\documentclass[twoside,fleqn]{ActaPS}
\usepackage[colorlinks=true, pdfstartview=FitV, linkcolor=black,
            citecolor=black, urlcolor=black]{hyperref}

\usepackage{latexsym}
\usepackage{amsthm}
\usepackage{calc}
\usepackage{epsfig}
\usepackage{subfig}
\usepackage{amsmath, amssymb, amsthm, euscript} 
\usepackage{psfrag}
\usepackage{caption}
\usepackage{txfonts}
\usepackage{textcomp}
\usepackage{pdfpages}

\usepackage{dsfont}

\usepackage{breakcites}
\usepackage{graphicx}
\usepackage{times,cite}
\usepackage{dsfont}
\usepackage{fancyhdr}

\usepackage[english]{babel}
\usepackage{changepage}
\usepackage{mathptmx}
\usepackage{listings}
\usepackage{slashbox}
\usepackage{enumitem}
\usepackage{nicefrac}
\usepackage{multirow}
\usepackage{makecell}
\usepackage{color}

\usepackage{palatino} % Use the Palatino font by default

\newcommand{\dn}{\downarrow}
\newcommand{\up}{\uparrow}
\newcommand{\Tr}{{\rm Tr}}

\renewcommand{\subsectionmark}[1]{}
\renewcommand{\thesection}{\arabic{section}}
\renewcommand{\thesubsection}{\thesection.\arabic{subsection}}

\renewcommand{\thefigure}{\thesection.\arabic{figure}}
\renewcommand{\thetable}{\thesection.\arabic{table}}

\def\shorttitle{Phase Transitions on Non-Euclidean Geometries}

\begin{document}

\pagerange{1}{100}

\vspace{-1.0cm}
\title{STUDY OF CLASSICAL AND QUANTUM PHASE TRANSITIONS\\ ON NON-EUCLIDEAN GEOMETRIES IN HIGHER DIMENSIONS}

\vspace{-0.2cm}
\author{Michal Dani\v{s}ka and Andrej~Gendiar\email{andrej.gendiar@savba.sk}}
{Institute of Physics, Slovak Academy of Sciences, D\'{u}bravsk\'{a} cesta 9\\
SK-845~11 Bratislava, Slovakia}

\abstract{The investigation of the behaviour of both classical and quantum systems on non-Euclidean surfaces near the phase transition point represents an interesting research area of the modern physics. However, due to the specific nature of the hyperbolic geometry, there are no analytical solutions available so far and the potential of analytic and standard numerical methods is strongly limited. The task of finding an appropriate approach to analyze the fermionic models on the hyperbolic lattices in the thermodynamic limit still remains an open question. In case of classical spin systems, a generalization of the Corner Transfer Matrix Renormalization Group algorithm has been developed and successfully applied to spin models on infinitely many regular hyperbolic lattices. In this work, we extend these studies to specific types of lattices. We also conclude that the hyperbolic geometry induces mean-field behaviour of all spin models at phase transitions. It is important to say that no suitable algorithms for numerical analysis of ground-states of quantum systems in similar conditions have been implemented yet. In this work we offer a particular solution of the problem by proposing a variational numerical algorithm Tensor Product Variational Formulation, which assumes a quantum ground-state written in the form of a low-dimensional uniform tensor product state. We apply the Tensor Product Variational Formulation to three typical quantum models on a variety of regular hyperbolic lattices. Again, as in the case of classical spin systems, we conjecture the identical adherence to the mean-field-like universality class irrespective of the original model. The main outcomes are the following: (1) We propose an algorithm for calculation and classification of the thermodynamic properties of the Ising model on triangular-tiled hyperbolic lattices. In addition, we investigate the origin of the mean-field universality on a series of weakly curved lattices. (2) We develop the Tensor Product Variational Formulation algorithm for the numerical analysis of the ground-state of the quantum systems on the hyperbolic lattices. (3) We study quantum phase transition phenomena for the three selected spin models on various types of the hyperbolic lattices including the Bethe lattice.}

\begin{minipage}{2.5cm}
\quad{\small {\sf KEYWORDS:}}
\end{minipage}
\begin{minipage}{10.0cm}
Phase Transitions and Critical Phenomena, Classical and Quantum Spin Models, Hyperbolic Lattice Geometry, Tensor Product States, Tensor Networks, Density Matrix Renormalization, Mean-field Universality
\end{minipage}

\fancyhead[LO]{Contents}
\tableofcontents

\fancyhead[LE,RO]{\thepage} \fancyhead[RE]{\shorttitle}
\fancyhead[LO]{\rightmark}

\newpage\setcounter{equation}{0} \setcounter{figure}{0} \setcounter{table}{0}
\section{Introduction}

The properties of both classical and quantum systems on non-Euclidean surfaces have been attracting researchers in various fields of modern physics. For example, experiments were performed with soft materials on conical geometry~\cite{cn} and magnetic nanostructures on various negatively curved surfaces~\cite{experiment1,experiment2,experiment3}.
In addition, the influence of non-flatness of the underlying surface on the thermal properties of the system can be important in specific applications. 

The main motivation of this PhD work is to investigate ground-state properties around phase transitions of strongly correlated systems, which are represented by a variety of Hamiltonians known in Solid-State Physics, when applied to negatively curved lattice geometries, often referred to as the so-called anti-de Sitter (AdS) space of the General Theory of Relativity. Here, wave functions of many-body interacting systems are intended to describe a non-trivial curved space, where time is excluded from consideration for the time being. The mutual relations among Solid-State Physics, General Theory of Relativity, and the Conformal Field Theory (CFT) enrich the interdisciplinary research, such as AdS-CFT correspondence known from the theory of Quantum Gravity~\cite{AdS-CFT1,AdS-CFT2,q-gravity1,q-gravity2}. 

In order to accomplish such a nontrivial task, the physical space can be considered to be discrete. The entire discrete space is occupied by interacting multi-state spin variables with the distances as small as the Plank length ($~10^{-35}$m) thus forming {\it a spin network}. The first elementary steps to tackle the given problem of the Quantum Gravity are studied. In particular, we analyze relations between Gaussian curvature and correlations of the interacting spin particles. The off-criticality represented by non-diverging correlation length at phase transition is one of the key features to understand the negatively curved (AdS) geometry. The final step of will be the determination of a relation between the entanglement von Neumann entropy and the Gaussian curvature, which are crucial issues for the {\it holographic principle} in Quantum Gravity. Therefore, we have chosen quantum Heisenberg, XY, and transverse-field Ising models as the reference spin systems. Our intention is to confirm a concept of the holographic entanglement entropy~\cite{tHooft,Susskind,Takayanagi}. It means that a non-gravitational theory is expected to live on the boundary of a subsystem $\partial {\cal A}$ of $(d+1)$-dimensional hyperbolic spaces. The entanglement entropy $S_{\cal A}$, associated with a reduced density matrix of ${\cal A}$, is a measure of the amount of information for the AdS/CFT correspondence. The entropy $S_{\cal A}$ is then related to a surface region $\partial {\cal A}$ in the AdS space. There is a duality in $(d+1)$-dimensional AdS and the $d$-dimensional system ${\cal A}$ in CFT.

We begin with the study of simple spin models on regular hyperbolic lattices constructed by tessellation of congruent $p$-sided polygons with coordination number $q$, which are denoted as $(p,q)$. The hyperbolic $(p,q)$ lattices satisfy the condition $(p-2)(q-2)>4$, exhibit constant negative curvature and their Hausdorff dimension is infinite if the thermodynamic limit is considered. On hyperbolic lattices the number of lattice sites $N$ grows exponentially as the lattice diameter increases linearly. Also, the boundary effects are not negligible in the thermodynamic limit $N \to \infty$ on the hyperbolic lattices and, therefore, the spin systems exhibit phase transitions exclusively in the center of the infinite hyperbolic lattice.
Due to these specific conditions, the standard numerical tools developed for either classical or quantum systems (such as, Monte Carlo simulations, transfer matrix exact diagonalization, the coordinate Bethe Ansatz, the algebraic Bethe Ansatz or the vertex operator approach) face significant difficulties when applied to study phase transitions on hyperbolic lattices in the thermodynamic limit.  

In case of the classical spin systems, the modified Corner transfer matrix renormalization group (CTMRG) algorithm was applied to an infinite series of hyperbolic $(p,4)$ lattices \cite{hctmrg-Ising-p-4, hctmrg-Ising-5-4, hctmrg-J1J2, hctmrg-clock-5-4}. Developing the original idea, we reformulate the CTMRG algorithm for use on the triangular $(3,q)$ as well as on weakly curved hyperbolic lattices, which represents a missing complementary study to the $(p,4)$ case. 

So far, an analogous algorithm designed for the ground-state analysis of quantum systems on hyperbolic surfaces has been missing. We expand a variational method, Tensor product variational formulation (TPVF)~\cite{TPVF54, TPVFp4} in order to find out an effective solution of the problem. Here, the quantum ground-state is approximated in the form of the tensor product state, which allows us to implement a generalization of the original CTMRG algorithm.

Our analyses of both the classical and the quantum spin systems confirm that the hyperbolic geometry causes that the mean-field universality behaviour at the phase transition point occurs, irrespective of the spin model used. We attribute this feature to the infinite Hausdorff dimension of the hyperbolic surfaces. Another key outcome of this work is an indirect analysis of the quantum spin models on the Bethe lattice, where the coordination number is fixed to be four. The Bethe lattice is attributed to the asymptotics of the $(p,4)$ lattices, where $p \to \infty$. These interesting outcomes have been published in Refs.~\cite{hctmrg-Ising-3-q, hctmrg-Ising-3-qn, TPVFp4}.

This review is structured into five chapters. Section 1 summarizes the most important aspects of the theory behind the phase transition phenomena, which are relevant in this study. The reader familiar with the basic theory of the phase transitions can directly proceed to the next chapter. Section 2 introduces the non-Euclidean geometry in general and the hyperbolic lattices in particular. The detailed description of the numerical algorithms CTMRG and TPVF for use on both the Euclidean and the hyperbolic lattices is provided in Section 3. We emphasize the details important for the practical implementation of the methods. Additional theoretical reasoning associated with the renormalization procedure can be found in references provided therein. The three Sections contain the theoretical part. The core of this work is represented by Sections 4 and 5, where the results of our numerical analyses are demonstrated. First, Section 4 analyzes phase transitions of the classical Ising model on the triangular $(3,q)$ lattice and weakly curved hyperbolic lattices. Second, we make use of the TPVF to perform a similar analysis for the quantum phase transition in the transverse-field Ising, XY and modified Heisenberg models on the series of the hyperbolic $(p,4)$ and $(4,q)$ lattices in Section 5. We estimate the properties of the respective quantum models on the Bethe lattice.

\newpage\setcounter{equation}{0} \setcounter{figure}{0} \setcounter{table}{0}
%\section{Introduction} %Uvod by teda mal byt v Preface
\section{Basic concepts}

\subsection{Classical phase transitions}
\label{Classical_Phase_Transitions}

%In physics the term \emph{phase} refers to a thermodynamic system throughout which the state variables (e. g. temperature, pressure, density, magnetization, ...) are uniform in thermal equilibrium.  
%with homogeneous macroscopic properties
In physics, the term \emph{phase} refers to a thermodynamic system throughout which the state variables (e. g. temperature, pressure, density, magnetization, ...) are spatially homogeneous. If a small change of external parameters 
%results in, establishes  
produces a new phase with qualitatively 
%considerably
different properties in comparison to the previous one, we talk about the \emph{phase transition}. This phenomenon is always hallmarked by a singularity in the free energy of the system or one of its derivatives. The phase transition is classified as "of $n$-th order" if there is a discontinuity in the $n$-th derivative of the free energy. %(or - in general - another appropriate thermodynamic potential). 
In this section we provide a brief introduction to these phenomena following the books \cite{Baxter} and \cite{Yeomans}. 
 
\begin{figure}[!b]
 \centering
 \includegraphics[width=2.8in]{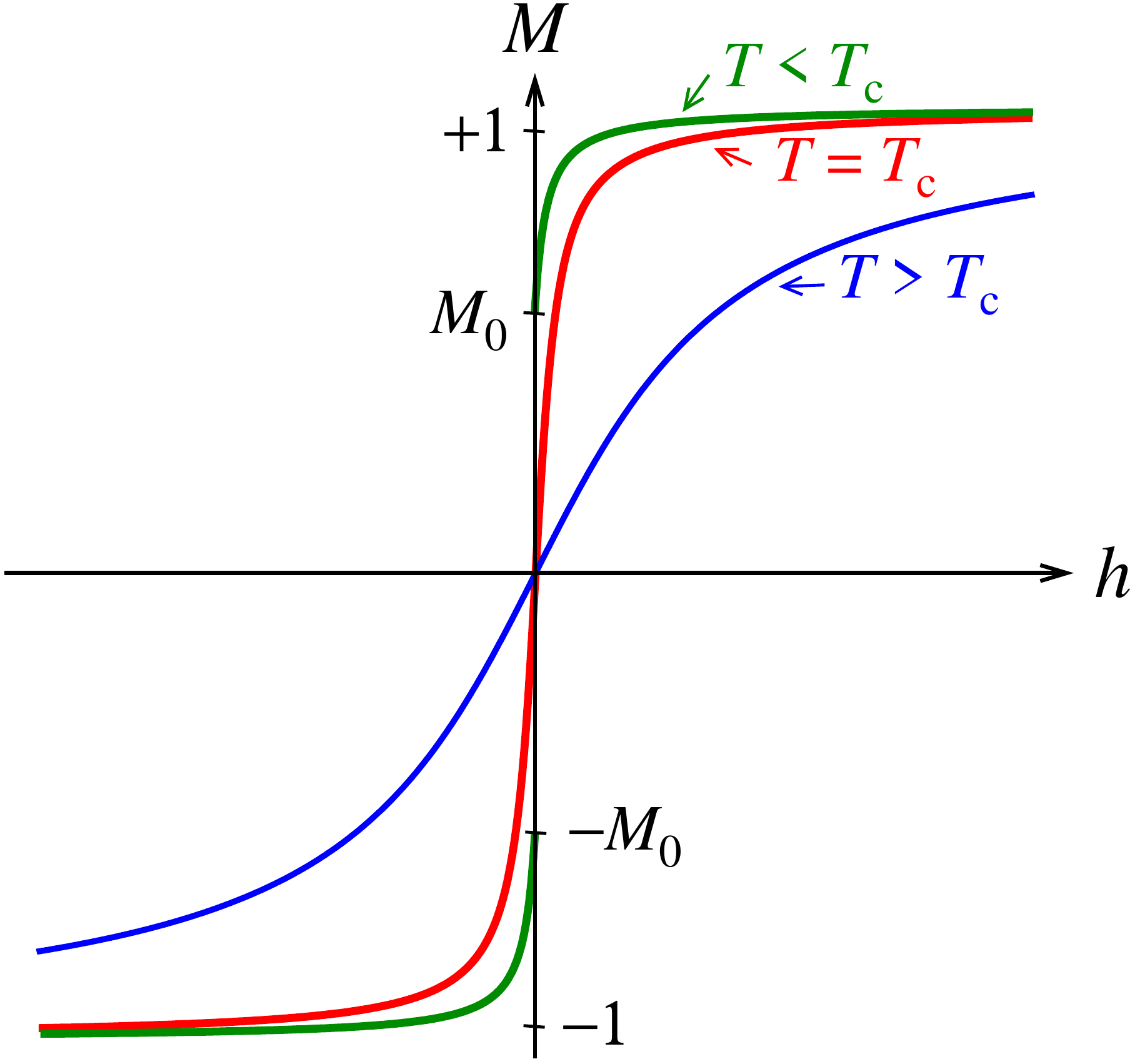}
   \caption{The magnetization $M(h,T)$ as a function of varying magnetic field $h$ at constant temperature $T<T_C$ (green), $T=T_C$ (red) and $T>T_C$ (blue).
}\label{magnetization_curves}
 \end{figure}

A common example of the phase transition is the abrupt change of properties of water at atmospheric pressure if its temperature $T$ rises over $100^{\circ}\textrm{C}$. Liquid water transforms into the gas form (steam) which results in sudden fall of the density.
Another important example is represented by phase transitions in ferromagnetic materials which can be authentically simulated even on very simple spin lattice models. %The phase diagram of a typical magnetic material is depicted in Fig.~??. 
%v popise diagramu popisat rovinu H, T
%If temperature $T$ is below the Curie temperature $T_{Curie}$ and the initially strong external magnetic field $h$ is monotonically decreased to zero, the magnetization $M(h,T)$ of the material also decreases, but to a nonzero value - the magnitude of which defines the \emph{spontaneous magnetization} $M_0(T)$.
The typical magnetization profiles of a magnetic material with respect to magnetic field $h$ for temperatures $T$ below, equal to and above the Curie temperature $T_{C}$ are depicted in Fig.~\ref{magnetization_curves}.
 Two phases of the ferromagnet are possible - one with positive magnetization $M(h,T) > 0$ if the magnetic field $h$ is parallel to the selected direction ($h>0$) or   one with negative magnetization $M(h,T) < 0$ at antiparallel magnetic field ($h<0$).
If the initially strong external magnetic field $h$ monotonically decreases to zero at given temperature $T$, the magnetization $M(h,T)$ of the material also decreases. The magnitude of magnetization at zero field defines the \emph{spontaneous magnetization} $M_0(T)$.
% but to a nonzero value - the magnitude of which defines the \emph{spontaneous magnetization} $M_0(T)$. 
%the \emph{spontaneous magnetization}, the absolute value of which is $M_0(T)$.
%amplitude, magnitude
 The term "spontaneous" reflects the fact that in the absence of the external field the magnetization is generated by the material itself. Orientation of the initial field plays the role of the symmetry-breaking mechanism which determines the orientation of the spontaneous magnetization after the field vanishes. Depending on whether the zero field is approached through positive or negative values $h$, we have
\begin{equation}
\lim\limits_{h \rightarrow 0^{+}} M(h,T) = M_0(T) \quad\textrm{or}\quad \lim\limits_{h \rightarrow 0^{-}} M(h,T) = -M_0(T). 
\end{equation}

\begin{figure}[tb]
 \centering
 \includegraphics[width=2.7in]{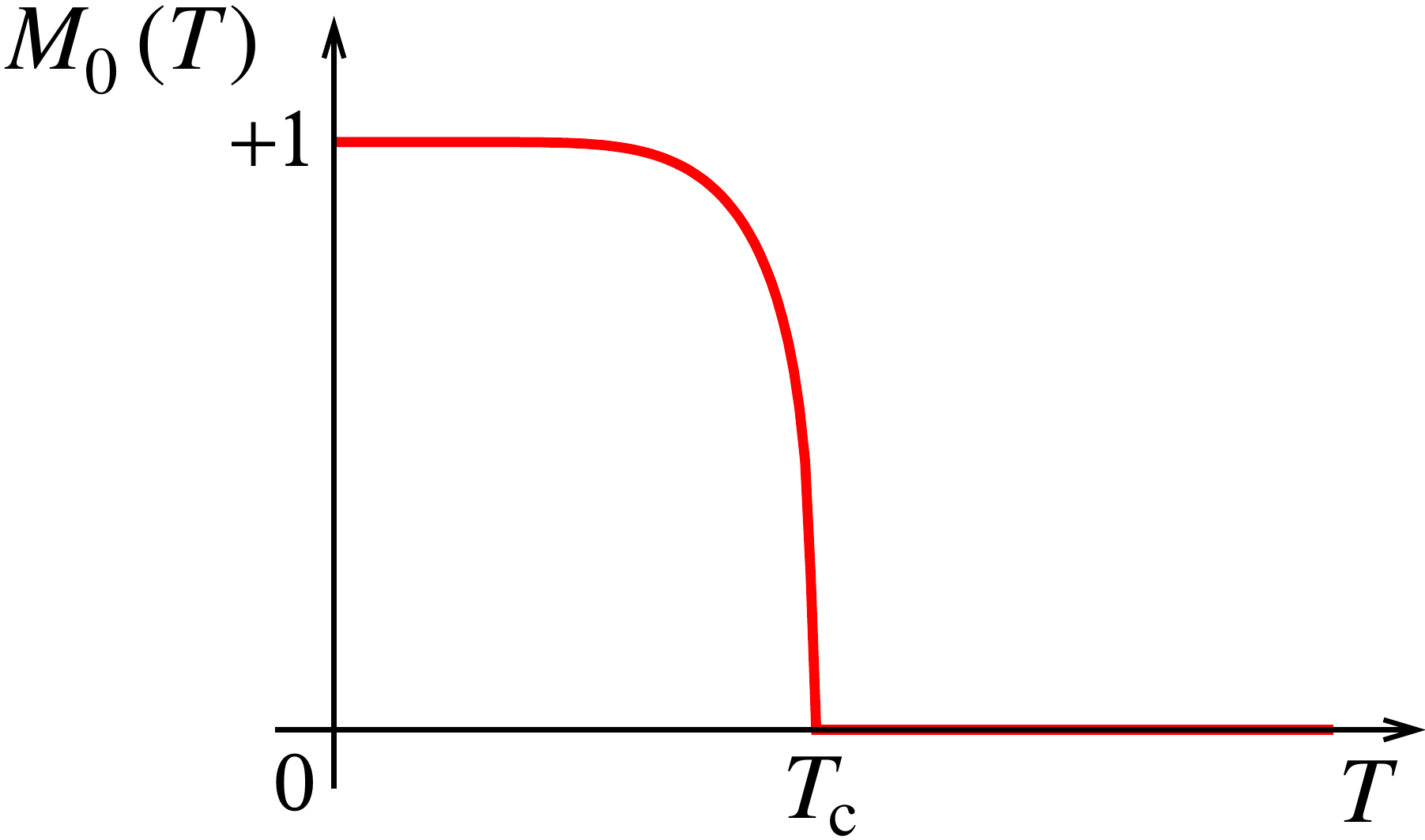}
   \caption{The spontaneous magnetization $M_0(T)$.
}\label{spontaneous_magnetization}
 \end{figure}

The temperature dependence of the spontaneous magnetization is depicted in Fig.~\ref{spontaneous_magnetization}.
 Whenever $T<T_{C}$, $M_0(T)$ is strictly positive. Therefore, at constant temperature $T<T_{C}$ and varying magnetic field $h$, the ferromagnet undergoes a phase transition at $h=0$ with discontinuity in the magnetization, changing suddenly from the negative value $-M_0(T)$ to the positive one $M_0(T)$ (or vice versa). Because the discontinuity occurred in magnetization, which can be calculated as the first partial derivative of the free energy with respect to $h$, it is the first-order phase transition. If $T \geq T_{C}$, $M_0(T)$ drops to zero and, thus, the magnetization $M(h,T)$ becomes a continuous function of $h$ at $h=0$ and analytic one if $T > T_{C}$. Therefore, there is no phase transition between the negative and the positive phase at $h=0$ and $T > T_{C}$. Although the magnetization is continuous at $T=T_{C}$, it is non-analytical (singular) due to infinite value of its first derivative $\frac{\partial M(h,T_{C})}{\partial h}$ (the magnetic susceptibility). This situation is equivalent to the absence of phase transition on the liquid-gas phase coexistence line above the critical temperature in the water phase diagram, where the difference in densities of both phases becomes continuous.

\begin{figure}[tb]
 \centering
 \includegraphics[width=2.5in]{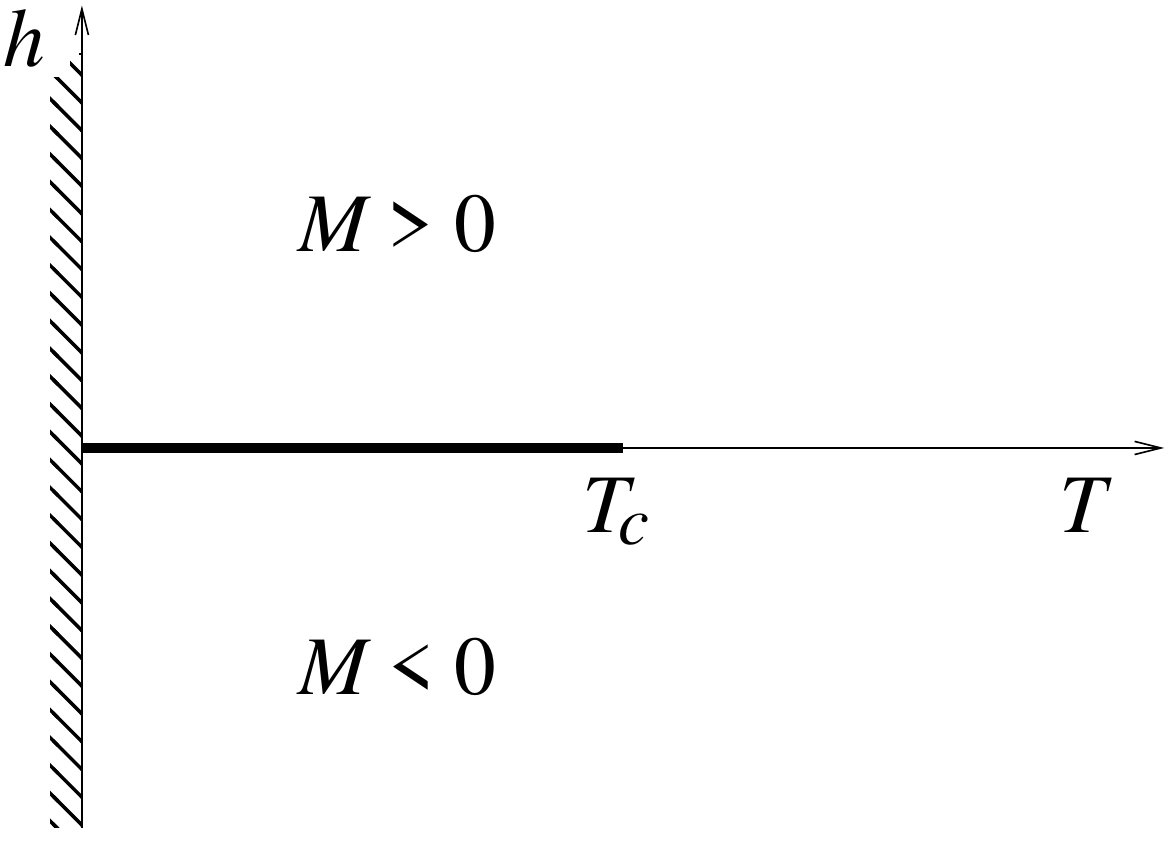}
   \caption{The phase diagram of the ferromagnetic material in the $(h,T)$ half-plane.
}\label{Phase_diagram}
 \end{figure}

The above mentioned observations are summarized in the phase diagram of the ferromagnet shown in Fig.~\ref{Phase_diagram}.
 The line $h=0$ represents the line of coexistence between phases, which separates the phase with the positive magnetization ($h>0$) from the negatively magnetized one ($h<0$). The magnetization is an analytic function of both $h$ and $T$ at all points of the $(h,T)$ half-plane, except those on the line segment $(h=0, 0 \leq T \leq T_{C})$, across which the phase transition occurs. The endpoint of this line segment $(h=0, T= T_{C})$ is denoted as the \emph{critical point}. If constrained to the phase coexistence line, two new phases can be defined - the ordered one with nonzero spontaneous magnetization $M_0(T)$ and the disordered one with $M_0(T)=0$. The two phases are separated at the critical point $T_{C}$ and the spontaneous magnetization
plays the role of the \emph{order parameter}, which identifies the ordered (disordered) phase by its nonzero (zero) value. The singular behaviour of $M_0(T)$ at the critical temperature $T_{C}$ is a hallmark of the phase transition between the ordered and the disordered phase. Note that this phase transition is generated by changing the temperature $T$ at constant field $h=0$. On the contrary, in case of the phase transitions between the negatively and the positively magnetized phases the field $h$ changes, while the temperature $T$ is held constant. 

\subsubsection{Basic notions from the classical statistical physics}

Let us consider a classical statistical system in external magnetic field $h$ at thermodynamic temperature $T$. The microstates of the system are labeled by index $r$ and their energies are $E(r,h)$. % which can occur in different microstates labeled by index $r$ and with corresponding energy $E_r(h)$. 
Then, the canonical partition function of the system is defined as
\begin{equation}
{\cal Z}(h,T) = \sum\limits_{r} \exp{\left(-\frac{E(r,h)}{k_B T}\right)} 
\label{Part_func_general_def}
\end{equation}
and the free energy as
\begin{equation}
{\cal F}(h,T) = -k_B T \ln{\cal Z}(h,T),
\label{free_energy_def}
\end{equation}  
where $k_B$ is the Boltzmann constant. The summands $\exp{\left(-\frac{E(r,h)}{k_B T}\right)}$ in \eqref{Part_func_general_def} are usually referred to as the statistical or the Boltzmann weight of the microstate $r$.  

  Now, complete
  % essential 
 information about the system can be in principle extracted from $\cal F$ (or $\cal Z$, equivalently) and its derivatives. Using the canonical probability of finding the system in the state $r$,
\begin{equation}
{\cal P}(r) = \frac{1}{\cal Z} \exp{\left(-\frac{E(r,h)}{k_B T}\right)},  
\end{equation}
 the thermal average of any thermodynamic function $X(r)$ is calculated as 
\begin{equation}
\langle X \rangle = \sum\limits_{r} X(r) {\cal P}(r) = \frac{\sum\limits_{r} X(r) \exp{\left(-\frac{E(r,h)}{k_B T}\right)}}{\cal Z} . 
\end{equation}
The \emph{internal energy}, defined as
\begin{equation}
E_{\rm int}(h,T) \equiv \langle E(r,h) \rangle = \frac{\sum\limits_{r} E(r,h)\exp{\left(-\frac{E(r,h)}{k_B T}\right)}}{\cal Z}, 
\end{equation}  
is a good example of such an averaged quantity. This formula can be   further rewritten into another convenient form
\begin{equation}
E_{\rm int}(h,T)= k_B T^2 \left( \frac{\partial \ln {{\cal Z}(h,T)}}{\partial T} \right)_h= - T^2 \frac{\partial}{\partial T} \left( \frac{{\cal F}(h,T)}{T} \right)_h,
\end{equation}
where the right bottom index $h$ in $\left( \frac{\partial }{\partial T} \right)_h$ explicitly identifies the variable which is held constant during the partial differentiation and the second equality follows from \eqref{free_energy_def}. Partial differentiation of the internal energy with respect to $T$ produces the \emph{specific heat} at constant external field
\begin{equation}
C_h(h,T) = \left( \frac{\partial E_{\rm int}(h,T)}{\partial T} \right)_h. 
\end{equation}  
If $E_{\rm int}$ is replaced by $- \cal F$ in the previous definition, we receive the formula for \emph{entropy} of the system
\begin{equation}
S(h,T) \equiv -k_B \sum\limits_r {\cal P}(r) \ln {\cal P}(r)=  -\left( \frac{\partial {\cal F}(h,T)}{\partial T} \right)_h . 
\end{equation}
As the entropy and the specific heat are given by the first and the second derivatives of the free energy $\cal F$, a discontinuity in these quantities witnesses, respectively, for the first- or second-order phase transition with respect to temperature change.

Analogously, phase transitions in magnetic materials induced by changes of the external magnetic field $h$ are classified by singularities in \emph{magnetization}
\begin{equation}
M(h,T) \equiv \langle m(r,h) \rangle = \frac{\sum\limits_{r} m(r,h)\exp{\left(-\frac{E(r,h)}{k_B T}\right)}}{\cal Z} = -\left( \frac{\partial {\cal F}(h,T)}{\partial h} \right)_T
\label{magnetization}
\end{equation}
and \emph{magnetic susceptibility}  
\begin{equation}
\chi(h,T) \equiv \left( \frac{\partial M(h,T)}{\partial h} \right)_T.
\label{susceptibility_def}
\end{equation}
Here $m(r,h)$ denotes magnetization of the $r$-th microstate at  magnetic field $h$. 
  
%The latter formula \eqref{susceptibility_def} serves as definition relation for $\chi(h,T)$, while the former one \eqref{magnetization} 

%A discontinuity in entropy or specific heat therefore witnesses for first- or second-order phase transition, respectively,   
%A discontinuity in entropy (first derivative of free energy $\cal F$) therefore witnesses for 

%As the free energy $\cal F$ is an extensive quantity which diverges with increasing system size,
%the asymptotic limit of the free energy per site
%\begin{equation}
%f(h,T) = \lim_{N \rightarrow \infty} \frac{{\cal F}(h,T)}{N},
%\end{equation}
%where $N$ denotes the total number of 
%is often investigated instead  
      
\subsubsection{The correlation function}

The characterization of phase transitions through averaged quantities such as magnetization or entropy represents the macroscopic approach. To be able to understand the transition phenomena on the microscopic level better, the concept of \emph{correlation functions} has been introduced. 

Let us consider a spin lattice system with $N$ spins. On each lattice site $i$ there is a spin variable $\sigma_i$, which can take two values, $\sigma_i=-1$ or $\sigma_i=1$. The set of microstates of the system consists of $2^N$ different configurations $\lbrace \sigma \rbrace \equiv \lbrace \sigma_1 ... \sigma_N\rbrace$ of the bivalent spin variables. The energy $E(\lbrace \sigma \rbrace,h)$ of the microstate (spin configuration) $\lbrace \sigma \rbrace$ is given by the Hamiltonian of the system ${\cal H}(\lbrace \sigma \rbrace,h)$. The Ising model with Hamiltonian 
\begin{equation}
E(\lbrace \sigma \rbrace,h)={\cal H}(\lbrace \sigma \rbrace,h)=
-J\sum\limits_{\langle i,j \rangle} \sigma_i \sigma_j - h \sum\limits_{i=1}^{N} \sigma_i ,
\end{equation}
where $\sum\limits_{\langle i,j \rangle}$ denotes summation over couples of nearest-neighbour lattice sites and $J$ the interaction strength, can be used as an example. The formula for the magnetization of the spin system is then
\begin{equation}
M(h,T) = \langle \sigma_1 + ... + \sigma_N \rangle /N.
\label{magnetization_def}
\end{equation} 

%The spin-spin correlation function between spins $\sigma_i$ and $\sigma_j$ defined as
The spin-spin correlation function between spins $\sigma_i$ and $\sigma_j$ is defined as 
\begin{equation}
g(\mathbf{r}_i,\mathbf{r}_j) \equiv 
\left\langle ({\sigma_i-\langle \sigma_i \rangle})
({\sigma_j-\langle \sigma_j \rangle})\right\rangle =
\langle \sigma_i \sigma_j \rangle - \langle \sigma_i \rangle \langle \sigma_j \rangle,
\label{cor-func_def}
\end{equation} 
where $\mathbf{r}_i$ is the position vector of the spin $\sigma_i$ on the lattice. Notice that $g(\mathbf{r}_i,\mathbf{r}_j)$ is only a specific member of a much wider class of correlation functions. Usually, the Hamiltonian ${\cal H}(\lbrace \sigma \rbrace,h)$ is translationally invariant, which yields $\langle \sigma_i \rangle = \langle \sigma_j \rangle, \forall i, j$ and consequently (after inserting into \eqref{magnetization_def})
\begin{equation}
\langle \sigma_i \rangle = M(h,T), \forall i. 
\end{equation}
As a result, the spin-spin correlation function depends only on the vector distance between the lattice sites $i$ and $j$
\begin{equation}
\mathbf{r}_{ij} \equiv \mathbf{r}_i - \mathbf{r}_j = r_{ij}\mathbf{e}_{ij},
\end{equation}
where $r_{ij}=\vert \mathbf{r}_{ij} \vert$ and $\mathbf{e}_{ij}=\mathbf{r}_{ij}/r_{ij}$ is a unit vector pointing in the direction of $\mathbf{r}_{ij}$. Therefore, $g(\mathbf{r}_{i},\mathbf{r}_{j})\equiv g(\mathbf{r}_{ij})$. %In this case, the quantity
%\begin{equation}
%G(\mathbf{r}_i,\mathbf{r}_j) =
%\langle \sigma_i \sigma_j \rangle 
%\end{equation}
%is an equivalent measure of correlation, since, at the given temperature $T$, the term $\langle \sigma_i \rangle \langle \sigma_j \rangle$ in \eqref{cor-func_def} only shifts $g(\mathbf{r}_i,\mathbf{r}_j)$ so that $g(\mathbf{r}_i,\mathbf{r}_j) \to 0$   
%as $r_{ij} \to \infty$.

The correlation function plays a crucial role in the concept of critical points. Away from the critical point 
($T_C$ in ferromagnets), 
both below and above it, any couple of spins becomes uncorrelated if their mutual distance is large enough, i. e., $g(\mathbf{r}_{ij}) \rightarrow 0$ if $r_{ij} \rightarrow \infty$. It is expected that the correlation function decays exponentially obeying the formula
\begin{equation}
%g(\mathbf{r}_{ij})=r_{ij}^{-\tau} \exp{\left(-r_{ij}/\xi\right)}
g(\mathbf{r}_{ij})=r_{ij}^{-\tau} \exp{\left(-\frac{r_{ij}}{\xi}\right)},
\label{cor_func_exp}
\end{equation}        
where $\xi$ is the \emph{correlation length} and $\tau$ is some number. The correlation length is a function of $h$, $T$ and the direction $\mathbf{e}_{ij}$, nevertheless, it is expected to become directionally independent near the critical point for large $r_{ij}$.

The critical point is by definition characterized by developing long-range correlations in the system which is hallmarked by diverging correlation length $\xi$. Hence, the necessary and sufficient condition for existence of a critical point (temperature) $T_C$ in the ferromagnet is
\begin{equation}
\lim\limits_{T \rightarrow T_C} \xi(h=0, T) = \infty,
\end{equation}
where the isotropicity of $\xi$ near criticality was utilized. As a result, the formula \eqref{cor_func_exp} breaks down. Instead, the correlation function decays as a power-like function
\begin{equation}
g(\mathbf{r}_{ij})=r_{ij}^{-d+2-\eta},
\label{crit_exp_eta_def}
\end{equation}
where $d$ is dimension of the underlying lattice and $\eta$ is a so-called critical exponent (see the next section for details).

% In particular, it is expected that
%\begin{equation}
%\xi(h=0, T) \propto {\vert t \vert}^{-\nu},
%\end{equation}
%where 
%\begin{equation}
%t=\frac{T-T_C}{T_C}
%\end{equation}
%is the so-called reduced temperature.  

\subsubsection{Critical exponents}     

Let us consider again the example of the ferromagnet in the following. It was argued in the previous sections that a critical point is inevitably coupled with singular behaviour of some thermodynamic functions in the form of discontinuities or divergences. It is expected that these singularities follow simple power-like formulae with non-integer exponents independent of $h$ and $T$ - the \emph{critical exponents}. 

Let us introduce a dimensionless measure of the deviation from the critical temperature $T_C$ in the form of the reduced temperature
\begin{equation}
t=\frac{T-T_C}{T_C}. 
\end{equation}
Hence, the critical point corresponds to $t=0$. The critical exponent $\lambda$ associated with a thermodynamic function $A(t) \equiv A(h = h_c \equiv 0, t)$ is defined by formula
\begin{equation}
\lambda = \lim_{t \rightarrow 0} \frac{\ln \vert A(t)\vert}{\ln \vert t \vert}.
\label{crit_coef_def}
\end{equation}
Equivalently, in the limit $t \rightarrow 0$ the thermodynamic function asymptotically obeys the power rule
\begin{equation}
\vert A(t) \vert \propto \vert t \vert^{\lambda},
\end{equation}  
as desired. The definition \eqref{crit_coef_def} applies to phase transitions induced by temperature
%thermal
 changes. If, instead, the transition generated by external magnetic field $h$
 %the transition generated
 is investigated through function $A(h) \equiv A(h , T=T_C)$, $t$ is replaced by $h$ in \eqref{crit_coef_def}, since $h_C \equiv 0$. 
 
 The most commonly used critical exponents and the associated thermodynamic functions are
\begin{align}
\label{C_h_exponent}
C_h(h=0, T) &\propto |t|^{-\alpha} & \text{if }  &t \to 0 ,  \\ 
M_0(T)      &\propto (-t)^{\beta} & \text{if }  &t \to 0^- , \\
\chi(h=0, T)&\propto {\vert t \vert}^{-\gamma} & \text{if }  &t \to 0 , \\
\xi(h=0, T) &\propto {\vert t \vert}^{-\nu} & \text{if } &t \to 0 ,\\
s(h=0, T) &\propto (-t)^{\mu} & \text{if } &t \to 0^- , \label{tension_exponent}\\
M(h, T=T_c) &\propto {\vert h \vert}^{1/\delta}{\rm sgn}(h) & \text{if }  &h \to 0 .
%g(\emph{r}) &\propto r^{-d+2-\eta} & \text{as } &t \to 0^{~} \label{corr_func} , \\
\end{align}
In addition, the critical exponent $\eta$ has already been introduced in equation \eqref{crit_exp_eta_def}.
The yet undefined quantity $s$ in the equation \eqref{tension_exponent} is the interfacial tension per unit area which represents the contribution of a unit area of the interface between the domains of coexisting positively and negatively magnetized phases at $h=0$ to the free energy $\cal F$. It is defined  within the ordered phase $(h=0, 0 \leq T \leq T_C)$ only.   

The above mentioned critical exponents are not mutually independent. Assuming the so-called \emph{scaling hypothesis}\footnote{See, e. g., \cite{Baxter} for more details.}, 
%which imposes a very specific formula for $M(h,T)$ near $T_C$,                    
one can obtain the following constraints:
\begin{align}
\gamma &= \beta \left( \delta - 1 \right) ,\\
\alpha + 2\beta + \gamma &= 2 ,\\
\left(2 - \eta \right) \nu &= \gamma ,\\
\mu + \nu &= 2 - \alpha  ,\\
d \nu &= 2 - \alpha.
\label{last_cond} 
\end{align}
The derivation of the last condition \eqref{last_cond} requires making further assumptions known as \emph{hyperscaling}. %The five scaling relations are in good agreement with experimental and theoretical results.
The importance of the five scaling relations, which are in good agreement with the experimental and theoretical results, rests in the fact that due to them, the knowledge of only two independent critical exponents is sufficient to determine all the remaining exponents.
%they make the knowledge sufficient

Now, we are ready to explain why the critical exponents are so important. It has been observed that quantities such as $M(h,T)$ and $T_C$ depend strongly on the details of interactions between spins or particles in the system in general. On the contrary, it is believed that the critical exponents are insensitive to details of the system Hamiltonian ${\cal H}({\lbrace \sigma \rbrace})$ and depend only on dimensionality of the system and symmetries of ${\cal H}({\lbrace \sigma \rbrace})$, which is known as the \emph{universality} assumption\footnote{Note that the scaling hypothesis and the universality idea represent two independent assumptions.}.
 Thus, the critical behaviour of a complicated realistic system can be correctly investigated on a model with drastically simplified Hamiltonian, provided that the dimensionality and symmetries of  
${\cal H}({\lbrace \sigma \rbrace})$ have been preserved.
%were not affected, were maintained
The set of systems represented by the same simple model forms a single \emph{universality class}. Each class is usually labeled by the simplest system.

\subsection{Spin models}

In this section we introduce
%present
 the most important lattice models of interacting systems which, due to their simplicity, were chosen as representatives of the corresponding universality classes. At the same time, the critical exponents uniquely assigned to each class are identified. Here, all the demonstrated models represent a set of spin variables positioned on vertices of a given lattice which differ only in the model specific Hamiltonian. %In classical case the spin denotes a variable $\sigma_i$, which can take two values $\sigma_i=\pm 1$.
%while 
%Similarly, its quantum counterpart is an observable with eigenvalues $\sigma_i^z=\pm 1$. Upravit podla Samaja. 
%the quantum spin   

At this point, it is important to emphasize that the phase transition may occur only on lattices which are of infinite size in each dimension. That is, the models have to be studied in the \emph{thermodynamic limit} $N_1 \rightarrow \infty, ..., N_d \rightarrow \infty$, where $N_1, ..., N_d$ denote number of lattice spins in the base directions of the $d$-dimensional lattice\footnote{This observation may be intuitively attributed to the presence of infinite functional series in the formula for the partition function $\cal Z$ which may generate a non-analytic function although all individual components are smooth functions. On the other hand, in case of finite lattices the finite series preserve the continuousness and differentiability of the summands and, thus, no singular behaviour may occur in the result.}.

\subsubsection{Classical spin models: Ising and mean-field}
\label{spin_models}

%The classical Ising model represents a set of bivalent spin variables $\sigma_i=\pm 1$ 
\paragraph{The Ising model}

The classical Ising model denotes a system governed by Hamiltonian
\begin{equation}
{\cal H}(\lbrace \sigma \rbrace,h)=
-J\sum\limits_{\langle i,j \rangle} \sigma_i \sigma_j - h \sum\limits_{i=1}^{N} \sigma_i,
\label{Ising_hamiltonian}
\end{equation}
where $N$ stands for the total number of spins in the system. The spin variable $\sigma_i$ can take only two values, $\sigma_i = +1$ if it is oriented 
%súhlasne
in the same direction as the magnetic field $h$ or $\sigma_i = -1$ if it points in the opposite direction. No other spin orientation is allowed. The coupling constant $J$ determines character of the spin-spin interaction. Positive value $J > 0$ favours ferromagnetic configuration with all spins pointing in the same direction, while if $J<0$, antiferromagnetic alignment represented by inverse orientation of the neighbouring spins is preferred.

The Ising model on one-dimensional chain can be solved analytically  using the transfer matrix formalism (see section \ref{transfer_matrix_section} and \cite{Baxter}). Although not difficult to solve, this case is not very interesting from physical point of view, because the ordered phase includes only a single point $(h=0, T=0)$, which simultaneously represents the critical point.
%lebo tam diverguje korelacna dlzka
The critical exponents for the 1D Ising model together with values for all the other models mentioned below are listed in Table~\ref{Tab_crit_coef}.

\renewcommand{\arraystretch}{1.3}
\begin{table}[tb]
\begin{center}
\begin{tabular}[tb]{|l|c|c|c|c|c|c|}
\hline
Representative & \multicolumn{6}{c|}{Universality class}\\
\cline{2-7}	
model & {$\alpha$} & {$\beta$} & {$\gamma$} & {$\delta$} & {$\nu$} & {$\eta$} \\
\hline
1D Ising${}^{(a)}$ & $1$ & $0$ & $1$ & $\infty$ & $1$ & $1$ \\
2D Ising${}^{(a),(b)}$ & $0$ & $1/8$ & $7/4$ & $15$ & $1$ & $1/4$ \\
3D Ising${}^{(b),(c)}$ & ${0.1137\; (t \rightarrow 0^{-})}\atop{0.1023 \; (t \rightarrow 0^{+})}$ & $0.3295$ & $1.24$ & $4.8$ & $0.63$ & $0.04$ \\
	mean-field${}^{(a)}$ & $0$ & $1/2$ & $1$ & $3$ & --- & --- \\
\hline
\end{tabular} 
\end{center}
\caption{Critical coefficients of selected universality classes and corresponding representative models. (After \cite{Baxter} ($a$), \cite{Yeomans} ($b$) and \cite{HoSRG}($c$)).}\label{Tab_crit_coef}
\end{table}  
\renewcommand{\arraystretch}{1.0}

The analytic solution of the Ising model on the two-dimensional square lattice has been found only in case $h=0$ \cite{Onsager, Baxter}. In the thermodynamic limit ($N_x, \: N_y \rightarrow \infty$) there is the only one critical temperature $T_C$ given by the relation
\begin{equation}
{\rm sinh}\left(\frac{2 J_x}{k_B T_C}\right)
{\rm sinh}\left(\frac{2 J_y}{k_B T_C}\right)
=1,
\end{equation}
where $N_x, N_y$ and $J_x, J_y$ denote the number of lattice sites and interaction strength in the $x$ and $y$ direction, respectively. 
Assuming the isotropic case with $J_x=J_y \equiv J$, we have
\begin{equation}
T_C = \frac{2 J}{k_B}\frac{1}{\ln(1+\sqrt{2})}.
\label{T_C_2D_Ising}
\end{equation} 
The 2D model with nonzero magnetic field or the three-dimensional one have not been solved analytically yet, however, they are precisely described through numerical calculations. 

\paragraph{The mean-field model}

As only a few lattice spin models can be solved exactly, a number of approximation methods were developed. One of the most widely used is the \emph{mean-field theory}, where the total effect of direct interaction of a selected spin with its coupling partners is mimicked  by 
%the impact of, influence
an averaged field generated by uniform contributions from all spins in the system. As an example, let us discuss the Ising model with $N$ spins at $h=0$, where each spin $\sigma_i$ is surrounded by $q$ neighbours. The total impact of all interactions affecting a single spin $\sigma_i$ is governed by Hamiltonian
\begin{equation}
{\cal H}_1 (\sigma_i) = -J \sigma_i \sum_{j \in (i,j)} \sigma_j,
\end{equation}   
where $\sum_{j \in (i,j)}$ denotes summation over the $q$ nearest neighbours of the spin $\sigma_i$. In the mean-field approach $\sum_{j \in (i,j)} \sigma_j$ is approximated by $\frac{q}{N}\sum_{j = 1}^{N} \sigma_j = q M$, where $M$ is the magnetization of the system. As a result, the mean-field Hamiltonian for the spin $\sigma$ takes the form 
\begin{equation}
{\cal H}_1^{MF} (\sigma_i) = -{J q M} \sigma_i. 
\end{equation}

The critical temperature $T_C$ can be obtained from the self-consistent equation for magnetization. Namely, as the system is translationally invariant, $M=\langle \sigma_i \rangle$ and, therefore,
\begin{equation} \label{MF_mag}
M = \frac{1}{{\cal Z}}\sum_{\sigma_i = \pm 1} \sigma_i \exp\left(\frac{q J M}{k_B T}\sigma_i \right)
= \frac{\exp\left(\frac{q J M}{k_B T}\right) -\exp\left(-\frac{q J M}{k_B T}\right) }{\exp\left(\frac{q J M}{k_B T}\right) +\exp\left(-\frac{q J M}{k_B T}\right) }   =  \tanh\left(\frac{q J M}{k_BT}\right). 
\end{equation} 
Using the identity $\frac{\rm d}{{\rm d} x} {\rm tanh}(x)=1-{\rm tanh}^2(x)$, it is easy to verify, that \eqref{MF_mag} has a nontrivial solution only iff 
\begin{equation}
\frac{{\rm d}}{{\rm d} M}\left[{{\rm tanh}\left(\frac{q J M}{k_B T}\right)}\right]_{M=0}=\frac{q J}{k_B T}\left[1-{{\rm tanh}^{2}\left(\frac{q J M}{k_B T}\right)}\right]_{M=0}=\frac{q J}{k_B T}>1.
\end{equation}
It follows from the fact that both sides of \eqref{MF_mag} take the same (zero) value for $M=0$,  derivative of the LHS is $1$, while the derivative of the RHS is a decreasing function starting at $\frac{q J}{k_B T}$ when $M=0$ and approaching zero as $M \rightarrow \infty$. 

The (spontaneous) magnetization $M(h=0, T)$ takes positive (nontrivial) values for $\frac{q J}{k_B T}>1$ and vanishes at $\frac{q J}{k_B T}=1$, which signalizes the critical temperature
\begin{equation}
T_C = \frac{q J}{k_B}
\label{T_C_mean_field}
\end{equation}
and the ordered phase for $(h=0, 0\leq T \leq T_C)$. Note the linear character of the dependence $T_C(q)$. Moreover, the critical temperature $T_C$ is not affected by details of the lattice layout provided that the coordination number $q$ is held constant. For example, the mean-field models on the two-dimensional triangular lattice and the three-dimensional cubic one ($q=6$ for both) share the identical value of $T_C$. The critical exponents of the mean-field universality class are $\alpha=0$, $\beta = 1/2$, $\gamma = 1$ and $\delta=3$. The exponents $\nu$ and $\eta$ are not defined in this case, since the equally strong interaction of a selected spin  with every other results in distance independent correlations. 

It can be shown that any classical statistical model with dimensionality $d \geq 4\equiv d_C$, where $d_C$ is the upper critical dimension, belongs to the mean-field universality class \cite{Yeomans}. This fact is of crucial importance within the framework of this thesis, as the minimal Hausdorff dimension of a space into which a hyperbolic lattice (see chapter \ref{chap:neeuklid}) can be embedded, is infinite. As a result, any model on the hyperbolic lattice exhibits mean-field behaviour in the vicinity of the critical point, irrespective of the original Hamiltonian.   
%The importance of the mean-field model within the framework of this thesis  

%Then the original Hamiltonian \eqref{Ising_hamiltonian} is replaced by
%\begin{equation}
%{\cal H}(\lbrace \sigma \rbrace,h)=-
%\sum\limits_{i=1}^{N}\sigma_i\left(h+{\frac{qJ}{2(N-1)}\sum\limits_{ \substack{j=1 \\
%j \neq i} }^{N} \sigma_j }\right)
%=
%-\frac{qJ}{N-1}\sum\limits_{(i,j) } \sigma_i \sigma_j - h \sum\limits_{i=1}^{N} \sigma_i,
%\end{equation} 
%where the symbol $(i,j)$ labels all the $N(N-1)/2$ distinct spin couples.  
\paragraph{Ising model on the Bethe lattice}

Applying the Ising Hamiltonian \eqref{Ising_hamiltonian} to the Bethe lattice is interesting for two reasons: First, it is exactly solvable and, second, the specific nature of the lattice causes the model to belong to the mean-field universality class, although no mean-field approximation is applied. Therefore, the critical exponents are identical, particularly, $\alpha=0$, $\beta=1/2$, $\gamma = 0$ and $\delta=3$. 

\begin{figure}[tb]
 \centering
 \includegraphics[width=2in]{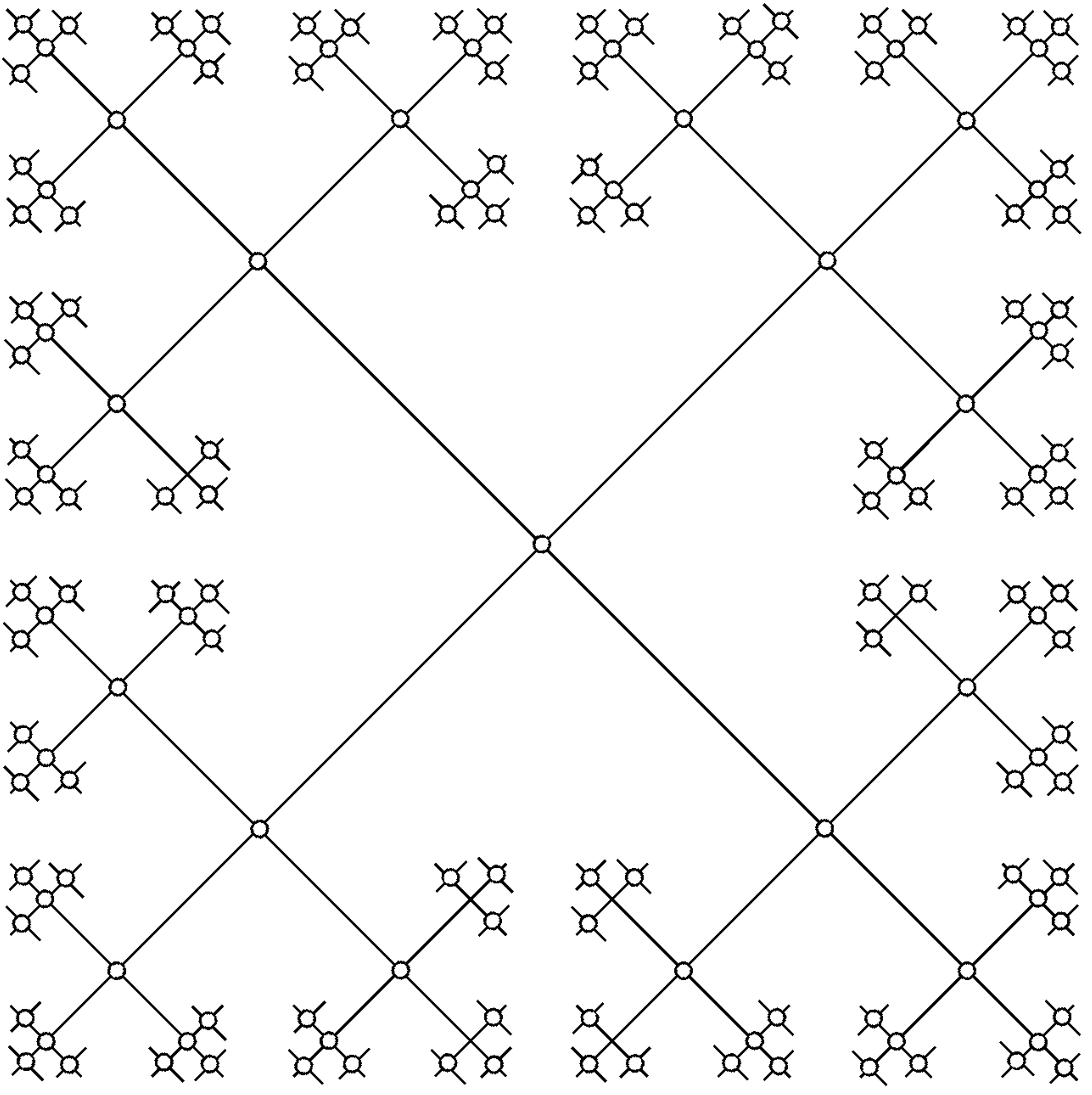}
   \caption{The Bethe lattice for $q=4$. Note that the interaction strength $J$ is identical for all spin pairs coupled by lattice edges, although the edge length was set different for better visualization.  
}\label{Bethe_lattice}
 \end{figure}

The Bethe lattice (cf. Fig.~\ref{Bethe_lattice})
 with the coordination number $q$ is constructed as follows: We start with a single central vertex and create $q$ links from it to its $q$ nearest neighbours, which form the first shell. Any next shell is constructed by connecting $q-1$ new vertices to each of the sites of the previous shell. The number of vertices in the $r$-th shell is $q(q-1)^{r-1}$, and the total number of sites in the lattice consisting of $r$ layers is 
\begin{equation}
n_q(r)=q \left[(q-1)^r-1\right]/(q-2).
\end{equation} 
The outermost shell forms the lattice boundary. The ratio of the number of boundary positions to the total number of lattice vertices tends to the nonzero value $(q-2)/(q-1)$ in the thermodynamic limit $r \to \infty$ and, therefore, the boundary effects cannot be removed by increasing the lattice size. In order to avoid this problem, we study only local properties of spins deep inside the lattice (far away from the boundary). The Bethe lattice is, by definition, formed by these deep interior vertices, which are all equivalent and have the coordination number $q$.

The dimension of the Bethe lattice is calculated as 
\begin{equation}
d = \lim\limits_{r \to \infty} \frac{\ln{n_q(r)}}{\ln r} = \infty  
\end{equation}      
which exceeds the critical dimension $d_C \equiv 4$. Hence, 
%in accord with the statement at the end of the section 
the critical behaviour of the Ising model on the Bethe lattice is governed by critical exponents with mean-field values as is confirmed by analytical calculations \cite{Baxter}. We emphasize that this mean-field-like critical behaviour is not induced by any mean-field approximations in the model, but by the infinite-dimensional lattice structure. 

The critical point of this model is positioned at $(h=0, T=T_C)$, where
\begin{equation}
T_C=\frac{2J}{k_B}\frac{1}{\ln\left[q/(q-2)\right]}.
\end{equation}

\subsubsection{Quantum spin models: Ising, XY and Heisenberg models}
\label{Quantum_spin_models}

Let us consider the Hamiltonian of a quantum system with $N$ spins in the form
\begin{equation}
{\cal H}(J_{xy}, J_z) = -\sum_{\langle i,j \rangle} 
\left[
J_{xy}\left(\boldsymbol{\sigma}^x_i\boldsymbol{\sigma}^x_j+
\boldsymbol{\sigma}^y_i\boldsymbol{\sigma}^y_j\right)
+J_z\boldsymbol{\sigma}^z_i\boldsymbol{\sigma}^z_j
\right]-h \sum_{i=1}^{N}\boldsymbol{\sigma}^x_i,
\end{equation} 
where the spin operators are in the z-representation given by the Pauli matrices 
\begin{equation}
\boldsymbol{\sigma}^x=
\left( \begin{matrix}
  0 & \phantom{-}1 \\
  1 & \phantom{-}0 
 \end{matrix} \right)
 \quad
\boldsymbol{\sigma}^y=
\left( \begin{matrix}
  0 & -i \\
	i & \phantom{-}0 
 \end{matrix} \right)
 \quad
\boldsymbol{\sigma}^z=
\left( \begin{matrix}
  1 & \phantom{-}0 \\
  0 & -1 
 \end{matrix} \right). 
\end{equation}
We assume arbitrary but fixed dimension of the system and periodic boundary conditions. Then, if $J>0$, the choice $J_{xy}=J_z=J$, $h=0$ defines the ferromagnetic Heisenberg model, $J_{xy}=J$, $J_z=0$, $h=0$ the ferromagnetic XY model and $J_{xy}=0$, $J_z=J$ with arbitrary $h$ the ferromagnetic Ising model in transverse field. If, instead, negative value $J<0$ is chosen, antiferromagnetic versions of the respective models are obtained.  

The class of Hamiltonians ${\cal H}(J_{xy}, J_z)$ exhibits some useful symmetries. Here, we focus on one of them only, but interested reader can find more information in, e. g., \cite{Samaj}. Let us suppose the underlying lattice is bipartite, i. e., the set of all lattice vertices can be factorized into two subsets $A$ and $B$ in such way that any couple of nearest neighbours $\langle i,j \rangle$ contains exactly one vertex from each of the two subsets. Then, the unitary transformation generated by operator $\mathbf{U}= \mathbf{U}^{\dagger}=\prod_{i \in A} \boldsymbol{\sigma}^z_i$ results in
\begin{equation}
\mathbf{U} {\cal H}(J_{xy},J_z)\mathbf{U}^{\dagger}={\cal H}(-J_{xy},J_z)=-{\cal H}(J_{xy},-J_z).
\label{UHU}
\end{equation}
This follows from the fact that the Pauli operator $\boldsymbol{\sigma}^z_i$, where $i \in A$ is arbitrary, but fixed,
%commute with all Pauli operators $\boldsymbol{\sigma}^x_j, \boldsymbol{\sigma}^y_j, \boldsymbol{\sigma}^z_j, j \in B$
commutes with all other operators $\boldsymbol{\sigma}^x_j, \boldsymbol{\sigma}^y_j, \boldsymbol{\sigma}^z_j, j \in A \cup B$ except $\boldsymbol{\sigma}^x_i$ and $\boldsymbol{\sigma}^y_i$. In the latter case we have
$\boldsymbol{\sigma}^z_i\boldsymbol{\sigma}^x_i
\boldsymbol{\sigma}^z_i=-\boldsymbol{\sigma}^x_i$ and $\boldsymbol{\sigma}^z_i\boldsymbol{\sigma}^y_i
\boldsymbol{\sigma}^z_i=-\boldsymbol{\sigma}^y_i$ which finalizes the proof idea. 

The equality between the first and the last term in \eqref{UHU} means that the energy spectra of the Hamiltonians ${\cal H}(J_{xy},J_z)$ and ${\cal H}(J_{xy},-J_z)$ are mutually related by reflection around the zero energy level $E=0$. Hence, the ground state $\Psi_0$ of the first system determines the most excited state $\Phi_{\rm MAX}$ of the second system via
\begin{equation}
\Phi_{\rm MAX}=\mathbf{U}\Psi_0
\end{equation}
and vice versa.
Another consequence of \eqref{UHU}, which will be used later in section~\ref{quantum_p4_model}, is that for $J>0$ the Hamiltonian ${\cal H}(J,-J)$ describes an antiferromagnetic model. Recall that, by definition, the  model is (anti)ferromagnetic if the sign of the expectation values  of local magnetization $\langle\boldsymbol{\sigma}_i^x\rangle,\langle\boldsymbol{\sigma}_i^y\rangle$ and $\langle\boldsymbol{\sigma}_i^z\rangle$ in the ground state is identical (opposite) for the nearest-neighbouring pairs of spins.  The Hamiltonian ${\cal H}(J,-J)$ can be obtained by the unitary transformation $\mathbf{U}$ of the antiferromagnetic Heisenberg model
\begin{equation}
{\cal H}(J,-J)=-{\cal H}(-J,J)=-\mathbf{U}{\cal H}(J,J)\mathbf{U}^{\dagger}=\mathbf{U}{\cal H}(-J,-J)\mathbf{U}^{\dagger},
\end{equation}
and therefore the ground-states $\Psi_0^{(J,-J)}$ and $\Psi_0^{(-J,-J)}$ of the two Hamiltonians obey
\begin{equation}
\Psi_0^{(J,-J)}=\mathbf{U}\Psi_0^{(-J,-J)}.
\end{equation}   
Using the same argumentation as in the text below \eqref{UHU}, we receive for $a=x$ or~$y$
\begin{equation}
\left\langle\Psi_0^{(J,-J)}\left| \boldsymbol{\sigma}_i^a \right| \Psi_0^{(J,-J)} \right\rangle = \left\langle\Psi_0^{(-J,-J)}\left| \mathbf{U}^{\dagger} \boldsymbol{\sigma}_i^a \mathbf{U}\right| \Psi_0^{(-J,-J)} \right\rangle= -\left\langle\Psi_0^{(-J,-J)}\left| \boldsymbol{\sigma}_i^a \right| \Psi_0^{(-J,-J)} \right\rangle
\end{equation}
 and
\begin{equation}
\left\langle\Psi_0^{(J,-J)}\left| \boldsymbol{\sigma}_i^z \right| \Psi_0^{(J,-J)} \right\rangle = \left\langle\Psi_0^{(-J,-J)}\left| \mathbf{U}^{\dagger} \boldsymbol{\sigma}_i^z \mathbf{U}\right| \Psi_0^{(-J,-J)} \right\rangle= \left\langle\Psi_0^{(-J,-J)}\left| \boldsymbol{\sigma}_i^z \right| \Psi_0^{(-J,-J)} \right\rangle.
\end{equation}
Therefore, the alternating sign structure of the local magnetization present in $\Psi_0^{(-J,-J)}$ is preserved also in $\Psi_0^{(J,-J)}$, which proves that ${\cal H}(J,-J)$ describes an antiferromagnetic system.

%if for example $i \in A, j \in B$ we have $\mathbf{U} \mathbf{U}^{\dagger}$  

\subsubsection{Quantum-classical correspondence}
\label{QCcorrespondence}

In this section we establish a mapping between the quantum transverse-field Ising model on the one-dimensional chain and the classical Ising model on the two-di\-men\-si\-o\-nal square lattice. %This specific example illustrates
%demonstrates
% a more general of correspondence between $d$-dimensional quantum and $d+1$-dimensional classical spin systems.  
In fact, it can be shown that $d$-dimensional quantum spin models can be mapped onto a system-specific $(d+1)$-dimensional classical spin model, which is known as \emph{quantum-classical correspondence}. This concept plays an important role in theoretical reasoning of the numerical algorithm Corner transfer matrix renormalization group   
(see section~\ref{CTMRG_general}) and, simultaneously, helps to determine the critical exponents of a quantum system by classification %of the universality class 
of its classical counterpart. 

%The hamiltonian of the quantum Ising model on a chain with $N$ spins in transverse field $h$ is
We start with the ferromagnetic quantum Ising model on the chain with $N$ spins in transverse field $h$ governed by Hamiltonian
\begin{equation}
{\cal H} = -J \sum_{i=1}^{N} 
\boldsymbol{\sigma}^z_i\boldsymbol{\sigma}^z_{i+1}-h \sum_{i=1}^{N}\boldsymbol{\sigma}^x_i \equiv
{\cal H}_A+{\cal H}_B,
\label{TFIM_hamiltonian}
\end{equation}
where $J>0$, ${\cal H}_A=-J \sum_{i=1}^{N} 
\boldsymbol{\sigma}^z_i\boldsymbol{\sigma}^z_{i+1}$, ${\cal H}_B=-h \sum_{i=1}^{N}\boldsymbol{\sigma}^x_i$ and periodic boundary conditions are imposed, i. e., $\boldsymbol{\sigma}^z_{N+1}\equiv\boldsymbol{\sigma}^z_{1}$.   

The partition function of (a quantum system) is defined as 
\begin{equation}
{\cal Z} = \Tr\left[{\exp(-\beta {\cal H})}\right],
\end{equation}
where $\beta=(k_B T)^{-1}$. Introducing a small imaginary time step $\Delta \tau=\beta/m$ with $m$ being a sufficiently large integer\footnote{In numerical practice $\Delta \tau \lesssim 10^{-2}$ is required.} and making use of the commutativity of $\cal H$ with itself, $\cal Z$ can be rewritten as
\begin{equation}
{\cal Z} = \Tr\left[{\exp(-\Delta\tau {\cal H})}\right]^m = %\Tr\underbrace{\left[{\exp(-\tau {\cal H})} ... {\exp(-\tau {\cal H})}\right]}_{m \,\, {\rm times}}.
%\Tr[\underbrace{{\exp(-\tau {\cal H})} \;...\; {\exp(-\tau {\cal H})}}_{m \,\, {\rm times}}].
\Tr\left[\underbrace{{\exp(- \Delta \tau {\cal H})} \;...\; {\exp(-\Delta \tau {\cal H})}}_{m \,\, {\rm times}}\right].
\label{SZ_part_func}
\end{equation}
Let us insert an identity operator in the form $ \mathbb{I} =  \sum\limits_{S^z_1=\up,\dn} ... \sum\limits_{S^z_N=\up,\dn} \left[\prod_{i=1}^{N}\left| S^z_i\right\rangle \left\langle S^z_i \right| \right] \equiv \sum\limits_{\lbrace \mathbf{S}^z \rbrace}\left| \mathbf{S}^z\right\rangle \left\langle \mathbf{S}^z \right|$ between any two consecutive factors in \eqref{SZ_part_func}, where $\left| S^z_i \right\rangle=\up,\dn$ are the eigenstates of the Pauli operator $\boldsymbol{\sigma}_i^z$ corresponding to eigenvalues $s_i^z=1$ and $s_i^z=-1$, respectively, $\left| \mathbf{S}^z\right\rangle = \left| S^z_1 \right\rangle \otimes \: ... \: \otimes \left| S^z_N \right\rangle$ and $\sum\limits_{\lbrace \mathbf{S}^z \rbrace}$ denotes summation over the complete set of $2^N$ base states $\left| \mathbf{S}^z\right\rangle$. After labeling each of the $m$ identities by index $l$ we receive
\begin{equation}
{\cal Z}  = \sum\limits_{\lbrace \mathbf{S}^{z,1} \rbrace} \:...\: \sum\limits_{\lbrace \mathbf{S}^{z,m} \rbrace}\prod_{l=1}^{m} \left\langle \mathbf{S}^{z,l} \right| {\exp(- \Delta \tau {\cal H})}  \left| \mathbf{S}^{z,l+1}\right\rangle.
\label{SZ_part_func2}
\end{equation}
As the operators ${\cal H}_A$ and ${\cal H}_B$ do not commute, the application of the Suzuki-Trotter expansion yields
\begin{equation}
\exp(- \Delta \tau {\cal H}) = \exp\left[- \Delta \tau ({\cal H}_A+ {\cal H}_B)\right] = \exp(- \Delta \tau {\cal H}_A)\exp(- \Delta \tau {\cal H}_B) + {\cal O}\left((\Delta \tau)^2\right),
\end{equation}    
where ${\cal O}\left((\Delta \tau)^2\right)$ denotes terms of order $(\Delta \tau)^2$ or higher, which vanish if $\Delta \tau \rightarrow 0$ (or, equivalently, $m \rightarrow \infty$). Because $\mathbf{S}^{z}$ are the eigenstates of the operator ${\cal H}_A$, we have
\begin{multline}
\left\langle \mathbf{S}^{z,l} \right|
\exp(- \Delta \tau {\cal H}_A)\exp(- \Delta \tau {\cal H}_B)
\left| \mathbf{S}^{z,l+1}\right\rangle
	=\\=
\exp\left(\Delta \tau J\sum_{i=1}^{N}s_i^{z,l} s_{i+1}^{z,l} \right)
\left\langle \mathbf{S}^{z,l} \right|
\exp(- \Delta \tau {\cal H}_B)
\left| \mathbf{S}^{z,l+1}\right\rangle
 .
\end{multline}  
The matrix elements on the RHS can be simplified by applying the identity relation $\left(\boldsymbol{\sigma}^{x}\right)^2=\mathbb{I}$ to the Taylor expansion of $\exp(\Delta \tau h \boldsymbol{\sigma}^{x})$, which gives
\begin{equation}
\exp(\Delta \tau h \boldsymbol{\sigma}^{x}) = \mathbb{I} {\rm cosh}(\Delta \tau h) + \boldsymbol{\sigma}^{x} {\rm sinh}(\Delta \tau h).
\label{Taylor_expHB}
\end{equation}
Inspired by structure of the partition function of a classical Ising model, we assume the partial matrix elements in the form
\begin{align}
\left\langle {S}_i^{z,l} \right|
\exp(\Delta \tau h \boldsymbol{\sigma}^{x}_i)
\left| {S}_i^{z,l+1}\right\rangle
=\Lambda \exp\left(\gamma s_i^{z,l} s_i^{z,l+1}\right).
\end{align} 
Expanding the LHS via \eqref{Taylor_expHB} 
%and using $\left\langle \up \right| \boldsymbol{\sigma}^{x} \left| \up \right\rangle = \left\langle \dn \right| \boldsymbol{\sigma}^{x} \left| \dn \right\rangle = 0$, $\left\langle \up \right| \boldsymbol{\sigma}^{x} \left| \dn \right\rangle = 1$ 
we receive
\begin{align}
	&\left\langle \up \right| \exp(\Delta \tau h \boldsymbol{\sigma}^{x}) \left| \up \right\rangle = \left\langle \dn \right| \exp(\Delta \tau h \boldsymbol{\sigma}^{x}) \left| \dn \right\rangle = {\rm cosh}(\Delta \tau h)=\Lambda\exp(\gamma),\\
	&\left\langle \up \right| \exp(\Delta \tau h \boldsymbol{\sigma}^{x}) \left| \dn \right\rangle =
\left\langle \dn \right| \exp(\Delta \tau h \boldsymbol{\sigma}^{x}) \left| \up \right\rangle =
{\rm sinh}(\Delta \tau h)=\Lambda\exp(-\gamma),
\end{align}  
which gives
\begin{equation}
\Lambda = \sqrt{{\rm sinh}(\Delta \tau h){\rm cosh}(\Delta \tau h)}
\qquad
\gamma = -\frac{1}{2} \ln \left[\tanh(\Delta \tau h)\right].
\end{equation}

Hence, in the limit $\Delta\tau \rightarrow 0 \: (m \rightarrow\infty)$ we obtain
\begin{align}
\left\langle \mathbf{S}^{z,l} \right| 
{\exp(- \Delta \tau {\cal H})}  
\left| \mathbf{S}^{z,l+1}\right\rangle
&=
\exp\left(\Delta \tau J\sum_{i=1}^{N}s_i^{z,l} s_{i+1}^{z,l} \right)
\prod_{i=1}^{N}
\left\langle {S}^{z,l}_i \right|
\exp(- \Delta \tau h \boldsymbol{\sigma}^{x}_i)
\left| {S}^{z,l+1}_i\right\rangle 
\nonumber\\
&=
\exp\left(\Delta \tau J\sum_{i=1}^{N}s_i^{z,l} s_{i+1}^{z,l} \right)
\prod_{i=1}^{N}
\Lambda \exp\left(\gamma s_i^{z,l} s_i^{z,l+1}\right)\\
&=
\Lambda^N
\exp\left(\Delta \tau J\sum_{i=1}^{N}s_i^{z,l} s_{i+1}^{z,l} 
+\gamma \sum_{i=1}^{N} s_i^{z,l} s_i^{z,l+1}\right)\nonumber
\end{align}  
which after inserting into \eqref{SZ_part_func2} yields
\begin{equation}
{\cal Z}=
\Lambda^{Nm}
\sum_{s^{z,l}_i=\pm 1}
\exp\left(\Delta \tau J\sum_{l=1}^{m}\sum_{i=1}^{N}s_i^{z,l} s_{i+1}^{z,l} 
+\gamma \sum_{l=1}^{m}\sum_{i=1}^{N} s_i^{z,l} s_i^{z,l+1}\right).
\end{equation}
Here $\sum_{s^{z,l}_i=\pm 1}$ denotes summation over the eigenvalues ${s^{z,l}_i}$ for all combinations $1\leq i \leq N, \: 1\leq l \leq m$, which replaces the original summation over the eigenstates $\left| S^{z,l}_i \right\rangle$.
 Finally, after replacing $s^{z,l}_i$ by the established notation $\sigma_{i,l}$, $\cal Z$ can be interpreted as the partition function of a classical Ising model on the two-dimensional infinite ($m \rightarrow \infty$) strip-lattice of width $N$ at temperature $T_{\rm classical}$ with Hamiltonian
\begin{equation}
{\cal H}(\lbrace {\sigma} \rbrace) = 
-J_1 \sum_{l=1}^{m}\sum_{i=1}^{N}\sigma_{i,l} \sigma_{i+1,l} 
-J_2 \sum_{l=1}^{m}\sum_{i=1}^{N} \sigma_{i,l} \sigma_{i,l+1},
\label{Class_2D_Ising_H}
\end{equation}
where
\begin{equation}
J_1 =  \Delta \tau J k_B T^{\rm classical}\qquad {\rm and} \qquad J_2 = \gamma k_B T^{\rm classical}.
\label{Class_2D_Ising_J}
\end{equation} %$J_1 =  \Delta \tau J k_B T_{\rm classical}$ and $J_2 = \gamma k_B T_{\rm classical}$. 
Note that, in general, the interaction strength in the mutually perpendicular axis directions labeled by indices $l$ and $i$ is different and $T \neq T^{\rm classical}$. Also, it can be shown that the quantum one-dimensional Ising model can be mapped to the classical two-dimensional one at zero field only. Thus, the role of the magnetic field $h$ in the quantum system is portrayed by the temperature $T^{\rm classical}$ in its classical counterpart.     

%Spomenut(asi skor v nasledujucej sekcii), ze ak $g<<1$ alebo $g>>1$, tak nemame korespondenciu s $d+1$ rozmernym sytemom ale len $d$-rozmernym, lebo sa to diagonalizovať. 

\subsection{Quantum phase transitions}

Until now, the phase transition phenomena have been discussed only within the context of classical statistical physics, where a special attention was paid to the second-order phase transition triggered by tuning the temperature $T$ around the critical temperature $T_C > 0$. However, as $T \rightarrow 0$, 
%the dying-out of the thermal effects allows to
the thermal effects die out and the so-far suppressed quantum fluctuations become important. As a result, the \emph{quantum phase transition} (QPT) may appear.
%quantum fluctuations
%a dovoluju prejavit sa kvantovym fluktuaciam, ktore doteraz prekryvali, upozadené.   

In order to briefly explain the concept of QPT (see, e.g., \cite{Batrouni, Sachdev} for more details), let us consider a lattice model at temperature $T=0$ with the Hamiltonian in the form
\begin{equation}
{\cal H}(g) = {\cal H}_A + g{\cal H}_B, 
\end{equation}
%where $g$ represents a continuously tunable dimensionless parameter and ${\cal H}_0$, ${\cal H}_1$ do not commute ($\left[ {\cal H}_0, {\cal H}_1 \right] \neq 0$). If $\left[ {\cal H}_0, {\cal H}_1 \right] = 0$, both ${\cal H}_0$ and ${\cal H}_1$ can be simultaneously diagonalized using the base of eigenstates they share and thus it is a problem of classical physics with no additional quantum effects as described in section~\ref{Classical_Phase_Transitions}.         
where $g \geq 0$ represents a continuously tunable 
%non-negative
 dimensionless parameter.  If ${\cal H}_A$ and ${\cal H}_B$ commute ($\left[ {\cal H}_A, {\cal H}_B \right] = 0$), both ${\cal H}_A$ and ${\cal H}_B$ can be simultaneously diagonalized using the base of eigenstates they share and, thus, it is a problem of classical physics as described in section~\ref{Classical_Phase_Transitions} with no additional quantum effects. A qualitatively new behaviour related to the quantum aspect of this problem arises only if 
\begin{equation}
\left[ {\cal H}_A, {\cal H}_B \right] \neq 0,
\label{Non_commutativity}
\end{equation}
 which we, therefore, assume to hold in the following. 
% to hold to be true/fulfilled from now on/in the following 

\begin{figure}[tb]
 \centering
 \includegraphics[width=4in]{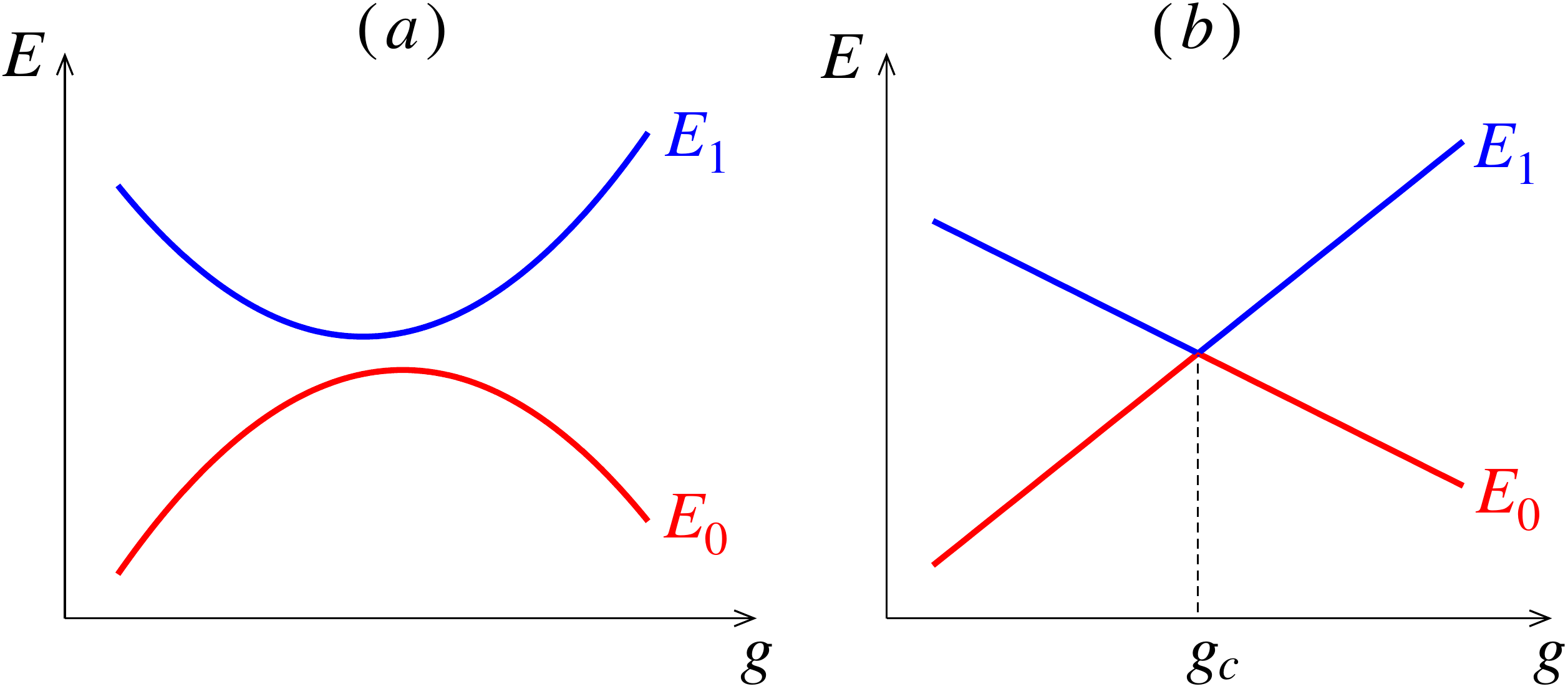}
   \caption{Possible energy profiles of the ground-state $E_0(g)$ and the first excited state $E_1(g)$ as functions of the tuning parameter $g$ for the quantum system of (a) finite and (b) infinite size.
}\label{QPT_energy_diagram}
 \end{figure}

%Let the two different ground-states (since $\left[ {\cal H}_0, {\cal H}_1 \right] \neq 0$) of the operators ${\cal H}_0$ and ${\cal H}_1$ be denoted $\Psi_0$ and $\Phi_0$, respectively. When $g \ll 1$, 
Let the ground-states of the Hamiltonians ${\cal H}_A$, ${\cal H}_B$ and ${\cal H}$ be denoted $\Phi_0^A$, $\Phi_0^B$ and $\Psi_0$ , respectively. We assume $\Phi_0^A \neq \Phi_0^B$, which, in general, is not guaranteed by the non-commutativity condition \eqref{Non_commutativity}, although it is implied by it in practice. 
When $g \ll 1$, ${\cal H} \approx {\cal H}_A $ and $\Psi_0 \approx \Phi_0^A$ with some quantum fluctuations caused by small, but nonzero term containing ${\cal H}_B$, while $g \gg 1$ results in ${\cal H} \approx {\cal H}_B $ and $\Psi_0 \approx \Phi_0^B$. Varying the value of $g$ between these two extreme limits, the energy profile of the ground-state $E_0(g)$ and the first excited state $E_1(g)$ of the Hamiltonian ${\cal H}$ is obtained. If the system is finite, the energy gap $E_1(g)-E_0(g)$ is always nonzero, although there could be a significant minimum at a specific $g$ (see Fig.~\ref{QPT_energy_diagram}(a)), and the groundstate $\Psi_0(g)$ changes smoothly from $\Phi_0^A$ to $\Phi_0^B$ as $g$ increases.
 On infinite lattice, however, the gap $E_1(g)-E_0(g)$ may vanish at $g_C$ (see Fig.~\ref{QPT_energy_diagram}(b)), which provides an opportunity for an abrupt change in the model ground-state $\Psi_0(g)$ by selecting an arbitrary state from the two-dimensional state space associated with the degenerated energy level $E_0(g)=E_1(g)$. If this occurs, we talk about the quantum phase transition. The (critical) point $g_C$ separates two phases - one with ${\cal H}_A$ dominating over $g{\cal H}_B$ at $0 \leq g < g_C$ and one, where the reverse is true at $g>g_C$. The ground-state $\Psi_0(g)$ plays the role of the order parameter which is in some sense closer to $\Phi_0^A$ than $\Phi_0^B$  if $g<g_C$, but the inverse relation holds if $g > g_C$. %Note that the critical point is also signaled by a singularity in the free energy ${\cal F}(g)$ which equals $E_0(g)$, since $T=0$.
Note that the singularity in $E_0(g)$, signaling the critical point $g_C$, appears also in the free energy ${\cal F}(g)$, since ${\cal F}(g)=E_0(g)$ at $T=0$.

Now, let us apply the mapping from a $d$-dimensional quantum model to its $(d+1)$-dimensional classical counterpart, where the extra dimension corresponds to the imaginary time $\tau$. As $g$ approaches $g_C$, the correlation length in the $d$ space directions diverges as
\begin{equation}
\xi(g) \propto \left| g-g_C \right|^{-\nu}, 
\label{xi_space}
\end{equation}  
but the divergence of the correlation length in the imaginary time $\xi_{\tau}$ may in general follow a slightly modified rule
\begin{equation}
\xi(g)_{\tau} \propto \left| g-g_C \right|^{-\nu z}, 
\end{equation}
which defines a new critical exponent $z$ that is unique to the quantum models.

%In this situation the ground-state $\Psi_0(g)$ plays the role of the order parameter, ktory oddeluje oblasti s relativnoou dominanciou HA a HB separovane bodom gC. 

\subsubsection{Ising model on the one-dimensional spin chain} 
 
As an example of the model with QPT, let us consider the quantum Ising model on the 1D chain with $N$ spins in the transverse magnetic field $h$ governed by Hamiltonian \eqref{TFIM_hamiltonian}, where the field $h>0$ plays the role of the tuning parameter, while $J$ is held constant. In section~\ref{QCcorrespondence}, the mapping of this model to the classical 2D Ising model in zero field was derived.  
We are interested in the critical phenomena and, therefore, assume the thermodynamic limit $N \rightarrow \infty$ at $T=0$. Hence, the  infiniteness of the corresponding classical model in both the space and the imaginary-time direction is guaranteed.

\paragraph{Critical field}
Inserting the expressions \eqref{Class_2D_Ising_J}, relating the parameters of the corresponding quantum and classical model into the formula \eqref{T_C_2D_Ising} which determines the critical temperature of the classical 2D Ising model, the equation for the critical magnetic field $h_C$ of the quantum model takes the form
\begin{equation}
{\rm sinh}\left(2 J \Delta\tau \right){\rm sinh}\left\lbrace-\ln\left[{\rm tanh}\left(h_C \Delta \tau \right)\right]\right\rbrace=1.
\end{equation}  
This can be simplified into
\begin{equation}
{\rm sinh}\left(2 J \Delta\tau \right) = 
{\rm sinh}\left(2 h_C \Delta\tau \right)
\end{equation}
and, thus, $h_C=J$.

\paragraph{Critical exponents}
It is known for the anisotropic 2D classical Ising model in zero field that the correlation length becomes directionally independent at the critical point. Therefore \eqref{xi_space} holds not only for space directions, but also for $\xi_{\tau}$ which yields $z=1$. %The non-analytic behaviour induced by tuning $h$ around $h_C$ in the quantum model is coupled to singularities characterizing the thermal phase transition in its classical counterpart. Hence, The classical 2D Ising model  
The magnetic field $h$ in the quantum model determines the temperature $T^{classical}$ in its classical counterpart through  \eqref{Class_2D_Ising_J}. 
%and thus the functional dependence $T^{classical}(h)$ is created. 
Inserting the functional dependence $T^{classical}(h)$ into the relations 
\eqref{C_h_exponent}-\eqref{tension_exponent} one can see that the
critical exponents describing the thermal phase transition ($\alpha$, $\beta$, $\gamma$) in the classical model are also related to the critical behaviour of the equivalent quantities in the quantum system, although now the control variable is $h$, not $T^{classical}$. As $T^{classical}(h)$ is not linear, critical exponents of the quantum model can, in general, differ from those in its classical counterpart. However, it turns out that in this case they preserve their original values. As a result, the critical behaviour at the QPT of the 1D quantum Ising model is ruled by
\begin{align}
C(h) \equiv \frac{\partial E_0(h)}{\partial h} &\propto |h-h_C|^{-\alpha} & \text{if }  &h \to h_C ,  \\ 
%\left\langle\boldsymbol{\sigma}^z(h)\right\rangle 
\left\langle S^z(h)\right\rangle\equiv \left\langle \Psi_0(h)\left|\boldsymbol{\sigma}^z\right|\Psi_0(h)\right\rangle     &\propto (h_C-h)^{\beta} & \text{if }  &h \to h_C^- , \label{crit_sz}\\
\chi(h) \equiv \frac{\partial \left\langle S^z(h)\right\rangle}{\partial h} &\propto {\vert h-h_C \vert}^{-\gamma} & \text{if }  &h \to h_C ,\\
\xi(h) &\propto {\vert h-h_C \vert}^{-\nu} & \text{if } &h \to h_C ,
\end{align}
where $\alpha=0$, $\beta=1/8$, $\gamma=7/4$ and $\nu=1$, which is identical to the classical 2D Ising model, cf. Table~\ref{Tab_crit_coef}.

%and also the remaining critical exponents of the classical 2D Ising model related to the thermal phase transition ($\alpha$, $\gamma$, $\mu$) migrate to the 1D quantum one with %preserved/
%the original values. 

%Pridat na vhodne miesto, ze fazove prechody nastavaju len pre $N$ nekonecne a je preto vhodne zaviest volnu enrgiu na $1$ site

\newpage\setcounter{equation}{0} \setcounter{figure}{0} \setcounter{table}{0}
\section{Non-Euclidean geometry}\label{chap:neeuklid}
\subsection{Euclidean geometry}\label{sect:Euclid}

The geometry of the world around us, which we are exposed to every day, is Euclidean. The mathematical description of the Euclidean geometry on a plane (surface)
%it 
%its mathematical theory 
is based on the following five axioms, which appeared %were presented
 for the first time in the Euclid's book \emph{the Elements} (about 300 B.C.). We present the axioms as formulated in the Coxeter's book \cite{Coxeter}:
 %in the formulation 
 %in their original form (as translated by Thomas Heath):    
%Euclidean geometry describes (zaoberá sa) the world we are familiar with and its is taught at schools. (In 2D )It is based on 5 postulates[orig-translation-from-greek-coxeter]:
%\begin{comment}
%povodne axiomy priamo pochadzajuce od Euklida
	\begin{enumerate}[label=(\Roman*)]
	\item A straight line may be drawn from any one point to any other point.
	\item A finite straight line may be produced to any length in a straight line.
	\item A circle may be described with any center at any distance from that center. 
	\item All right angles are equal.
	\item If a straight line meets two other straight lines, so as to make the two interior angles on one side of it together less than two right angles, the other straight lines will meet if produced on that side on which the angles are less than two right angles. 
\end{enumerate}

%The first four axioms have always been accepted by mathematicians.
The first four axioms have always been accepted by mathematicians, as they fulfill the essential requirements originally imposed on the axiom - to be so simple and obvious that no educated person could doubt its validity (cf. \cite{Greenberg}). %, p. 23).
%Marvin Greenberg Euclidean and Non-Euclidean geometries. p23
 However, the ‘‘non-self-evident’’ fifth axiom, which seems to be artificial, attracted the attention of mathematicians throughout centuries after Euclid's times.
%, who tried to deduce it as a theorem from the other four. Nevertheless, the only outcome of these attempts is a number of its various equivalent formulations.
All attempts to derive it as a theorem from the other four axioms and thus prove its redundancy within the axiomatic system failed. However, some people succeeded in reformulating it into an equivalent, but more ‘‘self-evident’’ form.
%All attempts to derive it as a theorem from the other four axioms and thus prove its redundancy within the axiomatic system failed 
%Although all attempts to prove its redundancy within the axiomatic system by deriving it as a theorem failed,   
%Although many of them tried to prove its redundancy derive it as a theorem from the other four axioms and , they only succeeded in re  
%Probably the best known version comes from the Scottish mathematician John Playfair's book \emph{Elements of Geometry} (1795), which states:
Probably the best known version comes from the Scottish mathematician John Playfair and his book \emph{Elements of Geometry} (1795), which states (cf. \cite{Greenberg}).% p. 21):
%..John Playfair, who in his book  \emph{Elements of Geometry} (1795) states:(ktorý ju vo svojej knihe ... uvádza nasledovne)
%\emph{In a plane, given a line and a point not on it, at most one line parallel to the given line can be drawn through the point.}    

\emph{For every line $\ell$ and for every point $P$ that does not lie on $\ell$, there exists a unique line $m$ through $P$ that is parallel to $\ell$.}

Due to this formulation, the fifth axiom is often referred to as the parallel postulate. In fact, the Playfair's version is not logically equivalent to the original one, but in the presence of the axioms I-IV, either of the two can be proved by assuming the other.
%As the postulate V (=) seems to be artificial and complicated, many mathematicians throughout centuries after Euclid tried to deduce it as proposition from postulates I-IV. However, they were only able to simplify it to various equivalent assumptions/ axioms/ postulates. The best known version is probably that of the Scottish mathematician John Playfair, which states:

%\emph{In a plane, given a line and a point not on it, at most one line parallel to the given line can be drawn through the point.}

%The "at most" clause is all that is needed since it can be proved from the remaining axioms that at least one parallel line exists.
%The statement is often written with the phrase, "there is one and only one parallel".

As the two-thousand-year long period of attempts to prove the parallel postulate as a theorem stalemated,
%bring to or cause to reach stalemate.
people started to think about the consequences of its replacement by its negation.
%začali uvažovať o tom, čo by sa stalo, keby piaty postulat neplatil o geometriách, kde piaty postulát neplatí.   
If one can find a geometry obeying the axioms I-IV and the negation of the fifth, this proves, that the parallel postulate cannot be derived as theorem from the other four. Otherwise, in any geometry based on the axioms I-IV the validity of the parallel axiom could be derived from the other four, which contradicts the existence of a geometry where its negation holds.

%napisat tu negaciu explicitne

Examples of a new type of geometry were indeed found independently by a Hungarian mathematician János Bolyai (1831) and a Russian Nikolai Lo\-ba\-chev\-sky (1829). %Among all non-Euclidean geometries (all which are not Euclidean - to nie je korektne, lebo v inych geometriach nemusia platit ani prve 4 postulaty). 
The two most common non-Euclidean geometries (curved two-dimensional surfaces) are the spherical geometry and the hyperbolic geometry. In the spherical geometry, a line has no parallels through a given point, while in hyperbolic (also called Bolyai-Lo\-ba\-chev\-skian) geometry for any given line $\ell$ and a point $P$ not on $\ell$ there are at least 2 distinct lines passing through $P$ and not intersecting $\ell$. We would like to emphasize that both the geometries describe spatially curved surfaces which are locally two-dimensional.  Considering three-dimensional space with Euclidean metrics, examples of the spherical geometry, such as the sphere or the ellipsoid, can be easily found. However, an infinite hyperbolic surface cannot be embedded into a space with finite Hausdorff dimension only\footnote{Examples of finite hyperbolic surfaces can be visualized in the three-dimensional space, nevertheless, a line of infinite length cannot be drawn there.}.

\iffalse
\subsubsection{Gaussian curvature}

If studying curved surfaces, the non-flatness is usually measured through the \emph{Gaussian curvature} $\kappa$.        
\fi

\subsection{Spherical geometry}
Although this thesis deals with systems on hyperbolic lattices, the spherical geometry will be discussed first, as it is easier to imagine due to its finiteness. Considering the essential properties, there is a sort of dual relationship between the spherical and the hyperbolic geometry.
The spherical geometry, as the name suggests, is the geometry of the sphere and related objects which are characterized by positive Gaussian curvature $\kappa>0$ at any point on the surface. The Gaussian curvature $\kappa$ of a regular sphere is constant and equal to $1/R^2$, where $R$ is the radius of the sphere. %\footnote{
%} (as the oscillating circle at any point on the sphere is the great circle and thus $\kappa_1=\kappa_2=1/R$). 
 Without loss of generality, we may assume $\kappa = 1$, as we can always measure the distance in the units of the sphere radius. The lines are represented as the great circles of the sphere. This agrees with the definition of a line as the set of points, where the shortest path from any point to another is the line segment between them. It is evident, that for any line $\ell$ and a point $P$ not on that line, there is indeed no line passing through $P$ and not intersecting $\ell$. 

The sum of angles $\alpha+\beta+\gamma$ of a triangle $\Delta$ in spherical geometry is always greater than $\pi$. For example, let us consider the triangle created as an intersection of the first octant of the Cartesian  coordinate system with the sphere centered in the origin of the coordinate system. The sides of the triangle are perpendicular to each other, hence $\alpha = \beta = \gamma = \pi/2$ and the sum of angles of the triangle $\alpha+\beta+\gamma = 3 \pi / 2 > \pi$. This follows from the simplified form of the Gauss-Bonnet formula
\begin{equation}
\alpha+\beta+\gamma -\pi = \iint_{\Delta} \kappa, 
\label{Gauss-Bonnet}
\end{equation}
where the integral is taken throughout the surface portion enclosed by the triangle $\Delta$. In spherical geometry, $\kappa$ is positive, which proves the statement $\alpha+\beta+\gamma -\pi>0$. Moreover, if the curvature $\kappa$ is constant, which is the case on the sphere, the area $A=\iint_{\Delta}$ of the triangle $\Delta$ is
\begin{equation}
A=\frac{\alpha+\beta+\gamma -\pi}{\kappa}.
\end{equation} 
 Hence,  the size of a triangle is uniquely defined by its angles. Similar triangles with identical angles and different sizes can exist only on Euclidean surfaces, where $\kappa=0$ and  $\alpha+\beta+\gamma = \pi$. 

The spherical surface has less space than the Euclidean one. Any attempt to flatten it results in tearing the spherical surface. Equivalently, we cannot create a sphere from a sheet of paper without cutting some paper away. There is no mapping from the spherical surface onto the Euclidean plane that preserves both angles and distance. However, the \emph{stereographic projection} can preserve the angles, although it disrupts the distances. Let us consider a unit sphere centered in the origin of the $xy$ plane. In this case, the northern (southern) hemisphere is mapped onto the outside (inside) of a unit circle in the plane. The south pole is projected onto the point $(0,0)$, while the north pole corresponds to the points in plane in infinity. All circles on the sphere are mapped onto circles in the plane and vice versa. In particular, a line\footnote{A line can be considered as a circle which passes through infinity.} in the plane is mapped onto a great circle on the sphere. Thus, there is a bijective mapping between the \emph{geodesics}\footnote{A geodesic is a generalization of the notion \emph{line} to curved spaces which represents the shortest route between two points in the space.} of the respective geometries. 

The disruption of the distance is expressed by the new metric induced by the stereographic projection. If we want to measure the Euclidean distance of two points on the unit sphere via their images on the xy plane, the metric
\begin{equation}
ds^2=\left(dx^2+dy^2\right)\frac{4}{(1+x^2+y^2)^2}\equiv\left(dx^2+dy^2\right)f(x,y)
\end{equation}  
must be applied. %Note the metric tensor $\mathbb{I}f(x,y)$ is isotropic, which confirms the angle preserving property of the mapping. 
The function $f(x,y)$ rapidly decreases if $(x,y)$ tends to infinity. The distance between the points $(0,0)$ and $(x \to \infty, 0)$ in the plane  is 
\begin{equation}
\int_0^{\infty}ds = \int_0^{\infty}\frac{2}{1+x^2}ds = \pi,
\end{equation}
which is the distance between the south and north pole of the sphere --- the pre-images of the two planar points in the stereographic projection.

\subsection{Hyperbolic geometry}
 
The hyperbolic surfaces exhibit negative Gaussian curvature  %$\kappa=\kappa_1\kappa_2<0$ at any point. The sign of the principal curvatures $\kappa_1$ and $\kappa_2$ is opposite, resulting in a saddle point. 
 $\kappa<0$ at any point. 
Hence, the entire surface is composed of saddle points only. Due to strong analogy with the spherical geometry, a surface with constant negative curvature $\kappa$ is called the \emph{pseudosphere}. 

\begin{figure}[tb]
 \centering
 \includegraphics[width=0.415\textwidth]{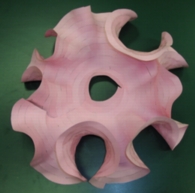}
 \includegraphics[width=0.45\textwidth]{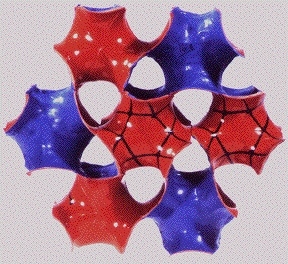}
   \caption{Examples of finite hyperbolic surfaces. After \cite{hypersurfcit1} (left) and \cite{hypersurfcit2} (right).
}\label{hypersurf}
 \end{figure}

Examples of hyperbolic surfaces of \emph{finite} size can be easily constructed in the three-dimensional space, cf. Fig.~\ref{hypersurf}. 
However, as the size increases, the surface curls more and more and 
%bend, loop
 the boundary parts start intersecting each other. The \emph{infinite} hyperbolic surface, therefore, cannot be placed in the three-dimensional space with Euclidean metric. On the contrary, it can be shown that the minimal Hausdorff dimension into which any infinite hyperbolic surface can be embedded is infinite.  

There is more space on the hyperbolic surface  than on the Euclidean plane. Any attempts to flatten it end up with a crunched object, portions of which overlap. As a consequence, for example, a circle of given diameter on the hyperbolic surface has larger area than its counterpart on the Euclidean plane - its area grows exponentially with increasing radius in comparison to a quadratic increase in the Euclidean case.

Developing the analogy with the spherical geometry, assuming $\kappa<0$ in the Gauss-Bonnet formula \eqref{Gauss-Bonnet} yields that the sum of angles of any triangle $\Delta$ on the hyperbolic surface $\alpha+\beta+\gamma$ is less than $\pi$ and the angles uniquely define the size of the triangle if $\kappa$ is constant. Note that, since $A=|\kappa|(\pi-(\alpha+\beta+\gamma))<|\kappa|\pi$, no triangle on the unit ($\kappa=-1$) pseudosphere can have area larger than $\pi$.

\begin{figure}[tb]
 \centering
 \includegraphics[width=2.3in]{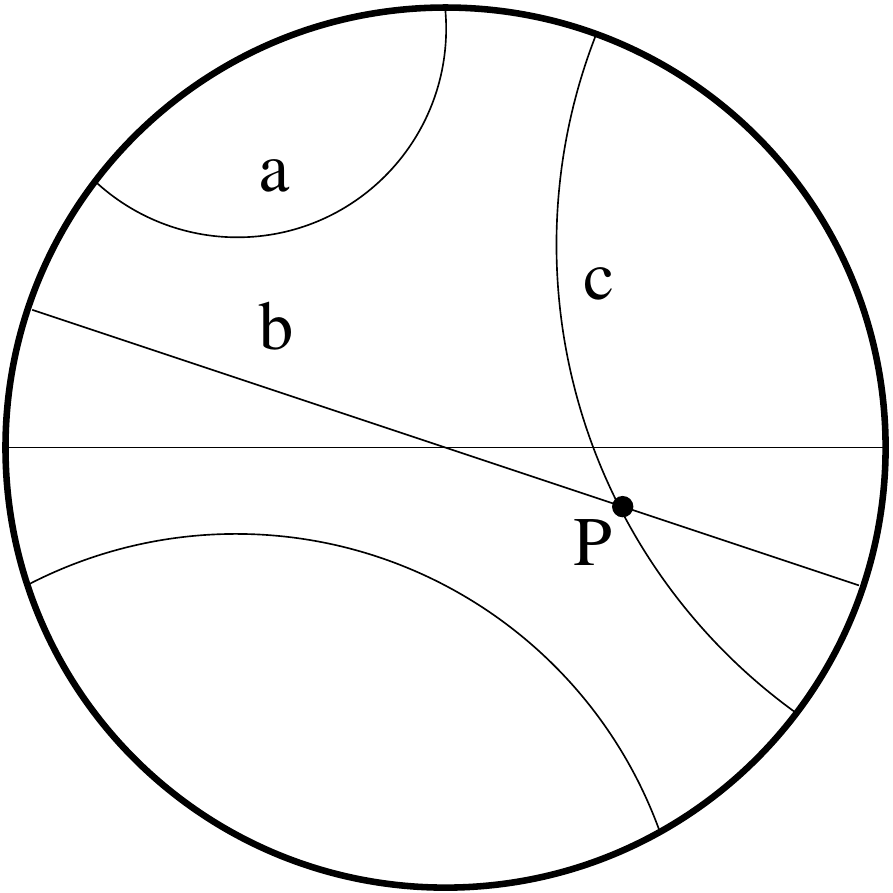}
   \caption{The Poincar\'e disc representation of the hyperbolic plane with several examples of the geodesics. Two distinct geodesics labeled  $b$, $c$ passing through the point $P$ are parallel to the geodesic $a$. The edge of the disc corresponds to infinity on the hyperbolic plane.
}\label{Poincare1}
 \end{figure}

Although the infinite hyperbolic surface with Euclidean metric cannot be embedded in the three-dimensional space, there is again an angle-preserving (and distance-distorting) mapping onto the Euclidean plane, or, to be more precise, onto a unit circle in the Euclidean plane. It is called \emph{Poincar\'e representation} \cite{Anderson} and the unit circle is referred to as the \emph{Poincar\'e disc}, cf. Fig.~\ref{Poincare1}.
% in some aspects. 
The points located in infinity on the hyperbolic plane are mapped onto the edge of the unit circle. The geodesics are represented as circles that meet the edge at right angle. Any non-intersecting such circles correspond to parallel lines on the hyperbolic surface. One can easily check that more than one parallel to a given line passing through a given point not on that line can be constructed. The metric of the Poincar\'e disc (assuming unit pseudosphere with $\kappa=-1$)
\begin{equation}
ds^2=\left(dx^2+dy^2\right)\frac{4}{(1-(x^2+y^2))^2}=\left(dx^2+dy^2\right)p(x,y)
\end{equation}         
strongly resembles the stereographic projection, as it
differs in the single minus sign in the denominator. However, the opposite sign changes the geometry completely. Near the edge of the circle $x^2+y^2 \approx 1$, $p(x,y) \to \infty$ and, therefore, the distance between the center and the edge of the disc
\begin{equation}
\int_0^{\infty}ds = \int_0^{\infty}\frac{2}{1-x^2}ds = \infty
\end{equation}
is infinite, as expected.   

%www.math.cornell.edu/%7Emec/Winter2009/Mihai/index.html
%https://en.wikipedia.org/wiki/Parallel_postulate#Equivalent_properties
%https://en.wikipedia.org/wiki/Playfair's_axiom
%http://aleph0.clarku.edu/~djoyce/java/elements/bookI/propI30.html
%CRC Concise Encyclopedia of Mathematics, Second Edition
%http://cs.unm.edu/~joel/NonEuclid/NonEuclid.html
%Coxeter Non-Euclidean geometry
%Marvin Greenberg Euclidean and Non-euclidean geometries.
%\end{comment}
        
\subsection{Regular tessellation}\label{sect:tessellation}

Our objective is to study regular lattice spin systems, %on hyperbolic surfaces.
where each lattice vertex corresponds to a position of a single spin and the lattice edges represent the bonds between neighbouring spins. The regularity means that the lattice is constructed by tessellation of congruent\footnote{The polygons are identical with fixed sizes of sides, i. e., the physical bond strength (the coupling constant $J$) is uniform throughout the system.} regular $p$-sided polygons and each spin has $q$ bonds to its $q$ nearest neighbours. Equivalently, each lattice vertex is shared by $q$ lattice polygons. %, see Fig.~??. 
The lattice of this structure will be denoted as $(p,q)$ and we refer to the integers $p$ and $q$ as the lattice parameter and coordination number, respectively.   

In Euclidean geometry, only the triangular $(3,6)$, square $(4,4)$ and hexagonal $(6,3)$ lattices can be formed. This can be justified by the following consideration. Assuming $q$ congruent non-overlapping regular polygons sharing one vertex with no space left, the vertex angle $\delta$ of the polygon must equal $2 \pi/q$. Each regular $p$-sided polygon can be divided into $p$ congruent triangles, the two vertices of which are the neighbouring vertices on the polygon circumference and the third one is the polygon center. The angle $\gamma$ of the triangle at the vertex coinciding with the polygon center is then $2 \pi/p$ and the remaining two angles are $\alpha=\beta=\delta/2=\pi/q$, thus giving the sum of the angles in the triangle $\alpha+\beta+\gamma=2 \pi(1/p+1/q)$. In Euclidean geometry, the sum of angles of the triangle must equal $\pi$, which means, that any $(p,q)$ lattice can be formed in the Euclidean plane if and only if $1/q+1/p = 1/2$ or, equivalently, $(p-2)(q-2) = 4$. As $p$, $q$ are positive integers greater than two, the only possible Euclidean lattices are $(3,6)$, $(4,4)$ and $(6,3)$.

In the spherical geometry, we have $\alpha + \beta + \gamma > \pi$, which results in $(p-2)(q~-2)<4$. This can be fulfilled only if $(p,q) \in \{(3,3),(4,3),(5,3),(3,4),(3,5)\}$. These lattices correspond to the "blown"\footnote{In the above-mentioned considerations, we assumed smooth surfaces, where a tangent plane exists at any point and the ratio of the circumference and the radius of an infinitesimal circle is $2\pi$. Hence, we do not receive the ordinary "angular" Platonic solids. Similarly, the hyperbolic $(p,q)$ lattices are considered as placed on the smooth hyperbolic surface. The interior of the $p$-sided polygon is not flat and there is no sharp edge between two neighbouring polygons as would be the case if real polygonal tiles were used in the tessellation.}
 versions of the five Platonic solids - the tetrahedron, the cube, the dodecahedron, the octahedron, the icosahedron. The "blown" version refers to a an object with identical structure of vertices and edges drawn on a (unit) sphere. 

\begin{figure}[tb]
 \centering
 \includegraphics[width=0.45\textwidth]{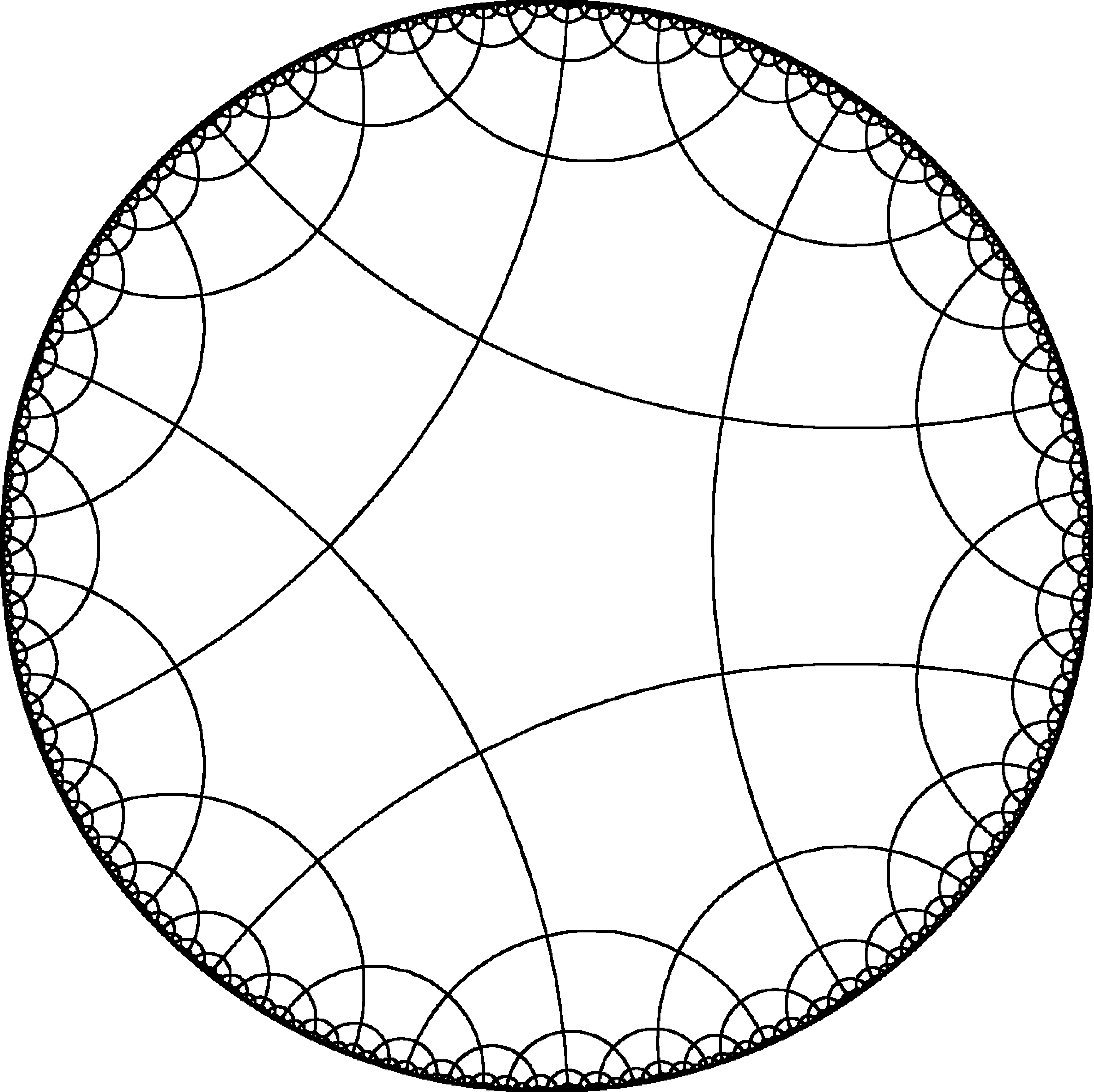}\qquad 
 \includegraphics[width=0.45\textwidth]{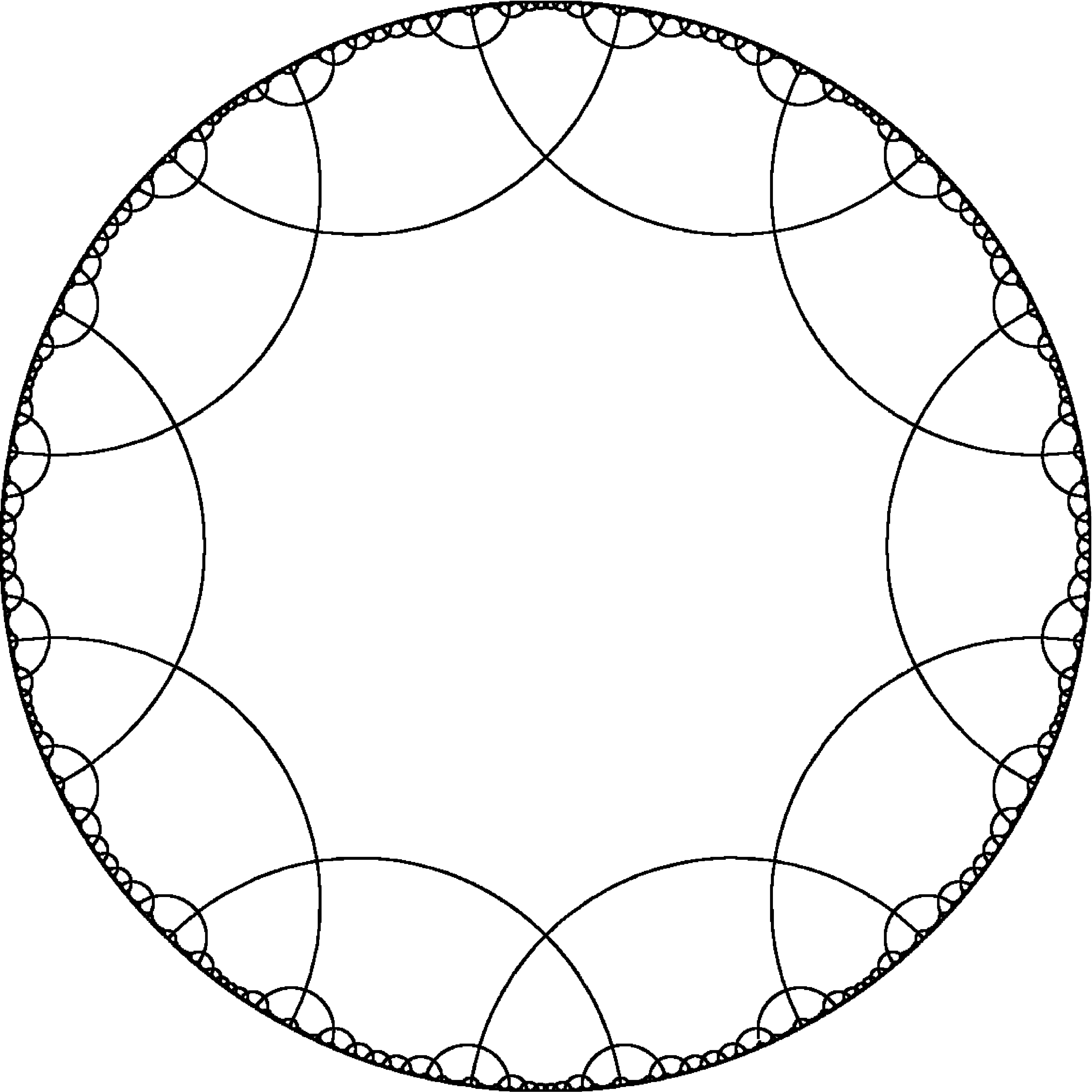}
\caption{Poincar\'e disc representation of the hyperbolic $(5,4)$ (left) and $(10,4)$ (right) lattices. All the polygons are of equal size and regular shape if the lattice is drawn on the pseudosphere. The projection of the pseudosphere onto the Poincar\'e disc, however, shows them deformed and progressively shrunk toward the disc boundary.}
\label{Lattice5_10_4}
\end{figure}

\renewcommand{\arraystretch}{1.3}
\begin{table}[!t]
\begin{center}
\begin{tabular}[!hbt]{|c|c|c|c|c|c|c|}
\hline
\backslashbox{$p$}{$q$} & $3$ & $4$ & $5$ & $6$ & $7$ & $...$\\ 
\hline
	$3$ & $S$ & $S$ & $S$ & {\bf E} & H & H\\ 
\hline
	$4$ & $S$ & {\bf E} & H & H & H & H\\ 
\hline
$5$ & $S$ & H & H & H & H & H\\ 
\hline
	$6$ & {\bf E} & H & H & H & H & H\\ 
\hline
$7$ & H & H & H & H & H & H\\ 
\hline
\vdots & H & H & H & H & H & H\\ 
\hline
\end{tabular} 
\end{center}
	\caption{The classification of regular $(p,q)$ lattices as Euclidean ({\bf E}), spherical ($S$) and hyperbolic (H) with respect to given combinations of the lattice parameter $p$ and the coordination number $q$.}
\label{pqclassification}
\end{table}   
\renewcommand{\arraystretch}{1.0}

In a similar way, we obtain the condition $(p-2)(q-2) > 4$ for the regular lattices on hyperbolic surfaces. On contrary to the two previous  geometries, there are infinitely many integer combinations $(p,q)$ which obey the relation. Thus, except the eight $(p,q)$ lattices realizable either on the Euclidean or the spherical surfaces, all the other $(p,q)$ lattices must be constructed on the hyperbolic surface. As an example, we show the $(5,4)$ and $(10,4)$ lattices in the Poincar\'e disc representation in Fig.~\ref{Lattice5_10_4}. Additional examples in the form of the triangular $(3,7)$ and $(3,13)$ lattices can be found in Fig.~\ref{Lattice3_7_13} in Chapter~\ref{chap:classical}. 
The classification scheme of the $(p,q)$ lattices into the respective geometries is shown in Table~\ref{pqclassification}. All the hyperbolic lattices fulfilling $(p-2)(q-2) > 4$ can be drawn on the unit pseudosphere ($\kappa=-1$), however, the polygon edge length must be rescaled accordingly. Equivalently, if the polygon edge length $l$ is held constant at the value $l=1$, the constant Gaussian curvature of the underlying hyperbolic and Euclidean surfaces obeys \cite{Mosseri}
\begin{equation}
\kappa_{p,q} = - \left[2 {\rm arccosh}\left(\frac{\cos(\pi/p)}{\sin(\pi/q)}\right)\right]^2.
\label{pqcurvature}
\end{equation}      

\newpage\setcounter{equation}{0} \setcounter{figure}{0} \setcounter{table}{0}
\section{Corner tensor networks}

In this chapter we describe numerical algorithms which form the computational background behind the core results of this thesis presented in chapters~\ref{chap:classical} and~\ref{chap:quantum}. First we introduce the mutually related concepts 
of the \emph{transfer tensor} and the \emph{corner transfer tensor} which play a crucial role in the \emph{Corner transfer matrix renormalization group} algorithm described in section~\ref{CTMRG_algorithm}. Next, we show how to modify the original algorithm, so that it can be implemented on hyperbolic  lattices and, finally, the \emph{Tensor product variational formulation} algorithm for quantum systems on the Euclidean and the hyperbolic surfaces is demonstrated.    

\subsection{Transfer approach to partition function analysis}

\subsubsection{Transfer tensor formalism}
\label{transfer_matrix_section}

Following the Baxter's book \cite{Baxter}, we explain the concept of the transfer tensor\footnote{In literature, it is more usual to refer to the transfer tensor as the \emph{transfer matrix}. This alternative terminology was established within the framework of one- and two-dimensional Euclidean lattices, where the transfer matrix applied to the vector of Boltzmann weights of the spin (row of spins) $i$ yields the vector of Boltzmann weights of the next spin (row of spins) $i+1$.
However, we prefer the tensor notation in order to unify the terminology with the next section, where the corner transfer tensor is introduced.} directly on a very simple model --- the classical  Ising model on the one-dimensional chain with $N$ spins. The Hamiltonian of the model is 
\begin{equation}
{\cal H} (\lbrace \sigma \rbrace) =-J \sum_{i=1}^{N}{\sigma_i\sigma_{i+1}} -h\sum_{i=1}^{N}{\sigma_i}, \label{1DIsingHamiltonian}
\end{equation}
where $\sigma_i = \pm 1$, $J>0$ is the ferromagnetic coupling and $h$ is the external field. We impose periodic boundary conditions, i. e., $\sigma_{1} \equiv \sigma_{N+1}$, which in combination with the site-independent values $J$ and $h$ create a translationally invariant system.
The formula for the partition function takes the form 
\begin{equation}
{\cal Z}_N = \sum_{\lbrace \sigma\rbrace}\exp\left[ K\sum_{i=1}^{N}{\sigma_i\sigma_{i+1}} + H\sum_{i=1}^{N}{\sigma_i}\right],
\label{1Dising_partfunc}
\end{equation}
where $\sum_{\lbrace \sigma\rbrace} = \sum_{\sigma_1}\sum_{\sigma_2} \dots \sum_{\sigma_N}$, $K=J/{k_B T}$ and $H=h/{k_B T}$.

Now, let us benefit from the sum in the argument of the exponential function. If we define
\begin{equation}
V(\sigma,\sigma') = \exp \left[ K\sigma\sigma' + \frac{H}{2}(\sigma + \sigma')  \right],  \label{transfer_matrix}
\end{equation}
the formula \eqref{1Dising_partfunc} for the partition function ${\cal Z}_N$ can be rewritten
in the product form
\begin{equation}
{\cal Z}_N = \sum_{\lbrace \sigma \rbrace}{V(\sigma_1,\sigma_2) V(\sigma_2,\sigma_3) V(\sigma_3,\sigma_4) \dots V(\sigma_N,\sigma_1)}. \label{partfunc_transfer_matrix}
\end{equation}
It is convenient to think of $V(\sigma,\sigma')$ as elements of a $2 \times 2$ matrix $\mathbf{V}$ defined as
\begin{eqnarray}
\mathbf{V} = \begin{pmatrix}
V(+1,+1) & V(+1,-1)\\ 
V(-1,+1) & V(-1,-1)
\end{pmatrix}  
=
\begin{pmatrix}
e^{K+H} & e^{-K}\\ 
e^{-K} & e^{K-H}
\end{pmatrix}.
\label{TM_ako_matica}
\end{eqnarray}
This way, the summations $\sum_{\sigma_2} ... \sum_{\sigma_N}$ in \eqref{partfunc_transfer_matrix} can be interpreted as consequent matrix multiplications producing the matrix element $\mathbf{V}^N(\sigma_1 , \sigma_1)$ and the last sum $\sum_{\sigma_1}$ as calculation of the trace. The formula \eqref{partfunc_transfer_matrix} thus simplifies into an elegant
% modifies
expression
\begin{equation}
{\cal Z}_N = \Tr(\mathbf{V}^N).
\label{traceV_N}
\end{equation}
The matrix $\mathbf{V}$ is the transfer matrix (tensor) of the 1D Ising model.
%The adjective \emph{transfer} originates in the following consideration. 
  
Our choice of the formula for $V(\sigma,\sigma')$ ensures that the matrix $\mathbf{V}$ is symmetric. As a result, $\mathbf{V}$ is diagonalizable and its eigenvectors can be chosen as mutually orthonormal, i. e.
\begin{equation}
\mathbf{V} = 
\mathbf{P}
\begin{pmatrix}
\lambda_1 & 0 \\
0 & \lambda_2
\end{pmatrix}
\mathbf{P}^{-1},
\label{TMdiag}
\end{equation}
where
\begin{equation}
\lambda_{1,2} = e^K \cosh{H} \pm \left({e^{2K}{\sinh^2{H}}+e^{-2K}}\right)^{\frac{1}{2}}%\frac{e^{K+h}+e^{K-h} \pm \left[ \left(e^{K+h}+e^{K-h}\right)^2-4\left(e^{2K}-e^{-2K}\right)\right]^{\nicefrac{1}{2}}}{2}
\label{eigenvalues1Dising}
\end{equation}
are the eigenvalues of $\mathbf{V}$ (labeled in the descending order, i.e., $\lambda_1 \geq \lambda_2$) and $\mathbf{P}$ is an orthogonal matrix\footnote{Orthogonal matrix $\mathbf{O}$ is a square matrix which obeys $\mathbf{O}^{\rm T}\mathbf{O}=\mathbf{O}\mathbf{O}^{\rm T}=\mathbb{I}$. Hence, $\mathbf{O}^{-1}=\mathbf{O}^{\rm T}$.} containing the mutually orthonormal eigenvectors as its columns. Thus,
\begin{eqnarray}
{\cal Z}_N = \Tr
\begin{pmatrix}
\lambda_1 & 0 \\
0 & \lambda_2
\end{pmatrix}^N
=
\Tr
\begin{pmatrix}
\lambda_1^N & 0 \\
0 & \lambda_2^N
\end{pmatrix}
=\lambda_1^N + \lambda_2^N 
\label{Zlambda}
\end{eqnarray}
and the free energy per site, ${ f}$, in the thermodynamic limit $N \to \infty$ is
\begin{equation}
{ f}=-k_B T \lim_{N \rightarrow \infty} \frac{\ln {{\cal Z}_N}}{N} = -k_B T  \lim_{N \rightarrow \infty}\left\lbrace\ln {\lambda_1} +\frac{1}{N} \ln \left[ 1 + \left({\frac{\lambda_2}{\lambda_1}}\right)^N \right]\right\rbrace.
\label{lnZ}
\end{equation}
For $T>0$ we have $\lvert {\lambda_2 /\lambda_1}\lvert < 1$ and, consequently, 
\begin{equation}
f=-k_B T \ln {\lambda_1},
\end{equation}
which is an analytic function. At $h=0$ and $T=0$ the correlation length $\xi = \left[ \ln(\lambda_1/\lambda_2)\right]^{-1}$ diverges, which is associated with the only critical point. There is no phase transition in the classical 1D Ising model for $T>0$ and real $h$.

\begin{figure}[tb]
 \centering
 \includegraphics[width=2.3in]{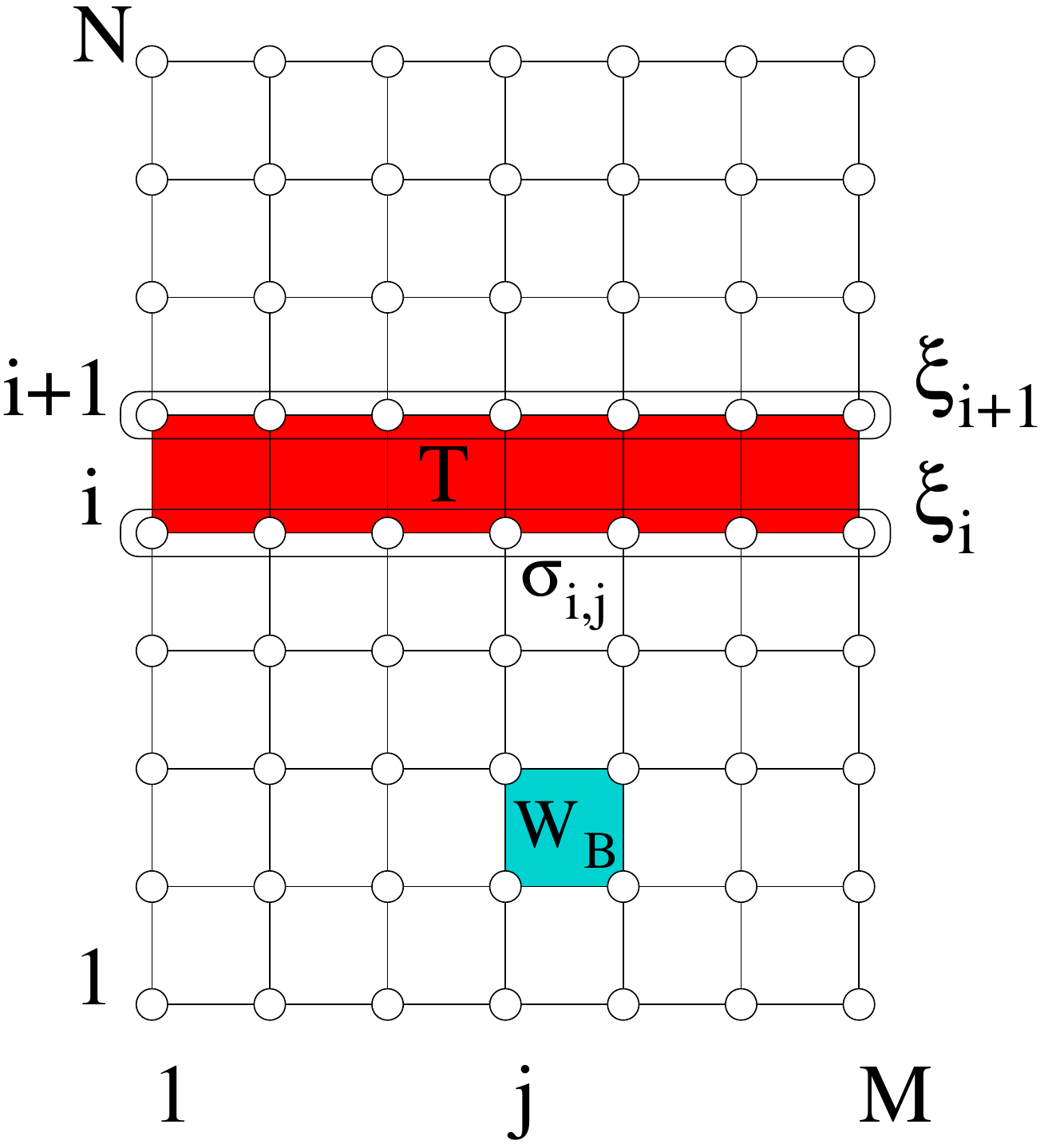}
   \caption{The $M \times N$ lattice with coloured areas represented by the Boltzmann weight tensor $\mathbf{W}_B$ (a single cell) and the transfer tensor $\mathbf{T}\left({\xi_i,\xi_{i+1}}\right)$ (an entire row of cells).
}\label{Transfer_matrix}
 \end{figure}

Now, let us apply the transfer tensor formalism to a more interesting ferromagnetic Ising model on a two-dimensional lattice with $M \times N$ spins located in the lattice vertices organized in $N$ rows and $M$ columns, cf. Fig.~\ref{Transfer_matrix}.
 We assume that the interaction strength $J>0$ is uniform throughout the entire lattice both in the horizontal and vertical directions. The Hamiltonian of the system is then given by
\begin{equation}
{\cal H}(\lbrace\sigma\rbrace) =-J \sum_{i=1}^{N}\sum_{j=1}^{M}\left({\sigma_{i,j}\sigma_{i+1,j}+\sigma_{i,j}\sigma_{i,j+1}} \right)-h\sum_{i=1}^{N}\sum_{j=1}^{M}{\sigma_{i,j}},
\label{2DIsing_hamiltonian}
\end{equation}
where $\sigma_{i,j}=\pm 1$ labels the spin in the $i$-th row and $j$-th column and $h$ is the homogeneous external field. We impose the periodic boundary conditions\footnote{In general, arbitrary, although identical, boundary conditions can be applied to spins $\sigma_{i,1}$ for $ 1 \leq i \leq N$ on the left lattice boundary and separately to spins $\sigma_{i,M}$ for $ 1 \leq i \leq N$ on the right lattice boundary. In that case, however, the expression $\prod_{j=1}^{M}\mathbf{W}_B\left(\sigma_{i,j},\sigma_{i,j+1},\sigma_{i+1,j},\sigma_{i+1,j+1}\right)$ in the formulae \eqref{Partfun_2DIsing} and \eqref{Transfer_matrix_2DIsing} has to be replaced by 
\begin{equation*}
\mathbf{W}_L\left(\sigma_{i,1},\sigma_{i+1,1}\right)\prod_{j=1}^{M-1}\mathbf{W}_B\left(\sigma_{i,j},\sigma_{i,j+1},\sigma_{i+1,j},\sigma_{i+1,j+1}\right)\mathbf{W}_R\left(\sigma_{i,M},\sigma_{i+1,M}\right),
\end{equation*}
 where the terms $\mathbf{W}_L\left(\sigma_{i,1},\sigma_{i+1,1}\right)$ and $\mathbf{W}_R\left(\sigma_{i,M},\sigma_{i+1,M}\right)$ represent the selected boundary conditions on the left and right boundary, respectively. For example, open boundary conditions on the left side result in
 \begin{equation*}
 \mathbf{W}_L\left(\sigma_{i,1},\sigma_{i+1,1}\right) = \exp\left[{\frac{h}{4 k_B T}{\left(\sigma_{i,1}+\sigma_{i+1,1}\right)}}+
\frac{J}{2 k_B T}{\left(
\sigma_{i,1}\sigma_{i+1,1}
\right)}\right].
\end{equation*}    
}, i.e., $\sigma_{N+1,j} \equiv \sigma_{1,j}$ and $\sigma_{i,M+1} \equiv \sigma_{i,1}$ for $1 \leq j \leq M$ and $1 \leq i \leq N$. 

Now, in order to rewrite the partition function ${\cal Z}_{N,M}$ of the $N \times M$ system into a matrix product form analogous to \eqref{partfunc_transfer_matrix}, 
%make further calculations more transparent
%(prehľadný/graspable?)
we introduce the Boltzmann weight tensor of the square-shaped cell bounded by four %defined
spins $\sigma_{i,j},\sigma_{i,j+1},\sigma_{i+1,j},\sigma_{i+1,j+1}$     
\begin{align}
&\mathbf{W}_B(\sigma_{i,j},\sigma_{i,j+1} \sigma_{i+1,j},\sigma_{i+1,j+1}) = \exp\left[
{\frac{h}{4 k_B T}{\left(\sigma_{i,j}+\sigma_{i,j+1}+\sigma_{i+1,j}+\sigma_{i+1,j+1}\right)}}
\right. &\nonumber\\
&\left.
+
\frac{J}{2 k_B T}{\left(
\sigma_{i,j}\sigma_{i,j+1}
+\sigma_{i,j+1}\sigma_{i+1,j+1}
+\sigma_{i+1,j+1}\sigma_{i+1,j}
+\sigma_{i+1,j}\sigma_{i,j}
\right)}\right].&
\label{WB}
\end{align}          
The expression in the argument of the exponential function represents contribution of the selected cell to the total Hamiltonian of the system ${\cal H}(\lbrace\sigma\rbrace)$, where the fractions $J/2$ and $h/4$ reflect that (assuming periodic boundary conditions) each bond and spin position is shared by $2$ and $4$ neighbouring lattice cells, respectively. Using \eqref{WB}, the formula for the partition function ${\cal Z}_{N,M}$ can be simplified into 
\begin{equation}
{\cal Z}_{N,M} = \sum_{\lbrace \sigma \rbrace }\prod_{i=1}^{N}\prod_{j=1}^{M} \mathbf{W}_B\left(\sigma_{i,j},\sigma_{i,j+1} \sigma_{i+1,j},\sigma_{i+1,j+1}\right),
\label{Partfun_2DIsing}
\end{equation}     
where the sum runs over all $2^{NM}$ possible spin configurations $\lbrace \sigma \rbrace$. Notice that the product $\prod_{i=1}^{N}\prod_{j=1}^{M} \mathbf{W}_B$ of the Boltzmann weights of all lattice cells gives the Boltzmann weight of the microstate (spin configuration) $\lbrace \sigma \rbrace$ of the entire system. Similarly, the Boltzmann weight of an arbitrary union of the lattice cells equals the product of the corresponding Boltzmann weight tensors $\mathbf{W}_B$.

%It happens to be beneficial if we have the Boltzmann weight of a row of cells at disposal. We therefore define the \emph{row transfer matrix} $\cal T$ by formula
It is useful if we introduce the \emph{(row) transfer tensor} ${\mathbf T}$ by formula
\begin{equation}
{\mathbf T}(\xi_i,\xi_{i+1})= \prod_{j=1}^{M}\mathbf{W}_B\left(\sigma_{i,j},\sigma_{i,j+1},\sigma_{i+1,j},\sigma_{i+1,j+1}\right),
\label{Transfer_matrix_2DIsing}
\end{equation}
where $\xi_i = \left\lbrace \sigma_{i,1} \sigma_{i,2} \dots \sigma_{i,M} \right\rbrace$ labels the $2^M$ configurations of $M$ grouped spins in the $i$-th lattice row. The transfer tensor $\mathbf{T}$ is constructed as the product of Boltzmann weights tensors of an entire row of cells, hence it represents
% the contribution of the $i$-th row to the Boltzmann weight  
the Boltzmann weight of the selected row at spin configurations $\xi_i$ and $\xi_{i+1}$. Considering the transfer tensor $\mathbf{T}$ in the matrix form allows us to modify the formula for the partition function ${\cal Z}_{N, M}$ into 
\begin{equation}
{\cal Z}_{N,M} = \sum_{\xi_1} \sum_{\xi_2}\dots \sum_{\xi_N}{{\mathbf T}(\xi_1,\xi_2) {\mathbf T}(\xi_2,\xi_3)\dots {\mathbf T}(\xi_N,\xi_1)}
=\Tr\left(\mathbf{T}^N\right),
\label{Z_TN}
\end{equation}
where we made use of the periodic boundary conditions in the vertical direction $\sigma_{N+1,j} \equiv \sigma_{1,j}$  for $1 \leq j \leq M$.
This expression is a formal analogue to \eqref{partfunc_transfer_matrix} and \eqref{traceV_N}, which have been developed in the one-dimensional case. Indeed, any $N \times M$ lattice can be considered as a 1D lattice consisting of $N$ "points", each of them representing one row of $M$ spins with $2^M$ configurations labeled by the variable $\xi_i$. %The row transfer matrix $\mathbf{T}$ thus ties the Boltzmann weights of spins $\Psi_{i}$ and $\Psi_{i+1}$ in the $i$-th and $i+1$-th row by
%connects, binds, establish relation/connection
%\begin{equation}
%\Psi_{i+1}=\mathbf{T}\Psi_{i}.
%\end{equation}   

Being able to calculate the largest eigenvalue $\lambda_1$ of the transfer matrix $\mathbf{T}$, the evaluation of the free energy per site $f$ of a system with finite number $M$ of spins in each row becomes straightforward, because
\begin{equation}
f =\frac{\cal F}{N M}=-\frac{k_B T}{N M} \ln \left(\Tr \mathbf{T}^N\right)=-\frac{k_B T}{N M} \left\lbrace N \ln\lambda_1 + \ln\left[1+\sum_{k=2}^{2^M}\left(\frac{\lambda_k}{\lambda_1}\right)^N\right]\right\rbrace.
\label{2Dtransfermatrixvolnanenergia}
\end{equation}
In the thermodynamic limit $N \rightarrow \infty$ taken along the vertical direction, the second term vanishes, and we receive
\begin{equation}
f =-\frac{k_B T}{M} \ln\lambda_1.
\end{equation}
The calculation of the largest eigenvalue $\lambda_1$ of the transfer matrix $\mathbf{T}$ is not a trivial task if the row length $M$ is large. 
%totiž
Namely, the numerical algorithms are significantly slowed down due to exponential increase of the matrix dimension ${\rm dim}(\mathbf{T})=2^M$ even if the complete diagonalization is not carried out. On the other hand, a well-known method, called \emph{Density matrix renormalization group} \cite{NishinoDMRG, DMRGbook}, can treat as large transfer matrices as $M>10^3$.    

\subsubsection{Corner transfer tensor formalism}

\begin{figure}[tb]
 \centering
 \includegraphics[width=2.3in]{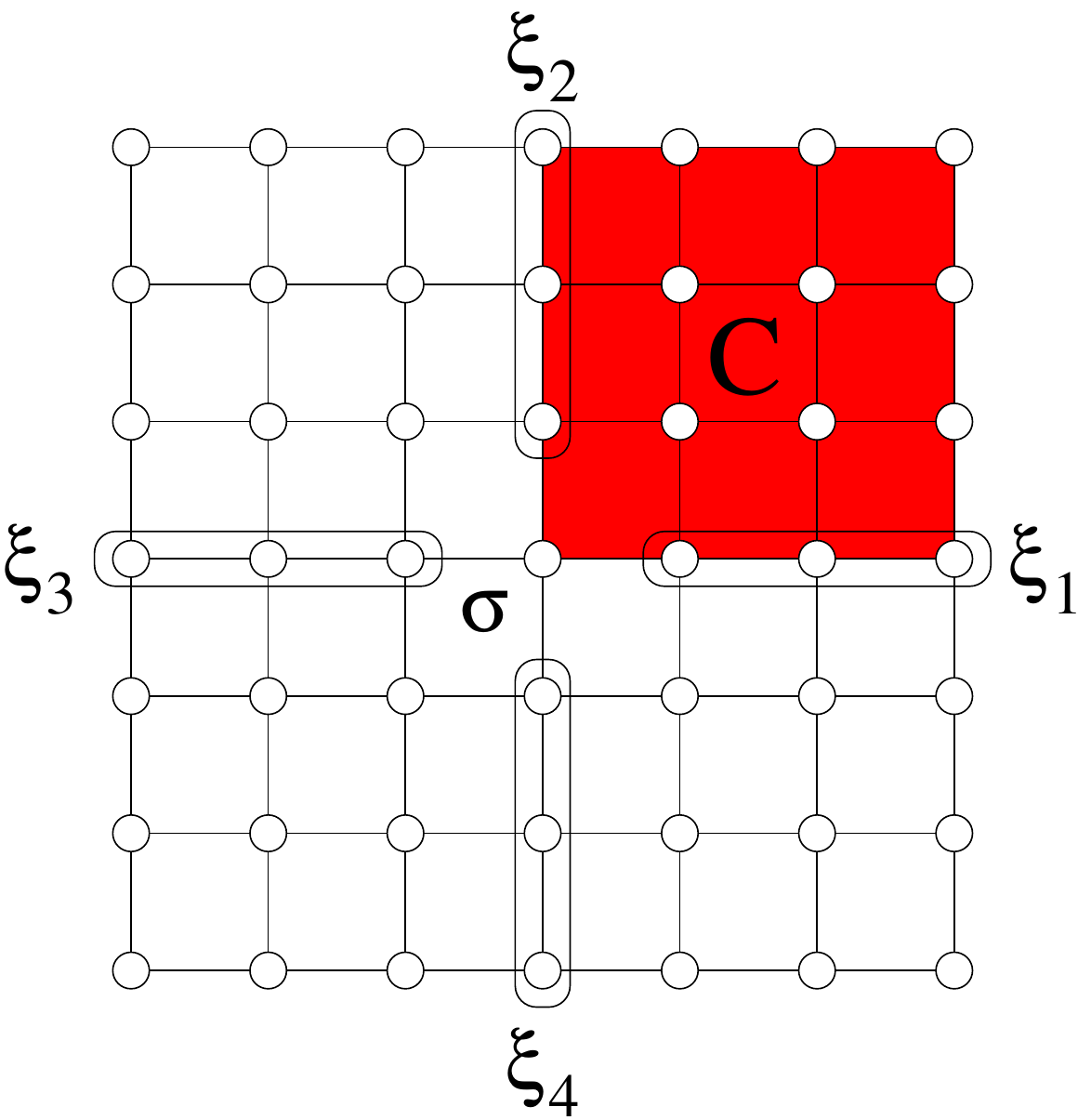}
   \caption{The square $(2L+1)\times(2L+1)$ lattice divided into four quadrants (corners) for $L=3$.  The multi-spin variables $\xi_1$, $\xi_2$, $\xi_3$, $\xi_4$ label the $2^L$ configurations of spins on the borders between two neighbouring corners, and $\sigma$ denotes the spin positioned in the lattice center. The upper-right corner represented by the corner tensor $\mathbf{C}(\sigma,\xi_1,\xi_2)$ is coloured in red. 
}\label{C4lattice}
 \end{figure}

\begin{figure}[tb]
 \centering
 \includegraphics[width=3.5in]{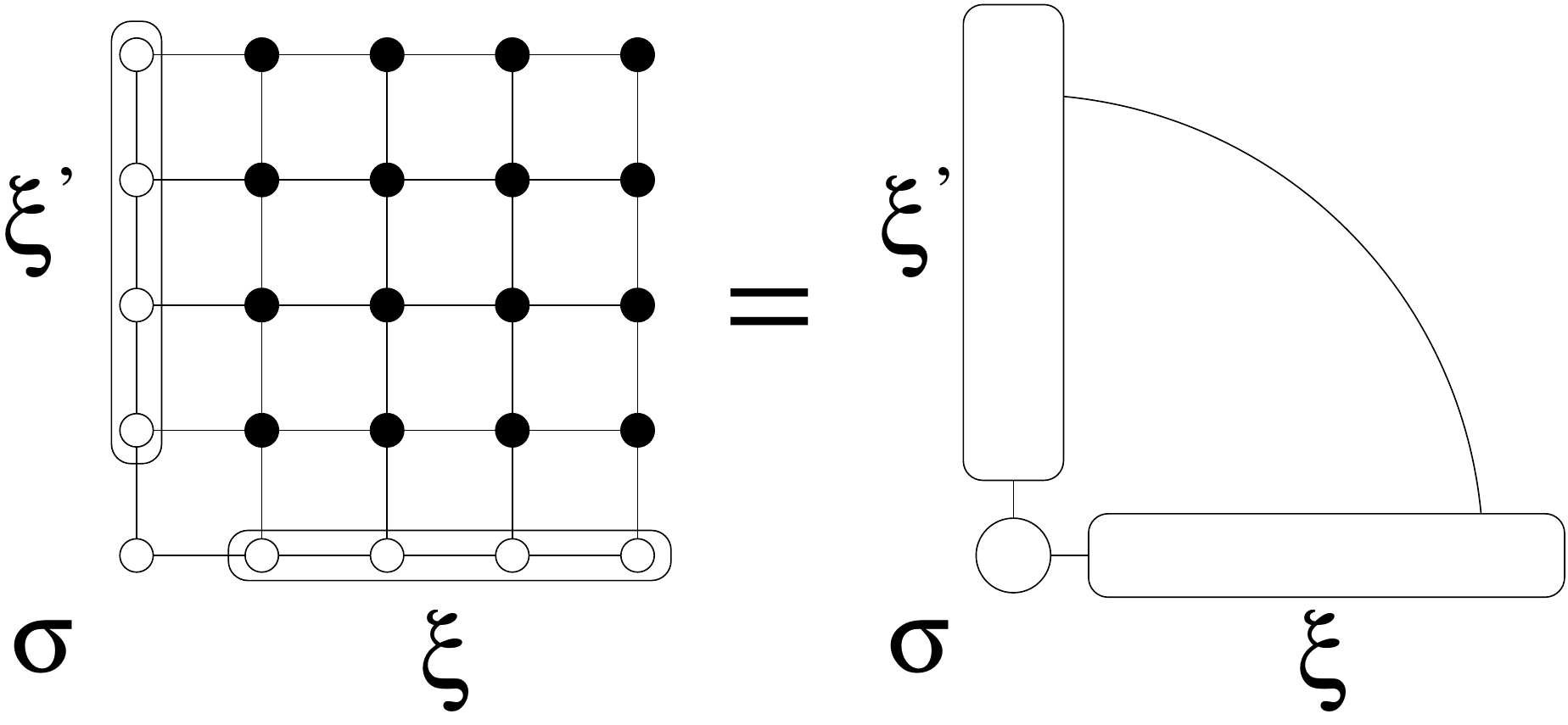}
   \caption{Graphical interpretation of the corner tensor $\mathbf{C}(\sigma, \xi, \xi')$ representing the upper-right lattice corner with $L=4$ spins on its interior border (left). Simplified graphical representation of the identical tensor (right) will be used in the following. 
}\label{Ctensor}
 \end{figure}

Let us again consider the 2D Ising model, however, unlike the previous case, open boundary conditions (OBC) are assumed. The lattice is a square, i. e., $M=N$, and its size $M=N=2L+1$ is considered to be odd so that we can divide the system into four equivalent quadrants  with respect to the central lattice spin% labeled $1-4$
\footnote{This concept can be further generalized, so that we can also consider $M \neq N$ for both even and odd $M$ and $N$.}%\footnote{In general the conditions $M=N$ and $M, N$ odd can be omitted if equivalence of the quadrants is not required. An arbitrary lattice point then defines $4$ non-identical quadrants which meet at this point.}
 (see Fig.~\ref{C4lattice}).
%We divide the system into $4$ equivalent quadrants with respect to the horizontal and vertical lines (passing through the lattice center).
%geodesic - najkratšia spojnica 2 bodov
 For each of these quadrants, we define the \emph{corner transfer tensor}\footnote{In literature, the corner transfer tensor is usually referred to as the corner transfer matrix. However, the matrix formalism requires a slightly modified definition
\begin{equation*}
\mathbf{C}\left( (\sigma, \xi), (\sigma', \xi') \right) = {\sum_{\lbrace\sigma\rbrace}}'\prod_{(i,j)} \mathbf{W}_B(\sigma_{i,j},\sigma_{i,j+1},\sigma_{i+1,j},\sigma_{i+1,j+1})\delta(\sigma,\sigma'),
\label{CTMatrix}
\end{equation*}
where the duplicated variable $\sigma'$ is created with the only intention of establishing $\mathbf{C}$ as a square matrix. We find this approach rather redundant and, therefore, we prefer the tensor notation \eqref{CTTensor} in the following.}
 $\mathbf{C}$ as
\begin{equation}
\mathbf{C}\left( \sigma, \xi, \xi' \right) = {\sum_{\lbrace\sigma\rbrace}}'\prod_{(i,j)} \mathbf{W}_B(\sigma_{i,j},\sigma_{i,j+1}, \sigma_{i+1,j},\sigma_{i+1,j+1}),
\label{CTTensor}
\end{equation}
where ${\sum_{\lbrace\sigma\rbrace}}'$ denotes summation over all configurations of spins inside the quadrant and on its outer border (represented by black filled circles in Fig.~\ref{Ctensor}), $\prod_{(i,j)}$ represents product over all the lattice cells within the quadrant, $\xi$, $\xi'$ label $2^L$ spin configurations of $L$ spins on each of the two  border lines of the corner with its neighbours and $\sigma$ labels the state of the central spin (cf. Fig.~\ref{Ctensor}).
 Note that due to the OBC some of the  fractions $J/2$ and $h/4$ in the formula \eqref{WB} for $\mathbf{W}_B$            
modify to $J$, $h$ or $h/2$ if the corresponding lattice cell is located on the lattice border, where the sharing of bonds and spin locations differs from situation inside the lattice. The corner transfer tensor thus represents the Boltzmann weight of the selected quadrant at the  configuration $\lbrace \sigma, \xi, \xi' \rbrace$ of spins on the inner border (if taking into account all possible configurations of the remaining spins in the corner).

  %It is often natural to think of the corner transfer matrix in the tensor form
%\begin{equation}
%{\cal C}(\sigma,\xi,\xi')=\mathbf{C}\left( \sigma \xi' \mid\sigma \xi \right),
%\end{equation}   
%which avoids the unphysical variable $\sigma'$ which was artificially created through the $\delta(\sigma, \sigma')$ term in \eqref{CTMatrix} with the only intention of establishing $\mathbf{C}$ as a square matrix. 

The structure of the corner transfer tensor allows us to rewrite the formula for the partition function into a convenient form
\begin{equation}
{\cal Z}_{N, N} =\sum\limits_{\sigma \xi_1 \xi_2\xi_3\xi_4}\mathbf{C}\left( \sigma, \xi_4, \xi_1 \right)\mathbf{C}\left(\sigma, \xi_1, \xi_2 \right)\mathbf{C}\left( \sigma, \xi_2, \xi_3 \right) \mathbf{C}\left( \sigma, \xi_3, \xi_4\right).
\end{equation}
%where the second index in ${\cal Z}_{N}$ was omitted due to equality $N=M$. 
This formula can be represented for homogeneous and isotropic spin systems in the simplified notation as
\begin{equation}
{\cal Z}_{N, N} = \Tr\left(\mathbf{C}^4\right).
\label{Z_C4}
\end{equation}
Therefore, the corner tensor $\mathbf{C}$ uniquely and exactly determines the partition function. In the previous section we derived the expression \eqref{Z_TN}, which establishes a similar relation between $\cal Z$ and the (row) transfer tensor $\mathbf{T}$. Now, instead of performing $N$ matrix multiplications or solving the eigenvalue problem, the multiplication of the four identical corner tensors $\mathbf{C}$ is required only. Note that the number of entries in the corner transfer tensor, $2^{2L+1}=2^{N}$, grows exponentially with the increasing system size $N$. It is, therefore, impossible to multiply the corner tensors $\mathbf{C}$ or even store them in memory if the system is large. Nevertheless, in the next section we %later
present a solution to this problem based on an appropriate renormalization technique.% in case we focus on the lowest energy states only. 

%We observed similar situation in the previous section, where it was shown that $\cal Z$ depends on the row transfer matrix $\mathbf{T}$ through \eqref{Z_TN}.    

\subsection{Corner transfer renormalization group}
\label{CTMRG_general}

In this section we introduce the Corner transfer matrix renormalization group (CTMRG) method in its original form as proposed by Nishino and Okunishi in \cite{NishinoCTMRG1, NishinoCTMRG2}. This algorithm provides highly accurate results for the classical spin models on large 2D square $N \times N$ lattices, provided that the structure of the model Hamiltonian is uniform and invariant to rotations of the system by $90^{\circ{}}$. The objective of the algorithm is to construct an "effective" corner transfer tensor ${\mathbf{C}}$ of the large system by an iterative sequence of step-by-step lattice expansions. A specific renormalization procedure
% of projecting onto a specific subspace 
 applied to each step guarantees that the number of entries of the "effective" tensor in the enlarged system does not exceed a preset bound. %Finally, the
Thus obtained corner transfer tensor ${\mathbf{C}}$ is used to evaluate the partition function $\cal Z$ via \eqref{Z_C4} or other quantities via similar formulae. %First we describe the CTMRG algorithm on 2D square lattice in detail, the theoretical reasoning
%background
% of the renormalization procedure is provided later.
 
%Let us assume the structure of the model Hamiltonian is uniform and invariant to rotations of the system by $90°$, which holds e. g. for the Ising model as described in previous sections. Consequently, the a single CTM matrix $\mathbf{C}$ is satisfactory for evaluation of $\cal Z$ via \eqref{Z_C4} and the Boltzmann weights $\mathbf{W}_B$ of all lattice cells are identical (except those with slightly different bond and vertex sharing on the lattice boundary). For tutorial purposes let us consider the Ising model with $\mathbf{W}_B$ given by \eqref{WB} in the following.    

\subsubsection{The algorithm}
\label{CTMRG_algorithm}
For tutorial purposes let us consider the 2D Ising model with the Hamiltonian \eqref{2DIsing_hamiltonian} and open boundary conditions. As required, the model is uniform and invariant by $90^{\circ}$ rotations, hence the Boltzmann weight tensors ${\mathbf W}_B$ of all lattice cells are identical and equal to \eqref{WB} (except those with slightly different bonds and vertices on the lattice boundary). From now on, in order to make the text shorter, we make no explicit difference between the tensors ${\mathbf W}_B$, $\mathbf{C}$ or $\mathbf{T}$ and the lattice structures they represent\footnote{This means that one can, for example, encounter sentences containing "we attach the {\it half-row} $\mathbf{T}^{(k)}$ to the {\it bottom side} of the corner $\mathbf{C}^{(k)}$". }.   

\begin{figure}[tb]
 \centering
 \includegraphics[width=4in]{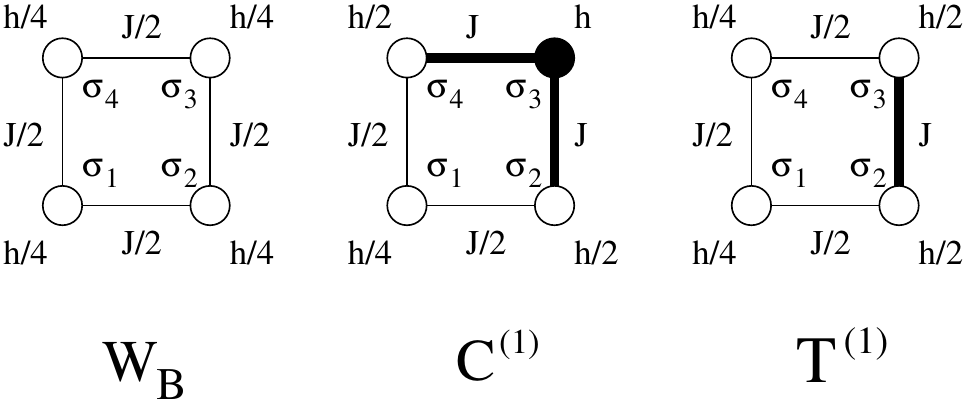}
   \caption{Graphical interpretation of the tensors ${\mathbf W}_B(\sigma_1, \sigma_2, \sigma_4, \sigma_3 )$, $\mathbf{C}^{(1)}(\sigma_1, \sigma_2, \sigma_4 )$ and $\mathbf{T}^{(1)}(\sigma_1, \sigma_2, \sigma_4, \sigma_3)$ with the correct sharing factors appearing in front of the terms $h\sigma_i$ and $J\sigma_i\sigma_{j}$ in the formulae \eqref{WB}, \eqref{WB_upper_right} and \eqref{HTM_init}, respectively. A thick edge labels a bond with the full contribution $2J/2=J$ and a filled circle denotes a spin variable, which is summed over in \eqref{CTM_init}. 
}\label{WCT_init}
 \end{figure}

\paragraph{I. Initialization}
The algorithm starts with a small system containing $3 \times 3$ spins on the square lattice, where each of the four corners is formed by a single cell. %The CTM matrix is initialized as
The corner tensor $\mathbf{C}^{(1)}$ of the initial upper-right $2 \times 2$ corner is then according to \eqref{CTTensor} given by  
\begin{equation}
{\mathbf C}^{(1)}(\sigma_1, \sigma_2, \sigma_4) = \sum_{\sigma_3}{ \mathbf W}^{\prime}_B(\sigma_1, \sigma_2, \sigma_4, \sigma_3),
\label{CTM_init}
\end{equation} 
where
\begin{align}
\begin{split}
{\mathbf W}^{\prime}_B(\sigma_1, \sigma_2, \sigma_4, \sigma_3)
= \exp & \left[
{\frac{h}{4 k_B T}{\left(\sigma_1+2\sigma_2+4\sigma_3+2\sigma_4\right)}}+
\right.\\
	&\hspace*{-0.2cm}+\left.\frac{J}{2 k_B T}{\left(
\sigma_1\sigma_2
+2\sigma_2\sigma_3
+2\sigma_3\sigma_4
+\sigma_4\sigma_1
\right)}\right]
\end{split}
\label{WB_upper_right}
\end{align}
is the Boltzmann weight tensor of the single corner cell. ${\mathbf W}^{\prime}_B$ is a modification of the original tensor ${\mathbf W}_B$ \eqref{WB}, which takes into account the different fractions of the bonds and the vertices on the lattice boundary in comparison to the lattice interior. The situation is illustrated in Fig.~\ref{WCT_init} (in the middle). %Spins $\sigma_2, \sigma_3, \sigma_4$ lie on the boundary, $\sigma_2,\sigma_4$
Spins $\sigma_2$ and $\sigma_4$ are shared only between two cells --- the current cell and the neighbouring one to the bottom and to the left, respectively. The sharing fraction of spins $\sigma_2$ and $\sigma_4$ is therefore $h/2=2h/4$. The spin $\sigma_3$ and the two bonds $\sigma_2\sigma_3$, $\sigma_3\sigma_4$ are not shared with other cells which is represented by the sharing fractions $J=2J/2$ and $h=4h/4$.    
%obrazky ako v minimovke, ale bez sum a C(...), T(...)

In CTMRG, the lattice expansion is carried out by attaching two half-rows of cells of matching length to the interior sides of the corner. The Boltzmann weight of these half-rows is represented by the (half-row) transfer tensor $\mathbf T$. The matching half-row attached to %the bottom of the 
the initial $2\times2$ corner is a single cell with one side on the boundary. %, cf. Fig~\ref{WCT_init} (right).
 The transfer tensor ${\mathbf T^{(1)}}$ corresponding to the {\it half-row} attached to the {\it bottom side} of the {\it right-upper} corner ${\mathbf C^{(1)}}$ is 
\begin{align}
\begin{split}
{\mathbf T^{(1)}}(\sigma_1, \sigma_2, \sigma_4, \sigma_3)
= \exp & \left[
{\frac{h}{4 k_B T}{\left(\sigma_1+2\sigma_2+2\sigma_3+\sigma_4\right)}}+
\right.\\
&\hspace*{-0.2cm}+\left.\frac{J}{2 k_B T}{\left(
\sigma_1\sigma_2
+2\sigma_2\sigma_3
+\sigma_3\sigma_4
+\sigma_4\sigma_1
\right)}\right],
\label{HTM_init}
\end{split}
\end{align}
i. e., the modification of ${\mathbf{W}_B}$ with the different prefactors at terms $\sigma_2$, $\sigma_3$, $\sigma_2\sigma_3$ which are related to the boundary as depicted in Fig.~\ref{WCT_init} (on the right)
\footnote{    
Correct matrix initialization is inevitable only if the lattice size is finite, i. e., $L < \infty$.  If, however, an infinite 2D Euclidean system is simulated, the initialization becomes irrelevant, because the ratio of the number of spins on the boundary to the number of spins in the entire lattice becomes zero in the thermodynamic limit. It is often useful to add a small magnetic field $g$ in the initial tensors $\mathbf{C}^{(1)}$ and $\mathbf{T}^{(1)}$ in order to enhance the symmetry breaking mechanism. On the contrary, proper initialization becomes
%is, gets
 essential on hyperbolic lattices, where the boundary is comparable in size with the interior. However, the boundary effects may get negligible even on hyperbolic lattices if, e. g., local quantities, such as the local magnetization $\langle \sigma\rangle$ on the central lattice site are evaluated.}.

\begin{figure}[tb]
 \centering
 \includegraphics[width=3.8in]{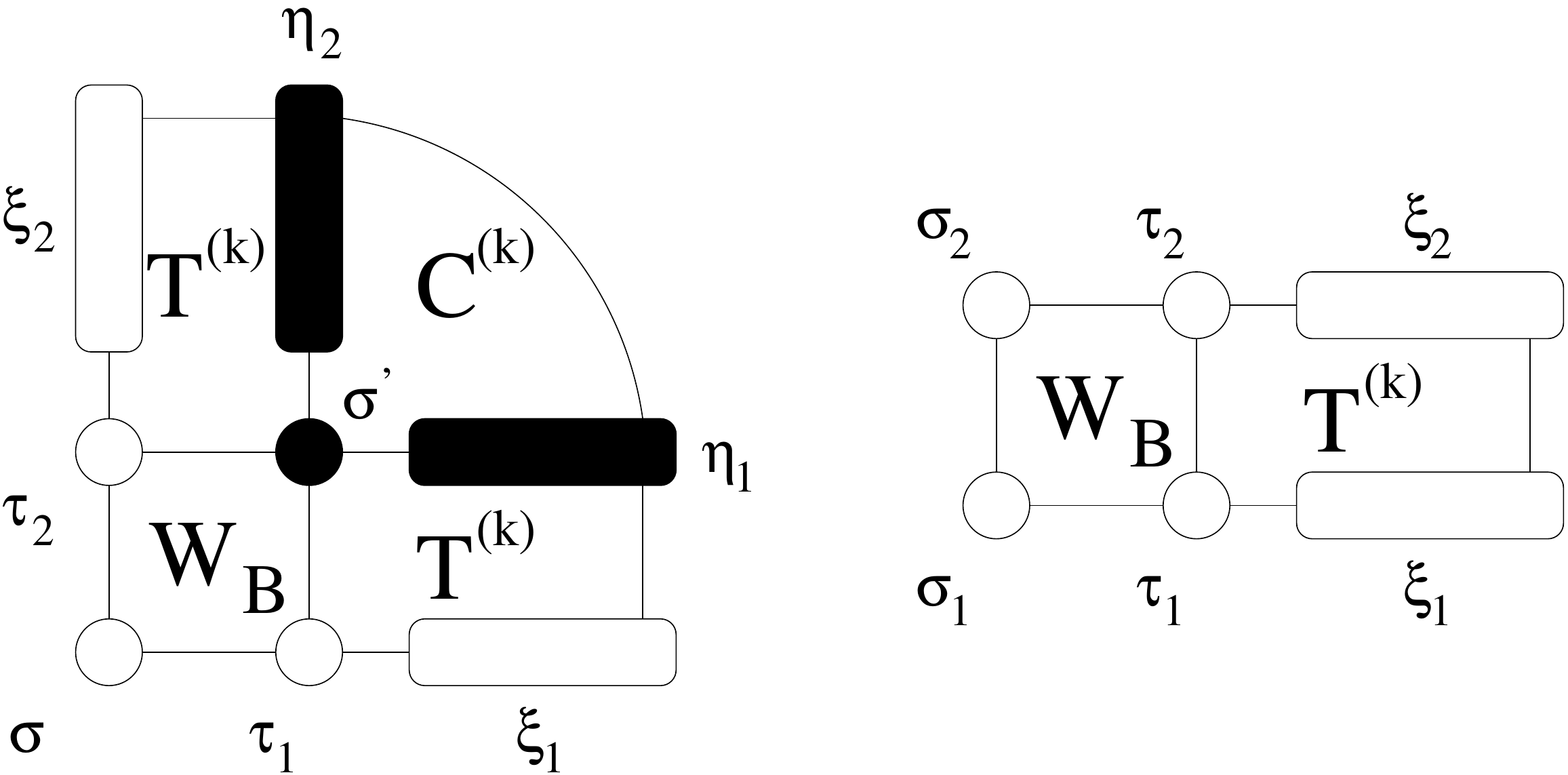}
   \caption{The expansion of the corner tensor ${\mathbf C}^{(k)}$ (left) and the transfer tensor ${\mathbf T}^{(k)}$ (right). Variables, which are summed over in \eqref{CTMexpanded}, are denoted as black-filled.    
}\label{CTexpansion}
 \end{figure}

\paragraph{II. Lattice and tensor expansion} 
Having initialized the tensors $\mathbf{C}$ and $\mathbf{T}$, the process of system expansion can start.
%be launched
%The process is schematically illustrated in Fig.~\ref{CTexpansion}.
 The lattice corner from the previous $k$-th step containing $k \times k$ cells or, equivalently $(k+1) \times (k+1)$ spins is extended by adding two half-rows of $k$ cells to its interior sides and a single cell at the central position. Boltzmann weights of the above-mentioned objects are represented by tensors ${\mathbf C}^{(k)}$,${\mathbf T}^{(k)}$ and $\mathbf{W}_B$, respectively. %The tildes denote renormalized versions of the respective tensors, as will be explained later. 
 As a result, we receive an enlarged corner with $(k+1)\times(k+1)$ cells. The corresponding corner tensor $\tilde{\mathbf C}^{(k+1)}$ is, therefore, given by
\begin{align}
\label{CTMexpanded}  
\begin{split}
\tilde{\mathbf C}^{(k+1)}(\sigma, \lbrace \tau_1 \xi_1 \rbrace, \lbrace \tau_2 \xi_2 \rbrace)
&= \sum_{\sigma', \eta_1, \eta_2}
{\mathbf T}^{(k)}(\sigma',\eta_2,\tau_2, \xi_2)
\mathbf{W}_B(\sigma, \tau_1, \tau_2, \sigma')\\
&\times 
{\mathbf C}^{(k)}(\sigma', \eta_1, \eta_2)
{\mathbf T}^{(k)}(\tau_1, \xi_1,\sigma', \eta_1),  
\end{split}       
\end{align}
where $\sigma$, $\tau_1$, $\tau_2$, $\sigma'$ are single-  and $\xi_1$, $\xi_2$, $\eta_1$, $\eta_2$ multi-spin variables. The tilde denotes the unrenormalized version of the tensor, as explained in the next section. The situation is illustrated in Fig.~\ref{CTexpansion} (left). %, where we follow the convention of depicting the variables, which are being summed as black filled objects.
 Due to the sum $\sum_{\sigma', \eta_1, \eta_2}$, the tensor elements \eqref{CTMexpanded} represent Boltzmann weight of the $(k+1)\times(k+1)$ corner at a given configuration of spins on the inner corner boundary, as required by the definition of the corner transfer tensor.

In the next iteration step, a half-row with $(k+1)$ cells will be required. This object is constructed by attaching a single cell to the interior side of the half-row represented by ${\mathbf T}^{(k)}$. The transfer tensor of the enlarged half-row is thus given by
\begin{align}
\begin{split}
\tilde{\mathbf T}^{(k+1)}(\sigma_1, \lbrace \tau_1 \xi_1 \rbrace,\sigma_2, \lbrace \tau_2 \xi_2 \rbrace)=
\mathbf{W}_B(\sigma, \tau_1, \sigma_2, \tau_2)
{\mathbf T}^{(k)}(\tau_1, \xi_1,\tau_2, \xi_2),  
\end{split} 
\label{TMexpanded}        
\end{align}           
as illustrated in Fig.~\ref{CTexpansion} (right). 
As a result, the total enlargement of the system, which is constructed from four corners, is equivalent to inserting two rows and two columns of cells of length $2(k+1)$ into the lattice center. 

Finally, we introduce the simplified notation, in which the recurrence expansion formulae take the form
\begin{equation}
\mathbf{C}^{(k+1)}
=
{\mathbf T}^{(k)}
\mathbf{W}_B
{\mathbf C}^{(k)}
{\mathbf T}^{(k)},
\label{C_schematic}
\end{equation}
\begin{equation}
\mathbf{T}^{(k+1)}=
\mathbf{W}_B
{\mathbf T}^{(k)},
\label{T_schematic}  
\end{equation}
where the tildes were omitted in order to emphasize the construction scheme, not the renormalization aspects. 

\paragraph{III. Renormalization}
The number of entries in the tensors $\mathbf{T}^{(k)}$ and $\mathbf{C}^{(k)}$ is $(2\times2^k)^2$ and $2\times2^{2k}$, respectively, where $2^k$ is the number of spin configurations of the  multi-spin variable labeling $k$ spins. The simple repeating of the expansion process described above leads to numerical overflows and enormous memory usage caused by exponential increase of the number of entries in the tensors $\mathbf{T}^{(k)}$ and $\mathbf{C}^{(k)}$.  It is, therefore, necessary to supply
%doplnit o
 each expansion step with a renormalization procedure, which reduces the number of entries of the newly created tensors $\tilde{\mathbf T}^{(k+1)}$ and $\tilde{\mathbf C}^{(k+1)}$ to an acceptable level. In CTMRG this is done by projecting the spin state space of the multi-spin variables onto a subspace with a significantly lower dimension. 
 
Let us denote the maximal acceptable
 %acceptable prípustný       
dimension of the multi-spin state space by the integer variable $m$. After the $k$-th expansion step, new $2\times \Omega_k$-dimensional multi-spin variables $\lbrace \tau\xi \rbrace $ are created, where $\Omega_k=\min\left({2^k},m\right)$ is the dimensionality of the variable $\xi$. Except for a few iterations at the beginning, $\Omega_k=m$, and the $2 \times m$-dimensional variable $\lbrace \tau\xi \rbrace $ has to be projected to an appropriate $m$-dimensional subspace. The instructions on how to construct the projection operator are summarized below.

\begin{figure}[tb]
 \centering
 \includegraphics[width=3.8in]{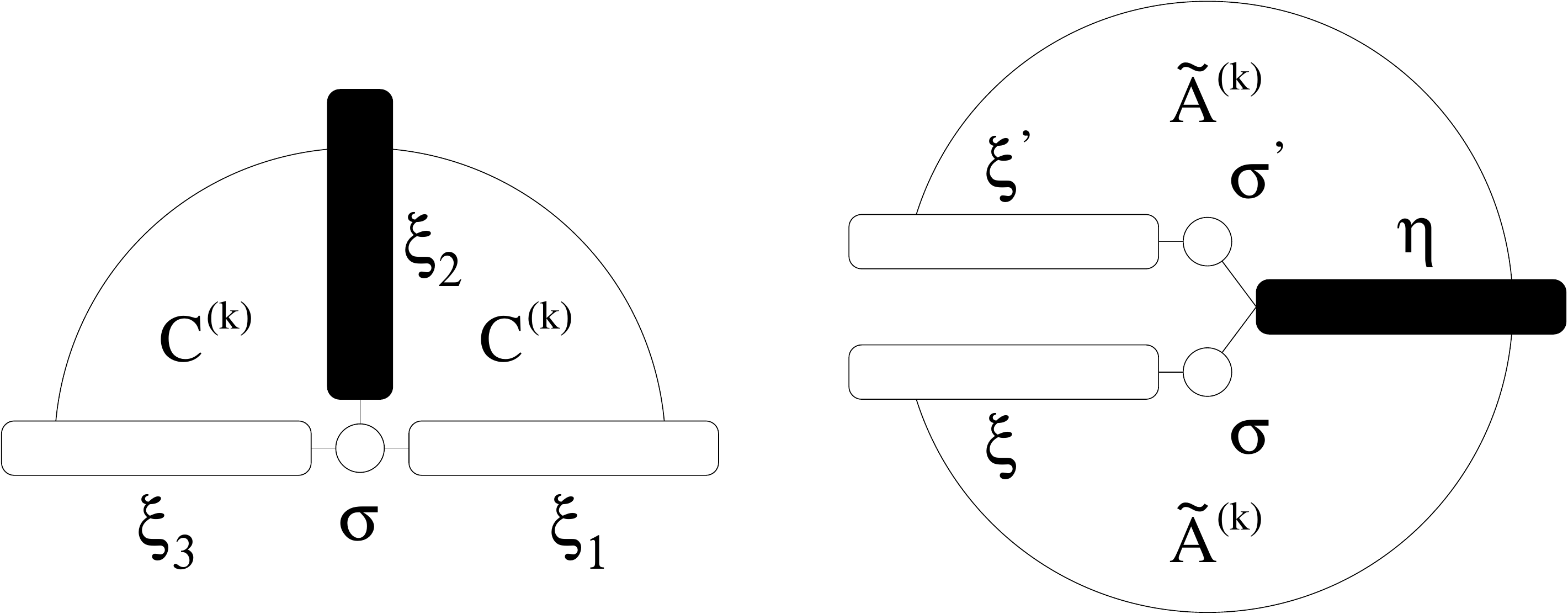}
   \caption{Graphical illustration of the tensor ${\mathbf A}^{(k)}$    (left) and the density matrix $\rho^{(k)}$ (right) corresponding to a cut in the lattice which starts in the lattice center and continues to the left boundary. 
}\label{A_DM}
 \end{figure}

First, we calculate a new tensor ${\mathbf A}^{(k)}$ by multiplying two corner tensors ${\mathbf C}^{(k)}$ of two neighbouring lattice quadrants from the previous step,  
\begin{equation}
{\mathbf A}^{(k)}(\sigma, \xi_1, \xi_3) = \sum_{\xi_2} {\mathbf C}^{(k)}(\sigma, \xi_1, \xi_2){\mathbf C}^{(k)}(\sigma, \xi_2, \xi_3), 
\label{nenormovanyA}
\end{equation}  
as illustrated in Fig.~\ref{A_DM} (left).
 In this way we defined an object, which represents the Boltzmann weight of one half of the lattice. The tensor ${\mathbf A}^{(k)}$ is normalized\footnote{ The normalization step may be omitted, as its only purpose is to guarantee the validity of the condition $\Tr{\rho}=1$. The projection operator is constructed from the normalized eigenstates of the density matrix $\rho$, which are not affected by scalar multiplication of $\rho$. It is, however, advisable to follow the normalization practice so that one can check the truncation error $\epsilon = \sum_{k=m+1}^{2m}{\omega_k}$, where $\omega_k$ are the eigenvalues of the density matrix labelled in the descending order.}
\begin{equation}
\tilde{\mathbf A}^{(k)}(\sigma, \xi, \eta) = \frac{{\mathbf A}^{(k)}(\sigma, \xi, \eta)}{\left|\left|{\mathbf A}^{(k)}\right|\right|}, 
\label{normovanyA}
\end{equation}
where the norm is
\begin{equation}
\left|\left|{\mathbf A}^{(k)}\right|\right| = \sqrt{\sum_{\sigma,\xi,\eta}{{\mathbf A}^{(k)}}^2(\sigma, \xi, \eta)}.
\end{equation}

Now, we construct the \emph{reduced density matrix} $\rho^{(k)}$ via
\begin{equation}
\rho^{(k)} \left(\lbrace \sigma, \xi \rbrace \left| \lbrace \sigma', \xi' \rbrace \right.\right)
=\sum_{\eta} \tilde{\mathbf A}^{(k)}(\sigma, \xi, \eta)\tilde{\mathbf A}^{(k)}(\sigma', \xi', \eta),
\label{def_DM}
\end{equation} 
as illustrated in Fig.~\ref{A_DM} (right). The elements of the density matrix $\rho \left(\lbrace \sigma, \xi \rbrace \left| \lbrace \sigma', \xi' \rbrace \right.\right)$ can be interpreted as the Boltzmann weights of a cut in the lattice at the spin configuration $\lbrace \sigma, \xi,\sigma', \xi' \rbrace$. The cut starts in the lattice center and continues along one of the main axes until the lattice border.

Without loss of generality, let us assume the usual situation, when ${\rm dim}\,\rho=2m$. The density matrix $\rho$ is constructed as a symmetric matrix, hence it is diagonalizable with orthonormal basis of eigenvectors. Let us label the eigenvalues $\omega_k$ of $\rho$ in the descending order, i. e., $\omega_1 \geq \omega_2 \geq ... \geq \omega_{2m}$ and the corresponding eigenvectors as $\Phi_k$. Then, the operator ${\mathbf P}$ of projection onto the $m$-dimensional state subspace is constructed as a $2m \times m$ matrix filled with eigenvectors $\Phi_1 ... \Phi_m$ as its columns, i. e.,
\begin{equation}
{\mathbf P} = 
\begin{pmatrix}
| 		&  	|		& \cdots	& |			\\
\Phi_1 	& \Phi_2 	& \cdots	& \Phi_{m}	\\
|		&	|		& \cdots	& |
\end{pmatrix}. 
\end{equation}  
The projection operator is applied to the tensors $\tilde{\mathbf C}^{(k+1)}$ and $\tilde{\mathbf T}^{(k+1)}$ given by equations \eqref{CTMexpanded} and \eqref{TMexpanded}, respectively. As a result, we obtain the renormalized tensors 
%(see Fig.??)
\begin{equation}
{\mathbf C}^{(k+1)}\left(\sigma, \xi_1,\xi_2 \right)=\sum_{\tau_1, \eta_1, \tau_2, \eta_2} {\mathbf P}\left({\lbrace \tau_1 \eta_1 \rbrace|\xi_1}\right)\tilde{\mathbf C}^{(k+1)}\left(\sigma, \lbrace \tau_1 \eta_1 \rbrace, \lbrace \tau_2 \eta_2 \rbrace\right){\mathbf P}\left({\lbrace \tau_2 \eta_2 \rbrace|\xi_2}\right)
\label{renormalizedC}
\end{equation}
and 
\begin{multline}
{\mathbf T}^{(k+1)}\left(\sigma_1, \xi_1, \sigma_2, \xi_2 \right)
=
\\
=
\sum_{\tau_1, \eta_1, \tau_2, \eta_2} 
{\mathbf P}\left({\lbrace \tau_1 \eta_1 \rbrace|\xi_1}\right)
\tilde{\mathbf T}^{(k+1)}\left(\sigma_1, \lbrace \tau_1 \eta_1 \rbrace,\sigma_2, \lbrace \tau_2 \eta_2 \rbrace\right)
{\mathbf P}\left({\lbrace \tau_2 \eta_2 \rbrace|\xi_2}\right),
\label{renormalizedT}
\end{multline}
whose multi-spin variables $\xi_1$, $\xi_2$ now live in the demanded $m$-dimensional space. The tensors ${\mathbf C}^{(k+1)}$ and ${\mathbf T}^{(k+1)}$ are used as an input in the following $(k+2)$-th iteration step, replacing ${\mathbf C}^{(k)}$ and ${\mathbf T}^{(k)}$ in the instructions above, which yields ${\mathbf C}^{(k+2)}$ and ${\mathbf T}^{(k+2)}$ and so on.

The choice of the classical density matrix, as an effective selector between the important and negligible states, is based on the quantum-classical correspondence. Namely, the quantum \emph{Density matrix renormalization group}  algorithm \cite{White1, White2} provides highly accurate results  for the ground-state or a few low excited states of quantum systems on a chain in the thermodynamic limit. The system is constructed from two identical blocks, which represent the left and the right part of the chain. In each step the blocks are iteratively expanded by adding a single site. The dimension of the effective Hilbert space is maintained within acceptable limits by projecting onto a suitable subspace. Here, the subspace is generated by the eigenvectors corresponding to the $m$ largest eigenvalues of the reduced density matrix of the quantum system. Applying the quantum-classical correspondence, the classical Density matrix renormalization group method \cite{NishinoDMRG} for two-dimensional classical lattice models was developed. The system gradually expands in the horizontal direction and the renormalization process is governed by the classical density matrix created from the normalized eigenvector ${\mathbf v}_1$ of the transfer matrix ${\mathbf T}$ corresponding to the largest eigenvalue $\lambda_1$. For large systems ${\mathbf v}_1 \approx \tilde{\mathbf A}$, cf. \eqref{nenormovanyA}, \eqref{normovanyA}, which is the idea of the CTMRG algorithm.

Note that as a result of multiple summations in the expansion-renormalization formulae, the tensor elements ${\mathbf C}^{(k+1)}\left(\sigma, \xi_1,\xi_2 \right)$ and ${\mathbf T}^{(k+1)}\left(\sigma_1, \xi_1, \sigma_2, \xi_2 \right)$ diverge exponentially as $k$ increases. In order to avoid this, normalization of tensors ${\mathbf C}^{(k+1)}$ and ${\mathbf T}^{(k+1)}$ before starting the next iteration step $(k+1)$ is necessary. As an example, one can use
\begin{equation}
{\hat{\mathbf C}}^{(k+1)}
=
\frac{{\mathbf C}^{(k+1)}}
{c_{k+1}}
\quad
{\rm and}
\quad
{\hat{\mathbf T}}^{(k+1)}
=
\frac{{\mathbf T}^{(k+1)}}
{t_{k+1}},
\label{CT_normalizacia}
\end{equation} 
where
\begin{equation}
c_{k+1}=\max_{\sigma, \xi_1,\xi_2} \left\{{\mathbf C}^{(k+1)}\left(\sigma, \xi_1,\xi_2 \right) \right\}
\quad
{\rm and}
\quad
t_{k+1}
=
\max_{\sigma_1, \xi_1, \sigma_2, \xi_2} \left\{{\mathbf T}^{(k+1)}\left(\sigma_1, \xi_1, \sigma_2, \xi_2 \right)\right\}.
\end{equation} 
\paragraph{IV. Calculation of the free energy and observables}

The CTMRG algorithm stops if the free energy per site $f$ and all important observables converged. Here, by the term convergence we mean a situation, when values of the respective quantities in two consecutive iterations differ by less than a preset tolerance constant $\epsilon$. For example, we demand $|f^{(k+1)}-f^{(k)}|<\epsilon$, where $f^{(k)}$ is the free energy per site $f$ in the iteration $k$. Below, we demonstrate how to calculate the free energy $f$ and quantities such as the local magnetization $\langle\sigma_{\ell}\rangle$ on a central lattice site $\ell$ or the nearest-neighbour correlation function $\langle\sigma_{\ell}\sigma_{\ell^{\prime}}\rangle$ using the tensors $\mathbf{C}$ and $\mathbf{T}$.  

Let us start with the free energy per site
\begin{equation}
f^{(k)}=\frac{{\cal F}^{(k)}}{N^{}} = - \frac{k_B T \ln{\cal Z}_{N,N}}{N^{}} = - \frac{k_B T \ln{\Tr\left({\mathbf C}^{(k)}\right)^4 }}{N^{}}, 
\label{fCk}
\end{equation}  
where $N = (2k+1)^2$ is the number of lattice vertices in the iteration $k$. Note that the free energy $\cal F$ is an extensive quantity, which diverges as the lattice increases and, thus, does not converge. In CTMRG the exact corner tensor in \eqref{fCk} is approximated by its renormalized version ${\mathbf C}^{(k)}$ at high accuracy. 

Since we have only the normalized tensors ${\hat{\mathbf C}}^{(k)}$ at disposal, $f^{(k)}$ is not calculated directly via \eqref{fCk}. Instead, the schematic form of the recurrence expansion formulae \eqref{C_schematic}, \eqref{T_schematic} and the normalization relations \eqref{CT_normalizacia} are used to determine the partition function ${\cal Z}_{N,N}$ as a product of the normalization constants $c_k$ and $t_k$. In particular, decomposing the corner tensor ${\hat{\mathbf C}}^{(k)}$ into the product form of its constituents, we have
\begin{align}
\begin{split}
{\hat{\mathbf C}}^{(k)} 
&=
\frac{{{\mathbf C}}^{(k)}}{c_k}
=
\frac{{\mathbf W}_B{\hat{\mathbf C}}^{(k-1)}\left({\hat{\mathbf T}}^{(k-1)}\right)^2}{c_k}
=
\frac{{\mathbf W}_B{{\mathbf C}}^{(k-1)}\left({{\mathbf T}}^{(k-1)}\right)^2}{c_k c_{k-1} t_{k-1}^2}
\\
&= 
\frac{{\mathbf W}_B\left[{\mathbf W}_B{\hat{\mathbf C}}^{(k-2)}\left({\hat{\mathbf T}}^{(k-2)}\right)^2\right]\left({\mathbf W}_B{\hat{\mathbf T}}^{(k-2)}\right)^2}{c_k c_{k-1} t_{k-1}^2}
\\
&=
\frac{{\mathbf W}_B^{(k-1)^2}\left({\mathbf T}^{(1)}\right)^{2(k-1)}{\mathbf C}^{(1)}}{\prod_{i=1}^{k}c_i t_{i}^{2(k-i)}}
=
\frac{{\mathbf W}_B^{k^2}}{\prod_{i=1}^{k}c_i t_{i}^{2(k-i)}}.
\end{split}
\end{align}  
The product ${\mathbf W}_B^{k^2}$ is the Boltzmann weight of a single corner on $(2k+1)\times(2k+1)$ lattice in the iteration $k$ with $(2k)^2$ cells. Hence
\begin{equation}
{\cal Z}_{N,N} = \Tr\left(\prod_{i=1}^{k}c_i t_{i}^{2(k-i)}{\hat{\mathbf C}}^{(k)}\right)^{4} = \left(\prod_{i=1}^{k}c_i t_{i}^{2(k-i)}\right)^4\Tr\left({\hat{\mathbf C}}^{(k)}\right)^{4}
\end{equation}
and
\begin{equation}
f^{(k)} = -\frac{k_B T}{(2k+1)^2}\left\{\ln\Tr\left({\hat{\mathbf C}}^{(k)}\right)^{4}+4\sum_{i=1}^{k}\left[\ln c_i + 2(k-i)\ln t_i \right]\right\}.
\end{equation} 
As a result, the complete set of the normalization constants $c_i$, $t_i$ for $1 \leq i \leq k$ must be stored in memory in order to calculate the free energy per site $f^{(k)}$ in iteration $k$. 

\begin{figure}[tb]
 \centering
\vspace*{-0.2cm}
 \includegraphics[width=2.3in]{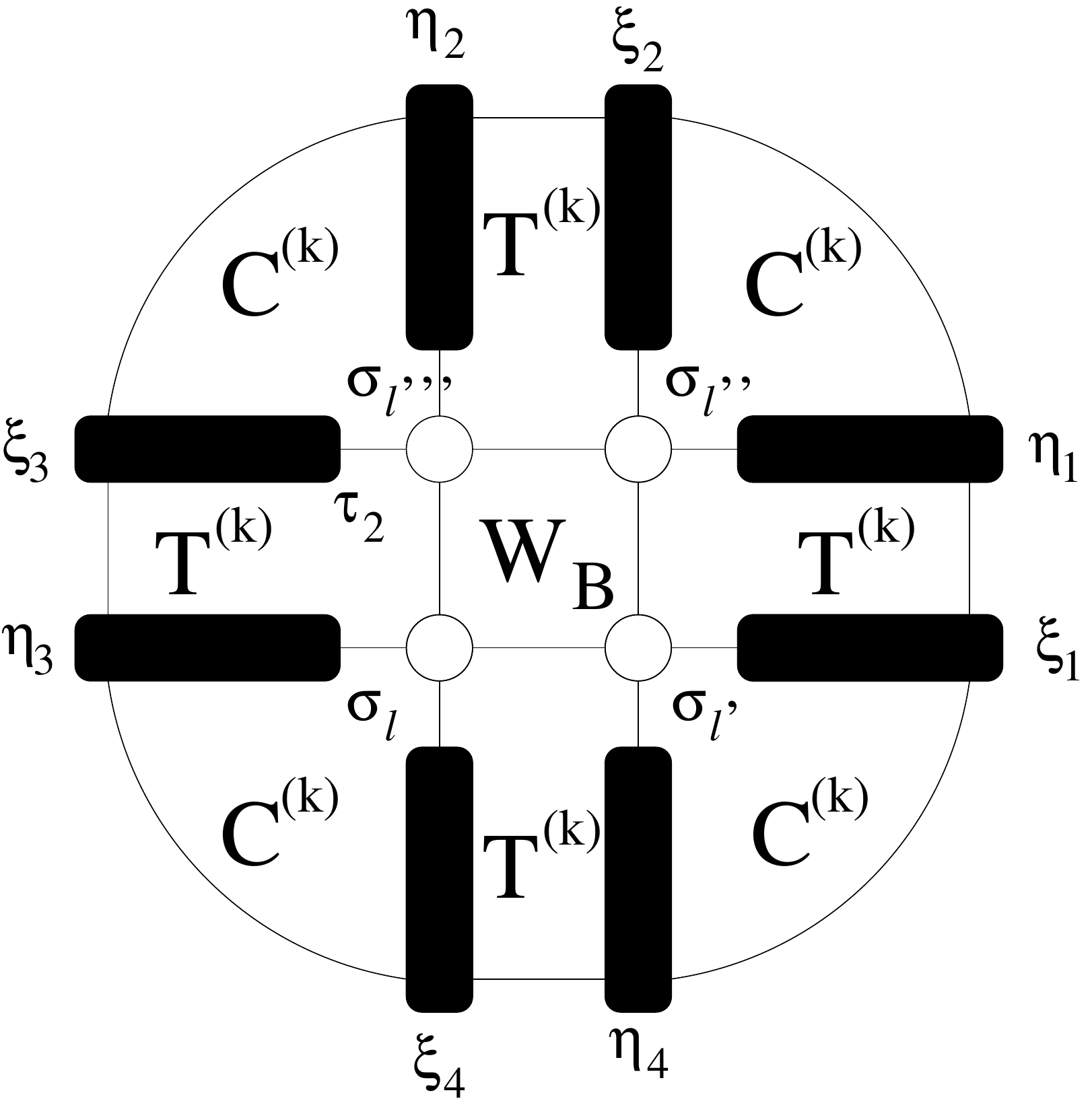}
   \caption{ The lattice structure required to calculate the correlation function $\langle\sigma_{\ell}\sigma_{\ell^{\prime}}\rangle$. %After the multi-spin variables $\xi and \eta$ are summed over, the Boltzmann weight 
Expectation values of any quantities depending on the spin variables    
$\sigma_{\ell},\sigma_{\ell^{\prime}},\sigma_{\ell^{\prime\prime}},\sigma_{\ell^{\prime\prime\prime}}$ can be easily calculated.
}\label{C4T4W}
\vspace*{-0.2cm}
 \end{figure}

The expected value of the local magnetization $\langle \sigma_{\ell}\rangle$ on the central lattice site $\ell$ is calculated as
\begin{equation}
\langle \sigma_{\ell}\rangle = \frac{\sum\limits_{\sigma_{\ell}}\sigma_{\ell}\sum\limits_{\xi_1,\xi_2,\xi_3,\xi_4}{\mathbf C}(\sigma_{\ell}, \xi_1, \xi_2){\mathbf C}(\sigma_{\ell}, \xi_2, \xi_3){\mathbf C}(\sigma_{\ell}, \xi_3, \xi_4){\mathbf C}(\sigma_{\ell}, \xi_4, \xi_1)}{\sum\limits_{\sigma_{\ell},\xi_1,\xi_2,\xi_3,\xi_4}{\mathbf C}(\sigma_{\ell}, \xi_1, \xi_2){\mathbf C}(\sigma_{\ell}, \xi_2, \xi_3){\mathbf C}(\sigma_{\ell}, \xi_3, \xi_4){\mathbf C}(\sigma_{\ell}, \xi_4, \xi_1)},
\label{C4_mag}
\end{equation}  
since the tensor product ${\mathbf C}^4$ gives the Boltzmann weight of the spin state $\sigma_{\ell}$. In the simplified notation, the formula  \eqref{C4_mag} takes the form
\begin{equation}
\langle \sigma_{\ell}\rangle = \frac{\Tr\left(\sigma_{\ell}{\mathbf C}^4\right)}{\Tr\left({\mathbf C}^4\right)}.
\label{exp_value_mag}
\end{equation}
Equivalently, using the definition \eqref{def_DM}, we can also write
\begin{equation}
\langle \sigma_{\ell}\rangle = \frac{\Tr\left(\sigma_{\ell}\rho\right)}{\Tr\left(\rho\right)},
\end{equation}
which reduces to $\Tr\left(\sigma_{\ell}\rho\right)$ if we normalize the density matrix so that $\Tr\left(\rho\right)=1$.
Such a simple calculation of the local magnetization can be performed on the central lattice site $\ell$ only, since in the corner tensor formalism the spin variable $\sigma_\ell$ is directly accessible, while spins on the remaining lattice sites are either summed over or incorporated into the multi-spin variables $\xi$, which do not take track of the original single-spin states. For the same reason, if we are interested in the nearest-neighbour correlation function $\langle\sigma_{\ell}\sigma_{\ell^{\prime}}\rangle$, it is necessary to construct the lattice as a central polygon (represented by the Boltzmann weight tensor ${\mathbf W}_B$) surrounded by an adequate number of the corners ${\mathbf C}$ and the half-rows ${\mathbf T}$, cf. Fig.~\ref{C4T4W}.
 
Now, all the four spins $\sigma_{\ell},\sigma_{\ell^{\prime}},\sigma_{\ell^{\prime\prime}},\sigma_{\ell^{\prime\prime\prime}}$ on the central polygon can be accessed directly. The nearest-neighbour correlation function $\langle\sigma_{\ell}\sigma_{\ell^{\prime}}\rangle$ is then calculated as
\begin{equation}
\langle \sigma_{\ell}\sigma_{\ell^{\prime}}\rangle = \frac{\Tr\left(\sigma_{\ell}\sigma_{\ell^{\prime}}{\mathbf W}_B{\mathbf T}^4{\mathbf C}^4\right)}{\Tr\left({\mathbf W}_B{\mathbf T}^4{\mathbf C}^4\right)},
\end{equation}
where the denominator determines the partition function $\cal Z$ of this system.
  
\iffalse  
\subsubsection{Theory behind CTMRG}

The main essence of the CTMRG algorithm rests in the way how the renormalization process is performed. The choice of the classical density matrix as an effective selector between the important and negligible states 
%stems from
is based on the quantum-classical correspondence. %Namely, it was shown by S. R. White \cite{White1, White2} that in one-dimensional quantum systems on a chain in the thermodynamic limit highly accurate results  for the ground-state or a few low excited states can be obtained if one proceeds as following: The large system is obtained by iterative expansion. In each step, two new sites are inserted into the lattice center. The dimension of the effective Hilbert space is maintained within   
Namely, the quantum \emph{density matrix renormalization group} (DMRG) algorithm \cite{White1, White2} provides highly accurate results  for the ground-state or a few low excited states for one-dimensional quantum systems on a chain in the thermodynamic limit.  The algorithm proceeds as following: The large system is obtained by iterative expansion. In each step, two new sites are inserted into the lattice center. The dimension of the effective Hilbert space is maintained within   
\fi

\subsection{CTMRG on hyperbolic lattices}
\label{CTMRG_hyp}

%A similar approach based on the generalized concept of the corner transfer tensor can be often successfully applied to any system if it is possible to divide it into a number of equivalent parts - the corners. The Ising model on the $(p,q)$ class of Euclidean and hyperbolic lattices with uniform bond strength $J$ and field $h$ serves as a positive example. If we consider $(p,q)$ lattices of infinite size, then any of them can be split into $q$ identical corners
%having been cut along its geodetic
% as depicted in Fig.~??. As a result, the above mentioned renormalization technique may be applied to all infinite $(p,q)$ lattices, which we discuss in detail in section ??.   

In section \ref{CTMRG_algorithm} the CTMRG algorithm was presented in its original form, as designed for systems on the two-dimensional Euclidean lattice with the square cells. %In $2007$ Ueda et all
Later, it was realized \cite{hctmrg-Ising-5-4} %realized
 that the algorithm can be naturally generalized to more complex lattices whenever a partitioning of the system into a set of equivalent "corners" is possible, and an expansion scheme for the corner (and the corresponding corner tensor) is supplied.
%provided, at siposal

Hyperbolic $(p,q)$ lattices constructed by tessellation of regular $p$-sided polygons allow us to satisfy both the above-mentioned conditions, as shown in the following. The modified CTMRG algorithm will be described through its application to the Ising model on several $(p,q)$ lattices. The description, however, contains all necessary instructions required to perform the calculations on arbitrary hyperbolic or Euclidean $(p,q)$ lattices. 

\subsubsection{The case $(4,q \geq 4)$}

\begin{figure}[tb]
 \centering
 \includegraphics[width=2.5in]{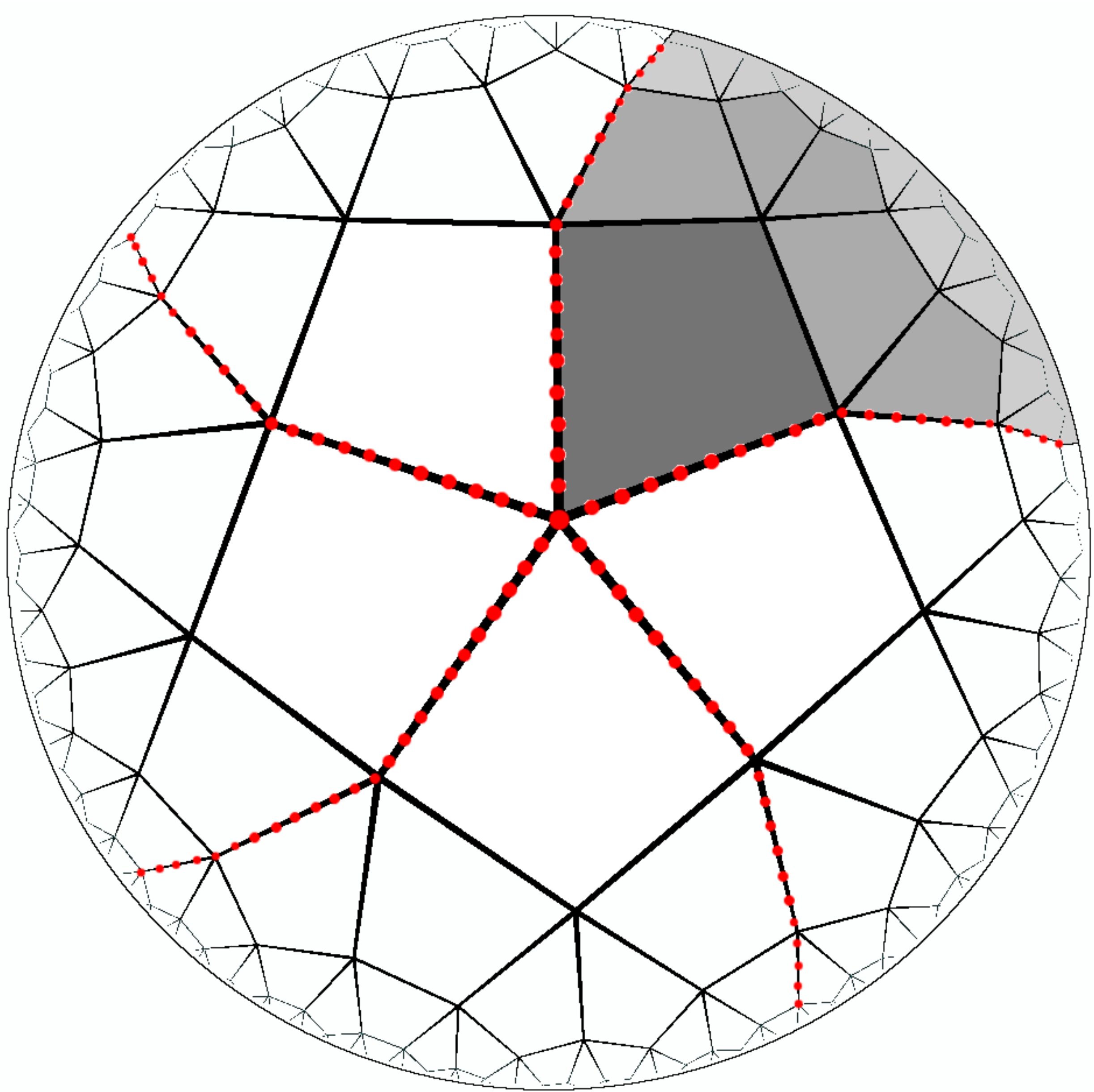}
	\caption{Decomposition of the hyperbolic $(4,5)$ lattice into $q=5$ corners in the Poincar\'{e} disc representation. Borders between the adjacent corners are highlighted by red-dashed curves; one of the five identical corners denoted by the red-dashed curves is explicitly filled in grey.  
}\label{5_4_corner}
 \end{figure}

Let us start with the class of $(4,q)$ lattices, where $q \geq 4$, which includes the well-discussed Euclidean $(4,4)$ lattice as a special case for $q=4$. Notice that the infinite $(4,q)$ lattice can be divided into $q$ equivalent corners at arbitrary vertex, as depicted in Fig.~\ref{5_4_corner}, where the $(4,5)$ lattice  is shown as an example.
%Notice, that any infinite $(4,q)$ lattice can be divided into $q$ equivalent corners at arbitrary vertex as depicted in Fig.~??. 
%As an example, lattices $(4,5)$ and $(4,6)$ are depicted in Fig.~??. 
The objective is to construct a sufficiently large lattice to describe the thermodynamic limit (at given $p$) by iterative corner expansions in analogy to the original CTMRG on the Euclidean square lattice. At each iteration $k$, the lattice is constructed by joining $q$ current corners ${{\mathbf C}^{(k)}}$ around the central lattice spin position. The partition function can be, thus,  calculated as
\begin{equation}
{\cal Z}^{(k)} = \Tr\left({{\mathbf C}^{(k)}}\right)^q. 
\end{equation}

Any lattice vertex is shared between $q$ lattice polygons, hence the fraction $\frac{h}{4 k_B T}$ in the formula \eqref{WB} for the Boltzmann weight tensor ${\mathbf W}_B$ on the $(4,4)$ lattice changes into $\frac{h}{q k_B T}$ on the general $(4,q)$ lattice. 
For the same reason, the original formulae \eqref{CTM_init} and \eqref{HTM_init} for the initial corner tensor $\mathbf{C}^{(1)}$ and the transfer tensor $\mathbf{T}^{(1)}$ are modified into the form 
\begin{align}
\begin{split}
{\mathbf C}^{(1)}(\sigma_1, \sigma_2, \sigma_4) = \sum_{\sigma_3}
 \exp & \left[
{\frac{h}{q k_B T}{\left(\sigma_1+\frac{q}{2}\sigma_2+q\sigma_3+\frac{q}{2}\sigma_4\right)}}+
\right.\\
	&\hspace*{-0.2cm}+\left.\frac{J}{2 k_B T}{\left(
\sigma_1\sigma_2
+2\sigma_2\sigma_3
+2\sigma_3\sigma_4
+\sigma_4\sigma_1
\right)}\right],
\end{split}
\end{align}

\begin{align}
\begin{split}
{\mathbf T}^{(1)}(\sigma_1, \sigma_2, \sigma_4, \sigma_3) = 
 \exp & \left[
{\frac{h}{q k_B T}{\left(\sigma_1+\frac{q}{2}\sigma_2+\frac{q}{2}\sigma_3+\sigma_4\right)}}+
\right.\\
&\hspace*{-0.2cm}+\left.\frac{J}{2 k_B T}{\left(
\sigma_1\sigma_2
+2\sigma_2\sigma_3
+\sigma_3\sigma_4
+\sigma_4\sigma_1
\right)}\right].
\end{split}
\end{align}
%As the lattice cells are the squares ($p=4$), the initial corner tensor $\mathbf{C}^{(1)}$ and the transfer tensor $\mathbf{T}^{(1)}$ are given by formulae \eqref{CTM_init} and \eqref{HTM_init}, respectively, from the original CTMRG algorithm on the $(4,4)$ lattice. However,
%The only difference rests in the fact that the  the formula \eqref{WB_upper_right} 
 The correct corner expansion scheme must satisfy that all interior vertices of the lattice constructed from the $q$ corners have the identical coordination number $q$. This condition is always fulfilled in the initialization step, where the only interior vertex is the central one, from which $q$ bonds forming the borders between the corners originate. 

\begin{figure}[tb]
 \centering
 \includegraphics[width=3.2in]{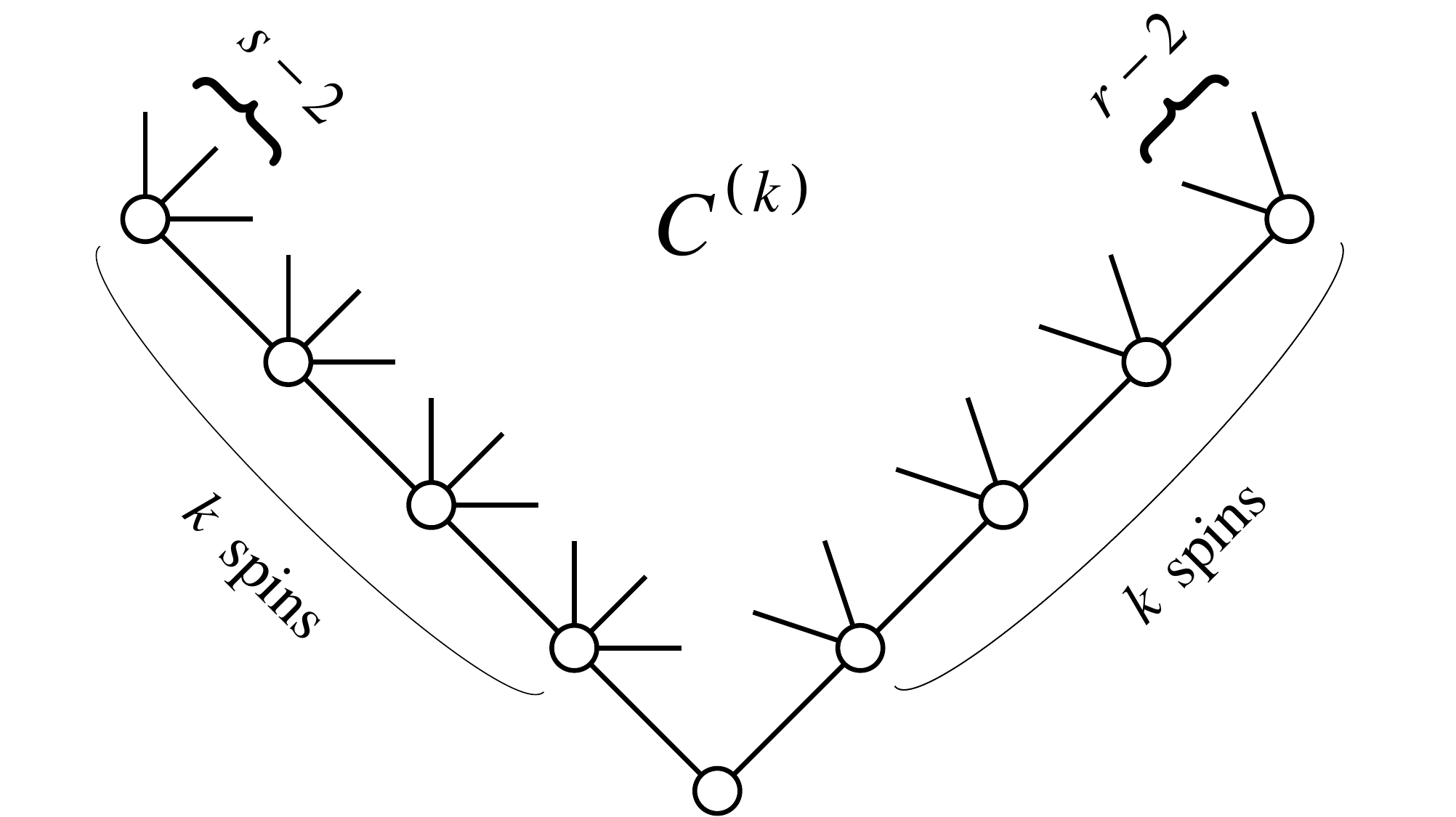}
   \caption{The generalized corner tensor ${\mathbf C}^{(k)}$ for the hyperbolic $(p,q)$ lattice with the coordination number $q = r+s-2 = 7$. There are $r=4$ bonds (the full lines) from each of the $k$ spins on the right boundary --- two bonds to its neighbours on the boundary and $r-2=2$ bonds to its neighbours in the interior of the corner. Similarly, there are $s=5$ bonds from each of the $k$ spins on the left boundary, where $s-2=3$ bonds tend to the interior of the corner.   
}\label{CT_hyperbolic}
 \end{figure}

Now, let us consider the corner in the $k$-th iteration with $k$ vertices on its left and right interior border. %labelled as $1^L, ..., k^L$ and $1^R, ..., k^R$, respectively (see Fig.~??). 
Let there be $r$ ($s$) bonds\footnote{The integers $r, s \geq 3$ may take arbitrary values as long as $r+s=q+2$. If $r=2$, then the vertices %$1^R, ..., k^R$ 
on the right interior border would form an unbranched line of length $k$ %. This is, however, not possible, because each corner is tessellated by squares and thus the maximal length of an unbranched line is $4$. Another argument is that the above mentioned spins %in the line $1^R, ..., k^R$ 
%are not
which is not connected to the interior of the corner. The lattice constructed from such $q$ corners would therefore consist of $q$ almost isolated segments which are connected only through the central vertex. Analogous reasoning holds for spins on the left side which proves $r,s \geq 3$.}
 from each of the right (left) border spins to its neighbours in the corner $\mathbf{C}^{(k)}$, see Fig.~\ref{CT_hyperbolic}.
  When the corners $\mathbf{C}^{(k)}$ are joined together so that the lattice is formed, the border spins and bonds from adjacent sides of the neighbouring corners merge together. After that, the number of bonds around any border spin is $r+s-2$. The lattice has to be uniform with $q$ bonds emerging from each spin, which yields the condition
\begin{equation}
r+s=q+2
\label{rsequality}
\end{equation}     
for admissible combinations of values $r$ and $s$. Similarly, we assume that there are $r$ ($s$) bonds from each of the right (left) border spins to its neighbours in the transfer tensor ${\mathbf T}^{(k)}$. Hence, when the transfer tensor ${\mathbf T}^{(k)}$ is attached to the corner ${\mathbf C}^{(k)}$ during the corner expansion, the coordination number of all spins on the line of contact is $q$, as required.   

\begin{figure}[tb]
 \centering
 \includegraphics[width=5in]{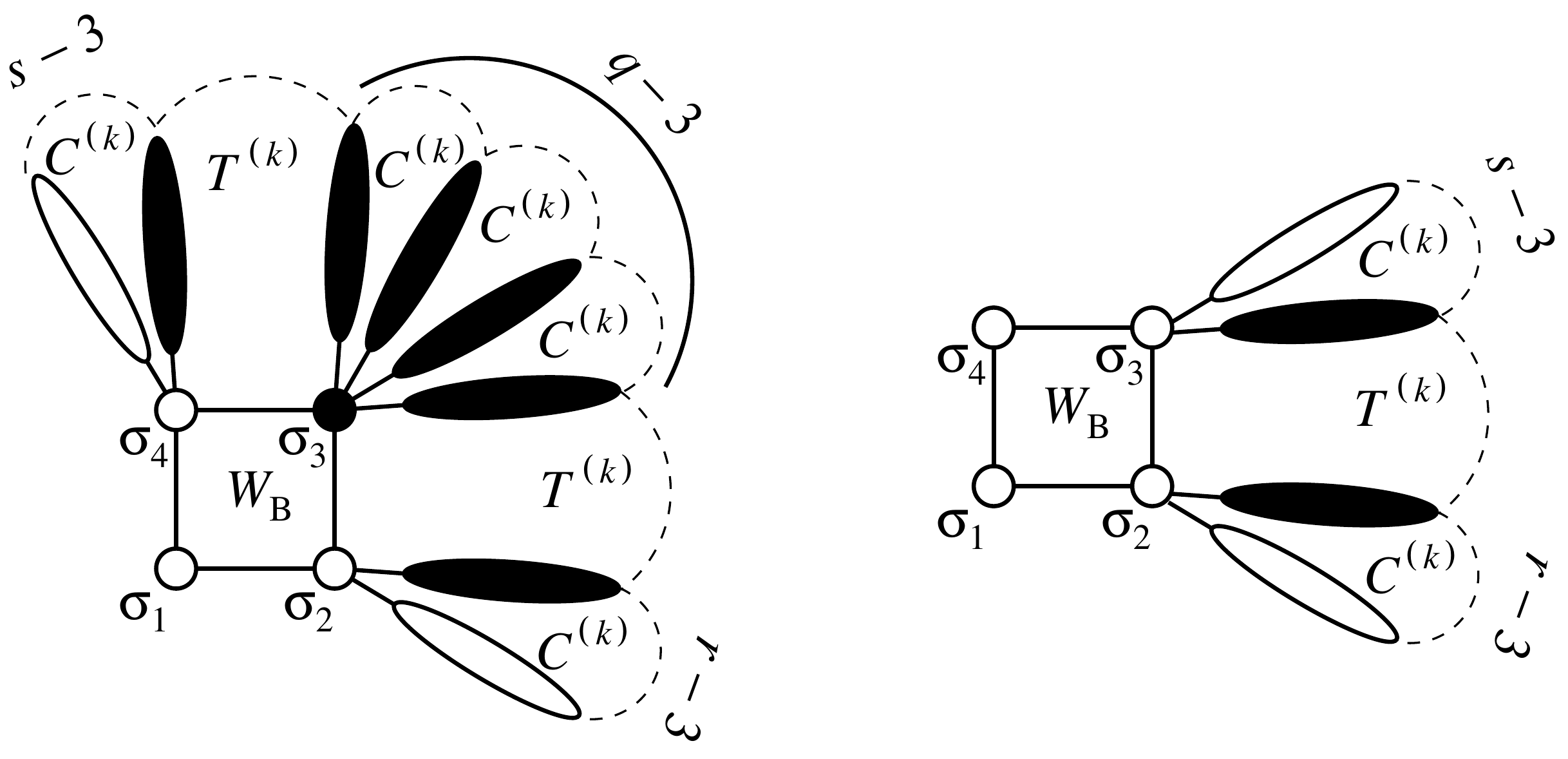}
   \caption{The expansion scheme for the corner tensor $\tilde{\mathbf C}^{(k+1)}$ (left) and the transfer tensor $\tilde{\mathbf T}^{(k+1)}$ (right) on the hyperbolic $(4,6)$ lattice. Here, $r+s=q+2=8$ and we make the symmetric choice $r=s=4$. There are $q-3=3$ corners ${\mathbf C}^{(k)}$ between the transfer tensors ${\mathbf T}^{(k)}$ in the left scheme for $\tilde{\mathbf C}^{(k+1)}$ and $r-3=s-3=1$ corner ${\mathbf C}^{(k)}$ on each of the outer sides in schemes for both $\tilde{\mathbf C}^{(k+1)}$ and $\tilde{\mathbf T}^{(k+1)}$. The black-filled objects represent the spin/multi-spin variables which are summed over in the recurrence expansion relations \eqref{C_hyperbol_expand} and \eqref{T_hyperbol_expand}. 
}\label{CTandT_hyperbolic}
 \end{figure}

The expansion process of the corner tensor ${\mathbf C}^{(k)}$ is illustrated in Fig.~\ref{CTandT_hyperbolic} (left).
 The scheme is analogous to the situation on the Euclidean $(4,4)$ lattice: The Boltzmann weight tensor ${\mathbf W}_B $ corresponding to a single cell is attached in the central position and two transfer tensors ${\mathbf T}^{(k)}$ are added on both sides. The difference is expressed by $q-3 \geq 1$ corners ${\mathbf C}^{(k)}$ (instead of a single one) between the two transfer tensors ${\mathbf T}^{(k)}$ and $r-3 \geq 0$ ($s-3 \geq 0$) corners ${\mathbf C}^{(k)}$ on the right (left) side of the resulting corner $\tilde{\mathbf C}^{(k+1)}$, where originally none were placed. These additional corners supply missing bonds so that there are $q$ bonds around the interior spin $\sigma_3$ and $r$ ($s$) bonds around the right (left) border spin $\sigma_2$ ($\sigma_4$). %This guarantees consistency of the expansion process, i. e. in the expanded corner there are $q$ bonds around each interior spin and $r$ ($s$) bonds around the border ones, as hold for the  
This guarantees consistency of the expansion scheme, i. e., the process started with the corner ${\mathbf C}^{(k)}$ (and the transfer tensor ${\mathbf T}^{(k)}$) with $q$ bonds around each interior spin and $r$ ($s$) bonds around the border ones, and created the enlarged corner $\tilde{\mathbf C}^{(k+1)}$ with identical properties.
%missing bonds to spins $\sigma_2, \sigma_3$ and $\sigma_4$. Now there are $q$ bonds around the interior spin $\sigma_3$ and $r$ ($s$) bonds around the right (left) border spin $\sigma_2$, ($\sigma_4$).       
As a result, the expansion recurrence formula for the corner tensor $\tilde{\mathbf C}^{(k+1)}$ takes the (schematic) form
%\begin{align}
%\begin{split}
%\mathbf{C}^{(k+1)}&=\mathbf{W}_B\left(\mathbf{C}^{(k)}\right)^{s-3} \mathbf{T}^{(k)}\left(\mathbf{C}^{(k)}\right)^{q-3}\mathbf{T}^{(k)}
%\left(\mathbf{C}^{(k)}\right)^{r-3}=\\&=\mathbf{W}_B
%\left(\mathbf{C}^{(k)}\right)^{2q-7}
%\left(\mathbf{T}^{(k)}\right)^{2},
%\end{split}
%\label{C_hyperbol_expand}
%\end{align}
\begin{equation}
\tilde{\mathbf C}^{(k+1)}=\mathbf{W}_B\left(\mathbf{C}^{(k)}\right)^{s-3} \mathbf{T}^{(k)}\left(\mathbf{C}^{(k)}\right)^{q-3}\mathbf{T}^{(k)}
\left(\mathbf{C}^{(k)}\right)^{r-3}=\mathbf{W}_B
\left(\mathbf{C}^{(k)}\right)^{2q-7}
\left(\mathbf{T}^{(k)}\right)^{2},
\label{C_hyperbol_expand}
\end{equation}
where \eqref{rsequality} was used in the second equality.                 

The expansion scheme of the transfer tensor $\mathbf{T}^{(k)}$  in the $k$-th iteration step on the $(4,6)$ lattice is illustrated in Fig.~\ref{CTandT_hyperbolic} (right). After attaching the Boltzmann weight tensor ${\mathbf W}_B $ to the transfer tensor $\mathbf{T}^{(k)}$, $r-3$ and $s-3$ corners $\mathbf{C}^{(k)}$ must be placed on the right and the left side of the tensor $\mathbf{T}^{(k)}$, respectively, in order to get $r$ ($s$) bonds around the spins $\sigma_2$ ($\sigma_3$). The expansion recurrence formula for the tensor $\tilde{\mathbf T}^{(k+1)}$ thus takes the form
\begin{equation}
\tilde{\mathbf T}^{(k+1)}=\mathbf{W}_B\left(\mathbf{C}^{(k)}\right)^{s-3} \mathbf{T}^{(k)}\left(\mathbf{C}^{(k)}\right)^{r-3}=
\mathbf{W}_B\left(\mathbf{C}^{(k)}\right)^{q-4}
\mathbf{T}^{(k)}.
\label{T_hyperbol_expand}
\end{equation}
It is evident that the transfer tensor $\mathbf{T}^{(k)}$ on the hyperbolic lattices has a much more complicated structure if compared to the simple row of cells on the Euclidean lattice. 

%Renormalizacia
The renormalization procedure follows the original idea described in section~\ref{CTMRG_algorithm}. If $q$ is even, the lattice can be divided into two identical halves, each consisting of $q/2$ corners. The Boltzmann weight tensor ${\mathbf A}^{(k)}$ of each half of the lattice in iteration $k$ thus equals the product of $q/2$ corner tensors $\mathbf{C}^{(k)}$
%\begin{equation}
%A^{(k)}(\sigma, \xi, \eta)=\sum_{\tau_1, ..., \tau_{q/2-1}}{\mathbf C}^{(k)}(\sigma, \xi,\tau_1){\mathbf C}^{(k)}(\sigma,\tau_1,\tau_2) ... {\mathbf C}^{(k)}(\sigma, \tau_{q/2-1},\eta)
%\end{equation}   
\begin{equation}
{\mathbf A}^{(k)}(\sigma, \xi, \eta)=\sum_{\tau_1, ..., \tau_{q/2-1}}\underbrace{{\mathbf C}^{(k)}(\sigma, \xi,\tau_1){\mathbf C}^{(k)}(\sigma,\tau_1,\tau_2) ... {\mathbf C}^{(k)}(\sigma, \tau_{q/2-1},\eta)}_{q/2}
\label{Aqtensor}
\end{equation}
or shortly ${\mathbf A}^{(k)}=\left({\mathbf C}^{(k)}\right)^{q/2}$. The normalized tensor $\tilde{\mathbf A}^{(k)}$ is then used to construct the density matrix $\rho^{(k)}$ via \eqref{def_DM}.
%\begin{equation}
%\rho^{(k)}(\{\sigma, \xi \}\vert \{\sigma', \xi'\})=\sum_{\eta} {\mathbf A}^{(k)}(\sigma, \xi, \eta){\mathbf A}^{(k)}(\sigma', \xi', \eta)
%\end{equation}
%after normalizing $A$'s so that $\Tr \rho = 1$.
A slightly modified approach is applied if $q$ is odd, because the lattice cannot be partitioned into two equally large parts constructed of whole corners only. In this case, $q=2u+1$, where $u$ is a positive integer. The density matrix must be calculated in the symmetrized form \cite{SchollwoeckRevModPhys, hctmrg-Ising-3-q}
\begin{equation}
\rho^{(k)}(\{\sigma, \xi \}\vert \{\sigma', \xi'\})=\frac{1}{2}\sum_{\eta} \left[\tilde{\mathbf A}^{(k)}(\sigma, \xi, \eta)\tilde{\mathbf B}^{(k)}(\sigma', \xi', \eta)+\tilde{\mathbf B}^{(k)}(\sigma, \xi, \eta)\tilde{\mathbf A}^{(k)}(\sigma', \xi', \eta)\right],
\label{def_DM_symmetrized}
\end{equation}
where ${\mathbf A}^{(k)}=\left({\mathbf C}^{(k)}\right)^u$ and ${\mathbf B}^{(k)}=\left({\mathbf C}^{(k)}\right)^{u+1}$ represent Boltzmann weights of the portions of the lattice containing $u$ and $u+1$ corners ${\mathbf C}^{(k)}$, respectively, and $\tilde{\mathbf A}^{(k)}$, $\tilde{\mathbf B}^{(k)}$ are their normalized versions so that $\Tr\left(\rho^{(k)}\right)=1$. The next steps are identical to the original CTMRG. The columns of the projection matrix $\mathbf P$ are filled with $m$ eigenvectors corresponding to the $m$ largest eigenvalues of the density matrix $\rho^{(k)}$ and the renormalized tensors ${\mathbf C}^{(k+1)},{\mathbf T}^{(k+1)}$ are created from $\tilde{\mathbf C}^{(k+1)},\tilde{\mathbf T}^{(k+1)}$ via \eqref{renormalizedC} and \eqref{renormalizedT}. 

\subsubsection{The case $(p\geq 4,4)$}

The expansion of the $(p,4)$ lattices, where the system is constructed from $q \equiv 4$ corners, is an analogous problem. The Boltzmann weight tensor ${\mathbf W}_B$ of the $p$-sided polygon is
\begin{equation}
{\mathbf W}_B(\sigma_1,\sigma_2, ..., \sigma_p) = \exp\left[
\frac{h}{4 k_B T}\sum_{i=1}^{p}\sigma_i
+
\frac{J}{2 k_B T}\sum_{i=1}^{p}\sigma_i\sigma_{i+1}
\right],
\label{WBp4}
\end{equation}
where the index $i$ labels the polygon vertices in the anti-clockwise order and $\sigma_{p+1}\equiv\sigma_1$. The initial tensors $\mathbf{C}^{(1)}$ and $\mathbf{T}^{(1)}$ are given by
\begin{align}
\begin{split}
{\mathbf C}^{(1)}(\sigma_1,\sigma_2, \sigma_p) &= \sum_{\sigma_3, ..., \sigma_{p-1}}\exp\left[
\frac{h}{4 k_B T}\left(\sigma_1+2\sigma_2+4\sum_{i=3}^{p-1}\sigma_i + 2\sigma_p \right)
+\right.\\&\left.+
\frac{J}{2 k_B T}\left(\sigma_1\sigma_{2} + 2\sum_{i=2}^{p-1}\sigma_i\sigma_{i+1} + \sigma_p\sigma_{1}\right)
\right]
\end{split}
\end{align}
and
\begin{align}
\begin{split}
{\mathbf T}^{(1)}(\sigma_1,\sigma_2, \sigma_p, \sigma_{p-1}) &= \sum_{\sigma_3, ..., \sigma_{p-2}}\exp\left[
\frac{h}{4 k_B T}\left(\sigma_1+2\sigma_2+4\sum_{i=3}^{p-2}\sigma_i + 2\sigma_{p-1}+\sigma_p \right)
+\right.\\&\left.+
\frac{J}{2 k_B T}\left(\sigma_1\sigma_{2} + 2\sum_{i=2}^{p-2}\sigma_i\sigma_{i+1} + \sigma_{p-1}\sigma_{p} +\sigma_p\sigma_{1}\right)
\right].
\end{split}
\end{align}

\begin{figure}[tb]
 \centering
 \includegraphics[width=5in]{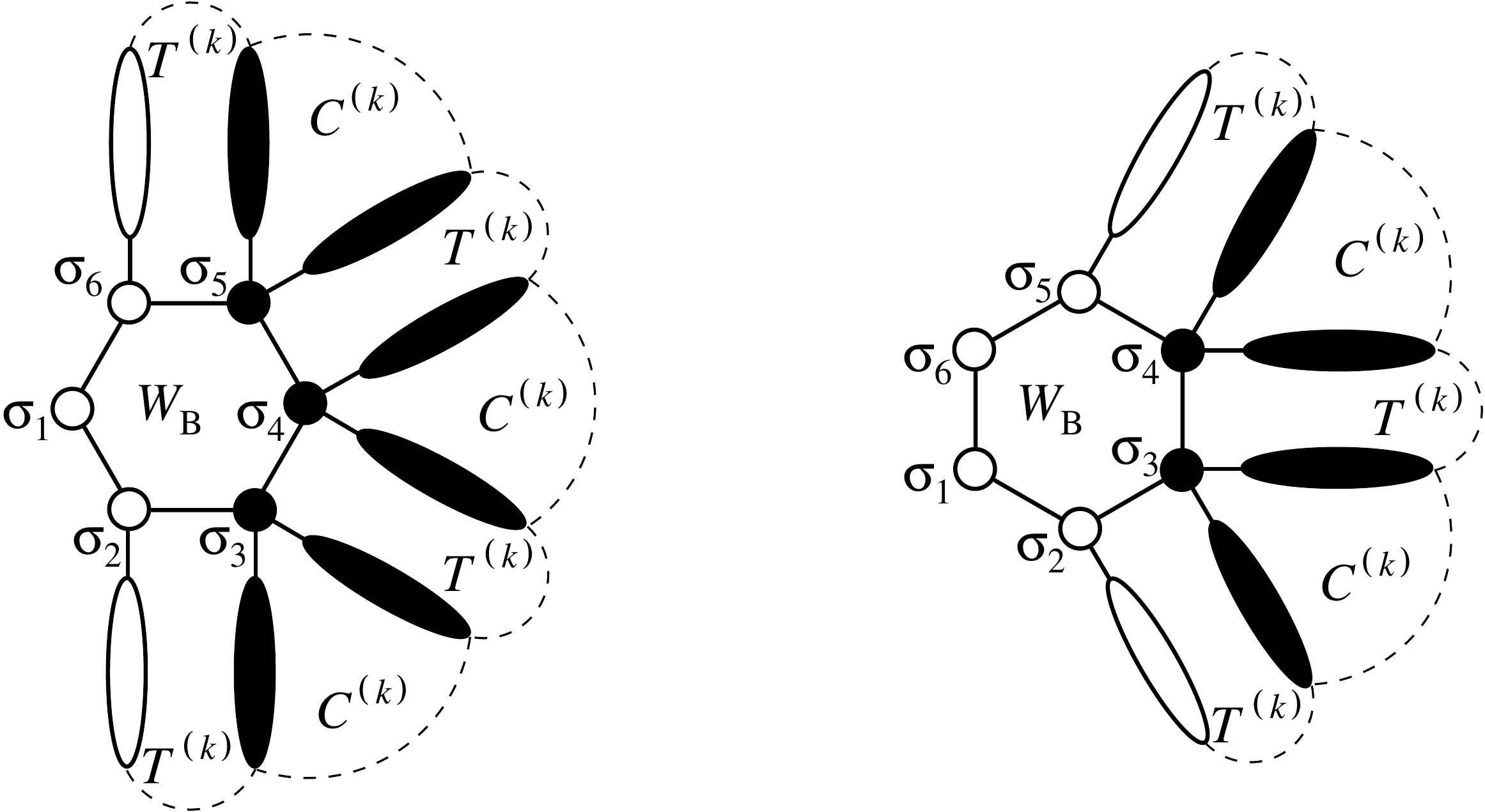}
   \caption{The expansion scheme for the corner tensor $\tilde{\mathbf C}^{(k+1)}$ (left) and the transfer tensor $\tilde{\mathbf T}^{(k+1)}$ (right) on the hyperbolic $(6,4)$ lattice. The black-filled objects represent the spin/multi-spin variables which are summed over in the recurrence expansion formulae \eqref{64expansionC} and \eqref{64expansionT}.   
}\label{CT64expansion}
 \end{figure}

 Without loss of generality, Fig.~\ref{CT64expansion} illustrates the expansion of both tensors $\mathbf{C}^{(k)}$ and $\mathbf{T}^{(k)}$ on the $(6,4)$ lattice.
 The only possible combination of values $r, s \geq 3$, fulfilling \eqref{rsequality}, is $r=s=3$ and, thus, there are $r-3=s-3=0$ corners attached to the right or left side of the expanded corner $\mathbf{C}^{(k)}$ or transfer tensor $\mathbf{T}^{(k)}$. If expanding the corner $\mathbf{C}^{(k)}$, the tensors $\mathbf{T}^{(k)}$ are attached to $p-2$ sides of the additional $p$-sided polygon so that we receive the demanded structure with a central spin and two lines of spins (with $r$ or $s$ bonds) on its sides. Similarly, if the tensor $\tilde{\mathbf T}^{(k+1)}$ is constructed, the attachment of $p-3$ tensors $\mathbf{T}^{(k)}$ to the polygon is necessary to create an object with a pair of single spins ($\sigma_1$, $\sigma_6$) and two lines of spins (with $r$ or $s$ bonds) on the sides. A single corner $\mathbf{C}^{(k)}$ is always inserted between neighbouring tensors $\mathbf{T}^{(k)}$ in order to create four bonds in total around the spin on the peak of the inserted corner $\mathbf{C}^{(k)}$. Hence, the expansion recurrence formulae for the transfer tensors take the form
\begin{align}
\tilde{\mathbf C}^{(k+1)}&=\mathbf{W}_B\left(\mathbf{T}^{(k)}\right)^{p-2} \left(\mathbf{C}^{(k)}\right)^{p-3}
\label{64expansionC}\\
\tilde{\mathbf T}^{(k+1)}&=\mathbf{W}_B\left(\mathbf{T}^{(k)}\right)^{p-3} \left(\mathbf{C}^{(k)}\right)^{p-4}.
\label{64expansionT}
\end{align}
As $q=4$, the construction of the density matrix $\rho$ and the subsequent renormalization process is identical to that in the original CTMRG algorithm on the Euclidean $(4,4)$ lattice. 

%Calculation of the free energy per site $f$ on a general $(p,q)$ lattice 

%VYPOCET Free energy - struktura s jednym polygonom v strede.
%PRIDAT VSADE Tildy k C-ckam 

\newpage\setcounter{equation}{0} \setcounter{figure}{0} \setcounter{table}{0}
\section{Tensor Product Variational Formulation}
\label{TPVF}

In the previous sections we described in detail the CTMRG algorithm for the \emph{classical} spin systems on both the Euclidean $(4,4)$ and the regular hyperbolic $(p,q)$ lattices. %, which is a highly accurate numerical tool for studying of \emph{classical} spin systems    
%The CTMRG algorithm described in the previous sections 
The objective of this section is to demonstrate that an analogous numerical analysis can also be performed in case of the \emph{quantum} spin systems.

Many analytical and computational techniques have been developed to study quantum spin models on the two-dimensional \emph{Euclidean} lattices. However, the task of finding an appropriate approach to analyze the quantum models on the \emph{hyperbolic} lattices still remains an open question. A remarkable demand for an appropriate numerical tool persists. For example, implementation of the Monte Carlo simulations fails due to exponential increase of the number of the lattice sites for models on the hyperbolic lattices with respect to the expanding lattice size from the lattice center~\cite{MC1,MC2}. 

Here we introduce a novel and sufficiently accurate numerical algorithm called \emph{Tensor Product Variational Formulation} (TPVF) \cite{TPVF54}, which combines an Ansatz for the quantum ground-state in the form of the Tensor Product State (TPS)~\cite{Orus} with the Corner transfer matrix renormalization group scheme. This algorithm can be used to study quantum spin systems in the thermodynamic limit on the regular hyperbolic $(p,q)$ lattices of constant negative Gaussian curvature.
%~\cite{Sadoc}
 Although the TPVF was originally designed %proposed
  in \cite{TPVA} for treating of quantum systems on the Euclidean $(4,4)$ lattice, we conjectured in ~\cite{TPVF54, TPVFp4} that TPVF is more suitable for models on the hyperbolic lattices.
  % due to off-critical and weakly correlated characteristics.
  This observation originates in the mean-field-like behaviour induced by the TPS Ansatz, which, as a consequence, cannot accurately approximate the correct ground state of those quantum models on the two-dimensional Euclidean lattice, which do not belong to the mean-field universality class, e. g., the transverse field Ising model. On the contrary, since the Hausdorff dimension of the hyperbolic lattices is infinite, spin models on these lattices belong to the mean-field universality class due to short range correlations, even though the mean-field approximation of the Hamiltonian is not applied, as discussed in Section~\ref{spin_models}.

%The mean-field-like behaviour of
%the phase transition exponents is caused by the non-Euclidean geometry of the underlying
%lattices~\cite{hctmrg-Ising-5-4,hctmrg-Ising-p-4,hctmrg-Ising-3-q}, and is not related
%to the analysis by the mean-field approximation.

\subsection{The model}
\label{TPVF_model}

The Tensor product variational formulation algorithm can approximate the ground-state of basic quantum spin models with the nearest-neighbour interaction on the Euclidean and the hyperbolic $(p,q)$ lattices. As an example, let us assume the quantum XY, Heisenberg and the transverse field Ising model (TFIM) on the $(p,4)$ lattices, which are formed by tessellation of regular $p$-sided polygons with the constant coordination number, which is equal to four. %The spin variables are located on the lattice vertices. 
We intend to study the quantum spin systems in the thermodynamic
limit, i.e., the number of the lattice vertices, where the spin variables are located, is infinite. 

\begin{figure}[tb]
 \centering
 \includegraphics[width=3in]{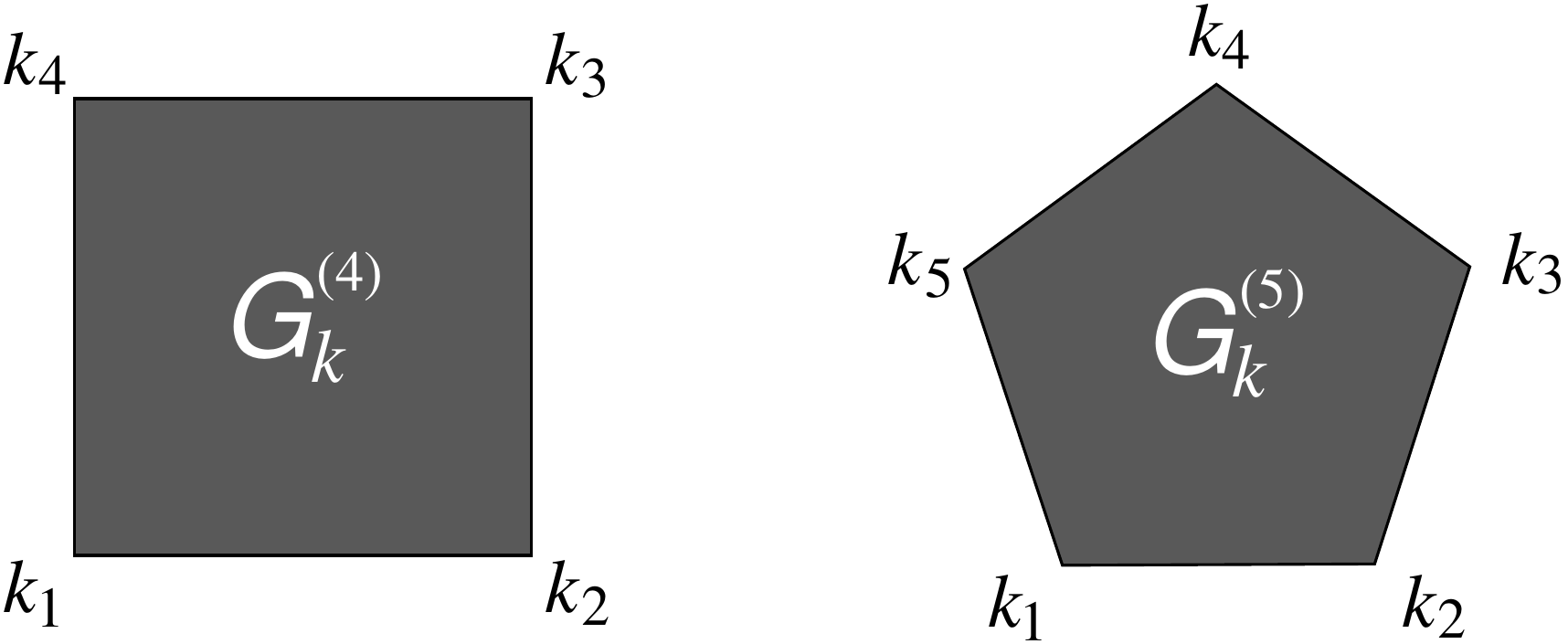}
\caption{Graphical representation of the local Hamiltonian $G_k^{(p)}$
with its particular shape in case of the square ($p=4$) on the left and the pentagon ($p=5$)
on the right.}
\label{45_shape_notation}
\end{figure}

The Hamiltonian ${\cal H}$
of the three models can be expressed in the following
compact form
\begin{equation}
{\cal H}\left(J_{xy}, J_z, h\right) = \sum\limits_{{\langle k \rangle}_p}^{~} G_{k}^{(p)}\left(J_{xy}, J_z, h\right)\, ,
\label{Hm1}
\end{equation}
where $G_{k}^{(p)}$ represents the local Hamiltonian of the $p$-sided polygon,
the lattice is constructed from, and $k$ marks the position of the polygon
on the lattice. The summation runs over all the positions of the polygons
${\langle k \rangle}_p$. The polygon on the $k^{\rm th}$ position is described
by the ordered set of spin indices $k_1$, $k_2$, ..., $k_p$, see Fig.~\ref{45_shape_notation},
where $k_i$ stands for the unique number which is assigned to the corresponding
vertex within the labeling scheme of the lattice vertices.
 The local Hamiltonian
has the expression
\begin{equation}
G_{k}^{(p)}\left(J_{xy}, J_z, h\right) = -\frac{1}{2}\sum\limits_{i=1}^{p} \left[
      J_{xy} \left( {\boldsymbol \sigma}_{k_i}^{x} {\boldsymbol \sigma}_{k_{i+1}}^{x} + {\boldsymbol \sigma}_{k_i}^{y} {\boldsymbol \sigma}_{k_{i+1}}^{y} \right)
      + J_{z} {\boldsymbol \sigma}_{k_i}^{z} {\boldsymbol \sigma}_{k_{i+1}}^{z}
      + \frac{h}{4} \left( {\boldsymbol \sigma}_{k_i}^{x} + {\boldsymbol \sigma}_{k_{i+1}}^{x}\right) \right] \, ,
\label{Hm2}
\end{equation}
where ${\boldsymbol \sigma}_{k_i}^x$, ${\boldsymbol \sigma}_{k_i}^y$, ${\boldsymbol \sigma}_{k_i}^z$ are the Pauli operators, and the spin
indices obey the cyclic condition $k_{p+1}\equiv k_1$. The $x$-component of the
external magnetic field is described by the variable $h$ and the constant
prefactors $\frac{J}{2}$ and $\frac{h}{8}$ reflect the sharing of the spin couplings
and the magnetic field, respectively, if the Hamiltonian is formed by the polygonal
tessellation in~\eqref{Hm1}. 
The spin couplings
$J_{xy}$ and $J_{z}$ specify the three models, as defined in Section~\ref{Quantum_spin_models}. %In particular, $J_{xy} = -J_{z} = 1$,
%and $h=0$ describe the Heisenberg model, $J_{xy} = 1$ and $J_{z} = h = 0$ specify
%the XY model, whereas $J_{xy} = 0$, $J_{z} = 1$ with an arbitrary $h$ lead to the
%transverse field Ising model. 
We assume the ferromagnetic versions of the models, so that a simpler TPS formulation with identical tensors $W_p$ can be used.
%{\color{red} TU POZOR, Heisenberg je antifero, ale to diskutovat az v casti s vysledkami.} 
% Since the ferromagnetic ordering of
%the models leads to a simpler TPS formulation, we have opted for the positive coupling,
%$J_{xy}=1$.%, and we consider dimensionless units throughout the entire work. 

Our objective is to obtain the ground-state of the system
\begin{equation}
      | \Phi_p\rangle = \lim\limits_{N\to\infty}
      \sum\limits_{\sigma_1^{~}\sigma_2^{~}\cdots\sigma_{N}^{~}}^{~}
      \Phi_p^{\sigma_1^{~}\sigma_2^{~}\cdots\sigma_{N}^{~}}
      |\sigma_1^{~}\sigma_2^{~}\cdots\sigma_N^{~}\rangle
\label{Phi}
\end{equation}
in the thermodynamic limit by a variational minimization of the ground-state energy
normalized per bond
\begin{equation}
{\cal E}_0^{(p)} = \min\limits_{\Phi_p}\lim\limits_{N_b\to\infty}\frac{1}{N_b}
            \frac{\langle \Phi_p | {\cal H} | \Phi_p \rangle}
                 {\langle \Phi_p            | \Phi_p \rangle} \, ,
\label{Eg}
\end{equation}
where $N$ stands for the total number of the lattice spins, $\sigma_j$, $j=1, ..., N$,
marks one of the two base states $\dn$ or $\up$ of the $j^{\rm th}$ lattice spin
and $N_b$ denotes the total number of the bonds (the nearest-neighbour pairs).

In order to simplify the numerical calculation, we approximate $| \Phi_p \rangle$
by a TPS $| \Psi_p \rangle$, which is given by the product of the identical
tensors $W_p$ of the same polygonal structure as each of the
local Hamiltonians $G_k^{(p)}$ has (cf. Fig.~\ref{45_shape_notation}). The $p$-rank tensors
depend on $p$ spin-$\frac{1}{2}$ variables labeled by indices $k_1, ..., k_p$ with two base states $\sigma_{k_i}^{~}=$ $\dn$ or $\up$.
The $p$ individual spin variables are grouped into a single one with $2^p$ base configurations denoted as $\{\sigma_k\}$ to simplify the notations if necessary.
It means that the tensor element $W_p(\{ \sigma_k \})\equiv W_p(\sigma_{k_1}^{~}\sigma_{k_2}^{~}\cdots
\sigma_{k_p}^{~} )$. For instance, there are 32 base spin configurations for
the pentagons, which can be represented in the arrow notation as $\{\dn\dn\dn\dn\dn\}$,
$\{\dn\dn\dn\dn\up\}$, $\{\dn\dn\dn\up\dn\}$, ..., $\{\up\up\up\up\up\}$.
Thus, the approximative ground state in the form of the polygonal TPS
\footnote{The auxiliary states in the TPS language are represented as
states with two degrees of freedom only. Such an approximation enhances the mean-field behaviour
around the criticality and is a compromise to make the calculations feasible numerically
due to the exponentially increasing complexity of the hyperbolic lattice structure.}\footnote{It is sufficient to consider the TPS $| \Psi_p\rangle$ and, equivalently, the tensor elements $W_p(\{\sigma_k\})$ as real numbers, since the Hamiltonian ${\cal H}\left(J_{xy}, J_z, h\right)$ contains no imaginary component and, thus, all its eigenstates, including $| \Phi_p \rangle$ can be set real.}~\cite{Orus}
has the following form in the thermodynamic limit
\begin{equation}
      | \Psi_p\rangle = \lim\limits_{N\to\infty}
      \sum\limits_{\sigma_1^{~}\sigma_2^{~}\cdots\sigma_{N}^{~}}^{~}
      \prod\limits_{{\langle k \rangle}_p}^{~} W_p(\{\sigma_k\}) 
      |\sigma_1^{~}\sigma_2^{~}\cdots\sigma_N^{~}\rangle \, ,
\label{Psi}
\end{equation}
where the sum runs over
the $2^N$ base spin states. Since the TPS $|\Psi_p\rangle$ has the product structure
of the identical tensors $W_p$\footnote{The TPS created from identical tensors $W_p$ can lead to a good approximation of the true ground state if the system is ferromagnetic. However, the situation gets complicated in the antiferromagnetic case. If $p$ is even, it is possible to construct a good TPS approximation in the product form of two tensors $W_p$ and $W^{\prime}_p$ with a chessboard-like arrangement on the lattice. However, odd $p$ leads to strong frustration of the system, since it is impossible to obtain a spin configuration with inverse orientation of all $p$ couples of the neighbouring  spins on the lattice polygon. Hence, the quality of the TPS approximation is disputable.
%Hence, a good TPS approximation cannot be created.
}, the variational problem in~\eqref{Eg} is in the
thermodynamic limit equivalent to the minimization of the local energy of an arbitrary
bond in the lattice center (to avoid boundary effects)
\begin{equation}
{E}_0^{(p)} \equiv \min\limits_{\Psi_p}
            \frac{2}{p}\frac{\langle \Psi_p | {G_\ell^{(p)}} | \Psi_p \rangle}
                 {\langle \Psi_p            | \Psi_p \rangle} \, > {\cal E}_0^{(p)} \, ,
\label{Egbond}
\end{equation}
where $\ell$ is the index of a polygon containing the selected central bond and
the factor $2/p$ reflects that each polygon contains $p$ bonds shared with neighbouring
polygons. Moreover, the product structure of $|\Psi_p\rangle$ enables us to express the
denominator
\begin{equation}
%\nonumber
{\langle \Psi_p | \Psi_p \rangle} = \sum\limits_{\sigma,\sigma^{\prime} }
              \prod\limits_{{\langle k \rangle}_p}^{~}
             W_p^{~}(\{\sigma^{\prime}_k\})  \delta_{\{\sigma^{\prime}_k\},\{\sigma^{~}_k\}}
             W_p(\{\sigma_k\})
            \equiv {\cal D}(W_p(\{\sigma\}))                                         
\label{denominator}
\end{equation} 
and the numerator
\begin{eqnarray}
\nonumber
{\langle \Psi_p | {G_\ell^{(p)}} | \Psi_p \rangle}  &=&  \sum\limits_{\sigma,\sigma^{\prime} }
                         \left[ W_p^{~}(\{\sigma^{\prime}_\ell\})
                                (G_\ell^{(p)})_{\{\sigma^{\prime}_\ell\},\{\sigma^{~}_\ell\}}
                                W_p^{~}(\{\sigma_\ell\}) \right.\\
           &\times &  \prod\limits_{{\langle k \rangle}_p \setminus \{\ell\}}^{~}	
         \left.  W_p^{~}(\{\sigma^{\prime}_k\})
                  \delta_{\{\sigma^{\prime}_k\},\{\sigma^{~}_k\}}
                  W_p^{~}(\{\sigma_k\}) \right] \equiv {\cal N}(W_p(\{\sigma\}))                                
\label{numerator}
\end{eqnarray} 
as sole functions of the tensor elements $W_p(\{\sigma\})$, where we removed the subscript
$k$ due to the uniform TPS. Here, $(G_\ell^{(p)})_{\{\sigma^{\prime}_\ell\},
\{\sigma^{~}_\ell\}}$ stands for the corresponding matrix element of the local Hamiltonian
$G_\ell^{(p)}$, $\delta_{\{\sigma^{\prime}_k\},\{\sigma^{~}_k\}}$ is the Kronecker symbol,
and ${\langle k\rangle}_p \setminus \{\ell\}$ denotes the set of all polygon indices except
for the index $\ell$. 

Consequently, the minimization over the set of variational parameters
$\Phi_p^{\sigma_1^{~}\sigma_2^{~}\cdots\sigma_{\infty}^{~}}$ in Eq.~\eqref{Eg}
is replaced by a much simpler problem
\begin{equation}
{E}_0^{(p)} =  \min\limits_{W_p(\{\sigma\})} 
               \frac{2}{p}\frac{{\cal N}(W_p(\{\sigma\}))}{{\cal D}(W_p(\{\sigma\}))} \, ,
\label{e0}
\end{equation}
where we minimize over $2^p$ tensor elements $W_p(\{\sigma\})$ only. 
This set can be further
significantly reduced if symmetries of the local Hamiltonian $G_{\ell}^{(p)}$ are taken into account, as discussed in %Section~\ref{tensor_symmetries}.  
the next section.
The optimization problem \eqref{e0} with the lower dimension is then solved by means of the TPVF algorithm described in Section~\ref{TPVF_algorithm}. 
%The calculation of the numerator ${\cal N}
%(W_p(\{\sigma\}))$ and the denominator ${\cal D}(W_p(\{\sigma\}))$ in Eq.~\eqref{e0}
%is carried out separately by means of a numerical algorithm described in section ??.

\subsection{The tensor symmetries}
\label{tensor_symmetries}

Rotational and spin-ordering symmetries of the local Hamiltonian $G_{\ell}^{(p)}$  are present in all the three spin models. As a typical example, let us consider the hexagonal lattice ($p=6$) and its particular base configuration of spins on the lattice polygon $\{\sigma^{*}\} = \{\up\dn\up\up\dn\dn\}$. Rotational symmetry requires that the tensor elements corresponding to the set of configurations $\{\dn\up\dn\up\up\dn\}$, $\{\dn\dn\up\dn\up\up\}$, $\{\up\dn\dn\up\dn\up\}$,  $\{\up\up\dn\dn\up\dn\}$, $\{\dn\up\up\dn\dn\up\}$ are identical to $W_{p=6}(\{\sigma^{*}\})$. Next, let us consider a spin-ordering operation, which reverses the order of the polygon spins. In particular, if the spins are labeled clockwise, the operation reorders them in the anti-clockwise direction. It means that the configuration $\{\up\dn\up\up\dn\dn\}$ is equivalent to $\{\dn\dn\up\up\dn\up\}$ by the spin-ordering symmetry and to all the rotations of the latter configuration ($\{\up\dn\dn\up\up\dn\}$, $\{\dn\up\dn\dn\up\up\}$, $\{\up\dn\up\dn\dn\up\}$, $\{\up\up\dn\up\dn\dn\}$, $\{\dn\up\up\dn\up\dn\}$) by the composition of the spin-ordering and the rotational symmetry. As a result, the 12 tensor elements $W_6(\{\sigma\})$ corresponding to the configuration $\{\sigma^{*}\}$ and its 11 equivalent configurations are represented by a single variational parameter, as they share the same value.

By performing a similar analysis on the set of all $2^p$ configurations $\{\sigma\}$ we can factorize it into $N_{\rm Ising}^{(p)}$ classes of equivalence with representatives $\theta_j$, where $j \in \{1, \dots, N_{\rm Ising}^{(p)}\}$. As a result, we have $W_p(\{\sigma\})=W_p(\theta_j)$ for all spin configurations $\{\sigma\}$ within the equivalence class labeled by $\theta_j$.  Thus, in case of a system with the rotational and the spin-ordering symmetry (as in the TFIM), there are only $N_{\rm Ising}^{(p)}$ free variational parameters $W_p(\theta_j)$ within the set of $2^p$ tensor elements $W_p(\{\sigma\})$. If there is no preferred spin alignment in the system (such as in the XY model, the Heisenberg model, as well as in the TFIM at and above the phase transition magnetic field), the spin-inversion symmetry appears. For instance, if $p=4$, the configuration $\{\up\up\up\dn\}$ is equivalent to $\{\dn\dn\dn\up\}$, which is obtained by flipping each spin, resulting in relation $W_4(\{\up\up\up\dn\})=W_4(\{\dn\dn\dn\up\})$. Such an additional symmetry results in consequent reduction of the set of the free variational parameters, the size of which drops to $N_{\rm Heis}^{(p)} < N_{\rm Ising}^{(p)}$. In this case, we denote the representatives of the equivalence classes as $\Theta_i$, where $i \in \{1, \dots, N_{\rm Heis}^{(p)}\}$\footnote{Any equivalence class labeled by $\Theta_i$ is a union of two equivalence classes $\theta_j$  at most.}. The numbers of the free variational parameters $N_{\rm Ising}^{(p)}$ and $N_{\rm Heis}^{(p)}$ with respect to the lattice parameter $p$ are summarized in Table~\ref{Tab_num_var_parameters}. In addition, one more variational parameter can be eliminated from each set of the free variational parameters by setting it to $1$, being the normalization condition in $W_p(\{\sigma\})$ and $| \Psi_p \rangle$, consequently.

\begin{table}[tb]
\begin{center}
\begin{tabular}{| c || r | r | r | r | r | r | r | r | }
\hline
$p$ & $4$ & $5$ & $6$ & $7$ & $8$ & $9$ & $10$ & $11$  \\
\hline
\hline
$N_{\rm Heis}^{(p)}$ & $4$ & $4$ & $8$ & $9$& $18$ & $23$ & $44$ & $63$ \\[0.04cm]
\hline
$N_{\rm Ising}^{(p)}$ & {\phantom{00}}$6$ & {\phantom{00}}$8$ & {\phantom{0}}$13$ & {\phantom{0}}$18$&
{\phantom{0}}$30$ & {\phantom{0}}$46$ & {\phantom{0}}$78$ & $126$ \\[0.04cm]
\hline
\end{tabular}
\end{center}
\caption{The numbers of the free variational parameters $N_{\rm Heis}^{(p)}$ (for the XY and the Heisenberg models) and $N_{\rm Ising}^{(p)}$ (for the TFIM) including the normalization parameter for the $(p,4)$ lattices.}
\label{Tab_num_var_parameters}
\end{table}

\begin{table}[tb]
\begin{center}
\begin{tabular}{|l|l|cccc|c|}
\hline
 
$\, j$ & $W_4^{~}(\theta_j^{~})$ & \multicolumn{4}{ c |} {$\{\sigma\} \equiv \{\sigma_{1}^{~}\sigma_{2}^{~}\sigma_{3}^{~}\sigma_{4}^{~}\}$} & $W_4^{~}(\Theta_i^{~})$ \\[0.1cm]
\hline

 $1$        & $W_4^{~}(\theta_1^{~})$ & $\{\dn\dn\dn\dn\}$ &                  &                  &                  & $W_4^{~}(\Theta_1^{~})$\\
 $2$        & $W_4^{~}(\theta_2^{~})$ & $\{\dn\dn\dn\up\}$ & $\{\dn\dn\up\dn\}$ & $\{\dn\up\dn\dn\}$ & $\{\up\dn\dn\dn\}$ & $W_4^{~}(\Theta_2^{~})$ \\
 $3$        & $W_4^{~}(\theta_3^{~})$ & $\{\dn\dn\up\up\}$ & $\{\dn\up\up\dn\}$ & $\{\up\up\dn\dn\}$ & $\{\up\dn\dn\up\}$ & $W_4^{~}(\Theta_3^{~})$ \\
 $4$ & $W_4^{~}(\theta_{4}^{~})$ & $\{\dn\up\dn\up\}$ & $\{\up\dn\up\dn\}$ &                  &         & \ $W_4^{~}(\Theta_4^{~})$ \\
 $5$        & $W_4^{~}(\theta_5^{~})$ & $\{\up\up\up\dn\}$ & $\{\up\up\dn\up\}$ & $\{\up\dn\up\up\}$ & $\{\dn\up\up\up\}$ & $W_4^{~}(\Theta_2^{~})$ \\
 $6$        & $W_4^{~}(\theta_6^{~})$ & $\{\up\up\up\up\}$ &                  &                  &                  & $W_4^{~}(\Theta_1^{~})$ \\[0.1cm]
\hline
\end{tabular}
\end{center}
\caption{The set of $2^4=16$ spin configurations for the $(4,4)$ lattice sorted into $N_{\rm Ising}^{(4)} = 6$ equivalence classes labeled by the representatives $\theta_j$. The last
column lists the reduced set of $N_{\rm Heis}^{(4)} = 4$ parameters $W_p(\Theta_i)$ in case the additional spin-inversion symmetry appears.}
\label{Tab_square}
\end{table}

\begin{table}[!tb]
\begin{center}
\begin{tabular}{|l|l|ccccc|c|}
\hline
 $\, j$ & $W_5^{~}(\theta_j^{~})$ & \multicolumn{5}{ c |} {$\{\sigma\}\equiv\{\sigma_{1}^{~}\sigma_{2}^{~}\sigma_{3}^{~}
 \sigma_{4}^{~}\sigma_{5}^{~}\}$} & $W_5^{~}(\Theta_i^{~})$ \\[0.1cm]
\hline
 $1$        & $W_5^{~}(\theta_1^{~})$ & $\{\dn\dn\dn\dn\dn\}$ &                     &                     &                     &                     & $W_5^{~}(\Theta_1^{~})$ \\
 $2$        & $W_5^{~}(\theta_2^{~})$ & $\{\dn\dn\dn\dn\up\}$ & $\{\dn\dn\dn\up\dn\}$ & $\{\dn\dn\up\dn\dn\}$ & $\{\dn\up\dn\dn\dn\}$ & $\{\up\dn\dn\dn\dn\}$ & $W_5^{~}(\Theta_2^{~})$ \\
 $3$        & $W_5^{~}(\theta_3^{~})$ & $\{\dn\dn\dn\up\up\}$ & $\{\dn\dn\up\up\dn\}$ & $\{\dn\up\up\dn\dn\}$ & $\{\up\up\dn\dn\dn\}$ & $\{\up\dn\dn\dn\up\}$ & $W_5^{~}(\Theta_3^{~})$ \\
 $4$ & $W_5^{~}(\theta_{4}^{~})$ & $\{\dn\dn\up\dn\up\}$ & $\{\dn\up\dn\dn\up\}$ & $\{\dn\up\dn\up\dn\}$ & $\{\up\dn\dn\up\dn\}$ & $\{\up\dn\up\dn\dn\}$ & \ $W_5^{~}(\Theta_4^{~})$ \\
 $5$        & $W_5^{~}(\theta_5^{~})$ & $\{\up\up\up\dn\dn\}$ & $\{\up\up\dn\dn\up\}$ & $\{\up\dn\dn\up\up\}$ & $\{\dn\dn\up\up\up\}$ & $\{\dn\up\up\up\dn\}$ & $W_5^{~}(\Theta_3^{~})$ \\
 $6$ & $W_5^{~}(\theta_{6}^{~})$ & $\{\up\up\dn\up\dn\}$ & $\{\up\dn\up\up\dn\}$ & $\{\up\dn\up\dn\up\}$ & $\{\dn\up\up\dn\up\}$ & $\{\dn\up\dn\up\up\}$ & \ $W_5^{~}(\Theta_4^{~})$ \\
 $7$        & $W_5^{~}(\theta_7^{~})$ & $\{\up\up\up\up\dn\}$ & $\{\up\up\up\dn\up\}$ & $\{\up\up\dn\up\up\}$ & $\{\up\dn\up\up\up\}$ & $\{\dn\up\up\up\up\}$ & $W_5^{~}(\Theta_2^{~})$ \\
 $8$        & $W_5^{~}(\theta_8^{~})$ & $\{\up\up\up\up\up\}$ &                     &                     &                     &                     & $W_5^{~}(\Theta_1^{~})$ \\[0.1cm]
\hline
\end{tabular}
\end{center}
\caption{The set of $2^5=32$ spin configurations for the $(5,4)$ lattice sorted into $N_{\rm Ising}^{(5)} = 8$ equivalence classes labeled by the representatives $\theta_j$. The last
column lists the reduced set of $N_{\rm Heis}^{(5)} = 4$ parameters $W_p(\Theta_i)$.}
\label{Tab_pentagon}
\end{table}

To be more specific, we demonstrate the factorization of the set of the $2^p$ base spin configurations in
%case $p=4$ and $p=5$
 the Tables~\ref{Tab_square} and \ref{Tab_pentagon} on the examples of the square ($p=4$) and pentagonal ($p=5$) lattices, respectively.
Each line in the Tables contains such
spin configurations, which are identical with respect to the rotational and spin-ordering
symmetry operations of the $p$-sided polygon. We count $N_{\rm Ising}^{(4)}=6$ or $N_{\rm Ising}^{(5)}=8$ variational parameters $W_p(\theta_j)$ used in the calculation of the transverse field Ising model. 
If, however, the spontaneous symmetry-breaking does not affect the solution, the
total number of the variational parameters decreases down to $N_{\rm Heis}^{(4)}=N_{\rm Heis}^{(5)}=4$ for both the $(4,4)$ and $(5,4)$ lattices. % In particular, the spin configuration
%probability for the pairs $\theta_0$ and $\theta_p$, $\theta_1$ and $\theta_{p-1}$, etc. becomes identical, %which
%can be formally generalized into the equations
%\begin{equation}
%\begin{tabular}{l}
%$W_p (\theta _j        ) = W_p (\theta_{p- j        }) \equiv a_{\min\{j,p-j\}}^{~}$,\\[0.3cm]
%$W_p (\theta_{j^\prime}) = W_p (\theta_{(p-j)^\prime}) \equiv 1$.
%\end{tabular}
%\label{coupled}
%\end{equation}
%There are the three equations in the upper expression defining the new variational parameters
%$a_0^{~}$, $a_1^{~}$, $a_2^{~}$, and one equation in the lower expression for
%$a_{2^{\prime}}^{~}$, which has already been eliminated from the set of the free parameters
%by putting $a_{2^{\prime}}^{~}\equiv1$ being the normalization condition in $W_p$.
Applying the normalization conditions $W_4(\Theta_4)=1$ or $W_5(\Theta_4)=1$, we find out that only three free variational parameters ($W_4(\Theta_1)$, $W_4(\Theta_2)$, $W_4(\Theta_3)$ or $W_5(\Theta_1)$, $W_5(\Theta_2)$, $W_5(\Theta_3)$) suffice to approximate the ground-state wave
function of the models on the square or pentagonal lattices, respectively, with no spontaneous symmetry-breaking phases. For the same reason\footnote{If setting, e. g., $W_4(\theta_6)=1$ and $W_5(\theta_8)=1$.},
if we consider the system without the spin-inversion symmetry, there are either five or
seven free variational parameters for $p=4$ or $p=5$, respectively.  

% Let function $\theta_j$ return all such configurations
%for which the integer number $j$ counts the number of the spins aligned upward as
%listed in Tabs.~\ref{Tab1} and \ref{Tab2}. 
%Each line in the Tables contains such
%spin configurations, which are identical with respect to the rotational
%symmetry operations of the $p$-sided polygon. %We make a difference
%between the spin configurations $\theta_j^{~}$ and $\theta_{j^\prime}^{~}$ when $j=2$
%(or $j=3$ for the case of $p=5$) as they are not rotationally equivalent. The prime
%symbol is used for such configuration (and its rotations), in which either the two
%($j=2$) or three ($j=3$) spins, aligned upward in each polygon, cannot be grouped
%together (or equivalently, if the alternating spin alignment is maximal).

\subsection{The algorithm}
\label{TPVF_algorithm}

The Tensor Product Variational Formulation algorithm consists of two parts. The first one evaluates the ratio in \eqref{e0}
by applying the CTMRG method separately to the numerator and the
denominator for a given set of
the variational parameters $W_p(\theta_j)$. The second part contains a multi-dimensional
minimizer, the Nelder-Mead simplex algorithm~\cite{gsl-page, gsl-manual,NelderMead},
which uses the first part to search for the optimized set of the variational parameters
$W_p^{*}(\theta_j)$, which minimize the ratio in \eqref{e0}.
The minimizer starts from an
initial simplex in the space of free variational parameters, one vertex of which is
specified by the initial tensor elements $W_p(\theta_j)$. The simplex undergoes an
iterative sequence of size changes and moves towards lower energies and stops if the
energy in \eqref{e0} converged.

\begin{figure}[tb]
\centerline{\includegraphics[width=3in]{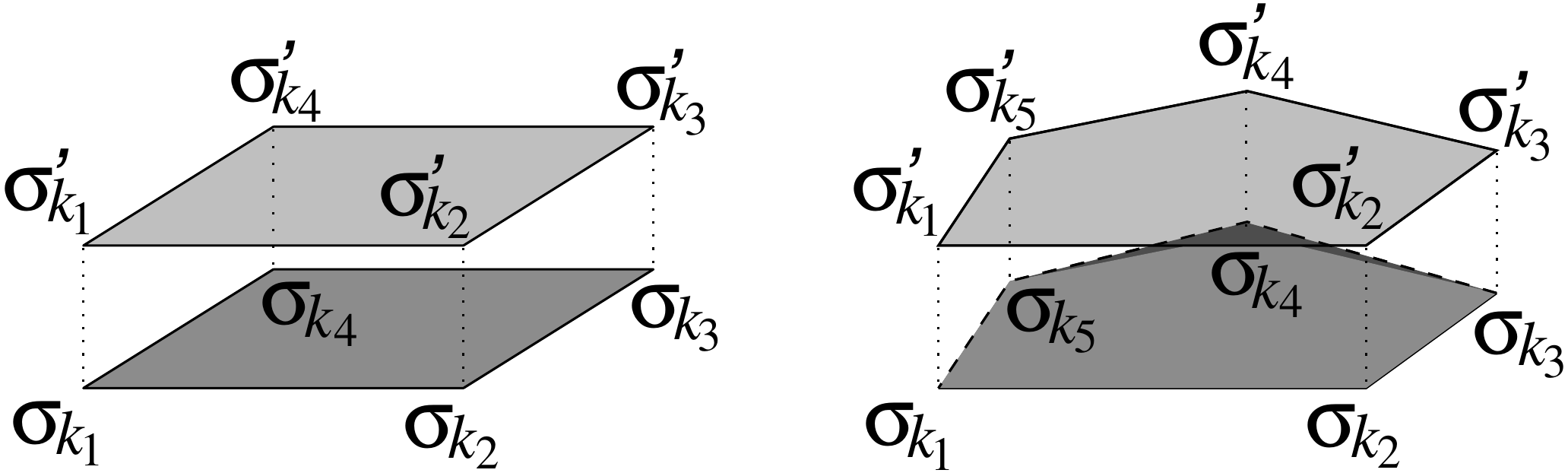}}
\caption{The double-layer tensor structure of ${Z}_4^{~}(\{\sigma_k^\prime
\sigma_k^{~}\})$ on the left and ${Z}_5^{~}(\{\sigma_k^\prime
\sigma_k^{~}\})$ on the right. The shaded polygonal
areas represent the tensors $W_p$.}
\label{Zlayers}
\end{figure}

The central
idea in calculation of the numerator ${\cal N}(W_p(\theta_j))$ and the denominator ${\cal D}(W_p(\theta_j))$ in \eqref{e0} is to replace the concept of the Boltzmann weight tensor ${\mathbf W_B}$ in the original CTMRG algorithm by the tensors $W_p$.
 In order to do this, let us introduce a double-layer tensor ${ Z}_p$
with the tensor elements
\begin{equation}
{Z}_p^{~}(\{\sigma_k^\prime\sigma_k^{~}\}) \equiv
W_p^{~}(\{\sigma^{\prime}_k\}) \delta_{\{\sigma^{\prime}_k\},\{\sigma^{~}_k\}}
W_p^{~}(\{\sigma_k\}) \, .
\end{equation}
Notice that there are $2^{2p}$ double-layer base spin configurations
$\{\sigma_k^\prime\sigma_k^{~}\}$. Figure~\ref{Zlayers} graphically depicts the
double-layer tensors ${Z}_p$ at the position $k$ for the square and pentagonal lattices.
 Thus, in the language of the classical statistical mechanics,
the general expression for the denominator ${\cal D}({W_p (\{\sigma\}}))=\langle\Psi_p
| \Psi_p\rangle$ in \eqref{denominator} corresponds to a tensor product object,
which is equivalent to the partition function of a (non-physical) classical Hamiltonian
given by the product of the tensors ${Z}_p$.

This generalization of the Boltzmann weight tensor enters the CTMRG algorithm and the consequent numerical calculation yields the denominator
${\cal D}(W_p(\theta_j))$ for the given set of the free tensor elements $W_p(\theta_j)$
according to \eqref{denominator}. A similar approach can be also used to determine
the numerator ${\cal N}(W_p(\theta_j))$, as it differs from ${\cal D}(W_p(\theta_j))$
only by the additional double-layer structure at the central position $\ell$ containing
the local Hamiltonian $G_\ell^{(p)}$.

%In particular, the tensors ${\cal Z}_p$ infinite TPS geometry is built up gradually by increasing
In particular, the  infinite TPS geometry is built up from gradually expanding corner tensors ${\mathbf C}^{(k)}$ and transfer tensors ${\mathbf T}^{(k)}$ which follow the identical initialization and expansion scheme for the selected $(p,q)$ lattice as in Section~\ref{CTMRG_hyp}. The only distinction rests in the replacement of the Boltzmann weight tensor ${\mathbf W_B}$ in the initialization and expansion formulae by the tensor ${Z}_p$.  

In order to grasp
 the additional structure at the central position $\ell$ in the numerator ${\cal N}(W_p(\theta_j))$, the $(p,4)$ lattices are constructed
from a central polygon surrounded by the alternating sectors represented by the corner tensors ${\mathbf C}^{(k)}$ and transfer tensors ${\mathbf T}^{(k)}$. The central polygon is represented by the
tensor ${Z}_p$ or $W_p^{~} G_\ell^{(p)}W_p^{~}$ in the structure of the denominator
${\cal D}({W_p (\theta_j)})$ and the numerator ${\cal N}({W_p (\theta_j)})$,
respectively. %Figure~\ref{Fig4} illustrates the situation for the case of
%${\cal D}({W_p (\{\sigma\})})$.
The construction scheme of the lattice is analogous to the situation, when the correlation function $\langle\sigma_{\ell}\sigma_{\ell^{\prime}}\rangle$ is evaluated in the classical CTMRG algorithm, as illustrated in Fig.~\ref{C4T4W} on the Euclidean $(4,4)$ lattice. 
Consequently, the relations \eqref{denominator} and
\eqref{numerator} for the denominator ${\cal D}({W_p (\theta_j)})$ and the numerator
${\cal N}({W_p (\theta_j)})$, formulated in the CTMRG language of the (corner)
transfer tensors, take the form\footnote{If we calculate the numerator  ${\cal N}({W_p (\theta_j)})$, %with double-layer structure at the central polygon containing the local Hamiltonian $G_\ell^{(p)}$
 we omit the $\delta_{\sigma^{\prime}, \sigma}$ term in the tensors $Z_p$ for those spin variables $\sigma$, which are located on the central polygon. This allows us to attach the tensors $\mathbf{C}$ and $\mathbf{T}$ to the double-layer structure at the central polygon containing the local Hamiltonian $G_\ell^{(p)}$.}
\begin{align}
\label{DWp}
{\cal D}({W_p (\theta_j)})&=
\lim\limits_{k\to\infty} {\rm Tr} \left( { Z}_p
\left[{\mathbf C}^{(k)}{\mathbf T}^{(k)}\right]^{p} \right),\\
\label{NWp}
{\cal N}({W_p (\theta_j)})&=
\lim\limits_{k\to\infty} {\rm \Tr} \left( W_p^{~} G_\ell^{(p)}W_p^{~}
\left[{\mathbf C}^{(k)}{\mathbf T}^{(k)}\right]^{p} \right).
\end{align}
On the $(4,q)$ lattices, a similar argumentation leads to formulae  
\begin{align}
{\cal D}({W_4 (\theta_j)})&=
\lim\limits_{k\to\infty} {\rm Tr} \left( {Z}_4
\left[\left({\mathbf C}^{(k)}\right)^{q-3}{\mathbf T}^{(k)}\right]^{4} \right),\\
{\cal N}({W_4 (\theta_j)})&=
\lim\limits_{k\to\infty} {\rm \Tr} \left( W_4^{~} G_\ell^{(4)}W_4^{~}
\left[\left({\mathbf C}^{(k)}\right)^{q-3}{\mathbf T}^{(k)}\right]^{4} \right).
\end{align}

\subsection{Remarks}

Here, we would like to turn the reader's attention to two important aspects of the TPVF algorithm.
%First, we would like to emphasize that the generalization of the CTMRG algorithm to quantum systems in the form of the Tensor product variational formulation is not based on the quantum-classical correspondence. The TPVF utilizes the formal analogy between the formulae for the quantity 
First, the generalization of the CTMRG algorithm to quantum systems in the form of the Tensor product variational formulation is not based on the quantum-classical correspondence. The TPVF utilizes the formal analogy between the formulae for the quantity 
%unnormalized energy
${\langle \Psi_p | {G_\ell^{(p)}} | \Psi_p \rangle}$ and the norm ${\langle \Psi_p | \Psi_p \rangle}$ in the quantum system and the partition function $\cal Z$ of the classical system if the quantum ground-state is approximated as the tensor product state $| \Psi_p \rangle$.    

Second, the TPS approximation \eqref{Psi} of the quantum ground-state may be a limiting factor regarding the accuracy of the TPVF on the Euclidean lattices near the critical point, where the correlation length $\xi$ diverges. The reason is the low dimension of the tensors $W_4$ in the TPS
approximation which suppresses the quantum long-range
correlations on the Euclidean $(4,4)$ lattice near the criticality. As a result, the TPS approximation \eqref{Psi} induces mean-field-like behaviour near the quantum phase transition irrespective of the true universality class the original model belongs in. We discuss this in more detail in Section~\ref{quantum_num_results} and in  \cite{TPVFp4, TPVF54}. 

An improvement of the numerical accuracy can be achieved if 
additional (non-phys\-ical) degrees of freedom are assigned to the spin variables $\sigma$.
% which, however, results in significant slow-down of the Nelder-Mead optimization algorithm. 
However, increasing the number of the free variational parameters prolongs the computational time of the Nelder-Mead optimization algorithm,
and may encounter numerical instability caused by trapping the system in a local
minimum of the energy, rather than approaching to the correct global minimum, which
corresponds to $E_0^{(p)}$.
On the other hand, a faster Nelder-Mead optimization with fewer parameters enables us to improve the
accuracy by increasing the number $m$ of states of the  multi-spin variables kept in the renormalization step in CTMRG.

On the contrary, it is expected that all the classical and quantum spin lattice models on various types of the
hyperbolic surfaces belong to the mean-field universality class, 
since the Hausdorff dimension of the hyperbolic lattices is infinite, which exceeds the critical values $d_C=4$ and $d_C=3$, respectively.
This was confirmed in studies~\cite{hctmrg-Ising-5-4,hctmrg-Ising-p-4,hctmrg-Ising-3-q} of classical spin models on the hyperbolic lattices, where the exponential decay of the density matrix spectra and the correlation function
result in the non-critical phase transition, since the correlation length, $\xi\lesssim1$, is
always finite, reaching its maximal value at the phase transition\cite{corrlen}. We assume similar
scenario also in case of the quantum systems. Hence, the mean-field approximation of the TPVF algorithm induced by the low-dimensional TPS \eqref{Psi} is not in conflict with the mean-field-like behaviour of quantum models on the hyperbolic lattice geometry.  For this
reason, we conjecture the TPFV analysis of the models
on the hyperbolic lattices is more accurate than on the Euclidean ones.

\newpage\setcounter{equation}{0} \setcounter{figure}{0} \setcounter{table}{0}
%\section{Classical spin systems on triangular-tiled hyperbolic lattices}
\section{Classical spin models on hyperbolic lattices}
\label{chap:classical}

In this chapter we first study the thermodynamic behaviour of the classical Ising model on the series of $(3,q)$ lattices constructed by tessellation of triangles. We assume $q \geq 6$, where $q=6$ represents the Euclidean triangular lattice and $q > 6$ corresponds to  the hyperbolic lattices. Later, we construct slightly curved surfaces by distributing exceptional lattice sites of the coordination number seven within the Euclidean $(3,6)$ lattice. The exceptional sites form a regular pattern with the typical distance between these sites proportional to an integer parameter $n$. This geometry allows us to study the influence of the increasing non-flatness of the underlying lattice on the thermal properties of the corresponding lattice model.    

In the past, classical spin models on the hyperbolic $(p,q)$ lattices with fixed coordination number $q = 4$ and various lattice parameters $p$ were investigated 
in reports~\cite{hctmrg-Ising-5-4,hctmrg-Ising-p-4,hctmrg-clock-5-4,hctmrg-J1J2}. For the Ising model on the $( p, 4 )$ lattices, the mean-field universality was
found~\cite{Shima,hctmrg-Ising-5-4}. 
%Here, we study effect of the varying coordination number $q$ at fixed lattice parameter $p=3$ on the thermodynamic properties.   
Thus, the study of models on the $(3,q)$ lattices addresses 
the complementary problem with the varying coordination number $q$ at fixed lattice parameter $p=3$.

\subsection{Ising model on the $(3,q)$ lattices}

\subsubsection{The model and expansion scheme}
\label{3qscheme}

\begin{figure}[!b]
 \centering
 \includegraphics[width=3.5in]{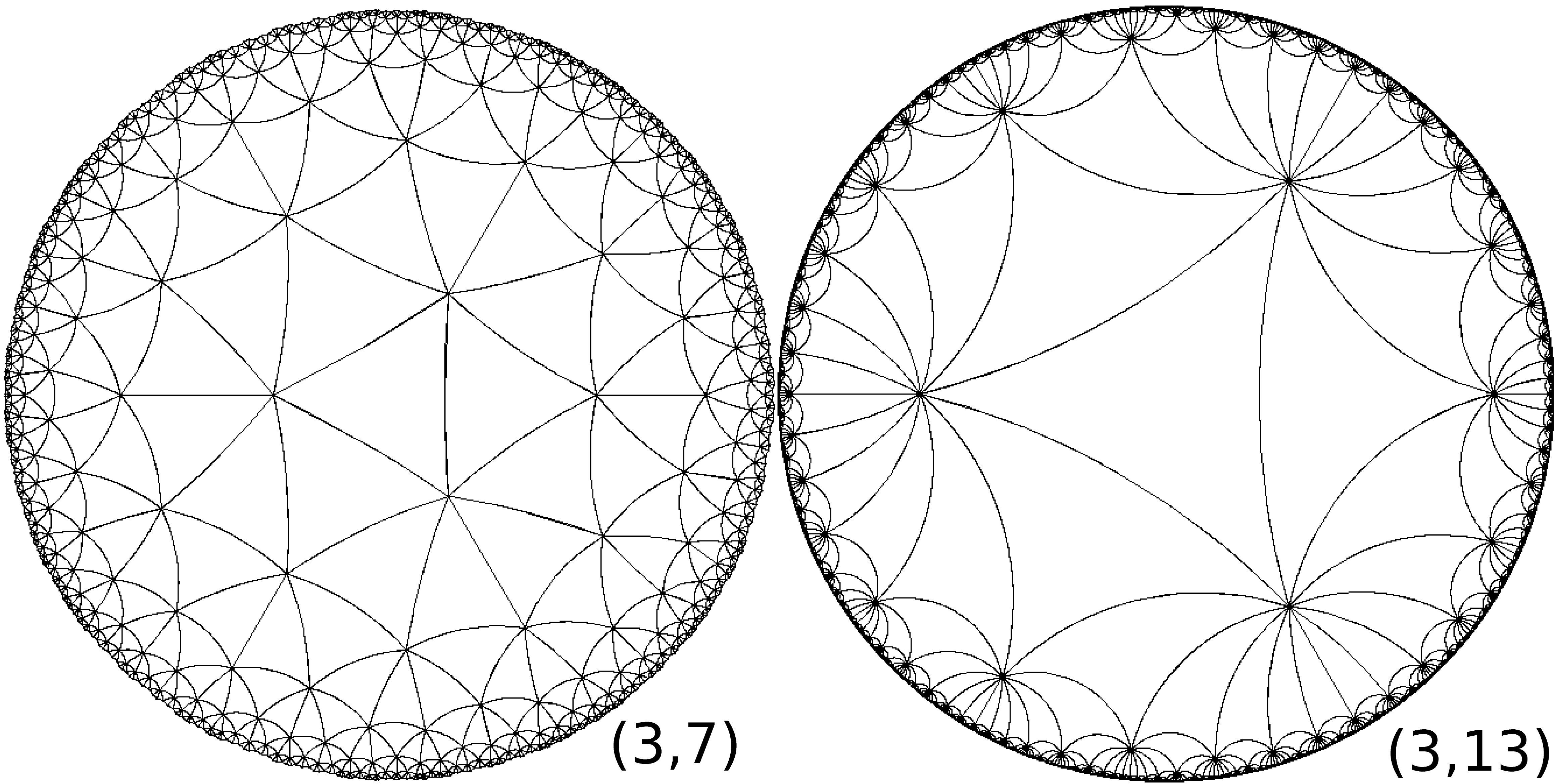}
\caption{Poincar\'e disc representation of the triangular hyperbolic lattices $(3,7)$ (left) and $(3,13)$ (right).}
\label{Lattice3_7_13}
\end{figure}

\begin{figure}[tb]
 \centering
 \includegraphics[width=4.5in]{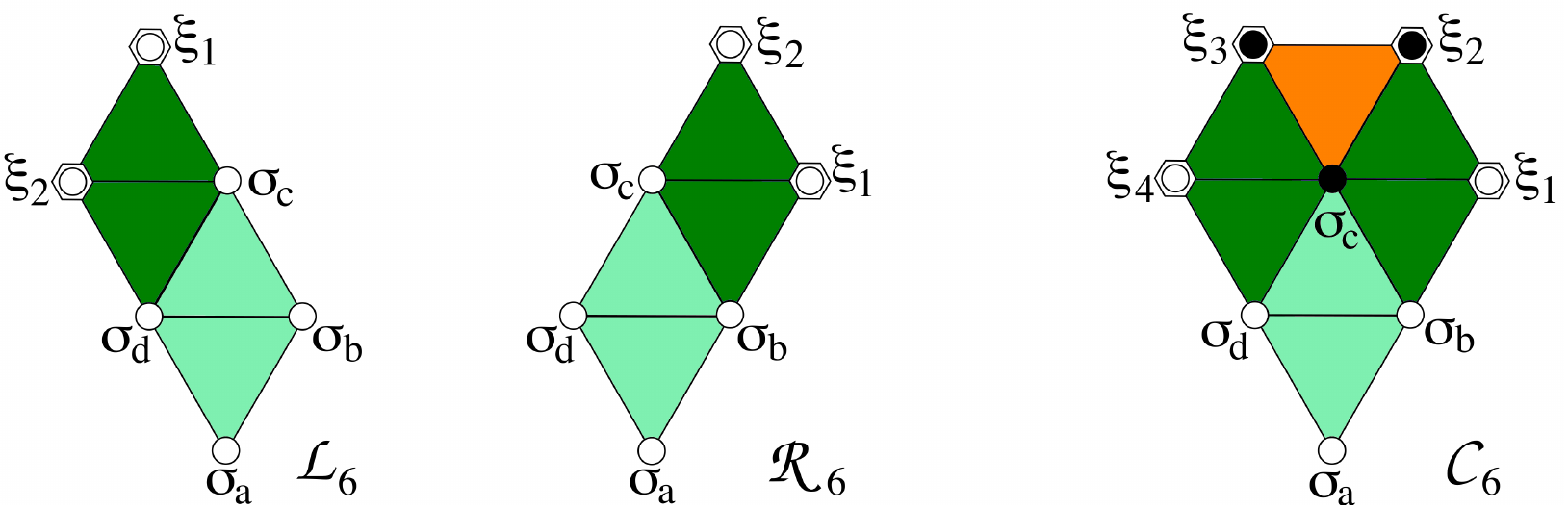}
\caption{Graphical representation of the extension process of
the left transfer tensor ${\tilde{\mathbf L}}^{~}_6$ (left), the right transfer tensor
${\tilde{\mathbf R}}^{~}_6$ (middle), and the corner tensor ${\tilde{\mathbf C}}^{~}_6$ (right)
on the $( 3, 6 )$ lattice. The transfer tensors ${{\mathbf L}}^{~}_6$ and ${{\mathbf R}}^{~}_6$ from the previous iteration are coloured in blue, while the corner tensor ${{\mathbf C}}^{~}_6$ in red. 
The black-filled symbols denote single- and multi-spin variables which are summed over in formulae \eqref{36lefttransfer}-\eqref{36corner}.}
\label{Lattice3_6_Rec_scheme}
\end{figure}

We consider the classical Ising model with Hamiltonian \eqref{Ising_hamiltonian}, where the spin variables $\sigma_i$ are located on the vertices of the $(3,q)$ lattices. As an example, the lattices $(3,7)$ and $(3,13)$ mapped onto the Poincar\'e disc are shown in Fig.~\ref{Lattice3_7_13}. 
Although the lattice polygon is a triangle, it is algorithmically more convenient to  assign the Boltzmann weight tensor ${\mathbf W_B}$ to a rhombus constructed from a pair of adjacent triangles $\sigma_a\sigma_b\sigma_d$ and $\sigma_b\sigma_c\sigma_d$ as shown in Fig.~\ref{Lattice3_6_Rec_scheme}.
The tensor  ${\mathbf W_B}$ for this pair of the triangles is then given by
\begin{align}
\begin{split}
 {\mathbf W_B}(\sigma_a, \sigma_b, \sigma_c, \sigma_d)=\exp
  &\left[\frac{J}{2k_{ B}T} (\sigma_a\sigma_b+\sigma_b\sigma_c
    +\sigma_c\sigma_d+\sigma_d\sigma_a+
        \right.
    \\
    &\left.	    
    +2\sigma_b\sigma_d) + 
    \frac{h}{qk_{ B}T} (\sigma_a+2\sigma_b+\sigma_c+2\sigma_d)
  \right] .
  \label{WB3q}
\end{split}
\end{align}
The factor $2$ in front of the $\sigma_b\sigma_d$ term arises from the fact that this bond is shared by the two adjacent triangles and, thus, entirely contained in the rhombus. Also, the factor $2$ at $\sigma_b$ and $\sigma_d$ reflects that the rhombus contains two of the $q$ triangles meeting at the corresponding lattice vertices.

In contrast to the general expansion process on the $(p,q)$ lattices, as described in section~\ref{CTMRG_hyp}, the transversal bond $\sigma_b\sigma_d$ in the tensor ${\mathbf W_B}$ makes it necessary to introduce two different kinds of transfer tensors --- the left tensor ${\mathbf L}_q$ and the right tensor ${\mathbf R}_q$.  Let us explain the recursive expansion scheme of the corner transfer 
tensor ${\mathbf C}_q$ and the tensors ${\mathbf L}_q$, ${\mathbf R}_q$ on the Euclidean $(3,6)$ lattice first. In this case, the expansions of the transfer tensors  ${\mathbf L}_6^{(k)}$, ${\mathbf R}_6^{(k)}$ and the corner tensor ${{\mathbf C}}_6^{(k)}$ in the iteration $k$ follow the formulae
\begin{eqnarray}
\label{36lefttransfer}
 {\tilde{\mathbf L}}_6^{(k+1)}(\sigma_d, \sigma_a, \sigma_b, \sigma_c, \xi_1, \xi_2)
 ={\mathbf W_B}(\sigma_a, \sigma_b, \sigma_c, \sigma_d)
  {\mathbf L}_6^{(k)}(\sigma_d, \sigma_c, \xi_1, \xi_2),\\
  \label{36righttransfer}
 {\tilde{\mathbf R}}_6^{(k+1)}(\sigma_c, \sigma_d, \sigma_a, \sigma_b, \xi_1, \xi_2)
 ={\mathbf W_B}(\sigma_a, \sigma_b, \sigma_c, \sigma_d)
  {\mathbf R}_6^{(k)}(\sigma_c, \sigma_b, \xi_1, \xi_2),\\
 \nonumber
 \label{36corner}
{\tilde{\mathbf C}}_6^{(k+1)}(\sigma_d, \sigma_a, \sigma_b, \xi_1, \xi_4)
 =\hspace{-0.2cm}\sum\limits_{\sigma_c,\xi_2,\xi_3}\hspace{-0.2cm}
       {\mathbf W_B}(\sigma_a, \sigma_b, \sigma_c, \sigma_d)
       {\mathbf L}_6^{(k)}(\sigma_d, \sigma_c, \xi_3, \xi_4)\\
 \times{\mathbf C}_6^{(k)}(\sigma_c,        \xi_2, \xi_3)
       {\mathbf R}_6^{(k)}(\sigma_c, \sigma_b, \xi_1, \xi_2) \, , 
\end{eqnarray}
as illustrated in Fig.~\ref{Lattice3_6_Rec_scheme}, where the position of the single- and multi-spin variables $\sigma$ and $\xi$, respectively, is also depicted. This recurrence scheme guarantees that at any iteration there are exactly $r=4$ and $s=4$ bonds stemming from all spins on the right and left boundary of the three tensors, respectively.  

\begin{figure}[tb]
 \centering
 \includegraphics[width=4.5in]{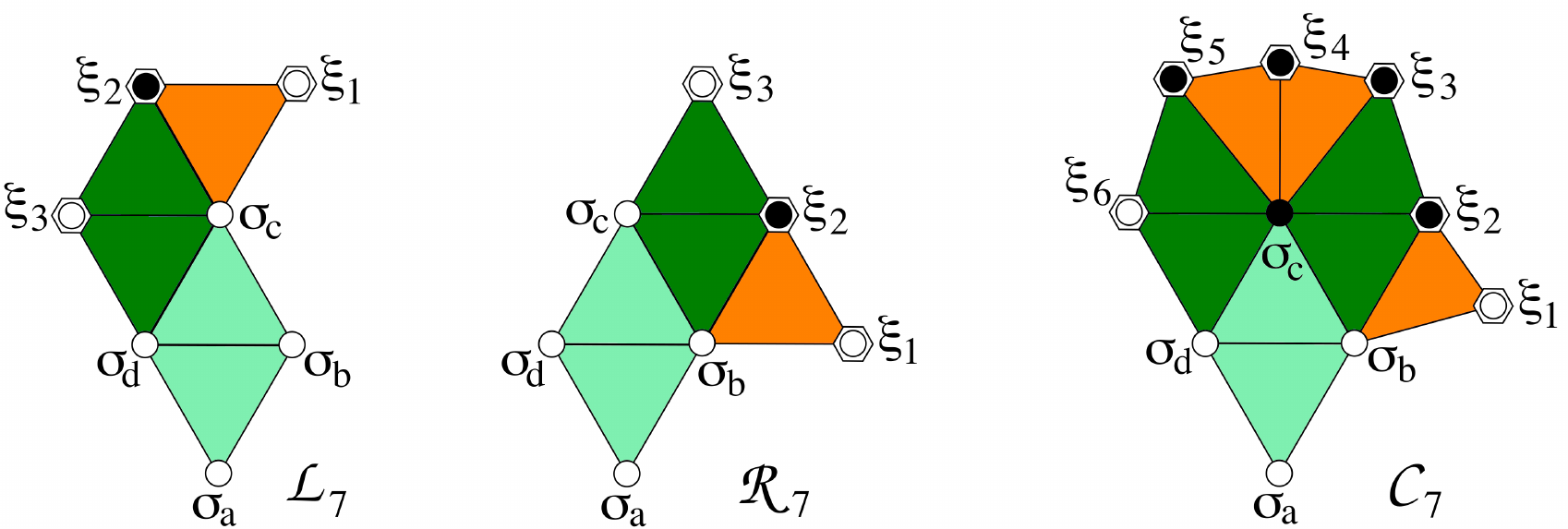}
 \includegraphics[width=4.5in]{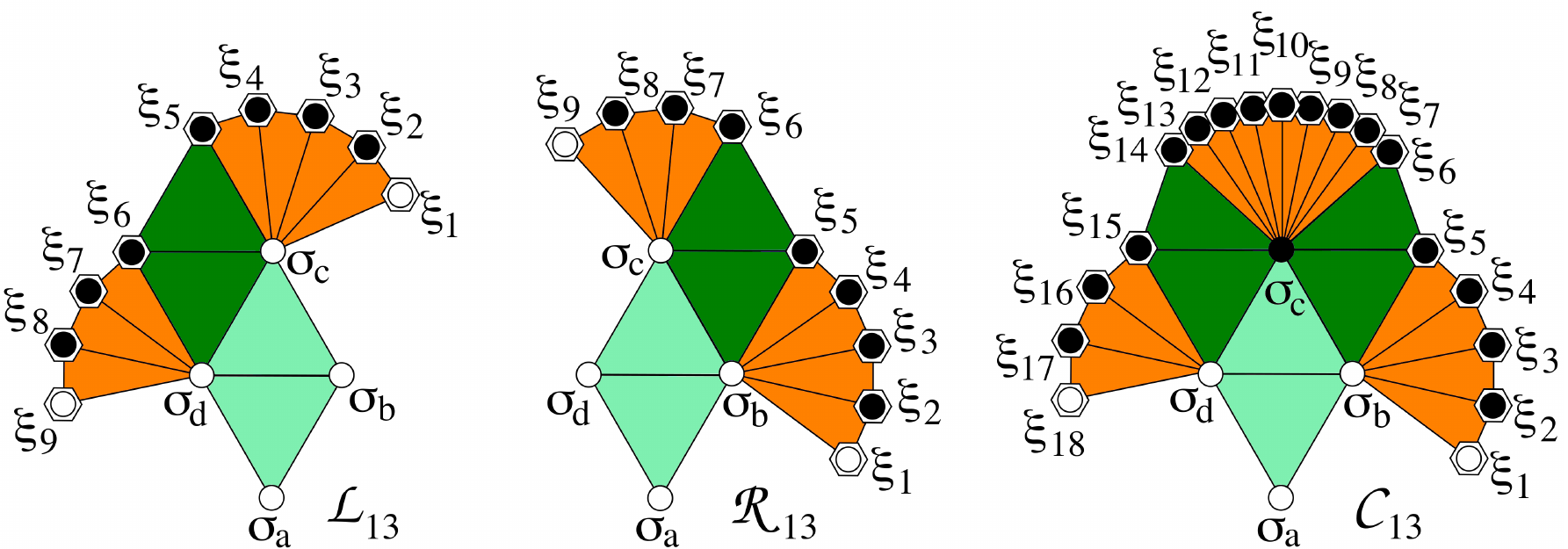}
\caption{The expansion process of
the tensors ${\tilde{\mathbf L}}^{~}_q$ (left),
${\tilde{\mathbf R}}^{~}_q$ (middle) and ${\tilde{\mathbf C}}^{~}_q$ (right)
on the $( 3, 7 )$ (top) and $( 3, 13 )$ (bottom) lattices, which follows the formulae \eqref{3qlefttransfer}-\eqref{3qcorner}.}
\label{3qlatticeexpand}
\end{figure}

The recursive expansion procedure is initialized in the following:
\begin{eqnarray}
&{\mathbf L}_6^{(1)}(\sigma_a, \sigma_b, \sigma_c, \sigma_d)
={\mathbf W_B^{\rm L}}(\sigma_a, \sigma_b, \sigma_c, \sigma_d),\\
&{\mathbf R}_6^{(1)}(\sigma_d, \sigma_a, \sigma_b, \sigma_c)
={\mathbf W_B^{\rm R}}(\sigma_a, \sigma_b, \sigma_c, \sigma_d),\\
&{\mathbf C}_6^{(1)}(\sigma_a, \sigma_b,\sigma_d)
=\sum_{\sigma_c} {\mathbf W_B^{\rm C}}(\sigma_a, \sigma_b, \sigma_c, \sigma_d),
\end{eqnarray}
where ${\mathbf W_B^{\rm L}}$,  ${\mathbf W_B^{\rm R}}$ and ${\mathbf W_B^{\rm C}}$ are modifications of the Boltzmann weight tensor \eqref{WB3q}, which reflect the specific situation in sharing of bonds and vertices between the rhombuses on the lattice boundary.
On the hyperbolic $(3,q)$ lattices, where $q \geq 7$, the recurrence expansion relations in the simplified notation take the form
\begin{eqnarray}
\label{3qlefttransfer}
  {\tilde{\mathbf L}}^{(k+1)}_q&=&{\mathbf W_B}\left({\mathbf C}^{(k)}_q\right)^{s-4}{\mathbf L}^{(k)}_q\left({\mathbf C}^{(k)}_q\right)^{r-4},\\
  \label{3qrighttransfer}
  {\tilde{\mathbf R}}^{(k+1)}_q&=&{\mathbf W_B}\left({\mathbf C}^{(k)}_q\right)^{s-4}{\mathbf R}^{(k)}_q\left({\mathbf C}^{(k)}_q\right)^{r-4},\\
 \label{3qcorner}
  {\tilde{\mathbf C}}^{(k+1)}_q&=&{\mathbf W_B}\left({\mathbf C}^{(k)}_q\right)^{s-4}{\mathbf L}^{(k)}_q\left({\mathbf C}^{(k)}_q\right)^{q-5}{\mathbf R}^{(k)}_q\left({\mathbf C}^{(k)}_q\right)^{r-4},
\end{eqnarray}
where the integers $r$ and $s$ are constrained by the condition \eqref{rsequality}. The recurrence scheme creates tensors with $r$ and $s$ bonds originating from all spins on the right and left boundary of the three tensors, respectively. In particular, we decided for the most symmetric combinations
\begin{align}
r&=\left\lceil \frac{q}{2} \right\rceil+1 
=: 
\min\left\{n \in \mathbb{Z} : n \geq \frac{q}{2}\right\}+1, 
\\
s&=\left\lfloor \frac{q}{2} \right\rfloor+1
=: 
\max\left\{n \in \mathbb{Z} : n \leq \frac{q}{2}\right\}+1,
\end{align}
as graphically depicted in Fig.~\ref{3qlatticeexpand} for the two representative lattices $(3,7)$ and $(3,13)$.

At each iteration, the reduced density matrix $\rho$ is created via \eqref{def_DM}, where the tensor $\mathbf A$ is calculated as \eqref{Aqtensor} if $q$ is even, or  via the symmetrized formula \eqref{def_DM_symmetrized} for odd $q$. After that, the three tensors are renormalized and subsequently normalized according to \eqref{renormalizedC}, \eqref{renormalizedT} and \eqref{CT_normalizacia}, respectively.

\subsubsection{Numerical results}

\paragraph{I. Magnetization and energy}

In order to suppress the influence of the system boundary on the thermodynamic properties and the phase transition analysis in case of the hyperbolic lattices, we concentrate on the
bulk properties of a sufficiently large inner region of the
lattice~\cite{Sakaniwa,hctmrg-Ising-p-4}. 
The local
%spontaneous 
magnetization $M(h,T)\equiv \langle \sigma_{\ell}\rangle$ of a spin on the central lattice position $\ell$ is an example. Generalizing the formula \eqref{exp_value_mag} for the case of the hyperbolic lattices, the magnetization is calculated as
\begin{equation}
M(h,T)=\frac{\Tr\left(\sigma_{\ell}{\mathbf C}^q\right)}{\Tr\left({\mathbf C}^q\right)}={\rm Tr} \left( \sigma_{\ell} \rho \right) / {\rm Tr} \,  \rho,
\end{equation}
where the second equality holds also for odd coordination numbers $q$ with density matrix $\rho$ in the symmetrized form \eqref{def_DM_symmetrized}.
Without loss of generality, we set the
coupling constant $J$ and the Boltzmann constant $k_{ B}$ to
unity, and all thermodynamic functions are evaluated in the unit
of $k_{ B}$.

\begin{figure}[!tbp]
 \centering
	\vspace*{-0.3cm}
 \includegraphics[width=4.2in]{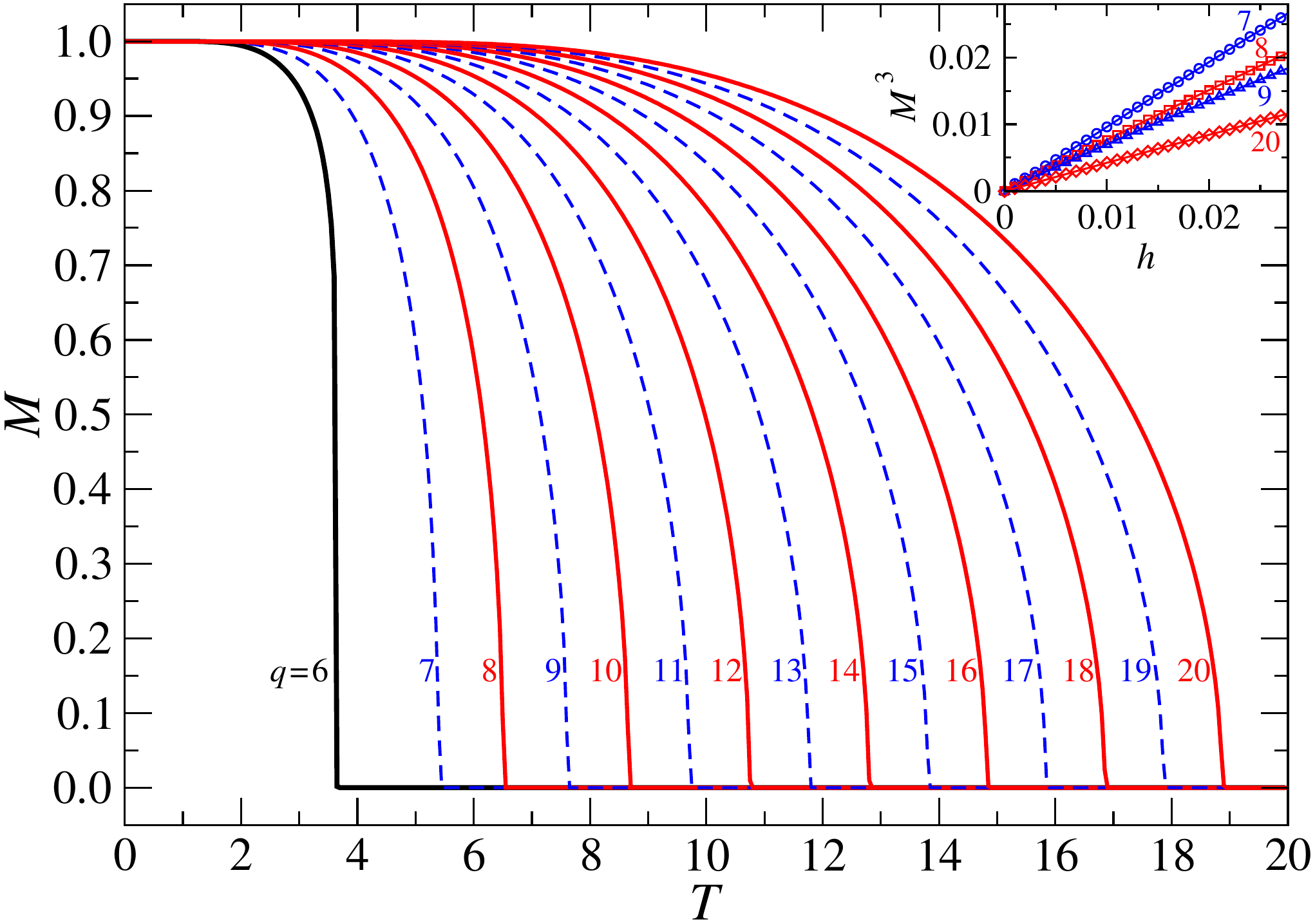}
	\vspace*{-0.2cm}
\caption{Spontaneous magnetizations $M_0(T)=M(h=0,T)$ with respect to temperature $T$ for
$6\leq q \leq 20$. The
full and the dashed curves, respectively, distinguish the even and odd values of $q$. The inset shows the linear behaviour of the cubic power of
the induced magnetization $M^3_{~}(h, T=T_{\rm pt}^{(q)})$ with respect the magnetic field $h$
around the transition temperatures $T_{\rm pt}^{(q)}$ for $q\geq 7$.}
\label{3qmagnetizacia}
\end{figure}

We first investigate the Euclidean $( 3, 6 )$ lattice. Keeping only
$m = 20$ states of the renormalized multi-spin variables $\xi$, the obtained spontaneous magnetization $M_0(T)=M(h=0,T)$ is shown in Fig.~\ref{3qmagnetizacia}. The estimated transition temperature $T_{\rm c}=3.641$ is
quite close to the exact value $T_{\rm c}=4/\ln 3\approx3.64096$~\cite{Baxter}. In the identical Figure, we also plot the temperature dependence of the spontaneous magnetization $M_0(T)$
for the hyperbolic $(3,q)$ lattices with coordination numbers $7\leq q \leq 20$. As we show later, the system is always off-critical whenever
$q \geq 7$, even at the transition temperature. We, therefore, use the
notation $T_{\rm pt}^{(q)}$ instead of $T_{\rm C}^{(q)}$ for $q \geq 7$ and
we also use $T_{\rm pt}^{(6)}$ for $q = 6$ in order to unify the notation.

\begin{figure}[!bhp]
 \centering
	\vspace*{-0.3cm}
 \includegraphics[width=4.2in]{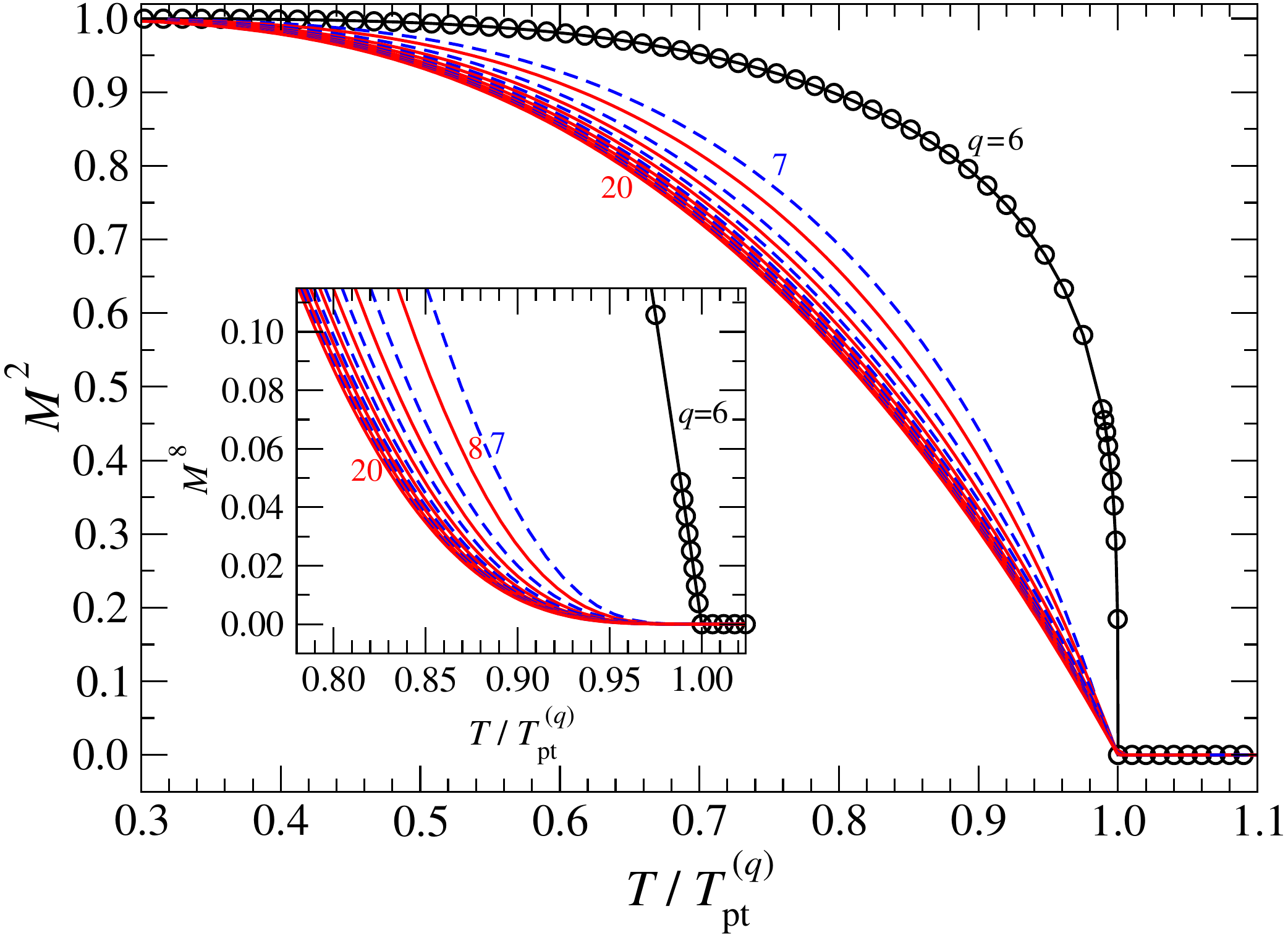}
	\vspace*{-0.3cm}
\caption{The squared spontaneous magnetization $M^2_{0}(T)$ with 
respect to the relative temperature $T/T_{\rm pt}^{(q \ge 6)}$ near the transition point.}
\label{3qsquaredM}
\end{figure}

If a small magnetic field $h$ is applied at the transition temperature 
$T_{\rm pt}^{( q \ge 7 )}$, the cubed induced magnetization $M^3_{~}(h, T=T_{\rm pt}^{(q)})$ is
always linear around $h = 0$. Thus, the model satisfies the scaling relation
$M(h,T=T_{\rm pt})\propto h^{1/\delta}$ with the mean-field critical exponent $\delta=3$. 
This value is in full agreement with the previous results for the 
hyperbolic $( p \ge 5, 4 )$ lattices~\cite{hctmrg-Ising-p-4}.

In order to observe the scaling relation of the spontaneous magnetization $M_0(T)$
in a unified manner, we plot the squared spontaneous magnetization $M^2_{0}(T)$ in Fig.~\ref{3qsquaredM}
with respect to the relative temperature $T/T_{\rm pt}^{(q \ge 6)}$.
The linearity of the curves near the transition point $T = T_{\rm pt}^{(q \ge 7)} $ agrees with the
mean-field behaviour $M(h=0,T)\propto(T_{\rm pt}^{(q)} - T)^\beta_{~}$ 
with  $\beta=\frac{1}{2}$ on hyperbolic lattices with $q \geq 7$. On the $( 3, 6 )$ lattice,
the exponent is $\beta=\frac{1}{8}$, as displayed in the inset, where the linearity is satisfied for $q=6$ only.

\begin{figure}[tb]
 \centering
 \includegraphics[width=4in]{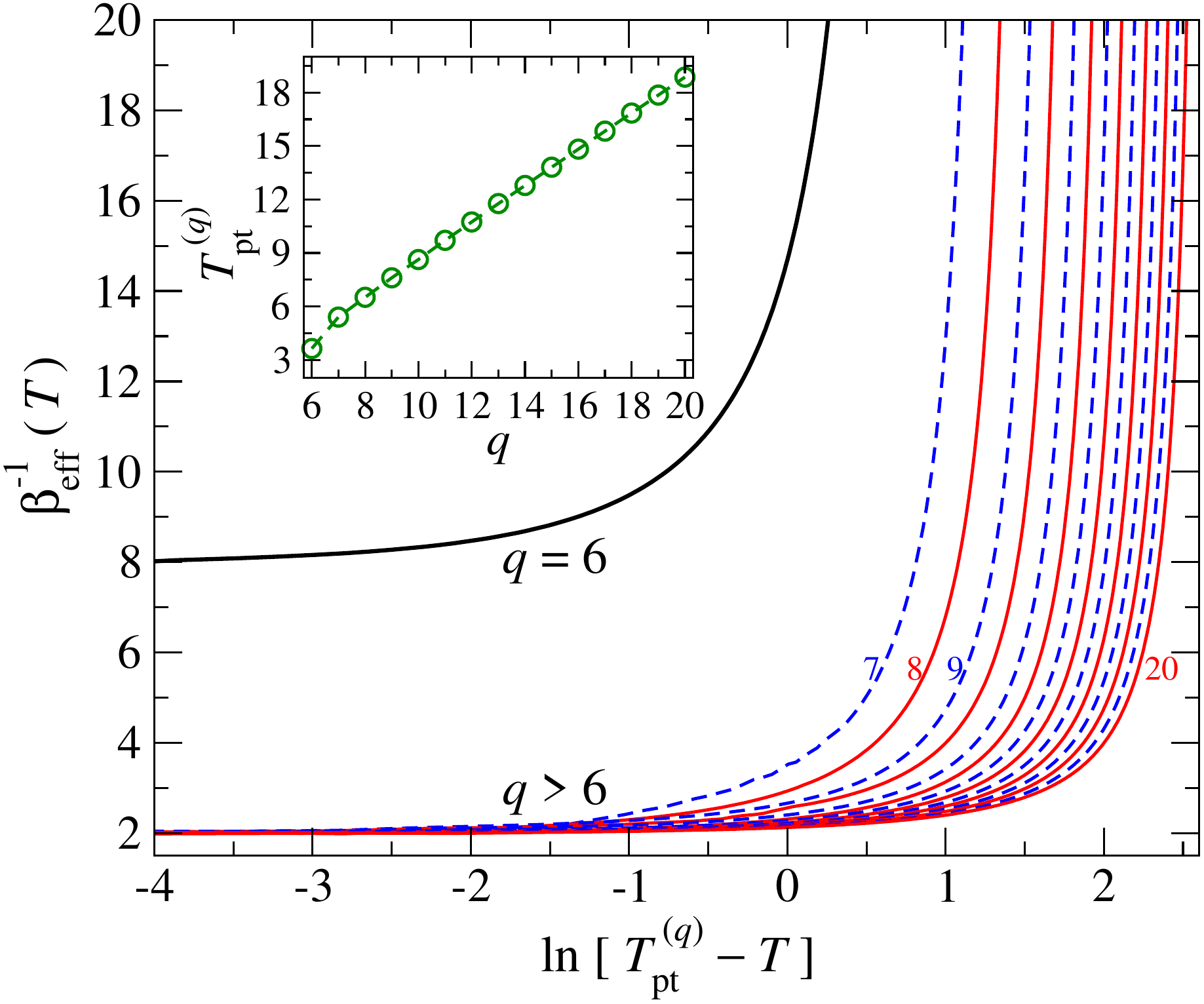}
\caption{Convergence rate of the effective critical exponent $\beta_{\rm eff}$. The inset shows the dependence of the phase transition temperatures $T_{\rm pt}^{(q)}$ on the coordination number $q$.}
\label{3qbeta}
\end{figure}

To detect the critical exponent $\beta$ in a more precise manner,
we calculate the effective exponent
\begin{equation}
\beta_{\rm eff}(T)=\frac{\partial \ln \left[ M \left( h=0,T<T_{\rm pt}^{(q)}
      \right) \right]} {\partial \ln \left[T_{\rm pt}^{(q)} - T\right]} \, 
\label{3qbeff}
\end{equation}
by means of the numerical derivative. The convergence of $\beta_{\rm eff}(T)$
with respect to $T_{\rm pt}^{(q)} - T$ is shown in Fig.~\ref{3qbeta}.  It is
apparent that the mean-field value $\beta=\frac{1}{2}$ is detected for any
$q \ge 7$, whereas we confirm $\beta=\frac{1}{8}$ on the flat $(3,6)$ lattice
only, which agrees with the two-dimensional Ising universality class. The linear increase of the transition temperature $T_{\rm pt}^{(q \ge 7)}$ 
with respect to $q$ is shown in the inset where the linearity appears already 
around $q\gtrsim8$. This agrees with the linear dependence $T_{\rm C}(q)$ \eqref{T_C_mean_field} observed in the mean-field model.

\begin{figure}[tb]
 \centering
 \includegraphics[width=4.5in]{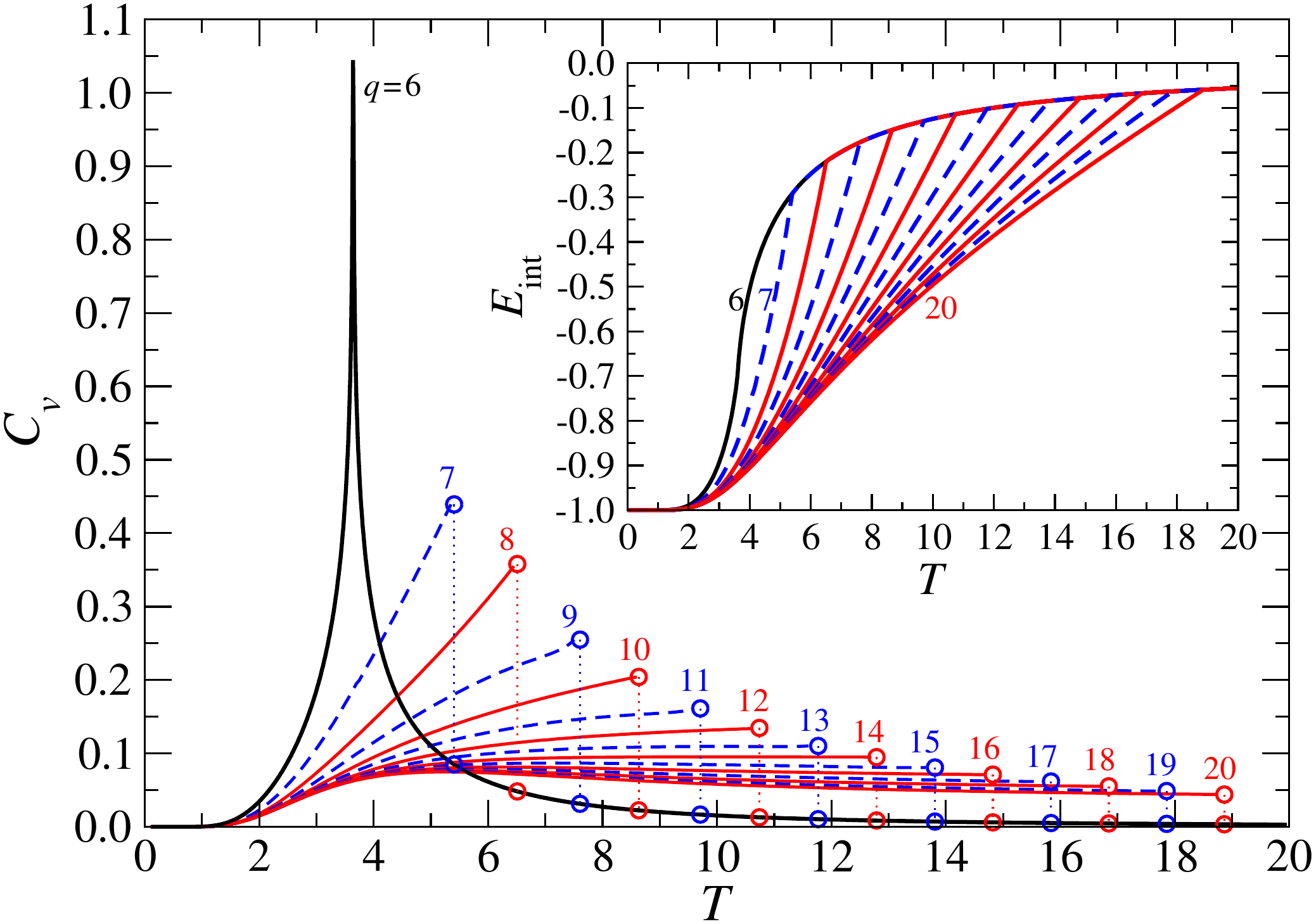}
\caption{Specific heat $C_h(h=0, T)$ as a function of temperature. The open circles connected
by the vertical dotted lines show the discontinuity. The inset shows the temperature 
dependence of the internal energy $E_{\rm int}^{~}(h=0, T)$. Both $C_h(h=0, T)$ and $E_{\rm int}(h=0, T)$ in the
paramagnetic region are almost independent on $q$.}
\label{3qenergy}
\end{figure}

Next, we investigate the temperature dependence of the internal energy per bond in absence of the magnetic field $h$ 
\begin{equation}
E_{\rm int}^{~}(h=0, T) =-J\langle\sigma_{\ell}\sigma_{\ell^{\prime}}\rangle
           =-J\,{\Tr}\left(\sigma_{\ell}\sigma_{\ell^{\prime}}\rho\right)/
           \Tr\left(\rho\right)
\end{equation}
and the specific heat per bond
\begin{equation}
C_h(h=0, T)=\frac{\partial E_{\rm int}^{~}(h=0, T)}{\partial T},
\end{equation}
where $\sigma_{\ell}$ and $\sigma_{\ell^{\prime}}$ denote two neighbouring spins located at the center of the lattice.
Figure~\ref{3qenergy} shows the results.
The internal energy $E_{\rm int}^{~}(h=0, T)$ is continuous for all the cases we
computed. The kink in $E_{\rm int}^{~}(h=0, T)$ at the
transition temperatures $T_{\rm pt}^{(q)}$ for $q \geq 7$ corresponds to the
discontinuity in $C_h(h=0, T)$~\cite{hctmrg-Ising-p-4,hctmrg-J1J2}.
For these cases the scaling exponent $\alpha$ is zero.

\paragraph{II. Entropy and correlation}

\begin{figure}[tb]
 \centering
 \includegraphics[width=4.45in]{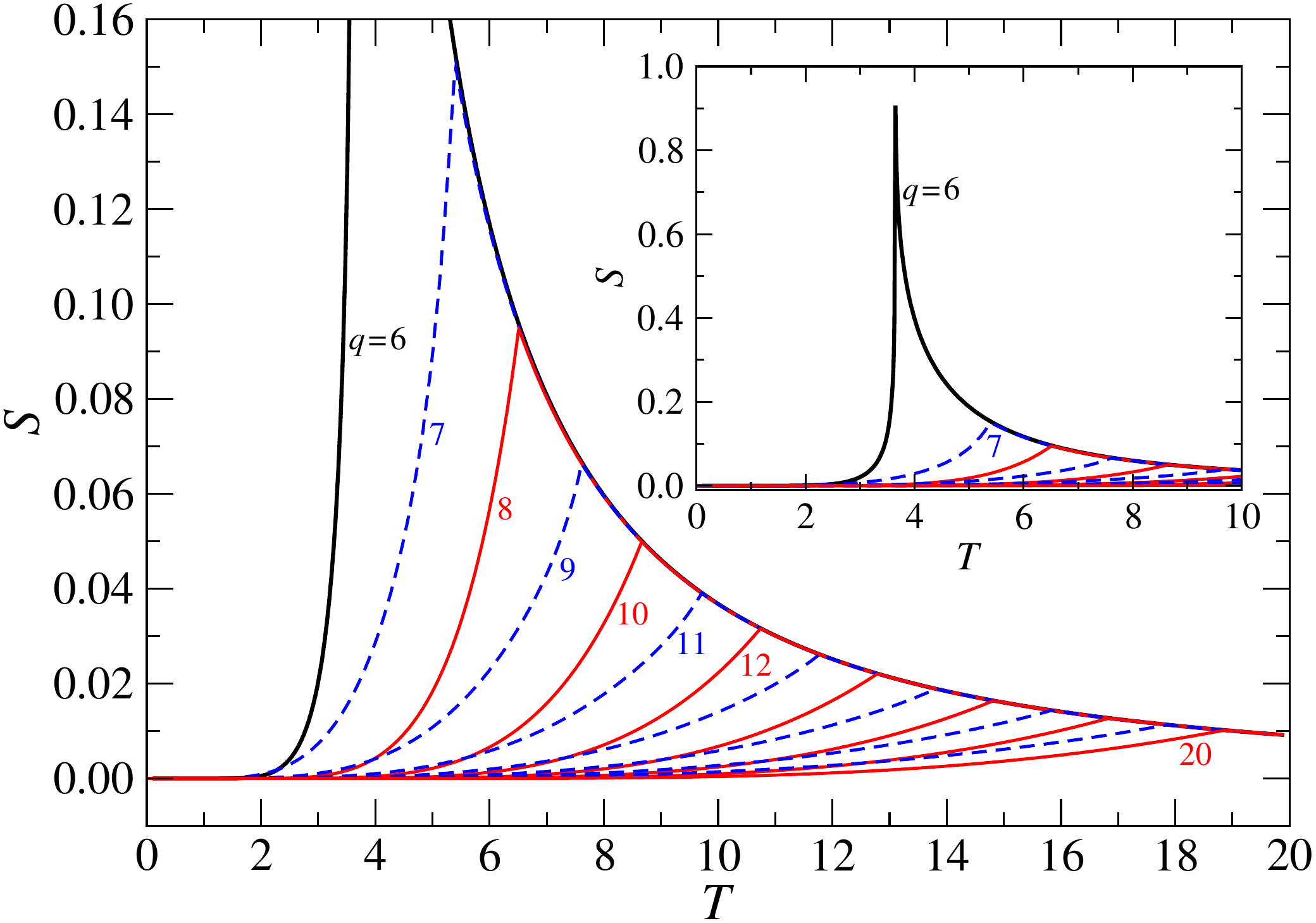}
\caption{Temperature dependence of the von Neumann entanglement entropy $S$.
The inset displays the dominant behavior of $S$ for the $( 3, 6 )$ lattice.}
\label{3qentropy}
\end{figure}

The von Neumann (or entanglement) entropy is defined via the reduced density matrix $\rho$
\footnote{If $q$ is odd, the symmetrized form \eqref{def_DM_symmetrized} of the density matrix $\rho$ is applied. In this case the entropy is considered to be less reliable than for even $q$'s,
and we regard such entropy as complementary information.}
as
\begin{equation}
 S=-{\Tr} \left(\rho\log_2\rho\right)
  =-\sum\limits_{i}\omega_i^{~}\log_2\omega_i^{~},
  \label{VonNeumannEntropy}
\end{equation}
where $\omega_i$ are the eigenvalues of $\rho$. Figure~\ref{3qentropy}
 shows the temperature dependence
of $S$ which remains finite for $q \ge 7$ even at the transition temperature $T_{\rm pt}^{(q)}$.
The entropies in the paramagnetic region are also almost independent on $q$
if $q \geq 7$ as observed also for $C_h^{~}(h=0, T)$ and $E_{\rm int}(h=0, T)$ in the previous section.

\begin{figure}[!thp]
 \centering
	\vspace*{-0.4cm}
 \includegraphics[width=4.13in]{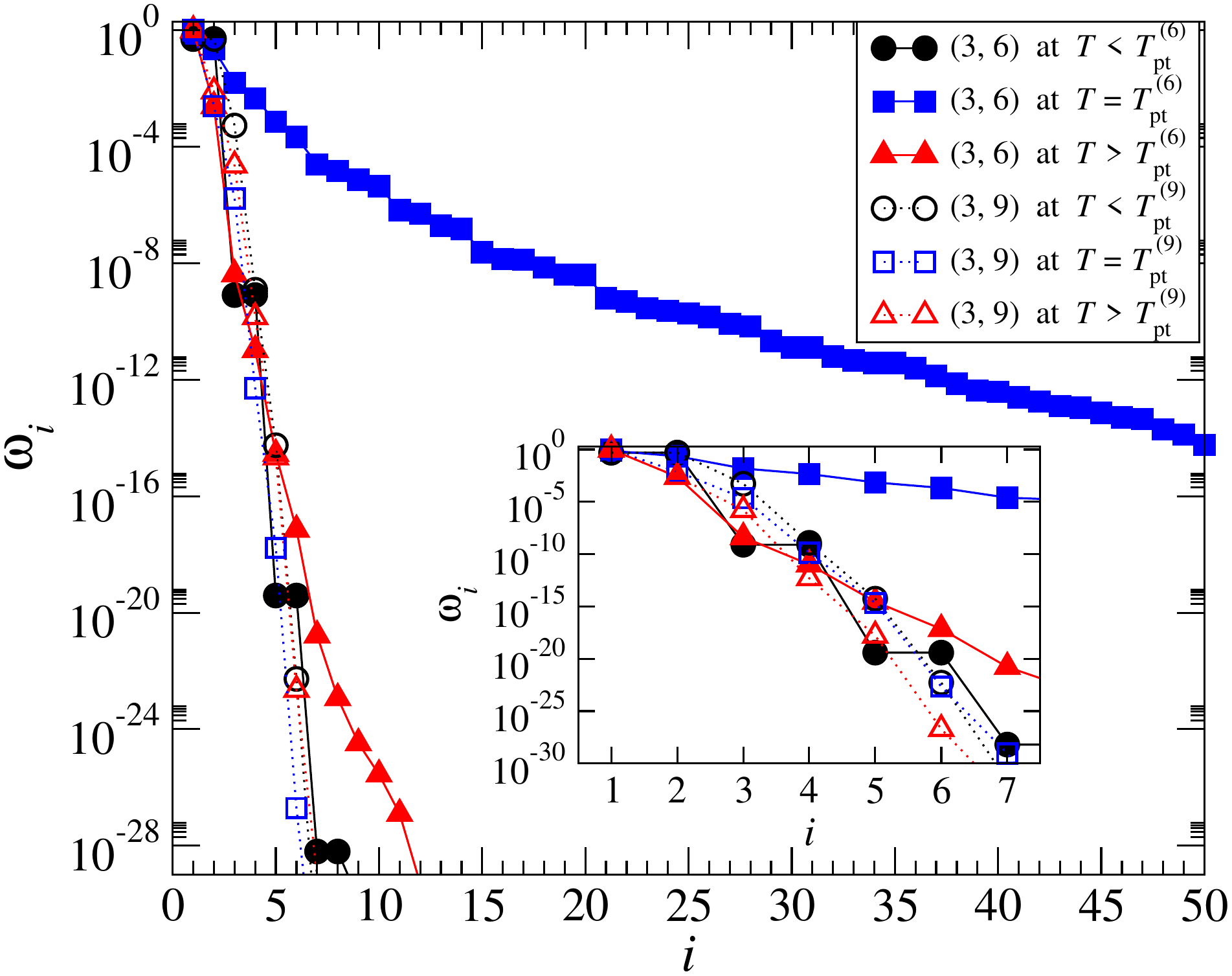}
	\vspace*{-0.2cm}
\caption{Decay of the density matrix spectra for the $( 3, 6 )$ lattice
(filled symbols) and the $( 3, 9 )$ lattice (open symbols).}
\label{3qomegadecay}
\end{figure}

The decay rate of the density matrix eigenvalues $\omega_i^{~}$ is shown in
fig.~\ref{3qomegadecay} 
\begin{figure}[!bhp]
 \centering
	\vspace*{-0.2cm}
 \includegraphics[width=4.13in]{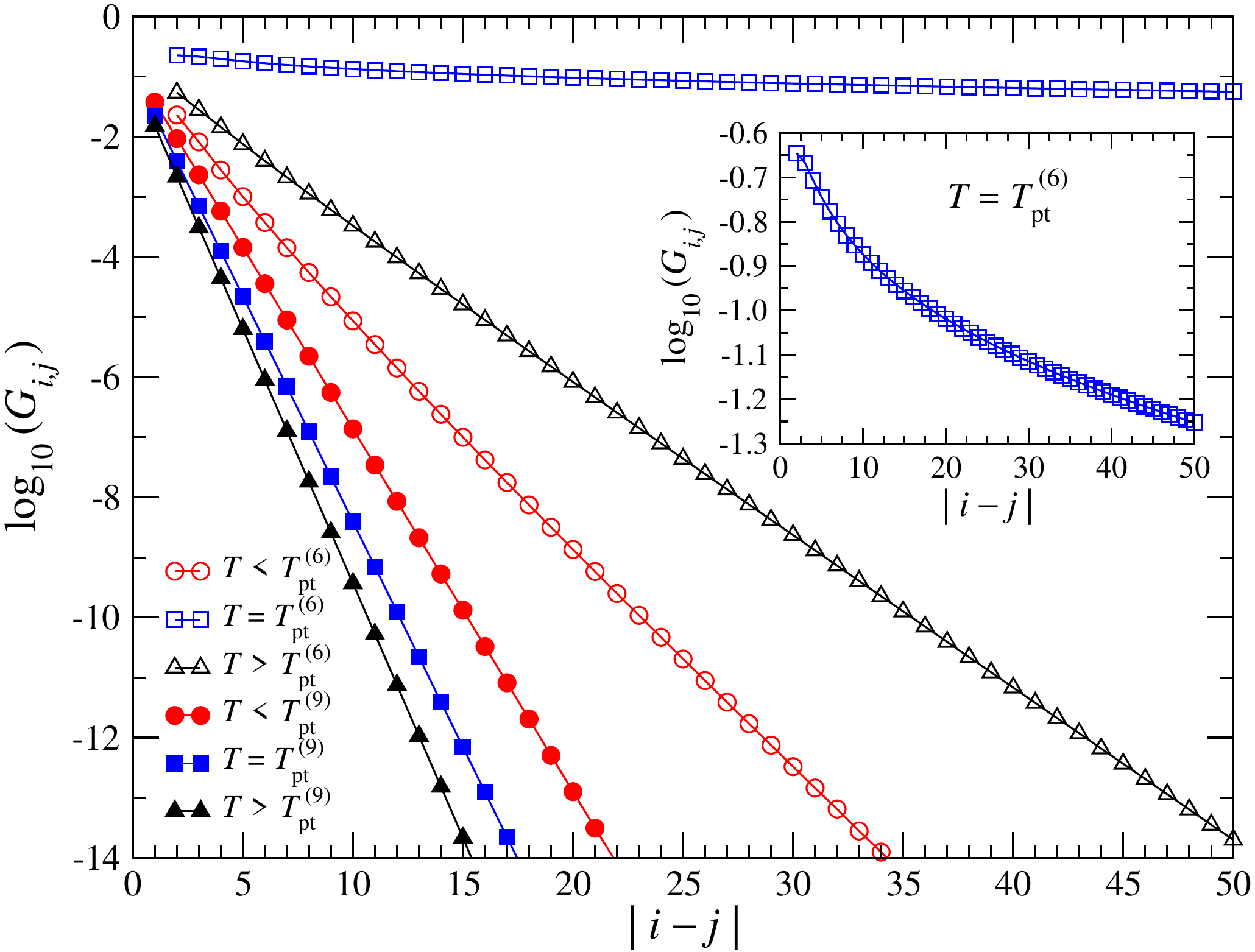}
	\vspace*{-0.2cm}
\caption{Decay of the correlation functions with respect to the spin
distance $\vert \mathbf{r}_{i}-\mathbf{r}_{j}\vert$. The open and the full symbols, respectively,
correspond to the  $( 3, 6 )$ lattice calculated at $T=3.0$, $3.641$,
and $5.0$, and the $( 3, 9 )$ lattice at $T=6.0$, $7.608$, and $9.0$. Note that $T_{\rm pt}^{(6)} \doteq 3.641$ and $T_{\rm pt}^{(9)} \doteq 7.608$.}
\label{3qGij}
\end{figure}
on a semilogarithmic scale for the Euclidean $(3, 6)$ and the hyperbolic $(3, 9)$ lattice. We confirm a power-law decay in $\omega_i^{~}$ only at the
transition point $T_{\rm pt}^{(6)}$ of the $( 3, 6 )$ lattice. The
eigenvalues $\omega_i^{~}$ decrease exponentially for $q\geq7$ at the
transition temperature.

The exponential decay of the density matrix spectra is also reflected in the
correlation function
\begin{equation}
g(\mathbf{r}_{i},\mathbf{r}_{j}) \equiv
g(|\mathbf{r}_{i}-\mathbf{r}_{j}|)={\rm Tr}\left( \sigma_i \sigma_j \rho \right)-
{\rm Tr}\left( \sigma_i\rho \right){\rm Tr}\left(\sigma_j \rho \right)
\label{cf}
\end{equation}
between two distant sites $i$ and $j$. We place the spin $\sigma_i$ at the
center of the system and $\sigma_j$ at the system boundary. %The distance is measured in units of the length of the lattice triangle edge, which yields $r_{ij}=|i-j|$   
Therefore, as the
lattice expands in the CTMRG algorithm, the distance between
these two spins increases progressively.

Figure~\ref{3qGij}
 depicts $\log_{10} \left[g(\mathbf{r}_{i},\mathbf{r}_{j})\right]$ as a function of the distance
$\vert \mathbf{r}_{i}-\mathbf{r}_{j}\vert$ for the Euclidean $( 3, 6 )$ and the hyperbolic $( 3, 9 )$ lattice. It is evident that the correlation functions always
decay exponentially on the $( 3, 9 )$ lattice regardless of the temperature.
We remark that an analogous exponential decay of $g(\mathbf{r}_{i},\mathbf{r}_{j})$ has been observed
for all $q \ge 7$ (not shown). On the $( 3, 6 )$ lattice, the correlation
function decays as a power law at the transition temperature $T_{\rm pt}^{(6)}$, as seen in the
inset.

%%%%%%%%%%%%%%%%%%%%%%%%%%
% AGAGAGA
%%%%%%%%%%%%%%%%%%%%%%%%%%

In the following, we compare the Gaussian curvature associated to the $(3, q)$
lattice with the correlation length at the transition temperature. There
are several ways to define the correlation length $\xi_q$~\cite{Baxter}.
For example, the decay rate of the correlation function directly provides $\xi_q$. 
This is straightforward, but the region of the distance for the fitting analysis
has to be valued carefully. Another possibility consists in using the largest
eigenvalue $\lambda_0(q)$ and the second largest one  $\lambda_1(q)$ of the
row-to-row transfer matrix ${\mathbf T}_q={\mathbf L}_q{\mathbf R}_q$,
where $\xi_q$ is determined from
\begin{equation}
\frac{1}{\xi_q} = \ln\left[\frac{\lambda_0(q)}{\lambda_1(q)}\right] \, .
\label{corlen}
\end{equation}
The relation can be generalized to the $( 3, q \ge 7 )$ lattices, in analogy
to our previous formulations for  the $( 5, 4 )$ lattice~\cite{corrlen}, 
via the construction of the row-to-row transfer matrix
\begin{equation}
    {\mathbf T}_q(\zeta_1 \sigma_c \zeta_4|\zeta_2 \sigma_d \zeta_3)=
    {\mathbf L}_q(\sigma_d, \sigma_c, \zeta_1, \zeta_2)
    {\mathbf R}_q(\sigma_c, \sigma_d, \zeta_3, \zeta_4)\, .
\label{TrMat}
\end{equation}
Using the notation of the recurrence scheme introduced previously,
we calculate the correlation length $\xi_q$ by use of Eq.~\eqref{corlen}.

\begin{figure}[!thp]
	\vspace*{-0.4cm}
\centerline{\includegraphics[width=4.15in]{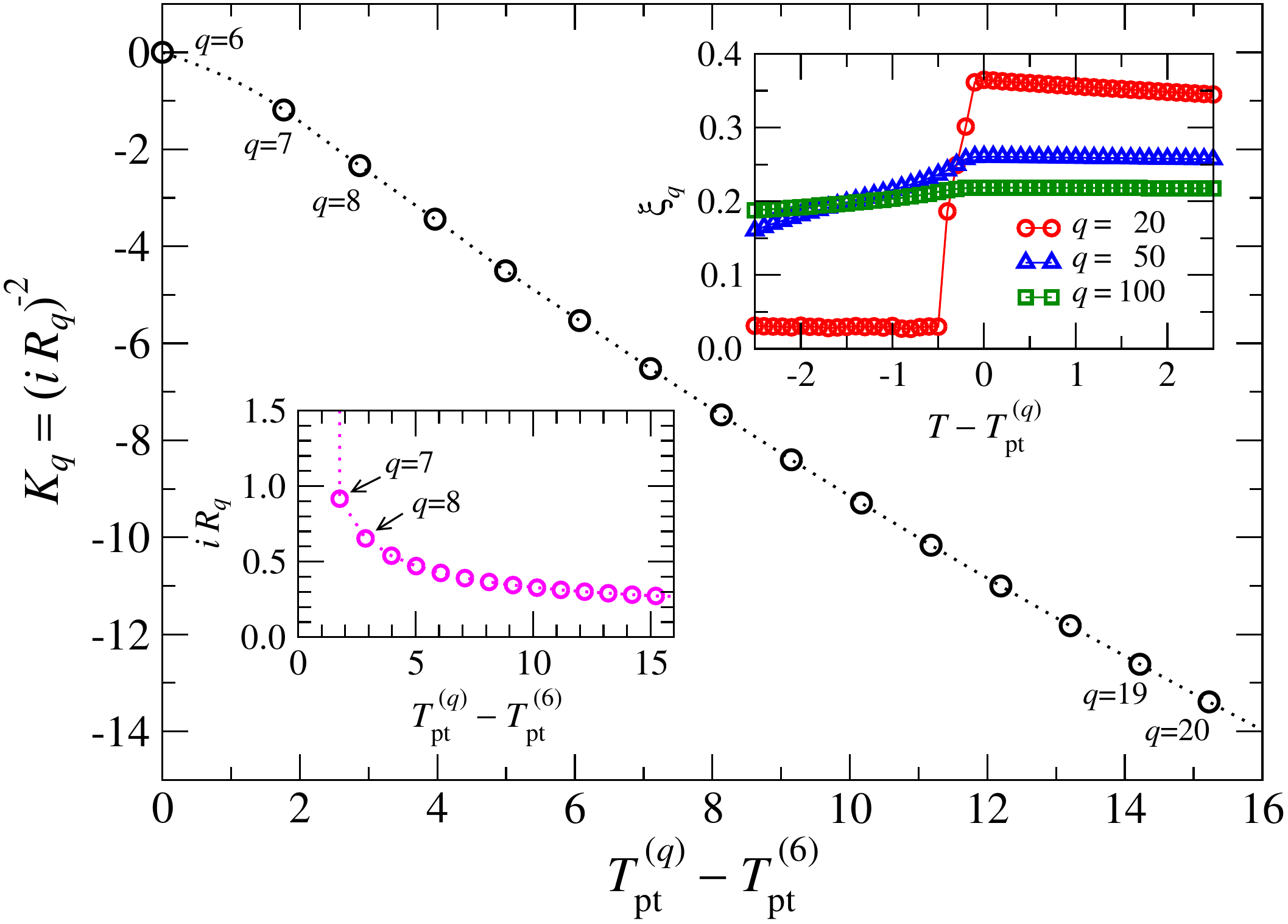}}
	\vspace*{-0.2cm}
\caption{Gaussian curvature $K_q$ with respect to the shifted phase transition
temperatures for $6\leq q\leq 20$. The inset on the left shows the related
radius of the curvature $iR_q$ via Eq.~\eqref{Kq-Rq} while that on the right
shows the correlation length in the vicinity of the phase transition.}
\label{fig12}
\end{figure}

The Gaussian curvature $K_q$ that corresponds to $( 3, q )$ lattice is given
by~\cite{Mosseri}
\begin{equation}
 K_q = \frac{1}{{(iR_q)}^2}=-4\,{\rm arccosh}
  \left[
    \frac{1}{2\sin\left(\frac{\pi}{q}\right)}
  \right]
\label{Kq-Rq}
\end{equation}
where $R_q$ is the curvature radius of the hyperbolic surface.
Recall that $K_q$ must be zero on the Euclidean flat space ($q=6$).
Figure~\ref{fig12} shows the relation between $K_q$ and  the 
shifted transition temperature $T_{\rm pt}^{(q)}-T_{\rm pt}^{(6)}$.
The lower-left inset shows complementary information about $R_q$.
The correlation function $\xi_q$ calculated around the phase transition
for three different $q$'s is plotted in the upper-right inset. Notice
that $\xi_q$ reaches its maximum at the phase transition which is not
well visible as $q$ increases.

\begin{figure}[!bhp]
	\vspace*{-0.2cm}
\centerline{\includegraphics[width=4.15in]{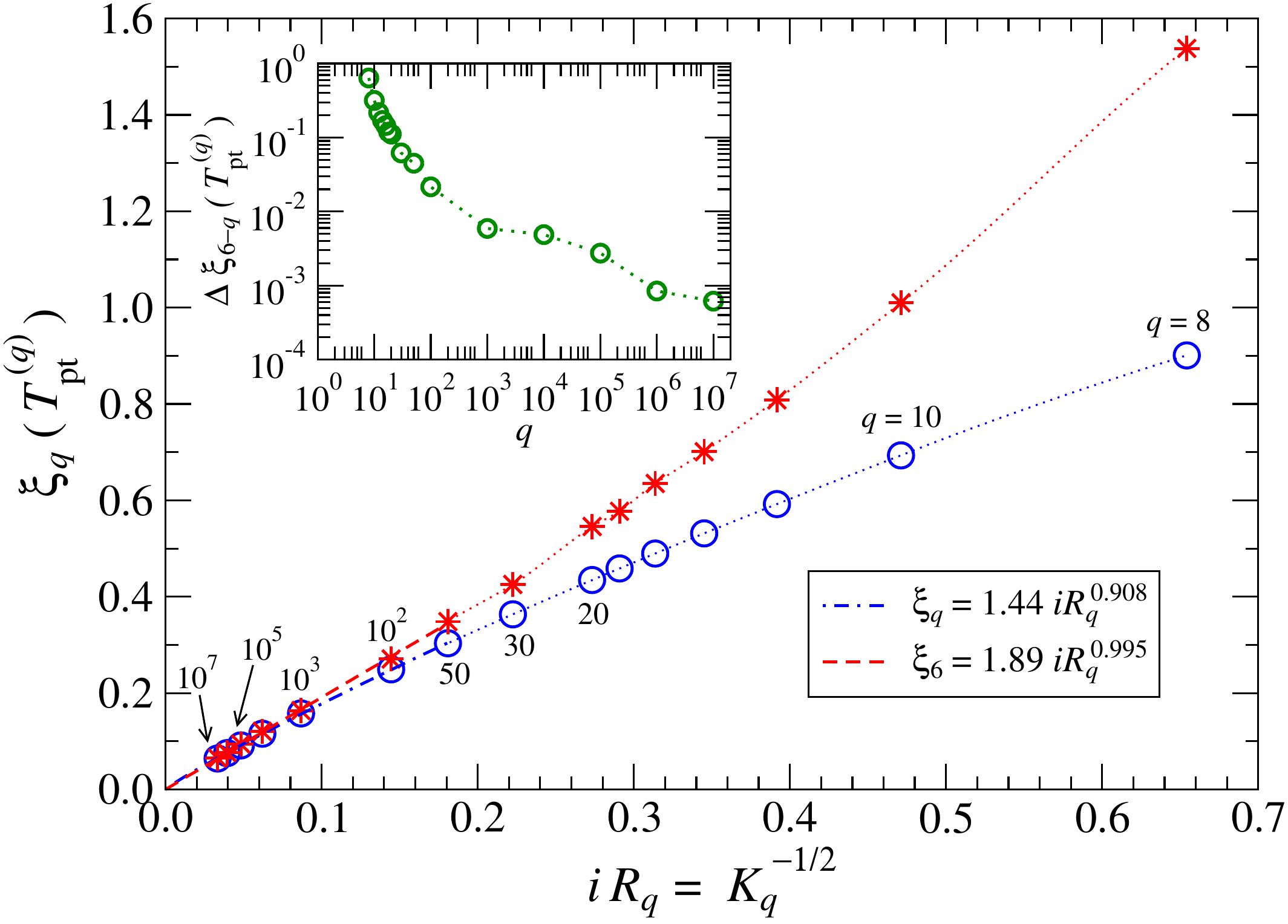}}
	\vspace*{-0.2cm}
\caption{Asymptotic scaling of the correlation length $\xi_q$ at the transition
temperatures $T_{\rm pt}^{(q)}$ with respect to $R_q$. The thin dotted lines
serve as a guide to the eye. The inset shows the difference $\Delta\xi_{6-q}
(T_{\rm pt}^{(q)})$ in Eq.~\eqref{deltaq} with respect to $q$ on a double 
logarithmic scale.}
\label{fig13}
\end{figure}

Figure~\ref{fig13} shows the dependence of the correlation length $\xi_q(T)$ at
the transition temperature with respect to the curvature radius $R_q$. In order
to collect these data, we performed extensive calculations up to 32 digits
numerical precision for the value of $q$ as large as $q = 10\,000\,000$ where
the corresponding Gaussian curvature $K_{10^7}$ is approximately  $900$. Note
that both quantities diverge on the $( 3, 6 )$ lattice, and therefore
$\xi_6(T_{\rm pt}^{(6)})$ and $R_6$ are not shown. Let us focus on the limit
$R_q\to0$ which corresponds to $q \rightarrow \infty$. Evidently, the correlation
length $\xi_q$ decreases to zero as $q$ tends toward infinity (the circles). 
Applying a least-square fit, we obtain the relation $\xi_q=1.44 (iR_q^{~})^{0.908}_{~}$
as shown by the thick dot-dashed curve. If we consider the error in the
calculation of the correlation length, we can conjecture that $\xi_q$ is
proportional to $R_q$.

Recall that the specific heat $C_v^{~}$, the internal energy $E_{\rm int}$,
and the entanglement entropy $S$ turned out to be weakly dependent on the
value of $q$ in the paramagnetic region $T>T_{\rm pt}^{(q)}$ for $q \ge 6$.
Thus, it can be conjectured that the disordered state is not modified by the
presence of the negative curvature. We, therefore, compare $\xi_{q\ge7}$ just
at the transition temperature $T_{\rm pt}^{(q)}$ with the correlation length
$\xi_6$ at the temperatures $T=T_{\rm pt}^{(q)}$. These values are plotted in
Fig.~\ref{fig13} by the asterisks. Since $T_{\rm pt}^{(q)}$ almost linearly
increases with $q$ for large values of $q$, the dotted line goes to the origin of the
graph. The circles and the asterisks in Fig.~\ref{fig13} are of the same order
for all $q$, and this fact supports our conjecture that $R_q$ represents the
only characteristic length of the hyperbolic lattice and that the phase
transition occurs at the temperature where $\xi_q$ is of the same order as
$R_q$. Note that $\xi_6(T_{\rm pt}^{(q)})>\xi_q(T_{\rm pt}^{(q)})$ is always
fulfilled as plotted in the inset of Fig.~\ref{fig13} where we show
the difference
\begin{equation}
\Delta\xi_{6-q}(T_{\rm pt}^{(q)})=
\left[\xi_6(T_{\rm pt}^{(q)})-\xi_q(T_{\rm pt}^{(q)})\right]\, .
\label{deltaq}
\end{equation}
The relation $\xi_6(T_{\rm pt}^{(q)})>\xi_q(T_{\rm pt}^{(q)})$ may be
explained by the effect of the negative curvature that prevents from a kind 
of {\it loop-back} of the correlation effect. Such suppression is also expected
to be present in higher-dimensional hyperbolic lattices and could be analytically
studied by means of the high temperature expansion.

We conjecture the reason why the correlation length remains finite even at the 
phase transition temperature $T_{\rm pt}^{(q)}$ for $q>6$, as follows. First of
all, the hyperbolic plane contains the typical length scale $R_q^{~}$, and it
might prevent scale invariance of the state expected at the criticality. A more
constructive interpretation could be obtained from the observation on the
row-to-row transfer matrix. The calculation of $\xi_q$ by means of Eq.~\eqref{corlen} 
requires diagonalization of the row-to-row transfer matrix ${\cal T}_q(\xi_1\sigma_a\xi_2|
\xi_1^\prime\sigma_a^\prime\xi_2^\prime)$ in Eq.~\eqref{TrMat}. 
The matrix corresponds to an area which connects (transfers) the
row of the neighboring spins $\{\xi_1\sigma_a\xi_2\}$ with the adjacent
ones $\{\xi_1^\prime\sigma_a^\prime\xi_2^\prime\}$. The shape of this area is 
very different from the standard transfer matrix on the Euclidean lattice,
which corresponds to a stripe of constant width. On the hyperbolic surfaces,
however, this distance between the spin rows is not uniform.
The distance is minimal at the center of the transfer matrix, i.e.,
between the two spins $\sigma_a$ and $\sigma_a^\prime$, and it increases 
exponentially with respect to the deviation  from the center to the direction of 
spin rows. Such a geometry~\cite{corrlen} could be imagined from the 
recurrence construction in Eq.~(6.8). As a consequence, the transfer matrix 
has an effective width, which is of the order of the curvature radius $R_q$. 
The region outside this width contributes as a sort of the boundary spins that
imposes mean-field effect to the bulk part. This situation is analogous to the Bethe 
lattice, being interpreted here as  ($\infty,q$)-lattices.~\cite{hctmrg-Ising-p-4}. 
Thus the Ising  universality could be observed only when the correlation 
length $\xi_q$ is far less than the curvature radius, $\xi_q \llless R_q$. As the
length $\xi_q$ increases toward the transition temperature, we expect a transient 
behavior to the mean-field 
behavior around the point when $\xi_q$ becomes comparable to $R_q$.

%%%%%%%%%%%%%%%%%%%%%%%%%%
% AGAGAGA
%%%%%%%%%%%%%%%%%%%%%%%%%%

\subsection{Ising model on the weakly curved lattices}

\subsubsection{The model}

\begin{figure}[tb]
\centering
\includegraphics[width=0.47\textwidth,clip]{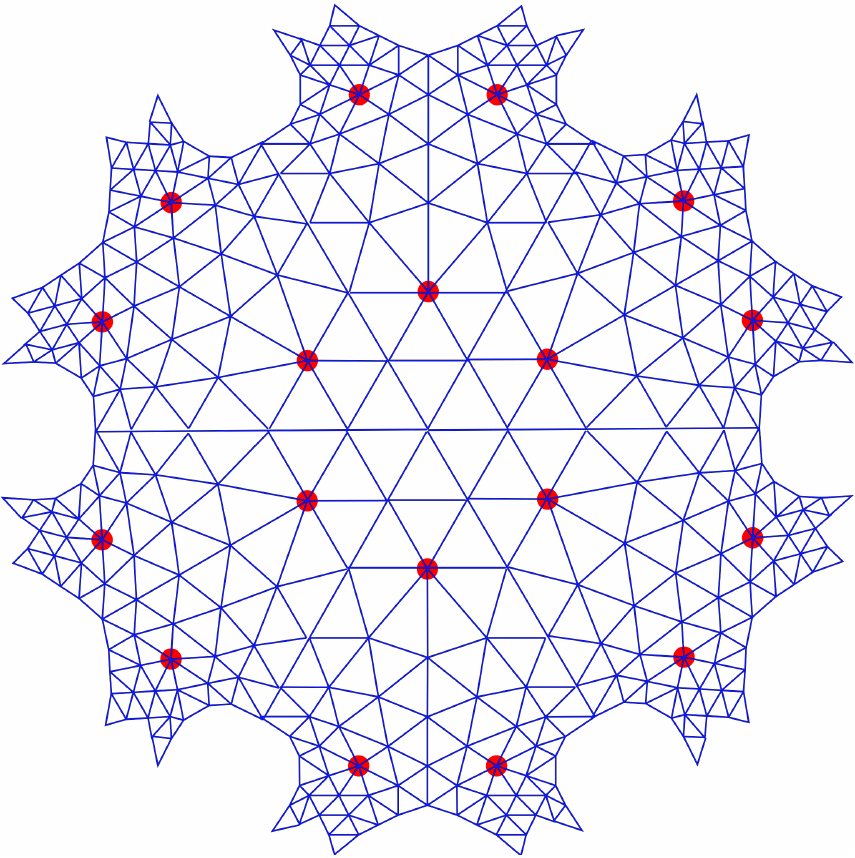}
\includegraphics[width=0.47\textwidth,clip]{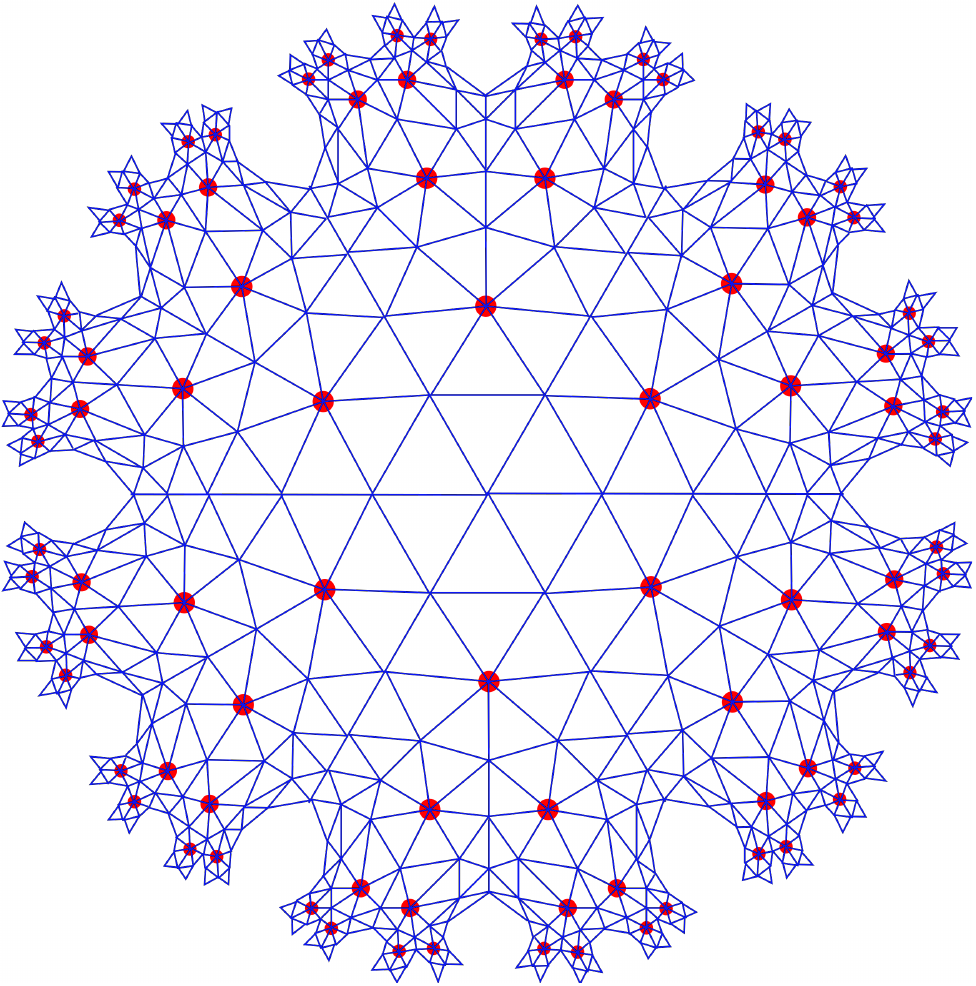}
\caption{The mixed lattices for
$n=1$ (left) and $n=2$ (right) in iteration $k = 5$. The filled
circles denote the exceptional lattice sites with the coordination
number seven, and number of the exceptional sites is $90$ in the left
and $18$ in the right, c.f. \eqref{SnM}.}
\label{3mixedlattice}
\end{figure}

In this section, we study the classical Ising model with Hamiltonian \eqref{Ising_hamiltonian} on a series of weakly curved lattices constructed by tessellation of triangles with non-constant coordination number $q$, which oscillates between integer values six and seven. The corresponding lattices are denoted as \emph{mixed} lattices and the vertices with coordination number seven are referred to as \emph{exceptional} sites. They are distributed regularly throughout the originally flat Euclidean $(3,6)$ lattice with the typical distance between nearest exceptional sites proportional to an integer $n$. Two examples of such lattice geometry are depicted in Fig.~\ref{3mixedlattice}. 
This concept is motivated by the fact that, assuming unit length of the lattice edges, although the $(3,7)$ lattice exhibits the least absolute value of the Gaussian curvature $\kappa$ among the hyperbolic $(3,q \geq 7)$ lattices, it is still far more curved than the Euclidean $(3,6)$ lattice. Indeed, the curvature radius $R=1/\sqrt{-K} \approx 0.917$ of the $(3,7)$ lattice (cf. \eqref{pqcurvature}) is of the order of the unit lattice edge length $l=1$, while $R=\infty$ on the Euclidean lattice. %The averaged curvature of the lattice with exceptional sites can, however, be controlled  
 However, surfaces with averaged curvature radii in between, i. e., $0.917 < R < \infty$, can be constructed by varying the parameter $n$ in the mixed lattices. As the integer $n$ increases, the flat triangular $(3,6)$ lattice is approached, which allows us to quantify the effect of the non-zero curvature to the order-disorder phase transition.
 
The mixed lattices are generated according to the extension scheme 
\begin{eqnarray}
\label{tm3qn}
\tilde{\mathbf L}^{(k+1)} &=& {\mathbf W_B} \, {\mathbf L}^{(k)}  \, , \nonumber\\
\tilde{\mathbf R}^{(k+1)} &=& {\mathbf W_B} \, {\mathbf R}^{(k)} \, , \\
\tilde{\mathbf C}^{(k+1)} &=& \left\{
\begin{array}{ll}
{\mathbf W_B} \, {\mathbf L}^{(k)} \left( {\mathbf C}^{(k)} \right)^2_{~} {\mathbf R}^{(k)}
\qquad ( {\rm at} ~ {\rm every} ~ n^{\rm th} ~{\rm step} ), \nonumber\\ 
{\mathbf W_B} \, {\mathbf L}^{(k)} \,\,\, {\mathbf C}^{(k)} \,\,\,\, {\mathbf R}^{(k)}
\qquad \,\,\,\,\, ( {\rm otherwise} ),
\end{array}
\right.
\end{eqnarray}
where the concept of the tensors ${\mathbf W}_B$, $\mathbf{L}$, $\mathbf{R}$ and $\mathbf{C}$ is analogous to the case of the $(3,q \geq 6)$ lattices in section~\ref{3qscheme}. These processes are almost identical to the extension scheme in \eqref{36lefttransfer}-\eqref{36corner} for the $(3,6)$ lattice, but when $k$ is a multiple of an integer parameter $n$, we insert an additional corner ${\mathbf C}^{(k)}$ in the extension process from ${\mathbf C}^{(k)}$ to $\tilde{\mathbf C}^{(k+1)}$. This process adds the exceptional lattice site with the coordination number seven whenever $(k\,\,{\rm mod}\,\,n)=0$. The tensors are initialized in the same manner as on the $(3,6)$ lattice. Note that we used the extension
process of ${\mathbf L}^{(k)}$ and ${\mathbf R}^{(k)}$ as in
\eqref{36lefttransfer} and \eqref{36righttransfer}. This restriction keeps the corner
${\mathbf C}^{(k)}$ symmetric to the spatial inversion; the
property is convenient for numerical calculations
by the CTMRG method. On the other hand, this simplification
introduces a slight inhomogeneity to the lattice, which
should be considered carefully.

\paragraph{I. Coordination number}
  
Examining the extension process in \eqref{tm3qn}, the total number ${\cal N}_n^{~}( k )$ of the lattice sites in the whole lattice
area $({\mathbf C}^{(k)})^6$ in iteration $k$ is calculated as \cite{hctmrg-Ising-3-qn}  
\begin{equation}
{\cal N}_n^{~}( k ) = 1+12\sum\limits_{j=1}^{k}
j\, 2^{M_n^{~}(nM_n^{~}(k,1)+n, \, j)}\, ,
\end{equation}
where $M_n(m,j)$ is the floor function
\begin{equation}
M_n(m,j)=\left\lfloor\frac{m-j}{n}\right\rfloor\equiv\max\left\{i\in{\mathbb Z}
\ \vert\ i\leq\frac{m-j}{n}\right\}\, .
\end{equation}
In the same manner, we can obtain the number of the exceptional
sites \cite{hctmrg-Ising-3-qn}
\begin{equation}
\label{SnM}
{\cal S}_n(k)=6\left[ 2^{M_n(k,1)} - 1 \right]
\end{equation}
for any set of $n$ and $k$. 

Considering the asymptotic limit $k \rightarrow \infty$, the ratio between
${\cal S}_n(k)$ and ${\cal N}_n(k)$ leads to the average density
of the exceptional sites
\begin{equation}
\lim\limits_{k \rightarrow \infty}^{~} \frac{{\cal S}_n(k)}{{\cal N}_n(k)}
= \frac{1}{2n(3n+1)} \, .
\end{equation}
%
%For sufficiently large $n$, the density becomes proportional to $n^{-2}_{~}$.
As a result, the averaged coordination number is
\begin{equation}
q_n^{~} = 6 + \frac{1}{2n(3n+1)} \, .
\label{qn}
\end{equation}
Note that $q_{\infty}^{~} = 6$ is the coordination number of the $( 3, 6 )$
lattice. Using the notation $q_n^{~}$ thus defined, we denote the
lattice constructed by \eqref{tm3qn} as the $( 3, q_n^{~} )$ lattice.

Length of the system lattice border ${\cal P}_n^{~}(k)$ in iteration $k$ is another essential
quantity that characterizes the geometry of the $( 3, q_n^{~} )$ lattice. 
The analytic formula of ${\cal P}_n^{~}(k)$ can be obtained as \cite{hctmrg-Ising-3-qn}
\begin{equation}
{\cal P}_{n}^{~}(k)=12\left[ k - n M_n^{~}(k,1) + n \sum\limits_{j=1}^{M_n^{~}(k,1)} 2^j_{~} \right] \, .
\end{equation}
It should be noted that the ratio of the boundary sites to the total number
of the lattice sites in the asymptotic limit
\begin{equation}
\lim\limits_{k \rightarrow \infty} \frac{{\cal P}_n(k)}{{\cal N}_n(k)} = 
\frac{2}{3 n + 1}
\end{equation}
is finite and inversely proportional to $n^{-1}$. Such a dominance of the boundary
sites over all lattice sites is a characteristic feature of the hyperbolic lattices. 
 The center of the $( 3, q_n^{~} )$ lattice, which represents our research target, the thermodynamic property of which we study, is, thus, surrounded by a wide system boundary.

\paragraph{II. Averaged curvature}

The hyperbolic nature of the $( 3, q_n^{~} )$ lattice
arises from the presence of the exceptional lattice sites which are distributed in
a sparse manner. Thus, when we consider the curvature of the $( 3, q_n^{~} )$
lattice, we have to take a certain average over the system. Apparently, such an
averaged curvature is dependent on the parameter $n$, and we write it as $\kappa_n^{~}$
in the following. Using \eqref{pqcurvature}, we evaluate the averaged curvature of the $(3,q_n)$ lattice with unit lattice edge length by
\begin{equation}
\kappa_n^{~} = - \left[2 {\rm arccosh}\left(\frac{\cos(\pi/p)}{\sin(\pi/q_n)}\right)\right]^2.
\label{Kn}
\end{equation}
Substituting
the asymptotic expression $q_n=6+1/6n^2$ from \eqref{qn} into
\eqref{Kn}, we obtain 
\begin{equation}
\kappa_n^{~} \, \sim \, - \frac{2}{3\pi} n^{-2}
\label{Kqn}
\end{equation}
with the dominant coefficient $2/3\pi \approx 0.212$ for large $n$.
Hence, the averaged curvature on the $(3,q_n)$ lattice $\kappa_n^{~} \propto -n^{-2}$.

\subsubsection{Numerical results}

We study the phase transition of the Ising model on the sequence
of the non-Euclidean $( 3, q_n^{~} )$ lattices, in particular,
\begin{equation}
( 3, q_1^{~} ), \quad ( 3, q_2^{~} ), \quad ( 3, q_3^{~} ), \quad
\cdots, \quad ( 3, q_{\infty}^{~} )\, .
\end{equation}
Without loss of generality, the coupling constant $J$ and the Boltzmann
constant $k_{ B}^{~}$ are set to unity. All thermodynamic functions
are considered in dimensionless units. The Boltzmann weight
tensor ${\mathbf W_B}$ of the elementary lattice rhombus characterized by spins $\sigma_a, \sigma_b, \sigma_c, \sigma_d$ is given by
\begin{align}
\begin{split}
 {\mathbf W_B}(\sigma_a, \sigma_b, \sigma_c, \sigma_d)=\exp
  &\left[\frac{J}{2k_{ B}T} (\sigma_a\sigma_b+\sigma_b\sigma_c
    +\sigma_c\sigma_d+\sigma_d\sigma_a+
        \right.
    \\
    &\left.	    
    +2\sigma_b\sigma_d) + 
    \frac{h}{6k_{ B}T} (\sigma_a+2\sigma_b+\xi\sigma_c+2\sigma_d)
  \right],
  \label{WB3qn}
\end{split}
\end{align}
which differs from \eqref{WB3q} (with $q=6$) only by the pre-factor $\xi$ in $\xi\sigma_c$. Normally, we set $\xi=1$, and $\xi$ is set to zero
when over-counting of interaction with external field $h$ happens at each
exceptional lattice point.  
 The reduced density matrix $\rho_n^{(k)}$ is calculated according to the standard definition \eqref{def_DM}, where we use the normalized tensor $\mathbf{A}^{(k)}_n = \left[{\mathbf C}^{(k)}\right]^3$ (cf. \eqref{Aqtensor}).
 
% \eqrefas {\color{red} a partial trace ${\rm Tr}^{\prime}$} of the six corner transfer tensors
%
%\begin{equation}
%\rho_n^{(k)} = {\rm Tr}^{\prime}\left[{\mathbf C}^{(k)}\right]^6_{~} %\, .
%\end{equation}
%

In our numerical calculations by CTMRG, we keep up to $m=200$ block spin
states, where we have
confirmed that all the data are converged with respect to $m$. As the 
iteration number $k$ increases, ${\mathbf C}^{(k)}$ approaches its thermodynamic limit during the
numerical calculations. Note that ${\mathbf C}^{(k)}$ possesses a minor 
dependence on $k$, since we keep inserting of the exceptional lattice sites
at every $n^{\rm th}$ extension step in accord with \eqref{tm3qn}.
We can either consider the cases where $k$ is multiple of $n$ or take the
average among the minor fluctuations. There is, however, no qualitative
difference in the two choices, and we have chosen the latter one. 
Again, we focus on the thermodynamic quantities deep inside
the system in order to suppress the boundary effects.

\begin{figure}[tb]
\centering
\includegraphics[width=4in]{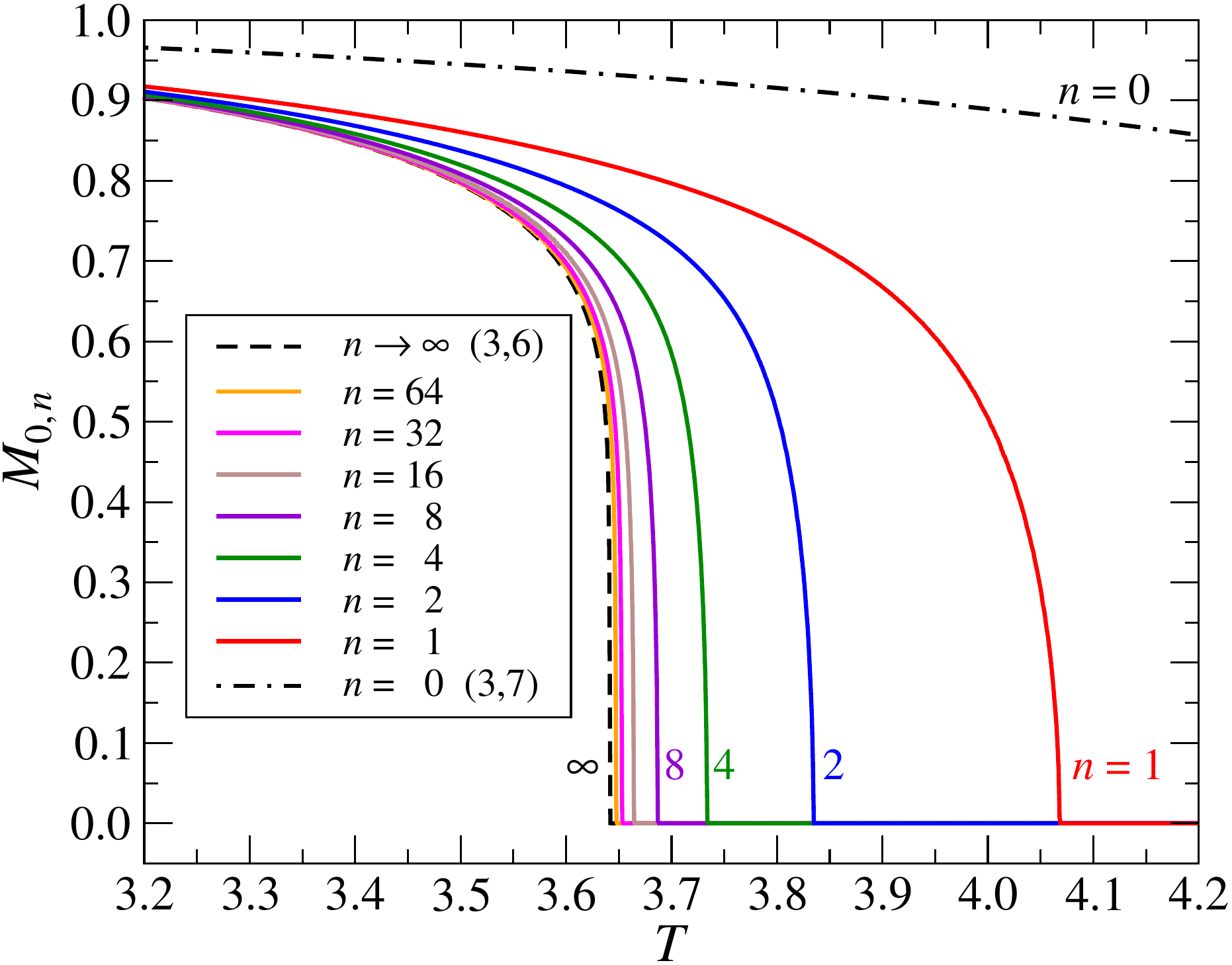}
\caption{Temperature dependence of the spontaneous
magnetization $M_{0,n}^{~}( T )$ on the $( 3, q_n^{~} )$ and $( 3, 7 )$
lattices.}
\label{Fig3qnmag}
\end{figure}

The spontaneous magnetization for the series of $(3,q_n)$ lattices
\begin{equation}
M_{0,n}^{~}( T ) \equiv M_{n}^{~}(h = 0, T ) =  \frac{{\rm Tr} \left( \sigma_{\ell}^{~} \rho_n^{~} \right)}{{\rm Tr}
\, \rho_n^{~}} 
\label{3qnmag}
\end{equation}
evaluated on the spin $\sigma_{\ell}$ at the center of the lattice system is displayed in Fig.~\ref{Fig3qnmag}. For comparison, we also
show the magnetization on the flat $( 3, 6 )$ lattice, denoted by $n \to \infty$,
as well as on the hyperbolic $( 3, 7 )$ lattice, denoted by $n = 0$.
Analogous notation by the subscript $n$ is also used for other thermodynamic quantities. 
The phase transition temperature $T_{{\rm pt} , n}^{~}$ monotonously decreases
with $n$ and approaches the analytically known value $T_{{\rm pt},\infty} \equiv T_{\rm pt}^{(6)} =
4 / \ln 3 \sim 3.64096$~\cite{Baxter} on the flat $( 3, 6 )$ lattice. Roughly
speaking, the difference $T_{{\rm pt} , n} - T_{{\rm pt} , \infty}$ is inversely proportional to $n$. 

\begin{figure}[tb]
\centering
\includegraphics[width=4in]{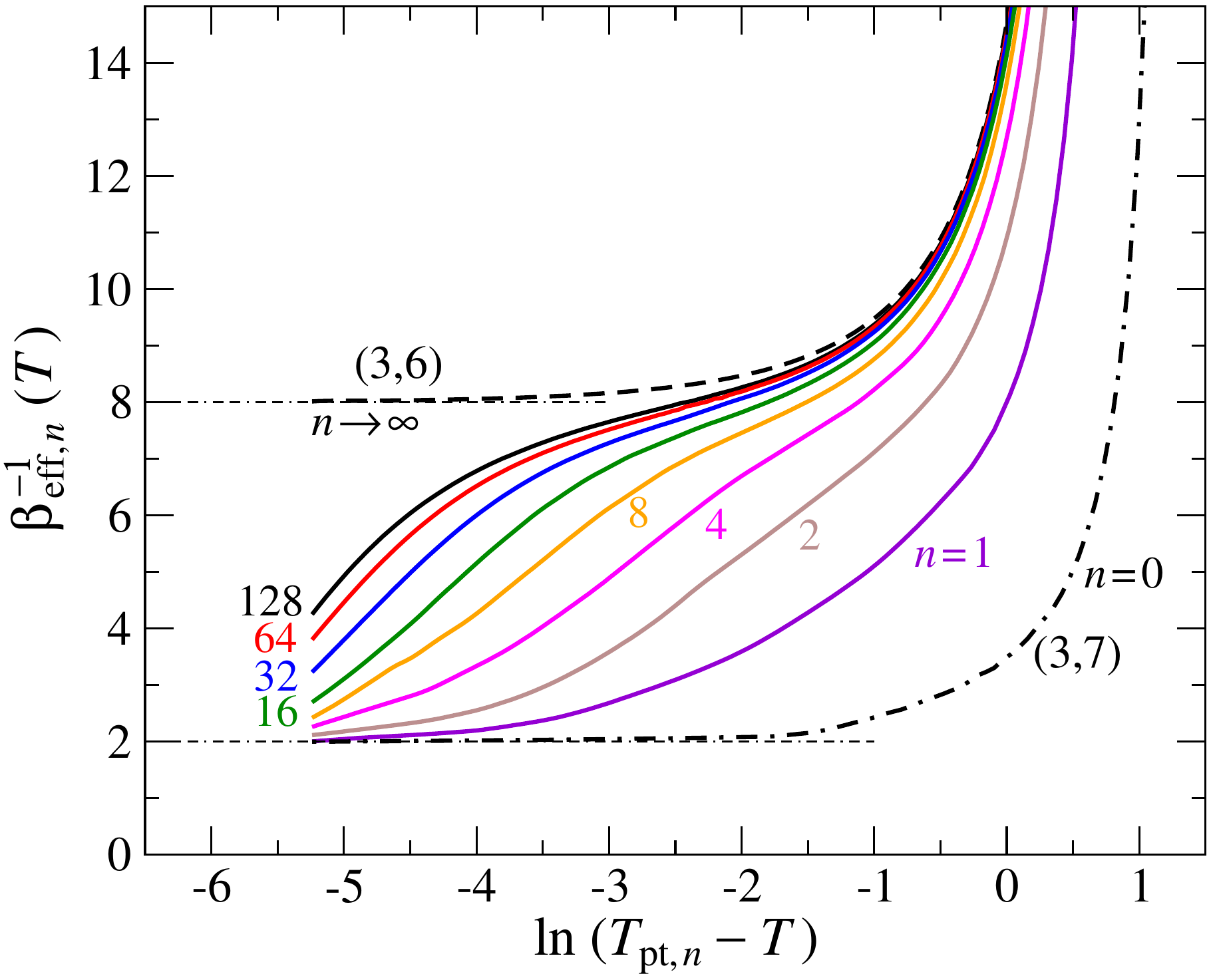}
\caption{Inverse of the effective magnetic exponent
$\beta_{{\rm eff} , n}^{~}( T )$ as a function of the logarithmic distance from the
transition temperature.}
\label{3qnbetaeff}
\end{figure}

%Just below the transition temperature, the power-law behaviour
%
%\begin{equation}
%M_n^{~}( T ) \propto {\left(T_n^{~} - T \right)}^{\beta_n}_{~} \, ,
%\end{equation}
%
%is expected. 
 In order to detect the magnetic exponent $\beta_n^{~}$ in the scaling relation $M_{0,n}(T) \propto (T_{{\rm pt} , n}-T)^{\beta_n}$, 
 we use the numerical derivative to calculate the effective exponent 
\begin{equation}
\beta_{{\rm eff}, n}^{~}( T ) = \frac{\partial \ln M_{0,n}^{~}( T )}{\partial \ln \left(T_{{\rm pt} , n} - T \right)} \, ,
\label{beff}
\end{equation}
within the ferromagnetic ordered phase $T \leq T_{{\rm pt} , n}$. 
Figure~\ref{3qnbetaeff} shows $\beta_{{\rm eff} , n}^{~}( T )$ thus obtained. 
When $T_{{\rm pt} , n} - T$ is relatively large, $\beta_{{\rm eff} , n}^{~}( T )$
follows the Ising universality value $\beta=\frac{1}{8}$,
however, in the neighbourhood of the transition temperature $T_{{\rm pt} , n}$,
the magnetic exponent $\beta_{{\rm eff} , n}^{~}$ for finite $n$ increases
and tends to $\beta_n=\frac{1}{2}$, the value which represents
the mean-field universality class.

\begin{figure}[tb]
\centering
\includegraphics[width=4in]{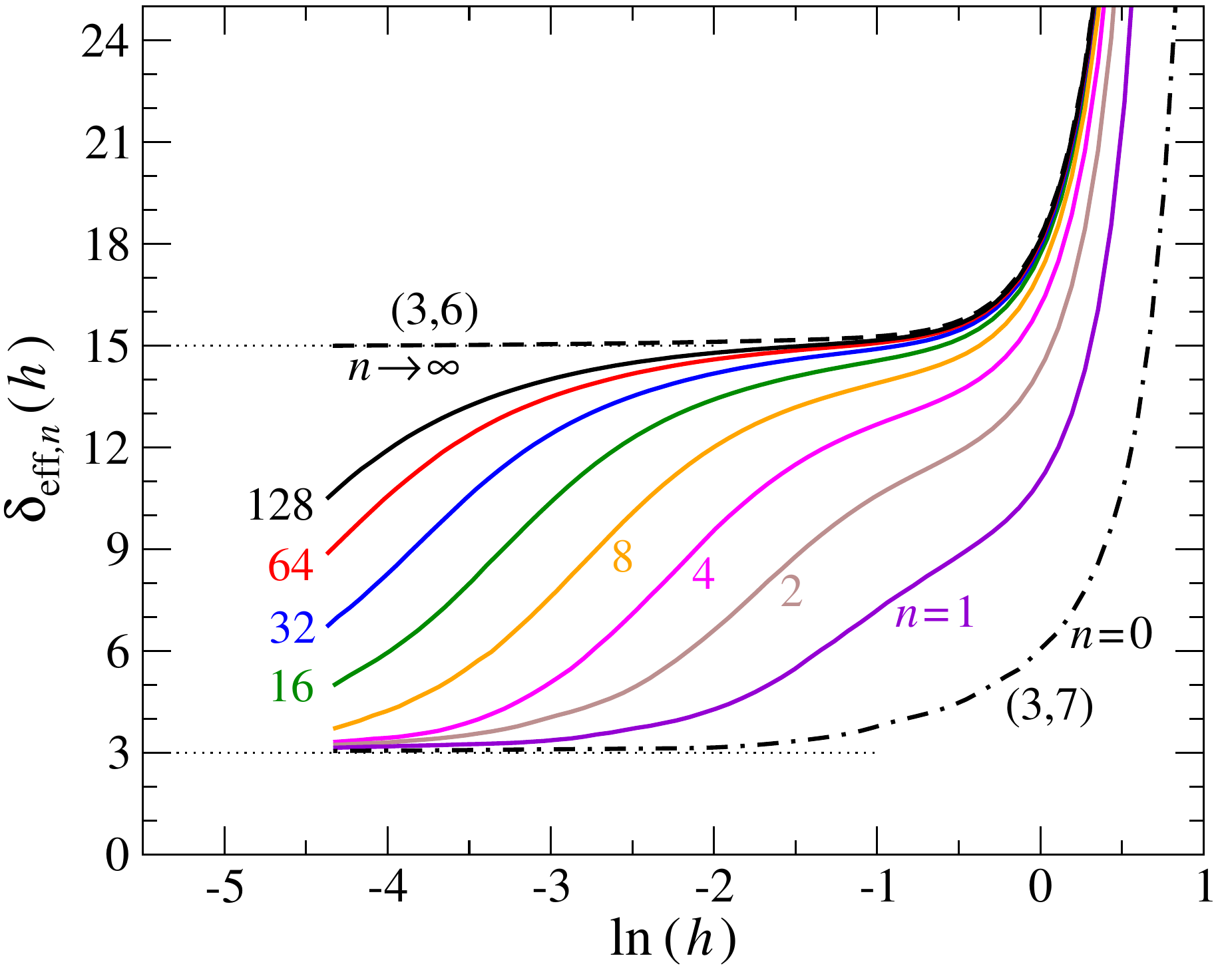}
\caption{Effective exponent $\delta_{{\rm eff}, n}^{~}( h )$ with
respect to the logarithm of the external magnetic field $h$ at the phase
transition temperature $T_{{\rm pt} , n}^{~}$.}
\label{3qndeltaeff}
\end{figure}

The critical exponent $\delta_n$ associated with the response of the magnetization to the uniform magnetic field $h$ at the phase transition temperature $T_{{\rm pt} , n}$ in the scaling relation $M_{ n}^{~}( h, T=T_{{\rm pt} , n} ) \propto h^{1/\delta_n}_{~}$ is evaluated in a similar manner. The effective exponent
\begin{equation}
\delta_{{\rm eff}, n}^{~}( h ) =
      {\left[
         \frac{\partial \ln M_n^{~}( h, T=T_{{\rm pt} , n}^{~} )}{\partial \ln h }
       \right]}^{-1}
\label{deff}
\end{equation}
obtained by numerical derivative in the limit $h\to0$ is shown in   Fig.~\ref{3qndeltaeff}. The observed behaviour qualitatively agrees with
that of the magnetic exponent $\beta_{{\rm eff} , n}$ depicted in Fig.~\ref{3qnbetaeff}.
The Ising universality value $\delta=15$ is obtained for the flat ($3,6$) lattice
only. It is obvious that the effective exponent $\delta_{{\rm eff} , n}^{~}( h )$ 
deviates from the Ising one when the external field becomes small, and
it again approaches the mean-field value $\delta_{{\rm eff} , n}^{~}(h \to 0)=3$ for any finite $n$.

\begin{figure}[tb]
\centering
\includegraphics[width=4in]{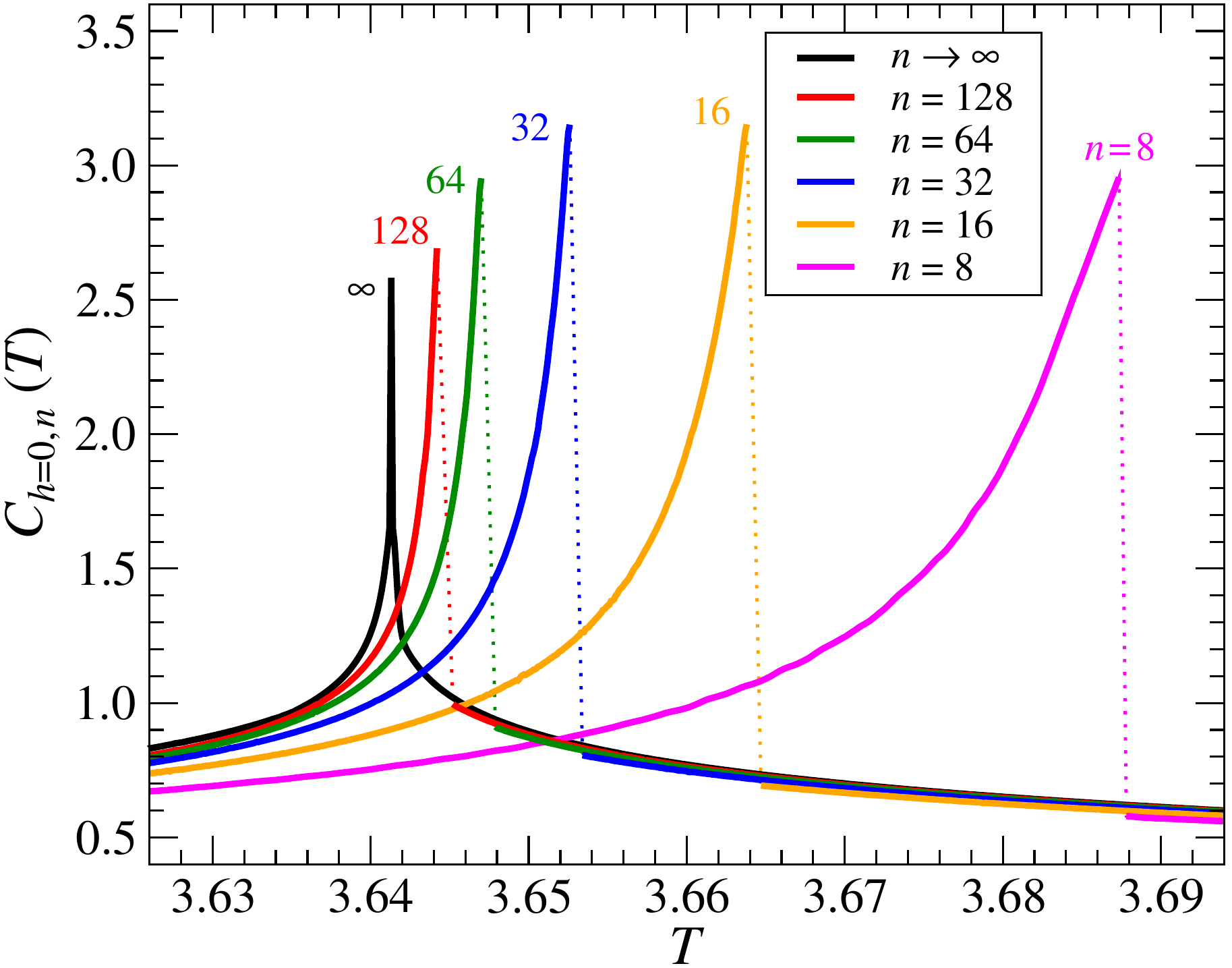}
\caption{The specific heat $C_{h,n}(h=0, T )$ on the $( 3, q_n^{~} )$ lattices.}
\label{3qnch}
\end{figure}

The internal energy per bond at the center of the system is
\begin{equation}
E_{{\rm int},n}^{~}(h=0, T ) =  -J\, \frac{{\rm Tr} \left( \sigma_{\ell} \sigma_{\ell'} \, \rho_{n}^{~} \right)}{
{\rm Tr} \, \rho_n^{~}} \, , 
\label{spc}
\end{equation}
where $\sigma_{\ell}^{~}$ and $\sigma_{\ell'}^{~}$ denote two neighbouring spins at the center.  Figure~\ref{3qnch} shows the
specific heat $C_{h,n}(h=0, T )=\partial E_{{\rm int},n}^{~}(h=0, T )/\partial T$, which is obtained by taking the numerical
derivative of $E_{{\rm int},n}^{~}(h=0, T )$ with respect to the temperature $T$. The maxima
of the specific heat for large $n$ are not obtained precisely, because
$E_{{\rm int},n}^{~}(h=0, T )$ around $T = T_{{\rm pt} , n}^{~}$ is very sensitive to a tiny numerical error.
The discontinuity in $C_{h,n}(h=0, T )$ for finite $n$ supports the fact that the
transition is of the mean-field nature. Note that the
specific heat, $C_{h,n}(h=0, T )$, in the disordered region $T \ge T_{{\rm pt} , n}$ for
various $n$ is close to $C_{h,\infty}(h=0, T )$ on the flat $( 3, 6 )$
lattice. This suggests a transient behaviour from the Ising universality to
the mean-field one which happens within the disordered phase.

\begin{figure}[!tbp]
\centering
\includegraphics[width=4in]{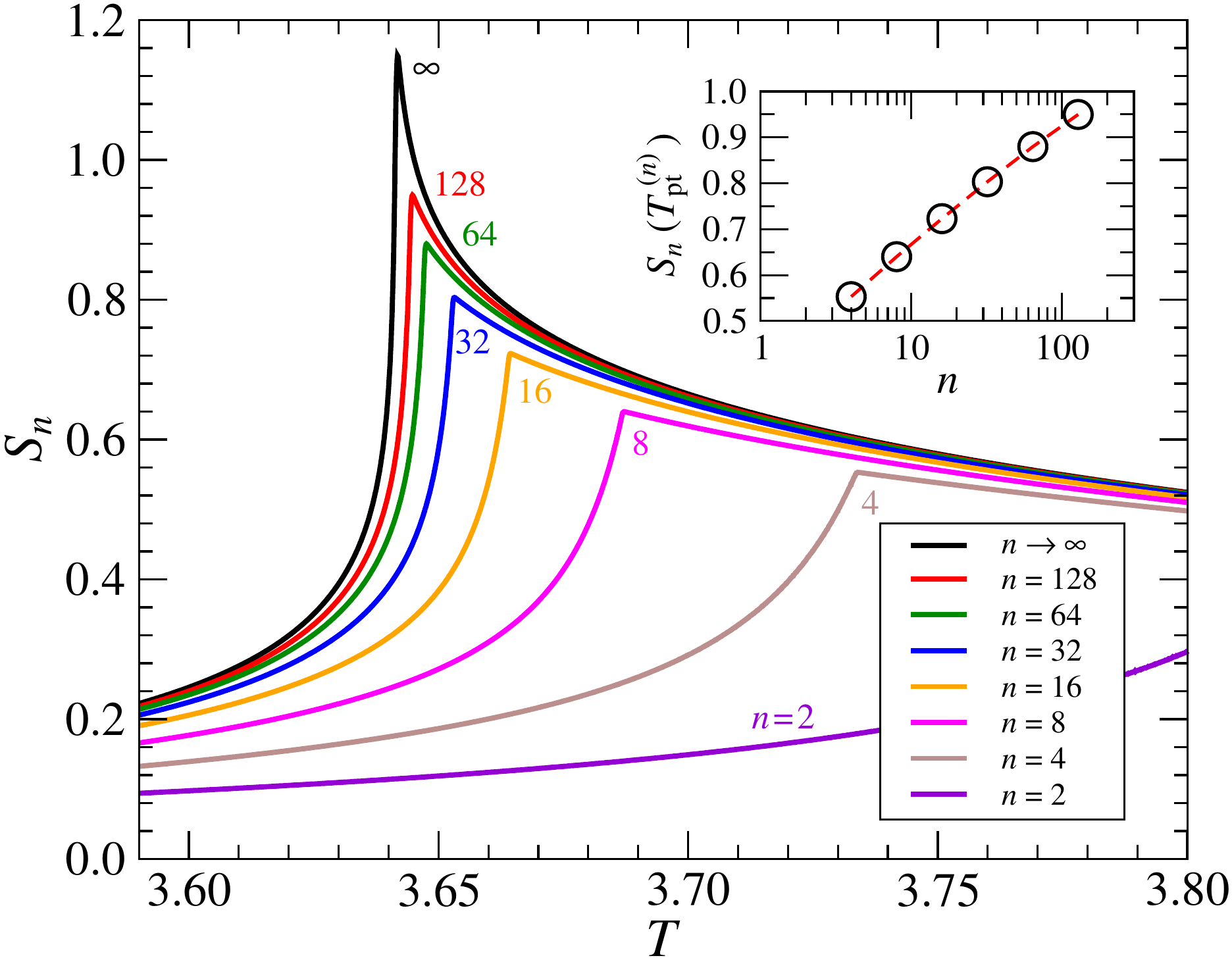}
\caption{Temperature dependence of the entanglement
entropy $S_n^{~}( T )$ with respect to $n$.}
\label{3qnentropy}
\end{figure}

As an independent measure of the phase transition, we look at the entanglement entropy 
$S_n^{~}$, which can be directly computed from the reduced density matrix spectrum
\begin{equation}
S_n^{~}( T ) = - 
     {\rm Tr} \left( \rho_{n}^{~} \,  \ln \rho_{n}^{~} \right) \, ,
\label{ent}
\end{equation}
where the reduced density matrices are normalized satisfying the condition
${\rm Tr}\,\rho_n=1$. Figure~\ref{3qnentropy} shows  $S_n^{~}( T )$, where the peak
values, $S_n^{~}( T_{{\rm pt}, n}^{~} )$, are shown in the inset. If the curvature radius
$R_n^{~}=\sqrt{-\kappa_n}$ controls the typical length scale at the transition temperature,
it is expected that $S_n^{~}( T_{{\rm pt}, n}^{~} )$ behaves as
\begin{equation}
S_n^{~}( T_{{\rm pt}, n}^{~} ) \, \sim \, \frac{c}{6} \, \ln \, R_n^{~} \, ,
\end{equation}
where $c$ is the central charge of the system. As shown in Fig.~\ref{3qnentropy},
the increase in $S_n^{~}( T_{{\rm pt}, n}^{~} )$ is close to the value $( \ln 2 ) / 12
= 0.05776$ when $n$ doubles, and the fitted value of the slope in the inset
gives $c \sim  0.48$. This value is consistent with $c = 1/2$ in the Ising
universality class. For this reason, our conjecture about the presence of
the typical length scale at $T_{{\rm pt},n}$, which is proportional to $n$
($R_n^{~}\propto 1/n$), is numerically supported.

\vfill
\phantom{where $c$ is the central charge of the system. As shown in Fig.~\ref{3qnentropy},
the increase in $S_n^{~}( T_{{\rm pt}, n}^{~} )$ is close to the value $( \ln 2 ) / 12
= 0.05776$ when $n$ doubles, and the fitted value of the slope in the inset
gives $c \sim  0.48$. This value is consistent with $c = 1/2$ in the Ising
universality class. For this reason, our conjecture about the presence of
the typical length scale at $T_{{\rm pt},n}$, which is proportional to $n$
($R_n^{~}\propto 1/n$), is numerically supported.
where $c$ is the central charge of the system. As shown in Fig.~\ref{3qnentropy},
the increase in $S_n^{~}( T_{{\rm pt}, n}^{~} )$ is close to the value $( \ln 2 ) / 12
= 0.05776$ when $n$ doubles, and the fitted value of the slope in the inset
gives $c \sim  0.48$. This value is consistent with $c = 1/2$ in the Ising
universality class. For this reason, our conjecture about the presence of
the typical length scale at $T_{{\rm pt},n}$, which is proportional to $n$
($R_n^{~}\propto 1/n$), is numerically supported.
where $c$ is the central charge of the system. As shown in Fig.~\ref{3qnentropy},
the increase in $S_n^{~}( T_{{\rm pt}, n}^{~} )$ is close to the value $( \ln 2 ) / 12
= 0.05776$ when $n$ doubles, and the fitted value of the slope in the inset
gives $c \sim  0.48$. This value is consistent with $c = 1/2$ in the Ising
universality class. For this reason, our conjecture about the presence of
the typical length scale at $T_{{\rm pt},n}$, which is proportional to $n$
($R_n^{~}\propto 1/n$), is numerically supported.}
\vfill
\newpage
\phantom{.}

\newpage\setcounter{equation}{0} \setcounter{figure}{0} \setcounter{table}{0}
%\section{Quantum spin systems on hyperbolic geometries}
\section{Quantum spin models on hyperbolic lattices}
\label{chap:quantum}

In this chapter we apply the Tensor Product Variational Formulation (TPVF) described in section~\ref{TPVF} to quantum spin systems in the thermodynamic limit on the $(p,q)$ hyperbolic lattices and the $(4,4)$ Euclidean lattice. First, we investigate the effect of the changing lattice parameter $p$
on the series of $(p,4)$ lattices with constant coordination number $q=4$, where $p \in \{5,6, \dots, 11\}$. We analyze the critical phenomena of the transverse field Ising model (TFIM), the XY and the modified Heisenberg model. In analogy to previous studies of classical spin models on these hyperbolic lattices \cite{hctmrg-Ising-5-4,hctmrg-Ising-p-4}, we expect fast convergence of the phase transition magnetic field of the quantum TFIM as well as the ground-state energies of the quantum XY and the modified Heisenberg models toward the asymptotic case $p \rightarrow \infty$, which represents the Bethe lattice \cite{hctmrg-Ising-p-4}. Numerical results presented in the following sections are in complete agreement with the expectations. The key feature of this work is the consequent indirect analysis of the quantum TFIM, XY, and Heisenberg models on the Bethe lattice with coordination number four, which has not been considered yet. %We also conjectured that TPVF is more suitable for models on the hyperbolic lattices than on the Euclidean one, as already mentioned in section~\ref{TPVF}.
\tabcolsep=1pt

An analogous study is performed on the complementary set of the $(4,q)$ lattices, where the transverse field Ising model is investigated. Here, we determine the phase transition fields on the respective lattices by maximizing the von Neumann entropy of the system. Assuming the results of the studies of the classical Ising model on the $(p,q)$ lattices \cite{hctmrg-Ising-3-q, Serina}, asymptotically linear dependence of the transition field on the coordination number $q$ was expected. However, the numerical results indicate the polynomial behaviour. Complete results for these quantum models on the series of the $(4,q)$ lattices will be published elsewhere. %\cite{TPVF4q}.

\subsection{Spin models on the $(p,4)$ lattices}

\subsubsection{The model}
\label{quantum_p4_model}

We study the ground-state properties and the phase transition of the quantum TFIM, XY, and modified Heisenberg models in the thermodynamic limit on a series of hyperbolic $(p,4)$ lattices with the lattice parameter $p \in \{5,6,\dots, 11\}$. Apart from the set, we include two additional cases: $p=4$ being the Euclidean square lattice and the asymptotic case $p\to\infty$, which is associated to the Bethe lattice. %Figure~\ref{Fig1} {\color{red} Obrazok je odkomentovany. Treba ho?} depicts the typical structure of the lattices. 
 The Euclidean $(4,4)$ lattice serves as the reference lattice, which allows us to compare the results obtained by TPVF with the outcomes of other numerical algorithms. Thus we can estimate the numerical inaccuracy of the TPVF algorithm, which varied from $1.2\%$ in the XY model to $3.7\%$ in TFIM at the phase transition. Analogous results for models on hyperbolic lattices are not available yet. We, however, expect significantly higher accuracy of the TPVF results on hyperbolic lattices, which are of our main interest, than on the Euclidean one, as already mentioned in section~\ref{TPVF}.

The Hamiltonian of the three models is given by formula \eqref{Hm1},  where the spin variables are positioned in the lattices vertices.  We consider the ferromagnetic Ising and the XY model with $J_{xy}=0$, $J_z=J$ and $J_{xy}=J$, $J_z=0$, respectively.  Without loss of generality, we set $J=1>0$. The modified Heisenberg model is specified by the choice $J_{xy}=J=1$, $J_z=-J=-1$, which corresponds to the unitary transformation of the  antiferromagnetic Heisenberg model with $J_{xy}=-1$, $J_z=-1$ if the coordination number $q$ is even, as discussed in section~\ref{Quantum_spin_models}.  The results for this specific type of the Heisenberg model may differ from the exact results, since the ground-state is antiferromagnetic if $p$ is even. However, in case $p=4$, a simple calculation confirms that the classical system corresponding to the 
optimal tensor ${ Z}_p$ in the role of the classical Boltzmann weight tensor ${\mathbf W_B}$ is ferromagnetic. Moreover, the relative error of the ground-state energy $E_0^{(4)}$ of this model if compared to the reference value \cite{HoSRG} is $2.2\%$ only. In case $p$ is odd, exact diagonalization for small lattice systems suggests that the  ground-state is ferromagnetic, although we have not managed to prove this property in general yet.

\subsubsection{Numerical results}
\label{quantum_num_results}

\paragraph{I. The transverse field Ising model}

The TFIM undergoes a quantum phase transition\footnote{We intentionally avoid the terms \emph{critical point}, \emph{critical field} and the corresponding index $C$ on hyperbolic lattices, since  the studies of the classical models on these lattices \cite{corrlen, hctmrg-Ising-3-q} conjecture that the correlation length $\xi$ remains finite at the transition. Analogous behaviour in case of quantum systems is also expected.}
 at a nonzero magnetic field $h_t^{(p)}>0$, where we explicitly emphasize its dependence on the lattice geometry. The nonzero spontaneous magnetization $\left\langle S^z(h)\right\rangle$ in the ordered phase at $h<h_t^{(p)}$ breaks the spin-inversion symmetry, which results in approximately twice larger set of the free variational parameters $N_{\rm Ising}^{(p)}$ in the TPVF algorithm if compared to $N_{\rm Heis}^{(p)}$ in the XY and Heisenberg models, cf.~table~\ref{Tab_num_var_parameters}. The computational time for a particular fixed field $h$ is, therefore, significantly prolonged. Moreover, in order to screen the vicinity of the phase transition field $h_t^{(p)}$, multiple calculations for a sequence of magnetic fields $h$ had to be performed. As a consequence, in order to restrict the total computational time, we have analyzed the TFIM on the hyperbolic lattices up to $p=10$ only. (Notice that the number of block spins states kept was $m=20$ for $p \in \{4,5, \dots, 8\}$, and only $m=4$ for $p \in \{9,10\}$, which was sufficient due to exponentially weak correlations caused by the hyperbolic lattice geometry~\cite{hctmrg-Ising-3-q}; any further increase of the states kept $m$ has not improved the numerical calculations significantly).

\begin{figure}[tb]
\centering
\includegraphics[width=4.5in]{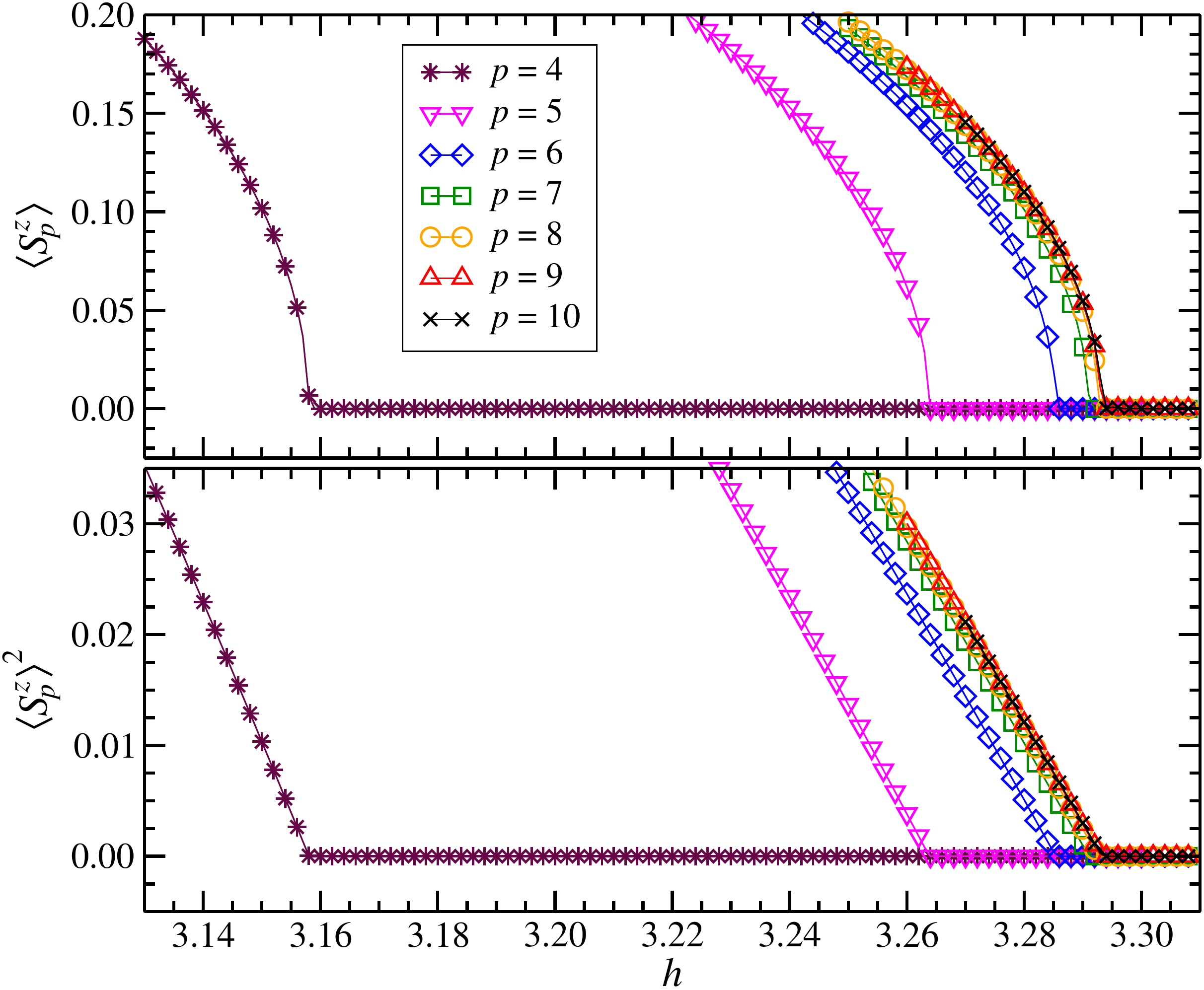}
\caption{The spontaneous magnetization $\langle S_p^{z}\rangle$ (the upper
graph) and its square ${\langle S_p^{z}\rangle}^{2}$ (the lower graph) in the vicinity of the phase transitions with respect to the magnetic field $h$ for $p \in \{4,5, \dots, 10\}$.}
\label{p4magnetizacia}
\end{figure}

We have analyzed the phase transition of the TFIM by the expectation value of the spontaneous magnetization $\langle S_p^{z} \rangle$ as well as by the magnetic susceptibility $\chi_p$. Solving the minimization problem in~\eqref{e0}, we received the optimal tensor elements $W^*_p(\{\sigma\})$, which uniquely define the approximative ground state $|\Psi_p^*\rangle$ via~\eqref{Psi}. Once $|\Psi_p^*\rangle$ has been constructed, we evaluated the spontaneous magnetization   
\begin{equation}
\langle S_p^{z} \rangle = \frac{\langle \Psi^*_p | \boldsymbol\sigma_{\ell}^{z} |\Psi^*_p \rangle}
                                    {\langle \Psi^*_p |                   \Psi^*_p \rangle} \, ,
\label{sz}
\end{equation}
where ${\ell}$ labels an arbitrary spin in the central polygon of the lattice in order to suppress boundary effects. Here, $\langle S_p^{z}\rangle$ denotes the order parameter of TFIM and specifies the
quantum phase transition at the phase transition field. The resulting dependence of the magnetization $\langle S_p^{z}\rangle$ with respect
to the magnetic field $h$ near the phase transition field $h_t^{(p)}$ is plotted in the upper graph of Figure~\ref{p4magnetizacia}. 
The quantum phase transition of the TFIM is characterized by a non-analytic behaviour of the magnetization curve, when $\langle S_p^{z}\rangle \to 0$ if approaching the phase transition field $h \to h_t^{(p)}$ from the ordered phase ($h < h_t^{(p)}$). 

The phase transition exponent $\beta_p$, which depends on the lattice geometry, describes the singularity through the scaling relation in the ordered phase, cf. \eqref{crit_sz}, 
\begin{equation}
\langle S_p^z (h) \rangle \propto {\left(h_t^{(p)} - h \right)}^{\beta_p} \, .
\label{powerlaw}
\end{equation}
Figure~\ref{p4magnetizacia} (the lower graph) shows the squared transversal magnetization $ {\langle S_p^{z}\rangle}^{2}$, where we point out the linearity of the squared magnetization if approaching the phase transition field $h_t^{(p)}$. Such a dependence confirms the mean-field exponent $\beta_p=\frac{1}{2}$ regardless of the lattice parameter $p$, which results in the mean-field-like behaviour of the TFIM if approaching the phase transition. The incorrect mean-field-like behaviour near the phase transition on the Euclidean lattice represented by the mean-field value $\beta_4=\frac{1}{2}$ is attributed to the exclusion of long-range correlations caused by the TPS approximation \eqref{Psi} which is built up by the tensors $W_4$ of the too low dimension. As a reference, the numerical TRG analysis~\cite{HoSRG} gives
correct $\beta_4^{\rm TRG} = 0.3295$ on the Euclidean $(4,4)$ lattice, which is also in agreement with Monte Carlo simulations.

\begin{figure}[tb]
\centering
\includegraphics[width=4in]{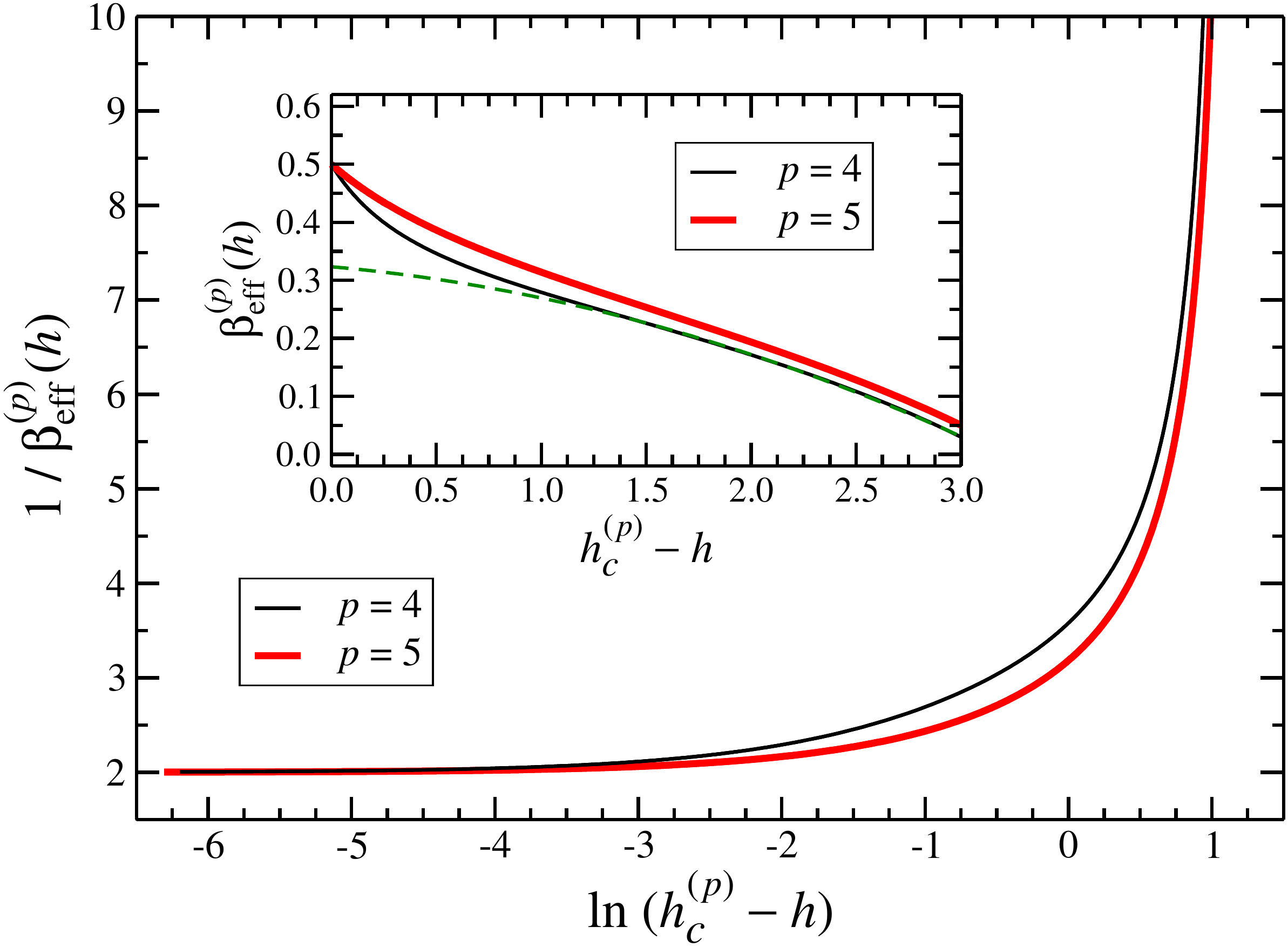}
\caption{The detailed dependence of the inverse effective magnetic
exponent $\beta^{(p)}_{\rm eff} (h)$ on the magnetic field in the logarithmic form for the Euclidean $(4,4)$ and the pentagonal $(5,4)$ lattice. The inset shows the effective
exponent at wider magnetic field scale. The green dashed line estimates behaviour of
the correct effective exponent for the Euclidean lattice.}
\label{p4-beff-fig}
\end{figure}

A more detailed analysis of the influence of the TPS approximation near the phase transition can be
visualized by evaluating the effective (magnetic-field dependent) exponent $\beta^{(p)}_{\rm eff}(h)$,
which converges to $\beta_p$ when approaching the phase transition field $h_t^{(p)}$
\begin{equation}
\beta_p = \lim\limits_{h \to h_t^{(p)}} \beta^{(p)}_{\rm eff} (h)
        = \lim\limits_{h \to h_t^{(p)}} \frac{\partial \ln\langle S_p^z (h) \rangle}
                                             {\partial \ln \left(h_t^{(p)} -  h \right)}.
\label{p4beta_eff}
\end{equation}
Figure~\ref{p4-beff-fig} shows the dependence of $\beta^{(p)}_{\rm eff}(h)$ on the magnetic field in case of the Euclidean $(4,4)$ and the pentagonal $(5,4)$ lattice. 
The effective exponent obviously converges to the
mean-field exponent $\beta_p=\frac{1}{2}$ for both lattice types if the phase transition
field is approached from the ordered phase, i.e., if $\ln(h_t^{(p)}-h) \to-\infty$. The
inset shows the same dependence on larger scales. The critical exponent on the square
lattice (the black curve for $p=4$) starts deviating at around $h>2.0$ from the expected
exponent (estimated by the blue dashed curve), which is known to converge to
$\beta_4^{\rm TRG}=0.3295$~\cite{HoSRG}. Knowing that $\beta^{(p)}_{\rm eff} (h)$ is expected to converge to $\beta_p=\frac{1}{2}$ at the transition field $h_t^{(p)}$, the value of $h_t^{(p)}$ can be determined at high precision. It is performed by varying $h_t^{(p)}$ in \eqref{p4beta_eff} so that $1/\beta^{(p)}_{\rm eff} (h)$ is as close as possible to the value $2$ as $\ln \left(h_t^{(p)} -  h \right) \to -\infty$.    

\def\arraystretch{1.6}
\tabcolsep=3pt
\begin{table}[tb]
\begin{center}
\begin{tabular}{| c | c | c | c | c | }
\hline
 $p$ 		&	$4$ & $5$ & $6$ & $7$  \\ \hline
 $h_t^{(p)}$& 	$3.158034$ & $3.263825$ & $3.285405$ & $3.291055$  \\ \hline
  $\Delta^{(p)}$ 	&	$1\times10^{-6}$ & $1\times10^{-6}$ & $1\times10^{-6}$ & $1\times10^{-6}$  \\ \hline\hline
 $p$ 		&	$8$ & $9$ & $10$ & $\infty$ \\ \hline
 $h_t^{(p)}$& 	$3.292647$ & $3.293113$ & $3.293263$& $3.29332$  \\ \hline
  $\Delta^{(p)}$ 	&	$2\times10^{-6}$ & $2\times10^{-6}$ & $5\times10^{-6}$ & $1\times10^{-5}$  \\ \hline
\end{tabular}
\end{center}
\caption{The phase transition fields $h_t^{(p)}$ of the TFIM including the estimated errors $\Delta^{(p)}$ with respect to the lattice parameter $p$.}
\label{Tab_p4_tranition}
\end{table}
\tabcolsep=1pt

\begin{figure}[tb]
\centering
\includegraphics[width=4.2in]{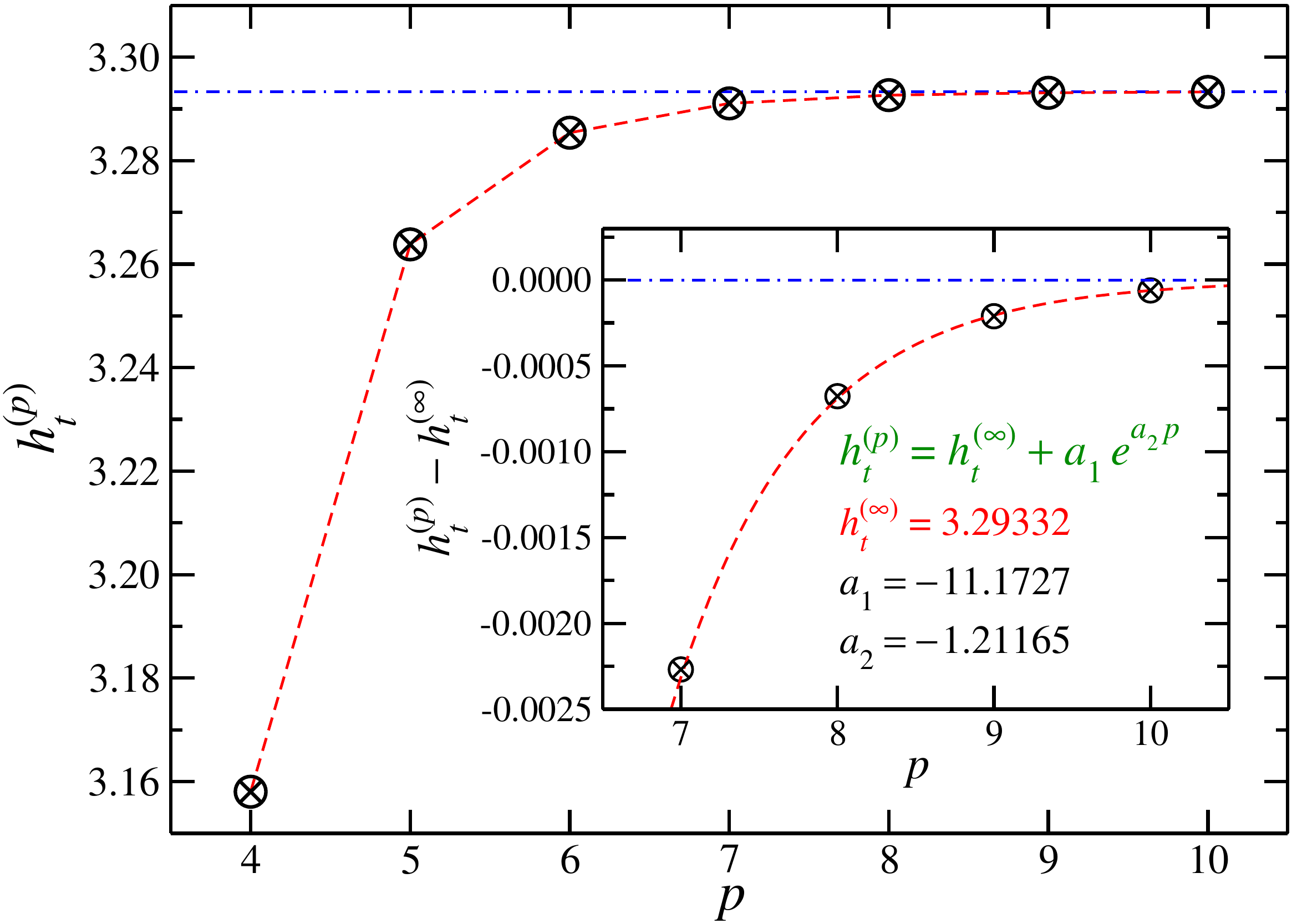}
\caption{The phase transition field $h_t^{(p)}$ of the TFIM with respect
to the lattice parameter $p$. The horizontal dot-dashed line represents the estimated
asymptotic value $h_t^{(\infty)}=3.29332$.}
\label{p4transition}
\end{figure}

The phase transition fields $h_t^{(p)}$, calculated according to the above mentioned method, are summarized in Table~\ref{Tab_p4_tranition} together with their errors $\Delta^{(p)}$. Notice that $\Delta^{(p)}$ represents only the error of the method providing that the calculated magnetization $\langle S_p^{z}\rangle$ is considered accurate. The most relevant value of the critical magnetic field for the TFIM on the Euclidean $(4,4)$ lattice by the TRG algorithm yields $h_t^{(4),{\rm TRG}} = 3.0439$~\cite{HoSRG}. The relative error of our result is thus $3.7\%$. The data are graphically plotted in Fig.~\ref{p4transition}, whereas the error bars are too small to be displayed. 

The monotonically increasing and rapidly saturating curve $h_t^{(p)}$  allows us to perform a meaningful extrapolation estimate of the transition field $h_t^{(\infty)}$ on the Bethe lattice.  
The fitting function is proposed in the form
 \begin{equation}
 h_t^{(p)} = h_t^{(\infty)} +a_1 \exp(a_2 p ) \, ,
 \label{fith_t}
\end{equation}
 where $h_t^{(\infty)}$, $a_1$, and $a_2$ are the fitting parameters, which were determined in the following way. First we defined a function $f(h)$, which returns the residual sum of squares ($RSS$) of the linear regression $\ln\vert h-h_t^{(p)}\vert = \ln\vert a_1\vert +a_2 p $. Then, $h_t^{(\infty)}$ was chosen as the argument, which minimizes the function $f(h)$. The corresponding linear regression $\ln\vert h_t^{(\infty)}- h_t^{(p)} \vert = \ln\vert a_1\vert + a_2 p $ specifies the parameters $a_1$ and $a_2$. If considering another way, $h_t^{(\infty)}$ is such a value that the curve $\ln\vert h_t^{(\infty)}- h_t^{(p)}\vert$ is as close as possible to a line, where the closeness is measured by the $RSS$. Applying this exponential fitting function to the critical magnetic fields $h_t^{(p)}$ for $p \in \{6, \dots, 10\}$
\footnote{We excluded $h_t^{(4)}$ from the fit, since TPVF is less accurate on the Euclidean lattice. The point $h_t^{(5)}$ was also excluded in order to restrict the fit to the tail of the curve.}
, we calculated the asymptotic phase transition field of the TFIM on the Bethe lattice $h_t^{(\infty)}=3.29332$ as listed in Table~\ref{Tab_p4_tranition}.

\begin{figure}[tb]
\centering
\includegraphics[width=4.5in]{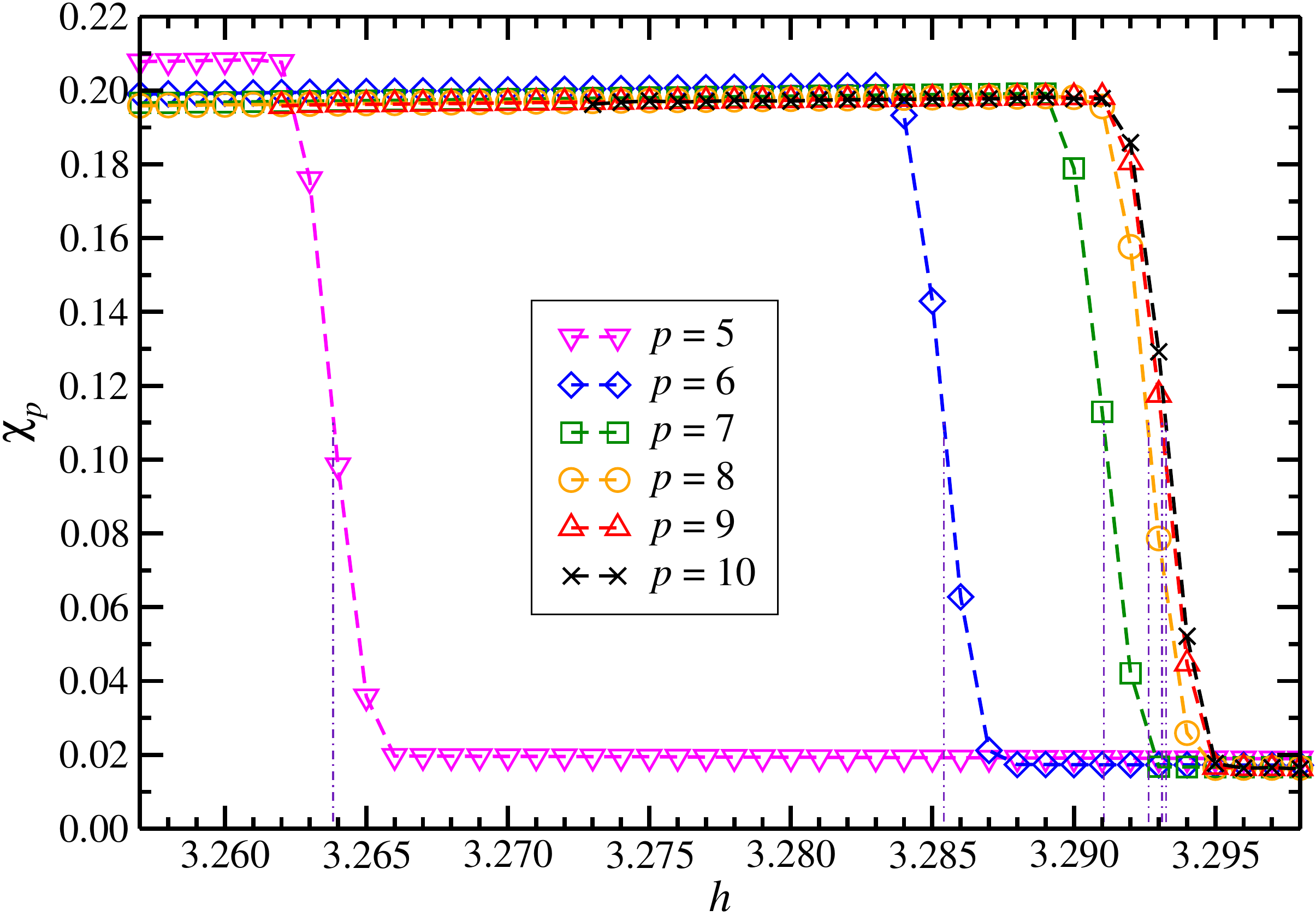}
\caption{The magnetic susceptibility $\chi_p$ of the TFIM as a function of the magnetic field $h$ for the hyperbolic lattices with $p \in \{5, \dots,10\}$. The vertical dot-dashed lines serve as guides for the eye and correspond to the phase transitions $h_t^{(p)}$.}
\label{p4susceptibility}
\end{figure}

Another independent way of obtaining (and confirming) the phase transition fields $h_t^{(p)}$ can be carried out by analyzing the magnetic susceptibility 
\begin{equation}
\chi_p=-\frac{\partial^2 E_0^{(p)}}{\partial h^2}\, .
\label{p4chi}
\end{equation}
The functional dependence of the susceptibility on the magnetic field $h$ is shown in Fig.~\ref{p4susceptibility}. 
A non-diverging discontinuity of $\chi_p$ occurs at the identical phase transition fields $h_t^{(p)}$, which we have determined above by the spontaneous magnetization analysis and are depicted by the vertical dot-dashed lines. The inaccuracy comes from performing the second derivative in~\eqref{p4chi} numerically, and the additional improvement rests in decreasing the spacing interval $\delta h$, i.e, in shrinking the distance between the magnetic fields, at which the ground-state energy is evaluated by TPVF. In the limit $\delta h \to 0$, the magnetic susceptibility undergoes a discontinuous jump at $h_t^{(p)}$. It is obvious that there is no significant difference between the phase transition magnetic fields $h_t^{(p)}$ obtained by the analysis of the transverse magnetization $\langle S_p^{z}\rangle$ and the magnetic susceptibility $\chi_p$.     

\begin{figure}[!thp]
\centering
\includegraphics[width=3.46in]{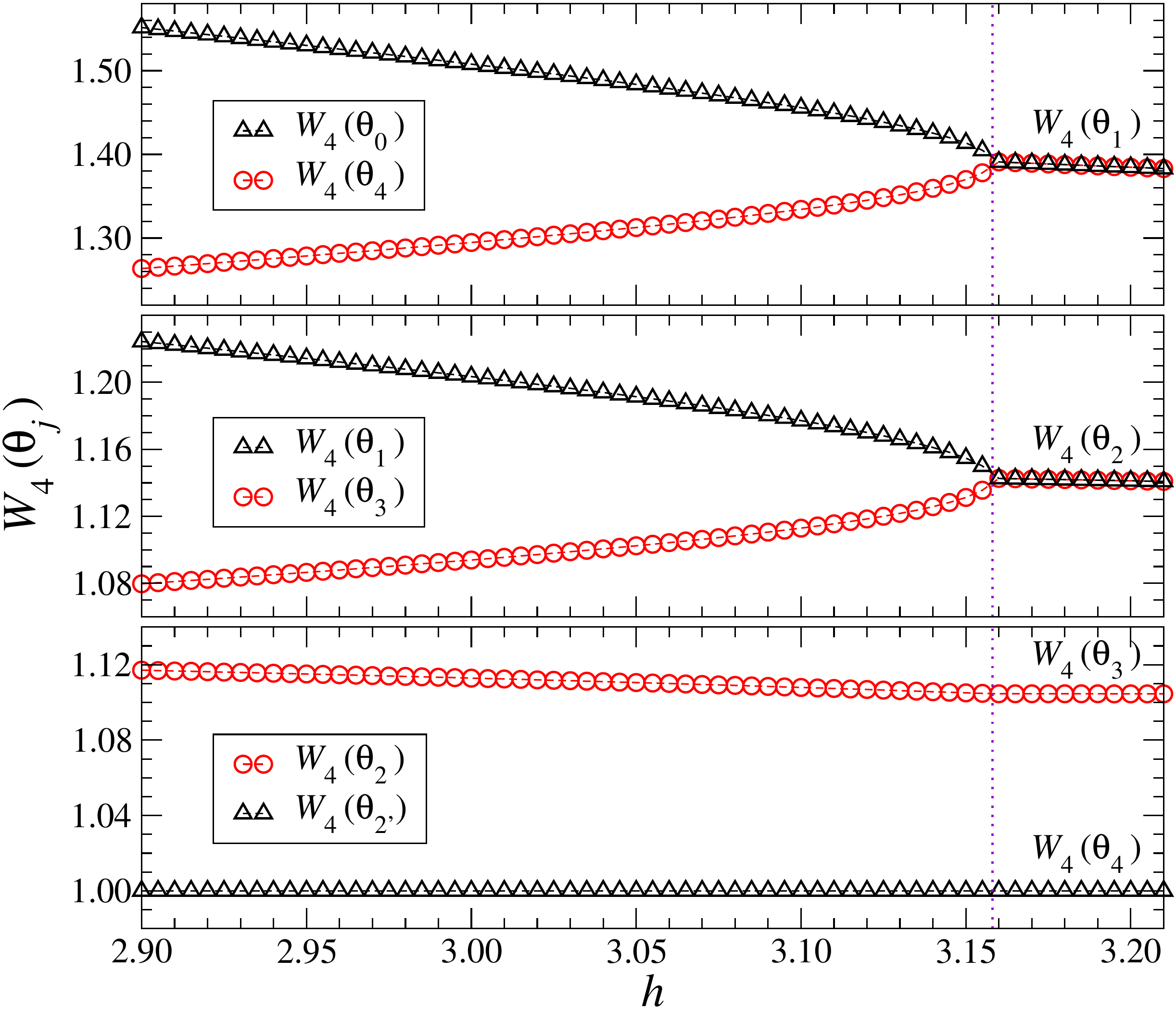}
\caption{The magnetic field dependence of the variational parameters
$W_4^{~}(\theta_j^{~})$ on the Euclidean $(4,4)$ lattice.  The dotted line marks the position of the transition field $h_t^{(4)}\approx3.158$.}
\label{44varparam}
\end{figure}
\begin{figure}[!bhp]
\centering
\includegraphics[width=3.46in]{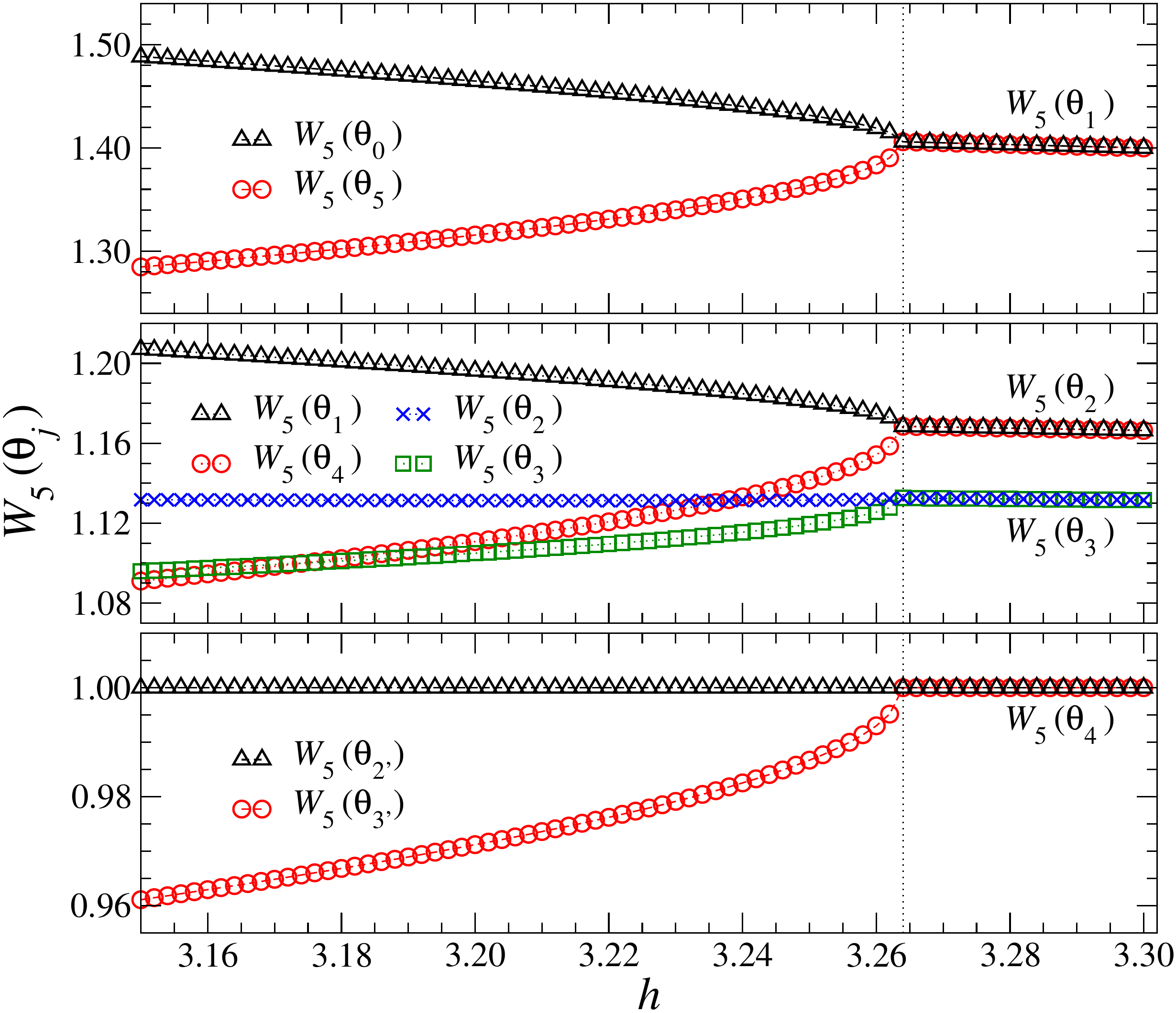}
\caption{The magnetic field dependence of the variational parameters
$W_5^{~}(\theta_j^{~})$ on the hyperbolic $(5,4)$ lattice. The dotted line marks the position of the transition field $h_t^{(5)}\approx3.264$.}
\label{54varparam}
\end{figure}

Except for the analysis of the phase transition by the spontaneous magnetization $\langle S_p^{z}\rangle$ and the magnetic susceptibility $\chi_p$, the field dependence of the set of the optimal free variational parameters $W^*_p(\theta_j)$ also provides helpful information about the phase transition $h_t^{(p)}$. The pairs of the optimal variational parameters $W^*_p(\theta_j)$ coupled by spin-inversion symmetry continuously collapse onto a single curve exactly at the phase transition $h_t^{(p)}$ determined by analysis of both the spontaneous magnetization and the magnetic susceptibility for all considered lattice geometries. However, due to the large number of the variational parameters $N_{\rm Ising}^{(p)}$, we only plot the $h$-dependence of $W^*_p(\theta_j)$ in Fig.~\ref{44varparam} and~\ref{54varparam} for the Euclidean $(4,4)$ and the pentagonal $(5,4)$ lattice, respectively.

%It is evident that above a certain magnetic
%field (depicted by the vertical dotted lines), specific pairs of the variational
%parameters collapse onto identical values, which yields four single curves in this region for both the lattices. These pairs represent the spin configurations
%which are equivalent with respect to spin inversion. The magnetic field, at which the
%collapse causes a singular behavior of $W_p^{~}(\theta_j^{~})$, corresponds to the
%quantum phase transition of TFIM at the identical magnetic field $h_t^{(p)}$, which was determined . 

In the ordered phase at $h<h_t^{(p)}$, the distinct optimized values of the
coupled parameters $W_p^{~}(\theta_j^{~})$, as specified in Tabs.~\ref{Tab_square} and
\ref{Tab_pentagon}, reflect the existence of the spontaneous sym\-metry-breaking
in the TFIM for both the lattice types. In the disordered phase at $h\geq h_t^{(p)}$,
the four-parameter description coincides with the variational parameters $W_4(\Theta_i)$ and $W_5(\Theta_i)$.
This confirms the relevance of the
additional symmetries in such systems, where the spontaneous sym\-metry-breaking mechanism
is not present, such as in the XY and Heisenberg systems at the zero magnetic field.

\paragraph{II. XY and modified Heisenberg models}

\def\arraystretch{1.6}
\tabcolsep=5pt
\begin{table}[tb]
\begin{center}
\begin{tabular}{| c | c | c | }
\hline
\multirow{2}{*}{$p$} & \multicolumn{2}{ c |}{$E_0^{(p)}$}  \\ \cline{2-3}
  & XY & Heisenberg  \\ \hline
$4$      & $-1.08456618$ & $-1.3089136$ \\ \hline
$5$      & $-1.08151200$ & $-1.2912704$ \\ \hline
$6$      & $-1.08097046$ & $-1.2925639$ \\ \hline
$7$      & $-1.08086301$ & $-1.2918936$ \\ \hline
$8$      & $-1.08084068$ & $-1.2919769$ \\ \hline
$9$      & $-1.08083585$ & $-1.2919403$ \\ \hline
$10$     & $-1.08083478$ & $-1.2919460$ \\ \hline
$11$     & $-1.08083453$ & $-1.2919437$ \\ \hline
$\infty$ & $-1.08083446$ & $-1.291944${\phantom{0}} \\ \hline
\end{tabular}
\end{center}
\caption{The ground-state energies per bond $E_0^{(p)}$ listed with respect to $p$ for the modified Heisenberg and XY models. The number of states $m$ of the multi-spin variables kept in renormalization process was $m=20$ for $4\leq p \leq 10$ and $m=10$ for $p=11$. The asymptotic estimate of $E_0^{(\infty)}$ corresponds to the model on the Bethe lattice.}
\label{Tab-p4-XYHeis}
\end{table}
\tabcolsep=2pt

We study the XY and the modified Heisenberg models at zero magnetic field, where these models are known to be critical in the Euclidean space. Therefore, there is no preferred direction (the spin alignment) in the system on the Euclidean lattice at $h \geq 0$, and the spin-inversion symmetry is present. We expect that the models on hyperbolic lattices also exhibit the spin-inversion symmetry. It enables us to reduce the number of the free variational parameters $W_p(\theta_j)$ within the TPVF minimization part down to $N_{\rm Heis}^{(p)}$ as listed in Table~\ref{Tab_num_var_parameters}. Despite the significant reduction, the number of the free parameters $N_{\rm Heis}^{(p)}$ still grows fast with respect to the increasing lattice parameter $p$. The computational time of the minimization algorithm is significantly prolonged due to (at least) linear dependence on the increasing number of the free variational parameters. Also, the algorithm may possibly be trapped in a local energy minimum and thus a series of initial conditions has to be tested in order to obtain the global energy minimum (or, at least, a sufficiently good approximation of it). For all these reasons, the calculations were stopped at $p=11$ with respect to the constraints of our computational resources and time.

The ground-state energies $E_0^{(p)}$ obtained by the TPVF algorithm for both the XY and the modified Heisenberg models are summarized in Table~\ref{Tab-p4-XYHeis}. The energies $E_0^{(p)}$ remained identical even if the larger set of $N_{\rm Ising}^{(p)}$ free variational parameters $W_p(\theta_j)$ in TPVF was used, whereby the optimal values of the parameters $W^{*}_p(\theta_j)$ coupled by spin-inversion symmetry were equal. These results witness the spin-inversion symmetry of the models on hyperbolic lattices.  Recall that $E_0^{(p)}$ represents only an upper estimate of the true ground-state energy ${\cal E}_0^{(p)}$. 

The energies $E_0^{(4)}$ calculated by TPVF on the Euclidean $(4,4)$ lattice for both the XY and the Heisenberg models are higher if compared to the results of the Monte Carlo simulations $E_0^{{\rm XY},{\rm MC}}=-1.09765$, $E_0^{{\rm Heis},{\rm MC}}=-1.33887$ \cite{XY,Heis} (the respective relative errors are $1.2\%$ and $2.2\%$). Again, because of the mean-field-like character of the TPS approximation, the TPVF algorithm is expected to be more accurate whenever a hyperbolic lattice geometry is considered~\cite{TPVF54, hctmrg-Ising-p-4}, since any quantum spin model on hyperbolic lattice belongs to the mean-field universality class.
% because the hyperbolic lattices exhibit the infinite Hausdorff dimension, which significantly exceeds the critical lattice dimension $D_c=3$~\cite{Yeomans, Baxter}.

\begin{figure}[tb]
\centering
\includegraphics[width=4.2in]{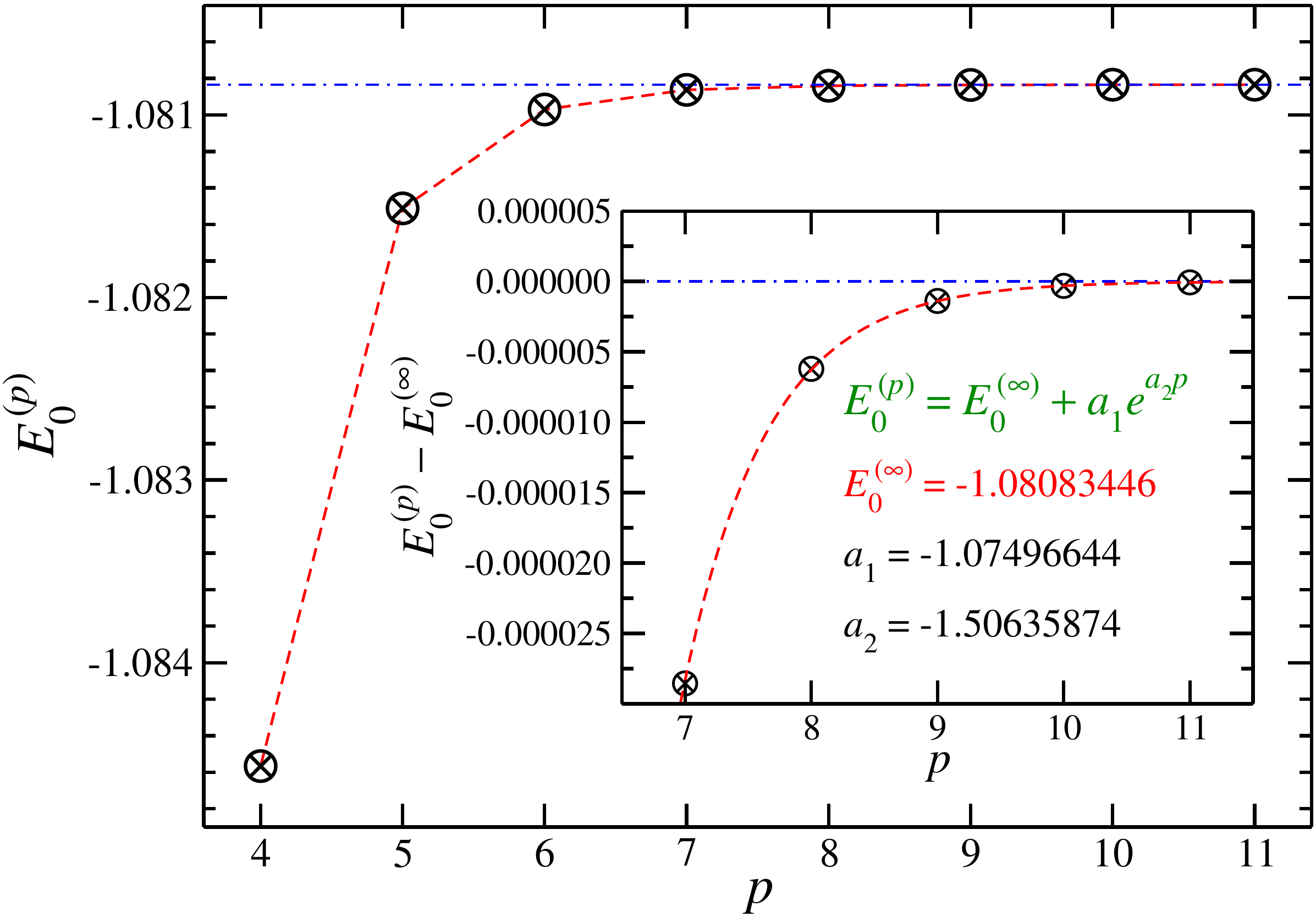}
\caption{The ground-state energy $E_0^{(p)}$ of the XY model
% (left) and the Heisenberg model (right) 
 with respect to the lattice parameter $p \in \{4,5,\dots, 11\}$. The inset shows the zoomed-in energy including the details of the fitting function.}
\label{p4XYenergy}
\end{figure}

Figure \ref{p4XYenergy} illustrates the monotonous and quickly saturating energy curve $E_0^{(p)}$ for the XY model with respect to the lattice parameter $p$. The inset depicts the tail of the curve in detail together with an exponential fit analogous to \eqref{fith_t} applied to the five energies $E_0^{(7)}, \dots, E_0^{(11)}$. 
%The case $p=4$, where the TPVF algorithm is not sufficiently accurate, was excluded from the extrapolation analysis. 
% The fitting function is proposed in the form analogous to . 
 The parameters of the fit $E_0^{(\infty)}$, $a_1$, and $a_2$ are listed in the inset of Fig.~\ref{p4XYenergy}, where the dot-dashed line represents the estimate of the ground-state energy per bond of the quantum XY model on the Bethe lattice $E_0^{(\infty)}=\lim\limits_{p \rightarrow \infty} E_0^{(p)}=-1.08083446$.
 
\begin{figure}[tb]
\centering
\includegraphics[width=4.2in]{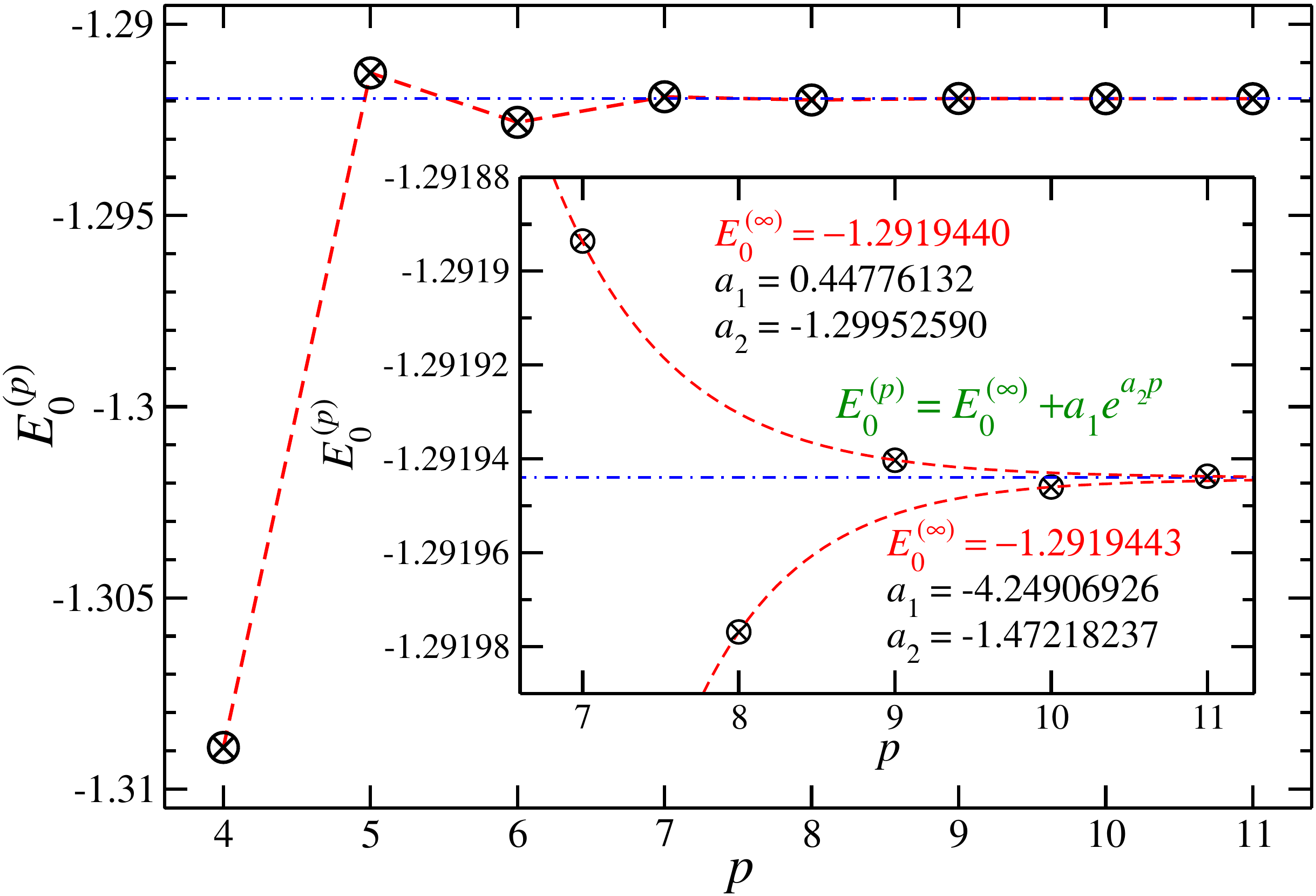}
\caption{The ground-state energy $E_0^{(p)}$ of the modified Heisenberg model with respect to the lattice parameter $p \in \{4,5,\dots, 11\}$. The fitting function parameters are shown in the inset.}
\label{p4Heisenergy}
\end{figure} 
 
Analogously, the ground-state energies $E_0^{(p)}$ of the Heisenberg model are plotted in Fig.~\ref{p4Heisenergy}. Again, rapid convergence of the energy to the asymptotic values is obvious from the data. We assume that the physical origin of the non-monotonic convergence (saw-like pattern) of $E_0^{(p)}$ may be attributed to the fact that the modified Heisenberg model on the lattices with even $p$ is a unitary transformation of the antiferromagnetic Heisenberg model, while this is not the case if $p$ is odd. However, a detailed analysis indicates that the exponential fitting function in~\eqref{fith_t} can successfully describe the data, if applied separately onto two sets: those with even $p \in \{6, 8, 10\}$ (the lower branch shown in the inset) and the odd $p \in \{5, 7, 9, 11\}$ (the upper branch). The fitting parameters of the two regressions are listed in the inset of Fig.~\ref{p4Heisenergy}. The lower and the upper branches yield the energies $E_0^{(\infty)}$ $-1.2919443$ and $E_0^{(\infty)}-1.2919440$, respectively. With respect to an independent application of additional analogous fits, we found $E_0^{(\infty)}=-1.291944$ (all the digits are valid) to be considered as the correct estimate of the ground-state energy per bond of the Heisenberg model (both the modified and the antiferromagnetic versions) on the Bethe lattice. 

We have not found any theoretical reasoning for the exponential convergence of the ground-state energies $E_0^{(p)}$ yet. However, if a power-law fitting function was applied instead, we obtained a less accurate fitting and greater $RSS$.

\subsection{Spin models on the $(4,q)$ lattices}

\subsubsection{The model}

In this section we apply the TPVF algorithm to study the quantum phase transition of the transverse field Ising model in the thermodynamic limit on the $(4,q)$ lattices, where $4 \leq q \leq 70$. Hence, we investigate the influence of the varying coordination number $q$ on the ground-state properties, which is a complementary problem to the previous study on the $(p,4)$ lattices. The number of the effective states which are kept after the renormalization procedure was set to $m=10$ in all calculations.

\begin{figure}[tb]
\centering
\includegraphics[width=4in]{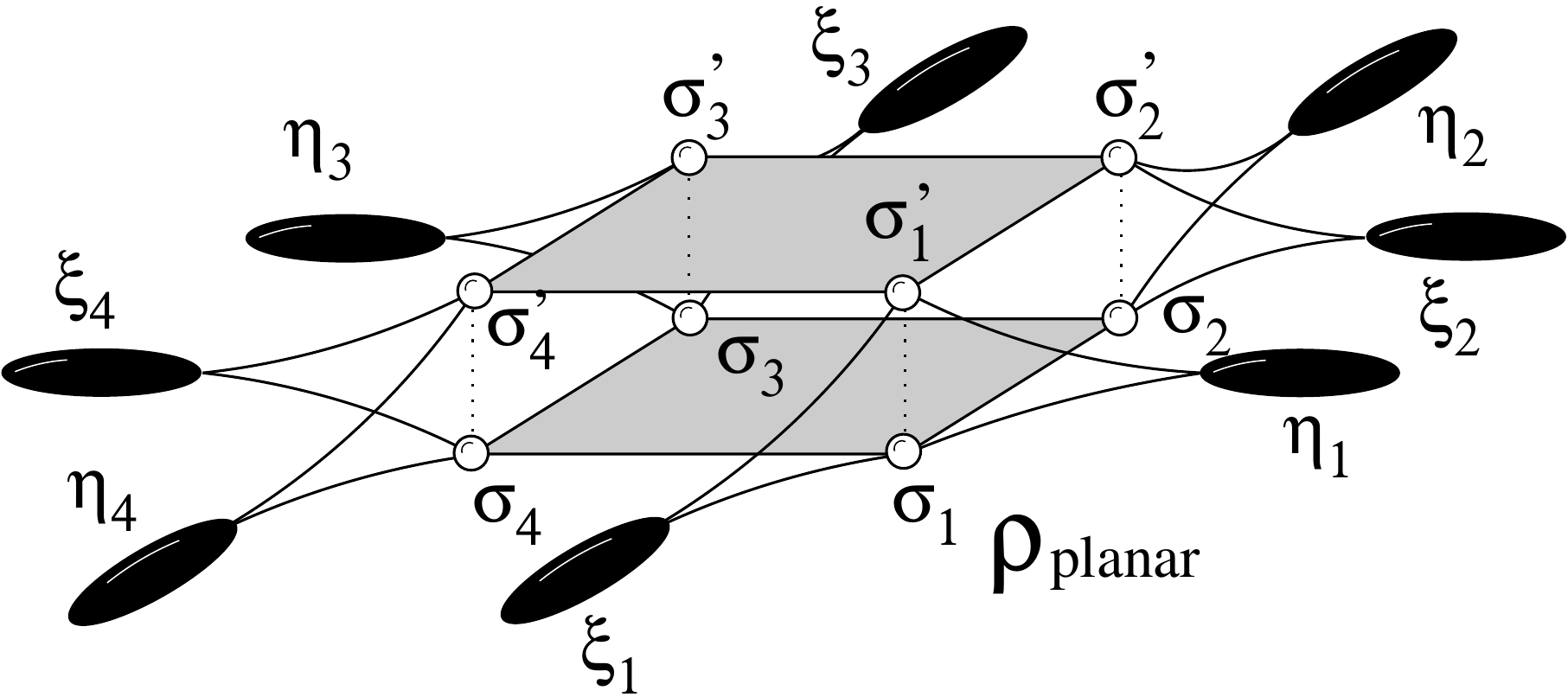}
\caption{Construction of the planar density matrix $\hat\rho$ according to \eqref{planarDMeq}. The black-filled objects represent the multi-spin variables $\xi_i, \eta_i$ which are summed over in \eqref{planarDMeq}. The double-layer structure in the center corresponds to the tensor $W_4^{~}(\{\sigma^{\prime}\})
W_4^{~}(\{\sigma\})$, which is surrounded by four tensors $\mathbf{C}$ and $\mathbf{T}$ in the identical alternating arrangement as in Fig.~\ref{C4T4W}.  }
\label{planarDM}
\end{figure}

In our analysis, we focused on the von Neumann entropy of the system given by formula \eqref{VonNeumannEntropy}. We introduce two kinds of the entropy - the linear entropy $S_{\rm linear}$ and the planar entropy $S_{\rm planar}$, which differ in the definition of the density matrix used in \eqref{VonNeumannEntropy}. The linear entropy $S_{\rm linear}$ is produced by assuming the standard reduced density matrix $\rho$, given by the relation \eqref{def_DM} equivalent to the partial trace of the tensor product $\mathbf{C}^4$, i.e.,
\begin{equation}
S=-\Tr(\rho \log_2 \rho).
\end{equation}
The planar entropy is
\begin{equation}
S=-\Tr(\hat\rho \log_2 \hat\rho),
\end{equation}
where $\hat\rho$ is the "planar" density matrix, calculated as 
\begin{align}
\begin{split}
\hat\rho\left(\{\sigma^{\prime}\} | \{\sigma^{~}\}\right) &= %Z_4(\{\sigma^{\prime} \sigma^{~}\})
W_4^{~}(\{\sigma^{\prime}\})W_4^{~}(\{\sigma\})\times\\
&\times\sum_{\substack{\xi_1,\xi_2,\xi_3,\xi_4\\\eta_1,\eta_2,\eta_3,\eta_4}}\prod_{i=1}^{4}\left[{\mathbf C}(\{\sigma^{\prime}_{i}\sigma^{~}_{i}\}, \xi_i, \eta_{i}){\mathbf T}(\{\sigma^{\prime}_{i}\sigma^{~}_{i}\}, \eta_i, \{\sigma^{\prime}_{{i+1}}\sigma^{~}_{{i+1}}\}, \xi_{i+1})\right],
\label{planarDMeq}
\end{split}
\end{align}
where $\xi_{4+1} \equiv \xi_1$, $\sigma_{4+1} \equiv \sigma_1$ and $\{\sigma\}=\{\sigma_1 \sigma_2 \sigma_3 \sigma_4\}$. The construction of the planar density matrix $\hat\rho$ is graphically illustrated in Fig.~\ref{planarDM}.
 This newly defined object represents a reduced quantum density matrix $\hat\rho=\Tr^{\prime} | \Psi_p \rangle \langle \Psi_p |$ of the TPS $| \Psi_p \rangle$, where the partial trace $\Tr^{\prime}$ is taken over the whole lattice but the four spins $\sigma_1$, $\sigma_2$, $\sigma_3$, $\sigma_4$ in the center. %Indeed, 
% In TPVF the scalar product $\langle \Psi_p | \Psi_p \rangle$
%is the lattice is constructed from a single tensor $Z_4$ 

\subsubsection{Numerical results}

\begin{figure}[!tb]
\centering
\includegraphics[width=4.5in]{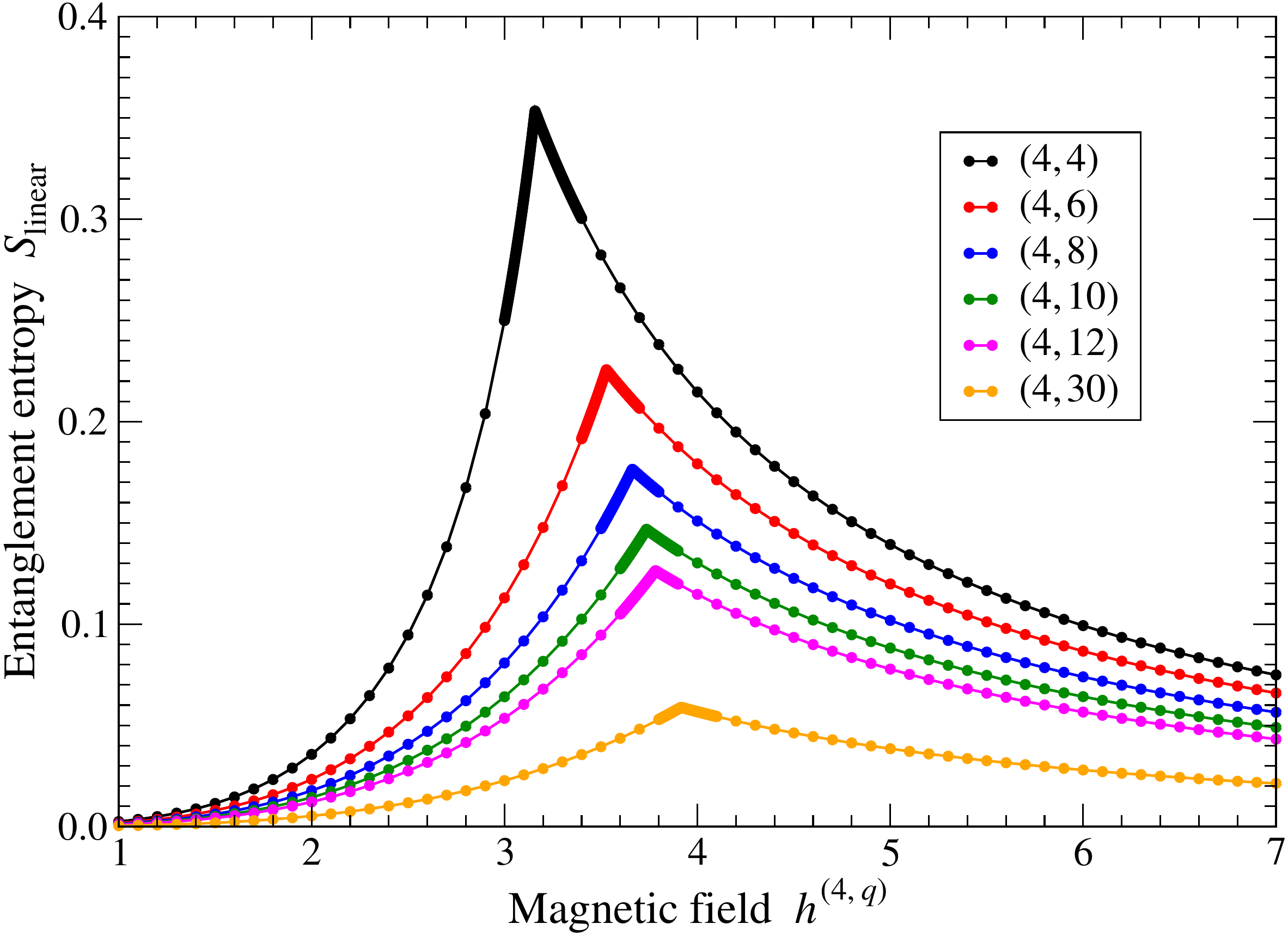}
\caption{Graph of the the linear entropy $S_{\rm linear}$ with respect to the magnetic field $h$ for $(4,q)$ lattices, where $4 \leq q \leq 30$. The sampling step is $\Delta h = 0.001$ around the phase transition, otherwise $\Delta h = 0.1$.}
\label{Entropy_vs_field}
\end{figure} 

\begin{figure}[!thp]
\centering
\includegraphics[width=3.9in]{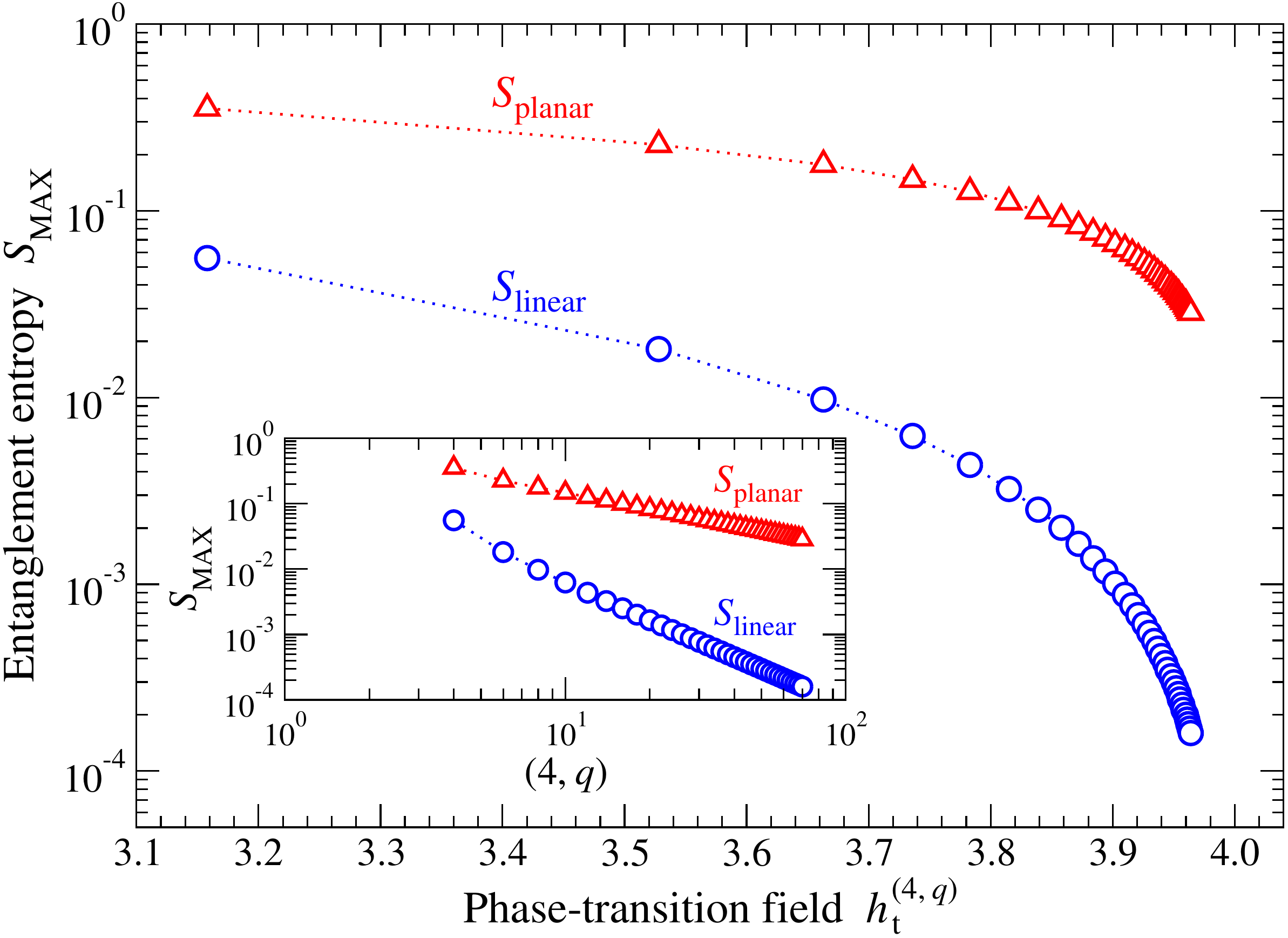}
\caption{The peak values of the two kinds of entropy $S_{\rm MAX}(q) \equiv S(h_{\rm t}^{(4,q)})$ with respect to $h_{\rm t}^{(4,q)}$ and the coordination number $q$ (in the inset) for $4 \leq q \leq 70$.   }
\label{Entropy_vs_q}
\end{figure}

We plot the curve of the linear entropy $S_{\rm linear}$  with respect to the magnetic field $h$ for selected $(4,q)$ lattices in Fig.~\ref{Entropy_vs_field}. The peak of the curve marks the phase transition field $h_{\rm t}^{(4,q)}$. Analogous graph with the identical position of the transition fields $h_{\rm t}^{(4,q)}$ can be obtained also for the planar entropy $S_{\rm planar}$. In order to perform more precise screening of the region near the phase transition fields $h_{\rm t}^{(4,q)}$, we sampled the magnetic field by step $\Delta h = 0.001$ there. %There is no singularity in $S_{\rm linear}$ for the Euclidean $(4,4)$ lattice, because of the mean-field approximation induced by the uniform TPS in TPVF. 
The peak of the curve shifts to the right and its maximum decreases as $q$ increases. To examine this dependency in a more precise manner, we located the maximum of the linear entropy $S_{\rm linear}(h)$ with respect to the magnetic field $h$. % using the golden-section method. 
 We determined the transition field $h_{\rm t}^{(4,q)}$ as the magnetic field $h$ which yielded the optimal value of $S_{\rm linear}(h)$. We plot the peak value of the two  entropies $S_{\rm MAX}(q) \equiv S(h_{\rm t}^{(4,q)})$ with respect to $h_{\rm t}^{(4,q)}$ and the coordination number $q$ in Fig.~\ref{Entropy_vs_q}. The apparent linearity of the curves $S_{\rm MAX}(q)$ in the $\log$-$\log$ scale suggests that the dependency $S_{\rm MAX}(q)$ has a polynomial character.
\begin{figure}[!bhp]
\centering
\includegraphics[width=3.9in]{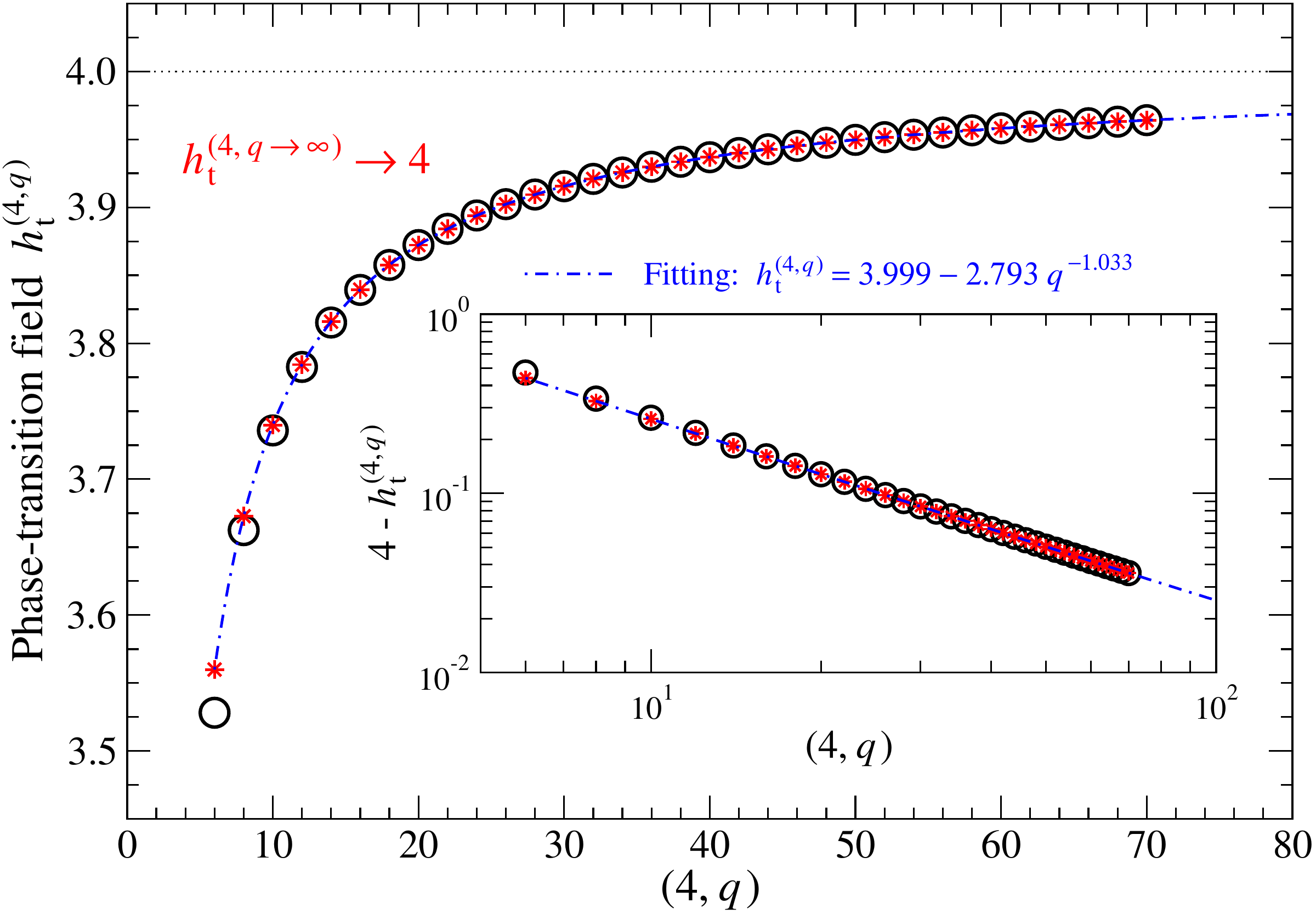}
\caption{The phase transition field $h_{\rm t}^{(4,q)}$ as a function of the coordination number $q$ for $4 \leq q \leq 70$. The open circles mark the calculated data, while the blue dot-dashed line and  the red stars illustrate the polynomial fitting function and the predicted values for the integers $q$, respectively. The inset illustrates the linearity $h_{\rm t}^{(4,q \to \infty)}-h_{\rm t}^{(4,q)} \equiv 4-h_{\rm t}^{(4,q)}$ with respect to $q$ in the $\log$-$\log$ scale, which supports the polynomial fit.}
\label{Scaling_ht_vs_q}
\end{figure} 

Figure~\ref{Scaling_ht_vs_q} depicts the phase transition field $h_{\rm t}^{(4,q)}$ as a function of the coordination number $q$. A detailed analysis, cf. the inset in Fig.~\ref{Scaling_ht_vs_q},  suggests that the best description of the data can be obtained by a polynomial fitting function. The parameters of the optimal fitting function are shown in the graph. Note that, according to the fit, as $q$ increases, the transition field $h_{\rm t}^{(4,q)}$ tends to the asymptotic value $h_{\rm t}^{(4,q \to \infty)} = 4$. %, which corresponds to the value of the lattice parameter $p=4$.
 The observed polynomial curve for $h_{\rm t}^{(4,q)}$ represents a new feature if compared to the results of the classical Ising model on the $(p,q)$ lattices, where a linear dependence of the transition temperature $T_{\rm pt}^{(q)} \propto q$ for large $q$ was detected \cite{Serina}.

\newpage\setcounter{equation}{0} \setcounter{figure}{0} \setcounter{table}{0}
\section{Conclusions and Perspectives}

In this work we focused on
numerical analysis of the phase transition phenomena of both classical and quantum spin systems on hyperbolic lattices.  
The hyperbolic lattices are constructed by tessellation of regular $p$-sided polygons with coordination number $q$, and we refer to them by using the notation $(p,q)$.

The presented task is highly non-trivial, since the number of the lattice sites increases exponentially with the diameter of the hyperbolic lattice. This exponential increase limits efficiency of the standard numerical tools such as the Monte Carlo simulations or exact diagonalization. Looking for an appropriate approach to deal with this challenging problem, we bet on a system-specific reformulation of the Corner transfer matrix renormalization group (CTMRG) algorithm, which was already successfully applied to classical spin systems on the $(p,4)$ lattices. 
In this work we presented a similar analysis in the complementary situation represented by the triangular $(3,q)$ lattices. In addition, we investigated the thermodynamic property of the Ising model on infinite sequence of weakly curved $(3,q_n)$ lattices, where $q_n$ represents the averaged coordination number and $n=0,1,2, ..., \infty$. As $n$ increases, the $(3,q_n)$ lattice flattens out approaching the triangular lattice $(3,6)=(3,q_{\infty})$. Next, we introduced the Tensor product variational formulation (TPVF) algorithm \cite{TPVF54}, which can be considered as a generalization of CTMRG to investigate quantum spin systems. Being interested in comparison of the phase transition phenomena in the classical and quantum case, we applied the TPVF algorithm to quantum spin models on the series of $(p,4)$ and $(4,q)$ hyperbolic lattices. The conclusions made from our studies of both classical and quantum systems are summarized in the following two Sections.

\subsection{Classical Ising model on triangular $(3,q)$ and weakly curved $(3,q_n)$ lattices}

First, we presented a detailed analysis of the phase transition phenomena for the Ising model on the Euclidean $(3,6)$ and the hyperbolic $(3, 7\leq q \leq 10^7)$ lattices. This work, which investigates the effect of the varying coordination number $q$ on the thermodynamic properties of the system, forms a supplement to the previous studies     
\cite{hctmrg-Ising-p-4, hctmrg-Ising-5-4, hctmrg-J1J2, hctmrg-clock-5-4}, where the influence of parameter $p$ was studied on the $(p,4)$ lattices. This task required a reformulation of the existing CTMRG algorithm, where different left and right transfer tensors were introduced, as explained in Section~\ref{3qscheme}. 

The phase transition temperatures $T_{\rm pt}^{(q)}$ were determined from
the analysis of the magnetization, internal energy, specific heat, and the
von Neumann entanglement entropy. We have shown that the transition temperature
$T_{\rm pt}^{(q)} \propto q$ as $q$ increases, which agrees with the mean-field behaviour \cite{Baxter}. On hyperbolic lattice, the
behaviour of the thermodynamic functions in the vicinity of the transition temperature
 is ruled
 by the critical
exponents $\alpha=0$, $\beta=\frac{1}{2}$, and $\delta=3$, which are characteristic for the mean-field
universality class. On the Euclidean $(3,6)$ lattice, the critical exponents $\alpha=0$, $\beta=\frac{1}{8}$, and $\delta=15$, known for the Ising universality class, are reproduced. The mean-field nature of the hyperbolic surfaces is
also characterized by the exponential decay of the reduced density matrix
eigenvalues and the correlation functions even at the transition temperature, which is the direct consequence of the finiteness of the correlation length. As a typical example of the non-diverging correlation length $\xi$ at the phase transition, the pentagonal $(5,4)$ lattice has been analyzed in detail~\cite{corrlen}. Due to finite values of the correlation length even at the transition point, the term \emph{critical} point on the hyperbolic lattices is not appropriate, since the critical point is always related to the divergence of the correlation length by definition.

In order to elucidate the origin of the mean-field universality induced by
the hyperbolic geometry, we have investigated the Ising model on the
slightly curved $( 3, q_n^{~} )$ lattices. On this lattice geometry, 
the Gaussian curvature can be easily manipulated which allows us to 
systematically approach the Euclidean (flat) geometry through an infinite series of weakly curved triangular lattices. Using a slight modification of the CTMRG method, as applied to the Euclidean $(3,6)$ lattice, we calculated the
thermodynamic functions deep inside the system
around the phase transition temperature. 
The curves of the spontaneous magnetization and the specific heat for the hyperbolic $(3,q_n)$ lattices continuously approach the curves for the Euclidean $(3,6)$ lattice as the curvature decreases to zero. The effective critical exponents $\beta_{\rm eff}$ and $\delta_{\rm eff}$ 
on the hyperbolic lattices follow the respective curves in the Euclidean case away from the transition point, however, they progressively bend to the mean-field values $\beta=\frac{1}{2}$, and $\delta=3$ as the transition point is approached.  

Assuming the previous studies \cite{hctmrg-Ising-p-4, hctmrg-Ising-5-4, hctmrg-J1J2, hctmrg-clock-5-4} including the results presented here, we conclude that classical spin systems on any hyperbolic lattice belong to the mean-field universality class. The mean-field-like behaviour observed in the hyperbolic geometry originates in the infinite Hausdorff dimension of the hyperbolic lattices which obviously exceeds the critical value $d_c=4$ \cite{Baxter, Yeomans}. We assume that the CTMRG method does not affect the critical behaviour, since it accurately reproduces all of the critical exponents on the 2D Euclidean lattices, as has been shown in \cite{hctmrg-Ising-5-4, hctmrg-Ising-p-4}.

\subsection{Quantum spin models on $(p,4)$ and $(4,q)$ lattices}

Generalizing the original idea proposed in \cite{TPVA} for the Euclidean lattice, we introduced the TPVF algorithm \cite{TPVF54} as a promising numerical tool for studying ground-states of quantum systems on the hyperbolic $(p,4)$ and $(4,q)$ lattices in the thermodynamic limit. Approximating the ground-state in the form of a uniform tensor product state (TPS), we receive a variational problem which is solved by a combination of a modified CTMRG and an optimization algorithm.  
The uniform TPS reduces the infinite number of the variational parameters in the thermodynamic limit down to $2^p$. Considering 
symmetries present in the Hamiltonian of the model, the number of the free variational parameters approximating the tensor product ground state is further significantly shrunk. 

First, applying the TPVF algorithm, we investigated three quantum spin-$\frac{1}{2}$ models (modified Heisenberg, XY, and transverse-field Ising model (TFIM)) on a series of hyperbolic $(p,4)$ lattices, where $p \in \{5, \dots,11\}$. The key feature of this study is the indirect analysis of the three models on the Bethe lattice with coordination number four, which is represented by the limit $p \to \infty$. This problem had not been addressed before. In order to assess accuracy of our results, the Euclidean square lattice ($p=4$) was also considered as a reference lattice, where highly precise results obtained through various numerical methods are available. The TPVF applied to the models on the square lattice is expected to be less accurate than on the hyperbolic lattices. This is caused by the too low dimension of the tensors in the TPS
approximation which suppresses the quantum long-range
correlations on the square $(4,4)$ lattice near the criticality. Thus, the TPVF algorithm itself is a source of an improved mean-field approximation, which is a new feature if compared to the classical case, where CTMRG produces correct results with no approximation.
Comparing our results with the reference study~\cite{HoSRG}, the ground-state energies $E_0^{(4)}$ of the XY and the modified Heisenberg model and the transition field $h_t^{(4)}$ in the TFIM on the Euclidean $(4,4)$ lattice deviate from the reference values by $1.2\%$, $2.2\%$ and  $3.7\%$, respectively.  
On the other hand, the mean-field-like behaviour, induced by the hyperbolic structure of the lattice (not the mean-field approximation of Hamiltonians), is natural, since the infinite Hausdorff dimension of the hyperbolic surfaces exceeds the critical dimensionality $d_c=3$ of quantum systems. Therefore, the improved mean-field approximation of the TPS is not in conflict with the mean-field universality induced by the hyperbolic geometry. 

The ground-state energies $E_0^{(p)}$ of the XY and the modified Heisenberg models have been studied in the absence of magnetic field on the series of the regular $(p,4)$ lattices with $4\leq p \leq 11$. The resulting dependence of the ground-state energy per bond $E_0^{(p)}$ on the lattice parameter $p$ differs considerably for the two models. While the energies $E_0^{(p)}$ of the XY model form a monotonically increasing and exponentially saturated sequence with increasing $p$, the modified Heisenberg model induces a saw-like dependence containing the separated upper (odd $p$) and the lower (even $p$) branches, both of them converging exponentially fast to the common asymptotic value $E_0^{(\infty)}$ which corresponds to the ground-state energy on the Bethe lattice with the coordination number four. The saw-like pattern in case of the modified Heisenberg model may be attributed to the fact, that if $p$ is even, the ground-state is antiferromagnetic, while if $p$ is odd, the ferromagnetic state is obtained. 

Within the identical series of hyperbolic $(p,4)$ lattices, we analyzed the phase transition magnetic fields $h_t^{(p)}$ of the TFIM for $4 \leq p \leq 10$ by the expectation value of the spontaneous magnetization $\langle S_p^z \rangle$, the associated magnetic exponent $\beta_p$, the magnetic susceptibility $\chi_p$, and the optimized variational parameters $W_p^{*}(\theta_j)$. The resulting phase transition magnetic fields $h_t^{(p)}$ form an increasing sequence, which exhibits exponential convergence to the asymptotic value $h_t^{(\infty)}$. Analogous behaviour had also been observed  
for the phase transition temperatures $T_{\rm pt}^{(p)}$ of the classical Ising model on the identical series of hyperbolic lattices in studies~\cite{hctmrg-Ising-5-4,hctmrg-Ising-p-4}. However, the physical interpretation of this phenomenon is still missing. The linearity of the squared spontaneous magnetization in the vicinity of the phase transition confirms the mean-field-like behaviour induced by the hyperbolic geometry, in which the associated magnetic exponents $\beta_p = \frac{1}{2}$. The mean-field approximation of the TPS results in the mean-field exponent $\beta_4 = \frac{1}{2}$ for the quantum TFIM on the Euclidean $(4,4)$ lattice, where the reference value is $\beta_4^{\rm TRG} = 0.3295$ \cite{HoSRG}.

Although the set of the phase-transition magnetic fields $h_t^{(p)}$ and the ground-state energies $E_0^{(p)}$ is restricted to $4 \leq p \leq 11$, which is far from the asymptotics $p\to\infty$, the fast convergence and the exponential character of $h_t^{(p)}$ and $E_0^{(p)}$ with increasing $p$ enables us to estimate the respective quantities of the quantum spin models on the Bethe lattice ($p \to \infty$). In particular, we conjecture that the phase transition field of the TFIM on the Bethe lattice is positioned at $h_t^{(\infty)} = 3.29332$ and the ground-state energies per bond of the XY and the Heisenberg  models, respectively, occur at $E_0^{(\infty)} = -1.08083446 $ and $-1.291944$. The latter value is common also to the antiferromagnetic Heisenberg model.

Finally, we presented the preliminary results of our studies of quantum models on the series of the $(4,q)$ lattices. In this case the phase transition field $h_{\rm t}^{(4,q)}$ of the transverse field Ising model was determined by maximizing the von Neumann entropy of the system. The calculated transition fields $h_{\rm t}^{(4,q)}$ suggest the polynomial character of the respective curve with respect to the coordination number $q$. Assuming the polynomial fit, the transition fields asymptotically converge to the value $h_{\rm t}^{(4,q \to \infty)} = 4$ as $q$ tends to infinity. This outcome has no analogy in the classical Ising model on the $(p,q)$ lattices, where, instead, the transition temperature $T_{\rm pt}^{(q)}$ grows linearly with increasing $q$ if $q$ is large. The polynomial dependence of the peak value of the entropy $S_{\rm MAX}\left(h_{\rm t}^{(4,q \to \infty)}\right)$ with respect to the coordination number $q$ was also detected. The origin of this polynomial behaviour has not been clarified yet.

\newpage
\addcontentsline{toc}{section}{Acknowledgement}
\begin{ack}
	
\medskip
This review summarizes the most important results obtained by myself and in close collaboration with other colleagues during my PhD study. The work is structured as follows: Section 1 provides brief review of the general theory of both the classical and quantum phase transitions. The critical behaviour of the elementary spin models is described. The concept of the non-Euclidean geometry is motivated in Section 2, where special attention is devoted to the hyperbolic surfaces. Section 3 concludes the theoretical part of this work. Here, we describe the numerical algorithms which represent the computational background of our analyses. One of the algorithms \cite{TPVF54} has been formulated during this PhD study.
The last two Sections form the core of this work, where we present the outcomes of numerical analyses of spin models on hyperbolic lattices. First, in Section 4, we investigate the phase transition phenomena in classical spin models, next, analogous situation for quantum systems is considered in Section 5. This part summarizes the results already published in \cite{hctmrg-Ising-3-q, hctmrg-Ising-3-qn, TPVFp4}.

At this point, I would like to express the deep gratitude to my supervisor Mgr. Andrej Gendiar, PhD. for his intensive support, encouragement and always positive mood, which made him more a good friend than only an academic fellow. I would like to thank him for all the time and effort he devoted to me and countlessly many pieces of advice, which helped me not only at writing this thesis. I thank him for his patience, contextual and grammar corrections and assistance in creating figures presented in this work. Without all of this, this work would never be completed.    

I would also like to thank to my office-colleague Mgr. Roman Kr\v cm\'ar, PhD. and RNDr. Ladislav \v Samaj, DrSc. for valuable discussions, which helped me to advance more quickly. I thank to Mgr. Jozef Genzor, PhD. for providing me with the latex template and the know-how related to the formal aspects of the graduating process.     
I thank to all other colleagues from our research department for all the help I received from them in many forms. I thank to my family for their support and last, but not least, I thank God.

This work has been supported by the projects EXSES APVV-16-0186, QETWORK APVV-14-0878, and VEGA Grant No. 2/0130/15.
\end{ack}

\newpage

\addcontentsline{toc}{section}{References}
\fancyhead[LO]{References}
\bibliographystyle{apalike}
\bibliography{main}

\end{document}